\documentclass[superscriptaddress,amsmath,prd, amssymb,aps]{revtex4-2}
\usepackage{graphicx}
\usepackage{dcolumn}
\usepackage{hyperref}
\hypersetup{colorlinks=true, citecolor=blue, urlcolor=blue, linkcolor=blue}
\usepackage{amsmath,amssymb}
\usepackage{mathrsfs}
\usepackage{esint}
\usepackage{comment}
\usepackage{tcolorbox}
\usepackage{booktabs}
\usepackage{caption}
\usepackage{ulem}

\newcommand{\myDelta}{\Delta\hspace{-2.3ex}\raisebox{0.3ex}{\Large$\sim$}}
\newcommand{\mydelta}{\Delta\hspace{-1.78ex}\raisebox{0.5ex}{\normalsize$\rule{0.25cm}{1pt}$}}

\begin{document}

\title{Relativistic second-order spin Hydrodynamics: A Kubo-type formulation for the quark-gluon plasma}
\medskip

\author{Duan She}
\email{sheduan@hnas.ac.cn}
\affiliation{Institute of Physics, Henan Academy of Sciences, Zhengzhou 450046, China}

\author{Yi-Wei Qiu}
\email{qiuyiwei1998@foxmail.com}
\affiliation{Institute of Particle Physics and Key Laboratory of Quark and Lepton Physics (MOE),
	Central China Normal University, Wuhan 430079, China}

\author{Defu Hou}
\email{houdf@mail.ccnu.edu.cn}
\affiliation{Institute of Particle Physics and Key Laboratory of Quark and Lepton Physics (MOE),
	Central China Normal University, Wuhan 430079, China}

\date{\today}

\begin{abstract}

Building upon Zubarev’s nonequilibrium statistical operator formalism, we derive a relativistic canonical-like second-order spin hydrodynamics for two power-counting schemes. We obtain comprehensive second-order expressions for dissipative fluxes, including the shear stress tensor, bulk viscous pressure, charge-diffusion currents, rotational stress tensor, boost heat vector, and spin tensor-related dissipative flux. By introducing novel transport coefficients and expressing them in terms of equilibrium correlation functions, we establish new Kubo-type formulas for second-order transport.  Our findings have significant implications for understanding the collective spin dynamics of strongly interacting matter and provide a robust theoretical basis for future experimental and theoretical studies.

\end{abstract}


 \maketitle

\section{Introduction}
\label{section1}

Recent measurements of hyperon spin polarization and vector meson spin alignment have offered new insights into the spin structure and transport properties of the quark-gluon plasma (QGP).  This has motivated theoretical investigations into spin-related phenomena within relativistic systems. Relativistic hydrodynamics provides a highly effective framework for describing low-frequency, long-wavelength phenomena in relativistic many-body systems. It is especially well suited for capturing the collective evolution of systems where macroscopic time and length scales far exceed microscopic mean free paths~\cite{Landau1987Fluid}. Relativistic hydrodynamics utilizes densities of conserved quantities, such as energy-momentum and particle number density, as its dynamical variables. The corresponding conservation laws serve as the foundation for the relativistic hydrodynamic equations. A key feature of this framework is the gradient expansion of physical quantities. The leading-order term corresponds to ideal hydrodynamics, while higher-order terms account for dissipative processes such as heat conduction, particle diffusion, and viscosity. Over the past decade, relativistic hydrodynamics has been successfully applied to describe the behavior of hot and dense matter produced in heavy-ion collisions at the RHIC and the LHC~\cite{CBM:2016kpk,Leupold:2011zz}. These experiments revealed the QGP, a novel state of matter characterized by near-perfect fluid behavior~\cite{Busza:2018rrf,Fukushima:2016xgg,Jeon:2015dfa}. The impressive success of hydrodynamic models in describing the evolution of the QGP has naturally led researchers to explore their applicability to spin transport phenomena. This pursuit has spurred the development of relativistic spin hydrodynamics~\cite{Hattori:2019lfp,Fukushima:2020ucl,Hongo:2021ona,Gallegos:2021bzp,Gallegos:2022jow,Peng:2021ago,Wang:2021wqq,Montenegro:2018bcf,Montenegro:2017rbu,Florkowski:2017ruc,Florkowski:2018fap,Li:2019qkf,Bhadury:2020puc,Montenegro:2020paq,Li:2020eon,Shi:2020htn,She:2021lhe,Hu:2021lnx,Hu:2021pwh,Hongo:2022izs,Singh:2022ltu,Daher:2022xon,Weickgenannt:2022zxs,Bhadury:2022ulr,Cao:2022aku}.

Relativistic spin hydrodynamics accounts for the evolution of the system's angular momentum, requiring simultaneous consideration of energy-momentum and angular-momentum conservation. Conservation of total angular momentum yields additional equations of motion that govern the dynamic evolution of the spin tensor. Systems with internal symmetries (e.g., baryon number conservation) necessitate the inclusion of corresponding charge conservation laws. Relativistic spin hydrodynamics, despite its rapid advancement, still requires theoretical clarification and development in several areas. These areas include the physical significance of pseudo-gauge transformations, the stability of spin hydrodynamic equations, and the calculation of new transport coefficients. Numerical solutions of spin hydrodynamic equations and their application to high-energy heavy-ion collisions remain pressing challenges. Relativistic spin fluids offer a macroscopic framework for describing spin polarization and vector meson spin alignment. Future numerical simulations will be crucial for refining the description of polarization phenomena observed in relativistic heavy-ion collisions.

First-order relativistic hydrodynamics exhibits unphysical modes that violate causality and lead to numerical instability at high momenta in practical calculations. This issue arises from the constitutive relation, which lacks a relaxation term, leading to a direct proportionality between the fluid's response and force. To address these issues, it is essential to introduce corresponding relaxation times, resulting in a second-order Israel-Stewart spin fluid dynamics framework. Previous studies~\cite{Weickgenannt:2022zxs,Weickgenannt:2022qvh} utilized spin-kinetic equations to derive second-order spin fluid dynamics equations. This approach explicitly includes spin-dependent collision terms and employs the moment method for dynamic equations. Reference~\cite{Biswas:2023qsw} derives the phenomenological form of second-order spin hydrodynamics through entropy current analysis. Reference~\cite{Dey:2024cwo} present a first-order spin hydrodynamic framework with spin chemical potential as leading order in gradient expansion using Zubarev's nonequilibrium statistical operator method. Reference~\cite{Tiwari:2024trl} develops a phenomenological second-order spin fluid with $\omega_{\mu\nu}\sim\mathcal{O}\left(\partial^{1}\right)$ using the Zubarev nonequilibrium statistical operator method. This study utilizes the Zubarev nonequilibrium statistical operator formalism~\cite{zubarev1974nonequilibrium,zubarev1997statistical,zubarev1979derivation,Hosoya:1983id,Horsley:1985dz,Harutyunyan:2018cmm,Harutyunyan:2017lrm,Harutyunyan:2021rmb}
to derive relativistic canonical-like second-order spin fluid dynamics equations with two power counting. This approach seeks to derive the complete second-order expressions for the shear stress tensor $\pi_{\mu\nu}$, bulk viscous pressure $\Pi$, charge-diffusion currents $\mathscr{J}_{a\mu}$, rotational stress tensor $\phi_{\mu\nu}$, boost heat vector $q_\mu$, and spin tensor-related dissipative flux $\varpi_{\lambda\mu\nu}$. The Zubarev formalism extends the Gibbs canonical ensemble to nonequilibrium states. It promotes the statistical operator to a nonlocal functional of thermodynamic parameters and their spacetime derivatives. Assuming sufficiently smooth variations of these parameters, the statistical operator is expanded in a gradient series to the desired order. Statistical averaging of relevant quantum operators then generates hydrodynamic equations for dissipative currents. A key advantage of the Zubarev formalism is its automatic generation of transport coefficients as Kubo-type relations, connecting them to specific correlation functions of the underlying field theory. These coefficients are expressed in terms of equilibrium correlation functions, simplifying subsequent calculations within standard thermal field theory.

This paper is organized as follows. Section~\ref{section2} provides a concise overview of Zubarev's formalism~\cite{zubarev1974nonequilibrium,zubarev1997statistical}. Section~\ref{section3} presents the derivation of second-order transport equations and transport coefficients for dissipative fluxes under the first power-counting scheme. Section~\ref{section4} presents the derivation of second-order transport equations and transport coefficients for dissipative fluxes under the second power-counting scheme. Section~\ref{section5} concludes with a summary of our findings and offers an outlook on future works. We work in flat space-time described by the metric tensor of the signature $g_{\mu\nu}=\mathrm{diag}\left(+,-,-,-\right)$ and the totally antisymmetric Levi-Civita tensor with the sign convention $\epsilon^{0123}=-\epsilon_{0123}=1$. We use natural units $\hbar=k_{B}=c=1$ throughout. In this manuscript, the symmetric and antisymmetric parts of a tensor $X^{\mu\nu}$ are denoted as $X^{\left(\mu\nu\right)}=\frac{1}{2}\left(X^{\mu\nu}+X^{\nu\mu}\right)$ and $X^{[\mu\nu]}=\frac{1}{2}\left(X^{\mu\nu}-X^{\nu\mu}\right)$, respectively. The fluid four-velocity $u^\mu$ satisfies the normalization condition $u^\mu u_\mu=1$. The projector orthogonal to $u^\mu$ is defined as $\Delta^{\mu\nu}= g^{\mu\nu}-u^{\mu}u^{\nu}$; by definition $\Delta^{\mu\nu}u_\mu=0$. Projection orthogonal to $u^\mu$ of a four-vector $X^\mu$ is defined as $X^{\left\langle \mu\right\rangle }=\Delta^{\mu\nu}X_{\nu}$. A traceless and symmetric projection operator orthogonal to $u^\mu$ is denoted as $\Delta_{\alpha\beta}^{\mu\nu}=\frac{1}{2}\left(\Delta_{\alpha}^{\mu}\Delta_{\beta}^{\nu}+\Delta_{\beta}^{\mu}\Delta_{\alpha}^{\nu}-\frac{2}{3}\Delta^{\mu\nu}\Delta_{\alpha\beta}\right)$.
Similarly, $\mydelta_{\alpha\beta}^{\mu\nu}=\frac{1}{2}\left(\Delta_{\alpha}^{\mu}\Delta_{\beta}^{\nu}-\Delta_{\beta}^{\mu}\Delta_{\alpha}^{\nu}\right)$
denotes the antisymmetric projection operator orthogonal to $u^\mu$ and $\myDelta_{\alpha\beta\gamma}^{\lambda\mu\nu}=\frac{1}{6}\left(\Delta_{\alpha}^{\lambda}\Delta_{\beta}^{\mu}\Delta_{\gamma}^{\nu}+\Delta_{\alpha}^{\mu}\Delta_{\beta}^{\nu}\Delta_{\gamma}^{\lambda}+\Delta_{\alpha}^{\nu}\Delta_{\beta}^{\lambda}\Delta_{\gamma}^{\mu}-\Delta_{\alpha}^{\mu}\Delta_{\beta}^{\lambda}\Delta_{\gamma}^{\nu}-\Delta_{\alpha}^{\nu}\Delta_{\beta}^{\mu}\Delta_{\gamma}^{\lambda}-\Delta_{\alpha}^{\lambda}\Delta_{\beta}^{\nu}\Delta_{\gamma}^{\mu}\right)$ denotes the totally antisymmetric projection operator orthogonal to $u^\mu$.

\section{The nonequilibrium statistical-operator formalism}
\label{section2}

We employ Zubarev's nonequilibrium statistical operator formalism to investigate a generic quantum system within the spin hydrodynamic regime. Our analysis commences with the operator-valued conservation laws pertaining to the energy-momentum tensor, charge currents, and total angular momentum tensor
\begin{align}
\partial_{\mu}\hat{N}_{a}^{\mu}=&0, \label{1}\\
\partial_{\mu}\hat{T}^{\mu\nu}=&0, \label{2}\\
\partial_{\lambda}\hat{J}^{\lambda\mu\nu}=&\partial_{\lambda}\hat{S}^{\lambda\mu\nu}+2\hat{T}^{[\mu\nu]}=0, \label{3}
\end{align}
where the conserved charges are indexed by $a=1,2,\cdots,l$, with $l$ representing the total number of conserved charges. In Eq.~\eqref{3}, we have decomposed the total angular momentum tensor $\hat{J}^{\lambda\mu\nu}$ into its orbital angular momentum $\hat{L}^{\lambda\mu\nu}=x^{\mu}\hat{T}^{\lambda\nu}-x^{\nu}\hat{T}^{\lambda\mu}$ and intrinsic spin angular momentum $\hat{S}^{\lambda\mu\nu}$.

The total angular momentum current can be decomposed into energy-momentum and spin tensor components in various ways. A given pair of these tensors can be transformed into another pair using a pseudogauge transformation~\cite{Daher:2022xon,Speranza:2020ilk}.  Among these, the Belinfante pseudogauge, based on the symmetric Belinfante energy-momentum tensor~\cite{BELINFANTE1939887,BELINFANTE1940449}, is particularly noteworthy. Alternatively, the phenomenological pseudogauge, employing an arbitrary energy-momentum tensor with symmetric and antisymmetric parts coupled to a spin tensor antisymmetric in its last two indices, is useful for constructing hydrodynamic frameworks~\cite{Hattori:2019lfp,Fukushima:2020ucl,Daher:2022xon,Weyssenhoff:1947iua}. At the microscopic level, canonical conserved currents, consisting of a symmetric-antisymmetric energy-momentum tensor and a fully antisymmetric spin tensor, can be derived. This choice is consistent with the Noether theorem applied to Dirac fermions in quantum field theory~\cite{Hongo:2021ona}. In contrast to the phenomenological approach, we will adopt the canonical-like conserved currents, which offer a more microscopic and fundamental description of the system. Reference~\cite{Buzzegoli:2024mra} provides further support for our choice of the canonical-like pseudogauge, demonstrating the thermodynamics of a field theory with axial current interactions is equivalent to Zubarev local equilibrium operator with the selection of the canonical pseudogauge. This shared understanding solidifies the canonical pseudogauge as a robust foundation for the development of spin hydrodynamics. 

Within the framework of spin hydrodynamics, the corresponding conservation laws are derived by calculating the statistical averages of the operators $\hat{T}^{\mu\nu}$, $\hat{N}_{a}^{\mu}$, and $\hat{S}^{\lambda\mu\nu}$ with respect to the complete nonequilibrium statistical operator. This operator is obtained by solving the quantum Liouville equation with an infinitesimal source term~\cite{zubarev1974nonequilibrium,zubarev1997statistical,zubarev1979derivation} and subsequently expanding it in a series of thermodynamic forces to the desired order. Statistical averaging of the dissipative currents then leads to the constitutive relations for these currents and explicit expressions for the transport coefficients in terms of equilibrium correlation functions of the system.

\subsection{Local-equilibrium statistical operator}

The thermodynamic state of a macroscopic quantum system is well characterized by the statistical operator $\hat{\rho}\left(t\right)$. In the Heisenberg picture, the Liouville-von Neumann equation governing its evolution is given by
\begin{eqnarray}
\frac{d\hat{\rho}\left(t\right)}{dt}=0,\quad\hat{\rho}\left(t\right)=\hat{\rho}\left(0\right)=\text{Const.}
\label{4}
\end{eqnarray}
Here, the operators acting on the system's quantum states are time dependent, while the statistical operator remains constant. The initial state of the system is denoted by $\hat{\rho}(0)$. In Eq.~\eqref{4}, the time derivative acts on both the operators and the thermodynamic parameters.

To be a valid statistical operator, $\hat{\rho}(t)$ must satisfy the normalization condition $\mathrm{Tr}\hat{\rho}=1$ in addition to the equation of motion given in Eq.~\eqref{4}. With this definition, the thermal expectation value of any quantum operator $\hat{F}\left(x\right)$ can be calculated using
\begin{eqnarray}
\langle\hat{F}(x)\rangle=\text{Tr}[\hat{\rho}\hat{F}(x)],
\label{5}
\end{eqnarray}
where $x\equiv\left(\boldsymbol{x},t\right)$ represents the four-dimensional spacetime coordinate.

In the framework of statistical mechanics, the equilibrium state of a system is uniquely determined by its conserved quantities. For a system coupled to a heat bath at temperature $T=\beta^{-1}$, a charge reservoir with chemical potentials $\mu_{a}$, and a spin reservoir with spin chemical potential $\omega_{\alpha\beta}$, the equilibrium state is described by the grand canonical ensemble:
\begin{eqnarray}
	\hat{\rho}_{\mathrm{eq}}=e^{\Omega-\beta\hat{K}},\qquad e^{-\Omega}=\mathrm{Tr}e^{-\beta\hat{K}}.
	\label{6}
\end{eqnarray}
Here, $\hat{K}=\hat{H}-\sum_{a}\mu_{a}\hat{\mathcal{N}}_{\mu}-\frac{1}{2}\omega_{\alpha\beta}\hat{\mathcal{S}}^{\alpha\beta}$,
where $\hat{H}$ is the Hamiltonian of the system, $\hat{\mathcal{N}}_{\mu}$ are the operators of the conserved charges, $\mu_a$ are the corresponding chemical potentials, $\hat{\mathcal{S}}^{\alpha\beta}$ is the operator of the spin, and $\omega_{\alpha\beta}$ is the corresponding spin chemical potential. Note that Eq.~\eqref{6} is defined in the rest frame of the system.

To generalize this equilibrium distribution to an arbitrary reference frame, we perform a Lorentz transformation of the Hamiltonian: $\hat{H}\to\hat{\mathcal{P}}^{\nu}U_{\nu}$, where $U^\nu$ is the four-velocity of the system in the chosen frame, $\hat{\mathcal{P}}^{\nu}$ is the four-momentum operator, and $\hat{H}\equiv\hat{\mathcal{P}}^{0}$ in the fluid rest frame. The operators $\hat{\mathcal{P}}^{\nu}$, $\hat{\mathcal{N}}_{a}$, and $\hat{\mathcal{S}}^{\alpha\beta}$ are expressed in terms of the energy-momentum tensor, charge currents, and spin tensor, respectively:
\begin{eqnarray}
\hat{\mathcal{P}}^{\nu}=\int d^{3}x\hat{T}^{0\nu}\left(x\right),\quad\hat{\mathcal{N}}_{a}=\int d^{3}x\hat{N}_{a}^{0}\left(x\right),\quad\hat{\mathcal{S}}^{\alpha\beta}=\int d^{3}x\hat{S}^{0\alpha\beta}\left(x\right).
\label{7}
\end{eqnarray}
Substituting Eq.\eqref{7} into Eqs.~\eqref{6} yields the Lorentz-covariant form of the Gibbs distribution:
\begin{align}
	\hat{\rho}_{\mathrm{eq}} & =\exp\biggl\{\Omega-\int d^{3}x\beta\Bigl[U_{\nu}\hat{T}^{0\nu}\left(x\right)-\sum_{a}\mu_{a}\hat{N}_{a}^{0}\left(x\right)-\frac{1}{2}\omega_{\alpha\beta}\hat{S}^{0\alpha\beta}\left(x\right)\Bigr]\biggr\},\label{8}\\
	e^{-\Omega} & =\mathrm{Tr}\exp\biggl\{-\int d^{3}x\beta\Bigl[U_{\nu}\hat{T}^{0\nu}\left(x\right)-\sum_{a}\mu_{a}\hat{N}_{a}^{0}\left(x\right)-\frac{1}{2}\omega_{\alpha\beta}\hat{S}^{0\alpha\beta}\left(x\right)\Bigr]\biggr\}.\label{9}
\end{align}

We consider a system that, while not in global thermodynamic equilibrium, exhibits local equilibrium within each sufficiently small (yet macroscopic) subsystem. This implies that each fluid element can be characterized by local values of hydrodynamic parameters: temperature $\beta^{-1}\left(x\right)$, chemical potentials $\mu_{a}\left(x\right)$, spin chemical potential $\omega_{\alpha\beta}\left(x\right)$, and a macroscopic four-velocity $u^{\nu}\left(x\right)$, which vary gradually in spacetime.

In this scenario, the global equilibrium distribution defined by Eqs.~\eqref{8} and~\eqref{9} is replaced by a local equilibrium statistical operator given by
\begin{align}
	\hat{\rho}_{l}(t) & =\exp\biggl\{\Omega_{l}(t)-\int d^{3}x\Bigl[\beta_{\nu}(x)\hat{T}^{0\nu}(x)-\sum_{a}\alpha_{a}(x)\hat{N}_{a}^{0}(x)-\frac{1}{2}\Omega_{\alpha\beta}\left(x\right)\hat{S}^{0\alpha\beta}\left(x\right)\Bigr]\biggr\},\label{10}\\
	e^{-\Omega_{l}(t)} & =\mathrm{Tr}\exp\biggl\{-\int d^{3}x\biggl[\beta_{\nu}(x)\hat{T}^{0\nu}(x)-\sum_{a}\alpha_{a}(x)\hat{N}_{a}^{0}(x)-\frac{1}{2}\Omega_{\alpha\beta}(x)\hat{S}^{0\alpha\beta}\left(x\right)\biggr]\biggr\},\label{11}
\end{align}
where
\begin{eqnarray}
\beta^{\nu}(x)=\beta(x)u^{\nu}(x),\quad\alpha_{a}(x)=\beta(x)\mu_{a}(x),\quad\Omega_{\alpha\beta}\left(x\right)=\beta\left(x\right)\omega_{\alpha\beta}\left(x\right).
\label{12}
\end{eqnarray}
The local equilibrium operator, $\hat{\rho}_{l}$, is constructed via the maximum entropy principle subject to constraints on the local values of energy-momentum, charge, and spin densities, namely $u_{\nu}\langle\hat{T}^{\mu\nu}\rangle$, $u_{\mu}\langle\hat{N}_{a}^{\mu}\rangle$ and $u_{\lambda}\langle\hat{S}^{\lambda\alpha\beta}\rangle$, respectively~\cite{zubarev1974nonequilibrium,zubarev1997statistical,zubarev1979derivation}. This local equilibrium distribution, as defined in Eq.~\eqref{10}, is also known as the relevant statistical operator in the literature~\cite{zubarev1974nonequilibrium,zubarev1997statistical}.

We define the operators for energy density, charge density, and spin density in the comoving frame as $\hat{\epsilon}=u_{\mu}u_{\nu}\hat{T}^{\mu\nu}$, $\hat{n}_{a}=u_{\mu}\hat{N}_{a}^{\mu}$ and $\hat{S}^{\alpha\beta}=u_{\lambda}\hat{S}^{\lambda\alpha\beta}$, respectively. The local values of the Lorentz-invariant thermodynamic parameters $\beta$, $\alpha_a$, and $\Omega_{\alpha\beta}$ are then determined by imposing matching conditions on the average values of $\hat{\epsilon}$, $\hat{n}_a$, and $\hat{S}^{\alpha\beta}$, as detailed in Refs.~\cite{zubarev1974nonequilibrium,zubarev1997statistical,zubarev1979derivation}
\begin{eqnarray}
\langle\hat{\epsilon}(x)\rangle=\langle\hat{\epsilon}(x)\rangle_{l},\quad\left\langle \hat{n}_{a}(x)\right\rangle =\left\langle \hat{n}_{a}(x)\right\rangle _{l},\quad\langle\hat{S}^{\alpha\beta}\left(x\right)\rangle=\langle\hat{S}^{\alpha\beta}\left(x\right)\rangle_{l},
\label{13}
\end{eqnarray}
where the local ensemble average of an operator $\hat{F}\left(x\right)$ is defined as
\begin{eqnarray}
\langle\hat{F}(x)\rangle_{l}=\operatorname{Tr}[\hat{\rho}_{l}(t) \hat{F}(x)].
\label{14}
\end{eqnarray}
Note that the conditions in Eq.~\eqref{13} serve to define the temperature, chemical potentials, and spin chemical potential as nonlocal functionals of~\cite{zubarev1972nonlocal}:
\begin{eqnarray}
\langle\hat{\epsilon}\left(x\right)\rangle\equiv\epsilon\left(x\right),\quad\langle\hat{n}_{a}\left(x\right)\rangle\equiv n_{a}\left(x\right),\quad\langle\hat{S}^{\alpha\beta}\left(x\right)\rangle\equiv S^{\alpha\beta}\left(x\right).
\label{15}
\end{eqnarray}

To establish a hydrodynamic description, thermodynamic parameters must be defined as local functions of energy, charge, and spin densities, analogous to their global equilibrium counterparts. This requires the assumption of statistical independence among fluid elements, each assumed to be in local equilibrium~\cite{Mori:1958zz}. Consequently, the local equilibrium expectation values $\langle\hat{\epsilon}\rangle_{l}$, $\langle\hat{n}_{a}\rangle_{l}$ and $\langle\hat{S}^{\alpha\beta}\rangle_{l}$ in Eqs.~\eqref{13} are evaluated at constant $\beta$, $\mu_a$, and $\omega_{\alpha\beta}$. These parameters are then determined by matching these local averages to their corresponding global values, $\langle\hat{\epsilon}\rangle$, $\langle\hat{n}_{a}\rangle$, and $\langle \hat{S}^{\alpha\beta}\rangle$, at each spacetime point. This procedure effectively assigns a fictitious local equilibrium state to every point, accurately reproducing local energy, charge, and spin densities.

It is instructive to express the relevant distribution in terms of the scalar fields $\hat{\epsilon}$, $\hat{n}_a$, and the tensor field $\hat{S}^{\alpha\beta}$. By transforming to the local rest frame of each fluid element, Eqs.\eqref{10} and \eqref{11} can be recast as:
\begin{align}
	& \hat{\rho}_{l}(t)=\exp\biggl\{\Omega_{l}(t)-\int d^{3}\widetilde{x}\beta(x)\Bigl[\hat{\epsilon}(x)-\sum_{a}\mu_{a}(x)\hat{n}_{a}(x)-\frac{1}{2}\omega_{\alpha\beta}\left(x\right)\hat{S}^{\alpha\beta}\left(x\right)\Bigr]\biggr\},\label{16}\\
	& e^{-\Omega_{l}(t)}=\mathrm{Tr}\exp\biggl\{-\int d^{3}\widetilde{x}\beta(x)\Bigl[\hat{\epsilon}(x)-\sum_{a}\mu_{a}(x)\hat{n}_{a}(x)-\frac{1}{2}\omega_{\alpha\beta}\left(x\right)\hat{S}^{\alpha\beta}\left(x\right)\Bigr]\biggr\},\label{17}
\end{align}
where $d^{3}\widetilde{x}=u^{0}\left(x\right)d^{3}x$ is the invariant volume element in the fluid's rest frame.

\subsection{Thermodynamic relations}

To derive the thermodynamic relations governing the local thermodynamic parameters, we begin with the relevant distribution given by Eq.~\eqref{10} or~\eqref{16}. Following Zubarev's formalism, we introduce the entropy operator as defined in Refs.~\cite{zubarev1997statistical,zubarev1979derivation}:
\begin{eqnarray}
\begin{aligned}
	\hat{S}(t) & =-\ln\hat{\rho}_{l}(t)=-\Omega_{l}(t)+\int d^{3}x\Bigl[\beta_{\nu}(x)\hat{T}^{0\nu}(x)-\sum_{a}\alpha_{a}(x)\hat{N}_{a}^{0}(x)-\frac{1}{2}\Omega_{\alpha\beta}\left(x\right)\hat{S}^{0\alpha\beta}\left(x\right)\Bigr]\\
	& =-\Omega_{l}(t)+\int d^{3}\widetilde{x}\beta(x)\Bigl[\hat{\epsilon}(x)-\sum_{a}\mu_{a}(x)\hat{n}_{a}(x)-\frac{1}{2}\omega_{\alpha\beta}\left(x\right)\hat{S}^{\alpha\beta}\left(x\right)\Bigr],
\end{aligned}
\label{18}
\end{eqnarray}
which allows us to express the relevant statistical operator as
\begin{eqnarray}
\hat{\rho}_{l}(t)=e^{-\hat{S}(t)}.
\label{19}
\end{eqnarray}

The thermodynamic entropy within a local equilibrium state is given by the expectation value of the entropy operator:
\begin{eqnarray}
S(t)=\langle\hat{S}(t)\rangle_{l}=-\Omega_{l}(t)+\int d^{3}\widetilde{x}\beta(x)\biggl[\left\langle \hat{\epsilon}(x)\right\rangle -\sum_{a}\mu_{a}(x)\left\langle \hat{n}_{a}(x)\right\rangle -\frac{1}{2}\omega_{\alpha\beta}\left(x\right)\langle\hat{S}^{\alpha\beta}\left(x\right)\rangle\biggr]\equiv\langle\hat{S}(t)\rangle,
\label{20}
\end{eqnarray}
where we have imposed the matching conditions of Eq.~\eqref{13}.

To derive the desired thermodynamic relations, we consider infinitesimal variations $\delta\epsilon\left(x\right)$, $\delta n_{a}\left(x\right)$, and $\delta S^{\alpha\beta}\left(x\right)$ in the local energy, charges, and spin densities, respectively. These perturbations induce corresponding infinitesimal changes in temperature, $\delta\beta\left(x\right)$, chemical potentials, $\delta\mu_{a}\left(x\right)$, and spin chemical potential, $\delta\omega_{\alpha\beta}\left(x\right)$. The resulting variation in $\Omega_{l}\left(t\right)$ is given by
\begin{eqnarray}
\delta\Omega_{l}\left(t\right)=\int d^{3}\widetilde{x}\biggl[\frac{\delta\Omega_{l}\left(t\right)}{\delta\beta\left(x\right)}\delta\beta\left(x\right)+\sum_{a}\frac{\delta\Omega_{l}\left(t\right)}{\delta\mu_{a}\left(x\right)}\delta\mu_{a}\left(x\right)+\frac{\delta\Omega_{l}\left(t\right)}{\delta\omega_{\alpha\beta}\left(x\right)}\delta\omega_{\alpha\beta}\left(x\right)\biggr],
\label{21}
\end{eqnarray}
where the square brackets contain Lorentz-invariant functional derivatives of $\Omega_{l}\left(t\right)$. From Eqs.~\eqref{16} and \eqref{17}, we obtain:
\begin{align}
&\frac{\delta\Omega_{l}(t)}{\delta\beta(x)}=\epsilon(x)-\sum_{a}\mu_{a}(x)n_{a}(x)-\frac{1}{2}\omega_{\alpha\beta}\left(x\right)S^{\alpha\beta}\left(x\right),\label{22}\\
&\frac{\delta\Omega_{l}(t)}{\delta\mu_{a}(x)}=-\beta(x)n_{a}(x),\label{23}\\
&\frac{\delta\Omega_{l}\left(t\right)}{\delta\omega_{\alpha\beta}\left(x\right)}=-\frac{1}{2}\beta\left(x\right)S^{\alpha\beta}\left(x\right).\label{24}
\end{align}
The infinitesimal change in the entropy can then be found from Eqs.\eqref{20}-\eqref{24},
\begin{eqnarray}
\begin{aligned}
\delta S(t)= & -\delta\Omega_{l}(t)+\int d^{3}\widetilde{x}\biggl[\delta\beta\Bigl(\epsilon-\sum_{a}\mu_{a}n_{a}-\frac{1}{2}\omega_{\alpha\beta}S^{\alpha\beta}\Bigr)\\
 & +\beta\Bigl(\delta\epsilon-\sum_{a}\mu_{a}\delta n_{a}-\frac{1}{2}\omega_{\alpha\beta}\delta S^{\alpha\beta}\Bigr)-\beta\sum_{a}n_{a}\delta\mu_{a}-\frac{1}{2}\beta S^{\alpha\beta}\delta\omega_{\alpha\beta}\biggr]\\
= & \int d^{3}\widetilde{x}\beta\left(x\right)\Bigl[\delta\epsilon\left(x\right)-\sum_{a}\mu_{a}\left(x\right)\delta n_{a}\left(x\right)-\frac{1}{2}\omega_{\alpha\beta}\left(x\right)\delta S^{\alpha\beta}\left(x\right)\Bigr].
\end{aligned}
\label{25}
\end{eqnarray}
Defining the invariant entropy density $s(x)$ as
\begin{eqnarray}
S(t)=\int d^{3} \widetilde{x} s\left(x\right).
\label{26}
\end{eqnarray}
Eq.~\eqref{25} becomes
\begin{eqnarray}
\int d^{3}\widetilde{x}\biggl\{\beta\left(x\right)\Bigl[\delta\epsilon\left(x\right)-\sum_{a}\mu_{a}\left(x\right)\delta n_{a}\left(x\right)-\frac{1}{2}\omega_{\alpha\beta}\left(x\right)\delta S^{\alpha\beta}\left(x\right)\Bigr]-\delta s\left(x\right)\biggr\}=0.
\label{27}
\end{eqnarray}
Since $\delta \epsilon\left(x\right)$, $\delta n_a\left(x\right)$ and $\delta S^{\alpha\beta}\left(x\right)$ are arbitrary variations, and the entropy density $s\left(x\right)$ is a local functional of $\epsilon\left(x\right)$, $n_a\left(x\right)$ and $S^{\alpha\beta}$[i.e., $s\left(\epsilon\left(x\right),n_{a}\left(x\right),S^{\alpha\beta}\left(x\right)\right)\equiv s\left(x\right)$], Eq.~\eqref{27} yields the following relation:
\begin{eqnarray}
T\left(x\right)\delta s\left(x\right)=\delta\epsilon\left(x\right)-\sum_{a}\mu_{a}\left(x\right)\delta n_{a}\left(x\right)-\frac{1}{2}\omega_{\alpha\beta}\left(x\right)\delta S^{\alpha\beta}\left(x\right),
\label{28}
\end{eqnarray}
which is the first law of thermodynamics for local variables.

To derive additional thermodynamic relations, we recall that the grand potential in global thermodynamic equilibrium is given by $\Omega_{l}=-\beta pV$, where $p$ is the pressure and $V$ is the system volume. In the regime of local equilibrium, $\Omega_{l}\left(t\right)$, as defined in Eq.~\eqref{17}, becomes a functional of $\epsilon\left(x\right)$, $n_{a}\left(x\right)$, and $S^{\alpha\beta}\left(x\right)$. Consequently, we introduce a scalar function $p\left(\epsilon\left(x\right),n_{a}\left(x\right),S^{\alpha\beta}\left(x\right)\right)\equiv p\left(x\right)$, representing the local pressure, such that
\begin{eqnarray}
\Omega_{l}(t)=-\int d^{3}\widetilde{x}\beta(x)p(x).
\label{29}
\end{eqnarray}
The function is determined by Eqs.~\eqref{17} and \eqref{29}, and the matching conditions of Eq.~\eqref{13}, which yield the temperature, chemical potentials, and spin chemical potential. From Eq.~\eqref{29}, we can express Eq.~\eqref{20} as
\begin{eqnarray}
S(t)=\int d^{3}\widetilde{x}\beta(x)\Bigl[\epsilon(x)+p(x)-\sum_{a}\mu_{a}(x)n_{a}(x)-\frac{1}{2}\omega_{\alpha\beta}S^{\alpha\beta}\Bigr].
\label{30}
\end{eqnarray}

Combining Eq.~\eqref{30} with Eq.~\eqref{26} leads to the well-known thermodynamic relation:
\begin{eqnarray}
\epsilon\left(x\right)+p\left(x\right)=T\left(x\right)s\left(x\right)+\sum_{a}\mu_{a}\left(x\right)n_{a}\left(x\right)+\frac{1}{2}\omega_{\alpha\beta}\left(x\right)S^{\alpha\beta}\left(x\right)\equiv w\left(x\right),
\label{31}
\end{eqnarray}
where $w$ is the enthalpy density. The Gibbs-Duhem relation can be derived from Eqs.\eqref{28} and \eqref{31}:
\begin{eqnarray}
\delta p\left(x\right)=s\left(x\right)\delta T\left(x\right)+\sum_{a}n_{a}\left(x\right)\delta\mu_{a}\left(x\right)+\frac{1}{2}S^{\alpha\beta}\left(x\right)\delta\omega_{\alpha\beta}\left(x\right).
\label{32}
\end{eqnarray}

The relevant statistical operator allows us to construct a complete set of thermodynamic variables, thereby establishing a fictitious local-equilibrium state. This is a crucial step in developing a comprehensive spin hydrodynamic description of the system, considering its energy-momentum tensor, charge current densities, and total angular momentum tensor.

\subsection{The nonequilibrium statistical operator}

While the statistical operator $\hat{\rho}_{l}$, defined in Eqs.~\eqref{10} and \eqref{11}, provides the local values of the macroscopic observables $u_{\nu}\langle \hat{T}^{\mu\nu}\rangle$, $u_{\mu}\langle \hat{N}_{a}^{\mu}\rangle$, and $u_{\lambda}\langle \hat{S}^{\lambda\alpha\beta}\rangle$, it fails to satisfy the Liouville equation~\eqref{4}, rendering it unsuitable for describing nonequilibrium thermodynamic processes.

To account for the irreversibility of thermodynamic processes, we introduce the statistical operator:
\begin{eqnarray}
\hat{\rho}_{\varepsilon}(t)\equiv\exp\left[-\varepsilon\int_{-\infty}^{t}dt^{\prime}e^{\varepsilon\left(t^{\prime}-t\right)}\hat{S}\left(t^{\prime}\right)\right],
\label{33}
\end{eqnarray}
this operator obeys the Liouville equation with an additional source term of order $\sim \varepsilon$, which explicitly breaks the time-reversal symmetry of the Liouville equation due to its retarded solution. To preserve irreversibility throughout the calculation, the thermodynamic limit must precede the limit $\varepsilon\to0^{+}$.

Consequently, the statistical average of any operator $\hat{F}(x)$ is determined according to the prescription outlined in Ref.~\cite{zubarev1997statistical}:
\begin{equation}
\langle\hat{F}(x)\rangle=\lim_{\varepsilon\rightarrow0^{+}}\lim_{V\rightarrow\infty}\text{Tr}[\hat{\rho}_{\varepsilon}(t)\hat{F}(x)],
\label{34}
\end{equation}
where $V$ represents the system volume. Recognizing that the statistical operator defined in Eq.~\eqref{33} incorporates memory effects, we anticipate that the ensuing equations of motion will possess causal properties, consistent with the findings of~\cite{zubarev1972nonlocal,morozov1998zubarev,Koide:2006ef,Koide:2008nw}. To ensure causality, we have constructed a causal nonequilibrium statistical operator by extending the relevant statistical operator.

Substituting the explicit expression for $\hat{S}\left(t\right)$ from Eq.~\eqref{18} into Eq.~\eqref{33} yields (with the index $\varepsilon$ suppressed for brevity):
\begin{equation}
\hat{\rho}(t)=Q^{-1}(t)\exp\biggl\{-\int d^{3}x\hat{Z}(\boldsymbol{x},t)\biggr\},\quad Q(t)=\mathrm{Tr}\exp\biggl\{-\int d^{3}x\hat{Z}(\boldsymbol{x},t)\biggr\},
\label{35}
\end{equation}
where
\begin{equation}
\hat{Z}(\boldsymbol{x},t)=\varepsilon\int_{-\infty}^{t}dt_{1}e^{\varepsilon\left(t_{1}-t\right)}\biggl[\beta_{\nu}\left(\boldsymbol{x},t_{1}\right)\hat{T}^{0\nu}\left(\boldsymbol{x},t_{1}\right)-\sum_{a}\alpha_{a}\left(\boldsymbol{x},t_{1}\right)\hat{N}_{a}^{0}\left(\boldsymbol{x},t_{1}\right)-\frac{1}{2}\Omega_{\alpha\beta}\left(\boldsymbol{x},t_{1}\right)\hat{S}^{0\alpha\beta}\left(\boldsymbol{x},t_{1}\right)\biggr].
\label{36}
\end{equation}
Using integration by parts, the local equilibrium term in Eq.~\eqref{36} can be separated as follows
\begin{equation*}
\begin{aligned}
\hat{Z}(\boldsymbol{x},t)= & \beta_{\nu}(\boldsymbol{x},t)\hat{T}^{0\nu}(\boldsymbol{x},t)-\sum_{a}\alpha_{a}(\boldsymbol{x},t)\hat{N}_{a}^{0}(\boldsymbol{x},t)-\frac{1}{2}\Omega_{\alpha\beta}\left(\boldsymbol{x},t\right)\hat{S}^{0\alpha\beta}\left(\boldsymbol{x},t\right)\\
&-\int_{-\infty}^{t}dt_{1}e^{\varepsilon\left(t_{1}-t\right)}\frac{d}{dt_{1}}\left[\beta_{\nu}\left(\boldsymbol{x},t_{1}\right)\hat{T}^{0\nu}\left(\boldsymbol{x},t_{1}\right)-\sum_{a}\alpha_{a}\left(\boldsymbol{x},t_{1}\right)\hat{N}_{a}^{0}\left(\boldsymbol{x},t_{1}\right)-\frac{1}{2}\Omega_{\alpha\beta}\left(\boldsymbol{x},t_{1}\right)\hat{S}^{0\alpha\beta}\left(\boldsymbol{x},t_{1}\right)\right].
\end{aligned}
\end{equation*}
The exponential factor $e^{\varepsilon\left(t_{1}-t\right)}$ ensures that the system asymptotically approaches the limiting behavior:
\begin{equation*}
\lim_{t_{1} \rightarrow-\infty} e^{\varepsilon\left(t_{1}-t\right)} \hat{F}\left(t_{1}\right)=0,
\end{equation*}
where $\hat{F}\left(t_{1}\right)$ is the term in square brackets in Eq.~\eqref{36}.

The conservation laws~\eqref{1},~\eqref{2}, and~\eqref{3} yield the relations $\partial_{\mu}\hat{T}^{\mu\nu}=\partial_{0}\hat{T}^{0\nu}+\partial_{i}\hat{T}^{i\nu}=0$,  $\partial_{\mu}\hat{N}_{a}^{\mu}=\partial_{0}\hat{N}_{a}^{0}+\partial_{i}\hat{N}_{a}^{i}=0$, and $\partial_{\lambda}\hat{S}^{\lambda\mu\nu}=\partial_{0}\hat{S}^{0\mu\nu}+\partial_{i}\hat{S}^{i\mu\nu}=-2\hat{T}^{[\mu\nu]}$, which lead to:
\begin{equation}
\begin{aligned}
	\partial_{0}\biggl(\beta_{\nu}\hat{T}^{0\nu}-\sum_{a}\alpha_{a}\hat{N}_{a}^{0}-\frac{1}{2}\Omega_{\alpha\beta}\hat{S}^{0\alpha\beta}\biggr)= &\hat{T}^{\mu\nu}\partial_{\mu}\beta_{\nu}-\sum_{a}\hat{N}_{a}^{\mu}\partial_{\mu}\alpha_{a}-\frac{1}{2}\hat{S}^{\lambda\alpha\beta}\partial_{\lambda}\Omega_{\alpha\beta}+\Omega_{\alpha\beta}\hat{T}^{[\alpha\beta]}\\
	&-\partial_{i}\biggl(\beta_{\nu}\hat{T}^{i\nu}-\sum_{a}\alpha_{a}\hat{N}_{a}^{i}-\frac{1}{2}\Omega_{\alpha\beta}\hat{S}^{i\alpha\beta}\biggr).
\end{aligned}
\label{37}
\end{equation}
Upon spatial integration, the final term in Eq.~\eqref{37} becomes a surface integral that vanishes when the boundary extends to infinity. This leaves us with
\begin{equation}
\begin{aligned}
	\int d^{3}x\hat{Z}(\boldsymbol{x},t)= & \int d^{3}x\biggl[\beta_{\nu}(\boldsymbol{x},t)\hat{T}^{0\nu}(\boldsymbol{x},t)-\sum_{a}\alpha_{a}(\boldsymbol{x},t)\hat{N}_{a}^{0}(\boldsymbol{x},t)-\frac{1}{2}\Omega_{\alpha\beta}(\boldsymbol{x},t)\hat{S}^{0\alpha\beta}(\boldsymbol{x},t)\biggr]\\
	&-\int d^{3}x\int_{-\infty}^{t}dt_{1}e^{\varepsilon\left(t_{1}-t\right)}\biggl[\hat{T}^{\mu\nu}\left(\boldsymbol{x},t_{1}\right)\partial_{\mu}\beta_{\nu}\left(\boldsymbol{x},t_{1}\right)-\sum_{a}\hat{N}_{a}^{\mu}\left(\boldsymbol{x},t_{1}\right)\partial_{\mu}\alpha_{a}\left(\boldsymbol{x},t_{1}\right)\\
	&-\frac{1}{2}\hat{S}^{\lambda\alpha\beta}\left(\boldsymbol{x},t_{1}\right)\partial_{\lambda}\Omega_{\alpha\beta}(\boldsymbol{x},t_{1})+\Omega_{\alpha\beta}\left(\boldsymbol{x},t_{1}\right)\hat{T}^{[\alpha\beta]}\left(\boldsymbol{x},t_{1}\right)\biggr],
\end{aligned}
\label{38}
\end{equation}
where the $4$-gradients are taken with respect to spacetime coordinates $\left(\boldsymbol{x},t_1\right)$. The initial term of this expression represents the local equilibrium component of the statistical operator. The integrand of the second term constitutes a thermodynamic "force" as it involves gradients of temperature, chemical potentials, spin chemical potential, and the velocity field. Consequently, the second term in Eq.~\eqref{38} is naturally associated with the nonequilibrium part of the statistical operator. By employing Eqs.~\eqref{35} and \eqref{38}, we can express the complete statistical operator as~\cite{Hosoya:1983id,Huang:2011dc}
\begin{equation}
\hat{\rho}(t)=Q^{-1} e^{-\hat{A}+\hat{B}}, \quad Q=\operatorname{Tr} e^{-\hat{A}+\hat{B}},
\label{39}
\end{equation}
with
\begin{align}
	\hat{A}(t)= & \int d^{3}x\Bigl[\beta_{\nu}(\boldsymbol{x},t)\hat{T}^{0\nu}(\boldsymbol{x},t)-\sum_{a}\alpha_{a}(\boldsymbol{x},t)\hat{N}_{a}^{0}(\boldsymbol{x},t)-\frac{1}{2}\Omega_{\alpha\beta}(\boldsymbol{x},t)\hat{S}^{0\alpha\beta}(\boldsymbol{x},t)\Bigr],\label{40}\\
	\hat{B}(t)= & \int d^{3}x\int_{-\infty}^{t}dt_{1}e^{\varepsilon\left(t_{1}-t\right)}\hat{C}\left(\boldsymbol{x},t_{1}\right),\label{41}\\
	\hat{C}(\boldsymbol{x},t)= & \hat{T}^{\mu\nu}(\boldsymbol{x},t)\partial_{\mu}\beta_{\nu}(\boldsymbol{x},t)-\sum_{a}\hat{N}_{a}^{\mu}(\boldsymbol{x},t)\partial_{\mu}\alpha_{a}(\boldsymbol{x},t)-\frac{1}{2}\hat{S}^{\lambda\alpha\beta}\left(\boldsymbol{x},t\right)\partial_{\lambda}\Omega_{\alpha\beta}(\boldsymbol{x},t)\nonumber\\
	&+\Omega_{\alpha\beta}(\boldsymbol{x},t)\hat{T}^{[\alpha\beta]}(\boldsymbol{x},t).\label{42}
\end{align}
The statistical operator defined in Eq.~\eqref{39} can be employed to derive transport equations for dissipative currents. The nonequilibrium component given by Eq.~\eqref{41} is treated as a perturbation. By retaining only the first-order terms in the Taylor expansion of $\hat{\rho}\left(t\right)$ with respect to the operator $\hat{B}(t)$, the conventional first-order dissipative hydrodynamic theory is recovered~\cite{Hosoya:1983id,Huang:2011dc}. Conversely, incorporating all second-order terms in the Taylor expansion leads to the well-established second-order dissipative hydrodynamic theory~\cite{Harutyunyan:2018cmm,Harutyunyan:2017lrm,Harutyunyan:2021rmb,Buzzegoli:2022kqx}. In this study, we extend this approach to formulate a second-order dissipative spin hydrodynamic theory within the Zubarev framework by including all second-order terms in the Taylor expansion.

\subsection{Second-order expansion of the statistical operator}

To derive the transport equations for dissipative currents, we expand the nonequilibrium statistical operator $\hat{\rho}$ to second order in a Taylor series with respect to the operator $\hat{B}$~\cite{Harutyunyan:2021rmb,Buzzegoli:2022kqx}:
\begin{eqnarray}
\hat{\rho}=\hat{\rho}_{l}+\hat{\rho}_{1}+\hat{\rho}_{2},
\label{43}
\end{eqnarray}
where the first-order correction is
\begin{eqnarray}
\begin{aligned}
	\hat{\rho}_{1} & =\int_{0}^{1}d\tau\Bigl(\hat{B}_{\tau}-\langle\hat{B}_{\tau}\rangle_{l}\Bigr)\hat{\rho}_{l}\\
	& =\int d^{4}x_{1}\int_{0}^{1}d\tau\Bigl[\hat{C}_{\tau}\left(x_{1}\right)-\bigl\langle\hat{C}_{\tau}\left(x_{1}\right)\bigr\rangle_{l}\Bigr]\hat{\rho}_{l},
\end{aligned}
\label{44}
\end{eqnarray}
and the second-order correction is
\begin{eqnarray}
\begin{aligned}
	\hat{\rho}_{2}= & \frac{1}{2}\int_{0}^{1}d\tau\int_{0}^{1}d\lambda\left[\widetilde{T}\bigl\{\hat{B}_{\lambda}\hat{B}_{\tau}\bigr\}-\bigl\langle\widetilde{T}\bigl\{\hat{B}_{\lambda}\hat{B}_{\tau}\bigr\}\bigr\rangle_{l}-\hat{B}_{\tau}\bigl\langle\hat{B}_{\lambda}\bigr\rangle_{l}-\hat{B}_{\lambda}\bigl\langle\hat{B}_{\tau}\bigr\rangle_{l}+2\bigl\langle\hat{B}_{\tau}\bigr\rangle_{l}\bigl\langle\hat{B}_{\lambda}\bigr\rangle_{l}\right]\hat{\rho}_{l}\\
	= & \frac{1}{2}\int d^{4}x_{1}d^{4}x_{2}\int_{0}^{1}d\tau\int_{0}^{1}d\lambda\Bigl[\widetilde{T}\bigl\{\hat{C}_{\lambda}\left(x_{1}\right)\hat{C}_{\tau}\left(x_{2}\right)\bigr\}-\bigl\langle\widetilde{T}\bigl\{\hat{C}_{\lambda}\left(x_{1}\right)\hat{C}_{\tau}\left(x_{2}\right)\bigr\}\bigr\rangle_{l}\\
	&-\bigl\langle\hat{C}_{\lambda}\left(x_{1}\right)\bigr\rangle_{l}\hat{C}_{\tau}\left(x_{2}\right)-\hat{C}_{\lambda}\left(x_{1}\right)\bigl\langle\hat{C}_{\tau}\left(x_{2}\right)\bigr\rangle_{l}+2\bigl\langle\hat{C}_{\lambda}\left(x_{1}\right)\bigr\rangle_{l}\bigl\langle\hat{C}_{\tau}\left(x_{2}\right)\bigr\rangle_{l}\Bigr]\hat{\rho}_{l},
\end{aligned}
\label{45}
\end{eqnarray}
where we introduced the abbreviated notation $\hat{X}_{\tau}=e^{-\tau\hat{A}}\hat{X}e^{\tau\hat{A}}$ for any operator $\hat{X}$ and $\int d^{4}x_{1}\equiv\int d^{3}x_{1}\int_{-\infty}^{t}dt_{1}e^{\varepsilon\left(t_{1}-t\right)}$, and $\widetilde{T}$ represents the antichronological time-ordering operator with respect to the parameters $\tau$ and $\lambda$.

The statistical average of an arbitrary operator $\langle\hat{X}\left(x\right)\rangle$ can be expressed as follows, based on the generic expansions provided above and the relations in Eqs.~\eqref{5},\eqref{43},\eqref{44}, and \eqref{45}:
\begin{equation}
\langle\hat{X}(x)\rangle=\langle\hat{X}(x)\rangle_{l}+\int d^{4}x_{1}\left(\hat{X}\left(x\right),\hat{C}\left(x_{1}\right)\right)+\int d^{4}x_{1}\int d^{4}x_{2}\left(\hat{X}\left(x\right),\hat{C}\left(x_{1}\right),\hat{C}\left(x_{2}\right)\right),
\label{46}
\end{equation}
where we have defined a two-point correlation function
\begin{equation}
\left(\hat{X}(x),\hat{Y}\left(x_{1}\right)\right)=\int_{0}^{1}d\tau\left\langle \hat{X}(x)\Bigl[\hat{Y}_{\tau}\left(x_{1}\right)-\bigl\langle\hat{Y}_{\tau}\left(x_{1}\right)\bigr\rangle_{l}\Bigr]\right\rangle _{l},
\label{47}
\end{equation}
and a three-point correlation function
\begin{eqnarray}
\begin{aligned}
	\left(\hat{X}(x),\hat{Y}\left(x_{1}\right),\hat{Z}\left(x_{2}\right)\right) & \equiv\frac{1}{2}\int_{0}^{1}d\tau\int_{0}^{1}d\lambda\biggl\langle\widetilde{T}\Bigl\{\hat{X}(x)\Bigl[\hat{Y}_{\lambda}\left(x_{1}\right)\hat{Z}_{\tau}\left(x_{2}\right)-\bigl\langle\widetilde{T}\hat{Y}_{\lambda}\left(x_{1}\right)\hat{Z}_{\tau}\left(x_{2}\right)\bigr\rangle_{l}\\
	&\quad-\bigl\langle\hat{Y}_{\lambda}\left(x_{1}\right)\bigr\rangle_{l}\hat{Z}_{\tau}\left(x_{2}\right)-\hat{Y}_{\lambda}\left(x_{1}\right)\bigl\langle\hat{Z}_{\tau}\left(x_{2}\right)\bigr\rangle_{l}+2\bigl\langle\hat{Y}_{\lambda}\left(x_{1}\right)\bigr\rangle_{l}\bigl\langle\hat{Z}_{\tau}\left(x_{2}\right)\bigr\rangle_{l}\Bigr]\Bigr\}\biggr\rangle_{l}.
\end{aligned}
\label{48}
\end{eqnarray}
From Eq.~\eqref{48}, we can directly obtain the pivotal symmetry relation
\begin{eqnarray}
\int d^{4} x_{1} d^{4} x_{2}\left(\hat{X}(x), \hat{Y}\left(x_{1}\right), \hat{Z}\left(x_{2}\right)\right)=\int d^{4} x_{1} d^{4} x_{2}\left(\hat{X}(x), \hat{Z}\left(x_{1}\right), \hat{Y}\left(x_{2}\right)\right),
\label{49}
\end{eqnarray}
which will be exploited in the following sections.

\subsection{Hydrodynamic equations: Canonical-like framework}

To isolate the dissipative processes related to viscous and diffusion currents, we decompose the energy-momentum tensor, charge currents, and spin tensor into their equilibrium and dissipative parts. The general form of these decompositions is
\begin{align}
&\hat{T}^{\mu\nu}=\hat{\epsilon}u^{\mu}u^{\nu}-\hat{p}\Delta^{\mu\nu}+\hat{h}^{\mu}u^{\nu}+\hat{h}^{\nu}u^{\mu}+\hat{\pi}^{\mu\nu}+\hat{q}^{\mu}u^{\nu}-\hat{q}^{\nu}u^{\mu}+\hat{\phi}^{\mu\nu},\label{50}\\
&\hat{N}_{a}^{\mu}=\hat{n}_{a}u^{\mu}+\hat{j}_{a}^{\mu},\label{51}\\
&\hat{S}^{\lambda\mu\nu}=u^{\lambda}\hat{S}^{\mu\nu}+u^{\mu}\hat{S}^{\nu\lambda}+u^{\nu}\hat{S}^{\lambda\mu}+\hat{\varpi}^{\lambda\mu\nu},\label{52}
\end{align}
where $\Delta^{\mu\nu}=g^{\mu\nu}-u^{\mu}u^{\nu}$ is the projection tensor onto the three-space orthogonal to the fluid four-velocity $u^\mu$. In accordance with the hydrodynamic gradient expansion, the energy density $\hat{\epsilon}$, particle number densities $\hat{n}_{a}$, and four-velocity $u^{\mu}$ scale as $\mathcal{O}\left(\partial^{0}\right)$. Conversely, the diffusion currents $\hat{j}_{a}^{\mu}$, shear stress tensor $\hat{\pi}^{\mu\nu}$, heat flux $\hat{h}^{\mu}$, boost heat vector $\hat{q}^{\mu}$, and rotational stress tensor $\hat{\phi}^{\mu\nu}$ scale as $\mathcal{O}\left(\partial\right)$. Importantly, $\hat{\pi}^{\mu\nu}$, $\hat{h}^{\mu}$, $\hat{q}^{\mu}$, $\hat{\phi}^{\mu\nu}$, $\hat{j}_{a}^{\mu}$, and $\hat{\varpi}^{\lambda\mu\nu}$ are all orthogonal to the fluid four-velocity $u_\mu$. Furthermore, $\hat{\pi}^{\mu\nu}$ is traceless:
\begin{eqnarray}
\begin{aligned}
&u_{\nu}\hat{h}^{\nu}=0,\,\,u_{\nu}\hat{j}_{a}^{\nu}=0,\,\,u_{\nu}\hat{\pi}^{\mu\nu}=0,\,\,u_{\mu}\hat{\pi}^{\mu\nu}=0,\,\,u_{\nu}\hat{q}^{\nu}=0,\,\,u_{\nu}\hat{\phi}^{\mu\nu}=0,\,\,u_{\mu}\hat{\phi}^{\mu\nu}=0,\\
&\hat{\pi}^{\mu\nu}=\hat{\pi}^{\nu\mu},\,\,\hat{\phi}^{\mu\nu}=-\hat{\phi}^{\nu\mu},\,\,\hat{\pi}_{\,\,\,\mu}^{\mu}=0,\,\,u_{\lambda}\hat{\varpi}^{\lambda\mu\nu}=0,\,\,u_{\mu}\hat{\varpi}^{\lambda\mu\nu}=0,\,\,u_{\nu}\hat{\varpi}^{\lambda\mu\nu}=0.
\end{aligned}
\label{53}
\end{eqnarray}

Due to the totally antisymmetric nature of the spin current, the spin density $\hat{S}^{\mu\nu}$ is subject to the Frenkel condition, $\hat{S}^{\mu\nu}u_{\mu}=0$, in addition to the intrinsic antisymmetry, $\hat{S}^{\mu\nu}=-\hat{S}^{\nu\mu}$. As a result, $\hat{S}^{\mu\nu}$ possesses only three dynamical degrees of freedom, making it a leading-order quantity within the hydrodynamic gradient expansion, scaling as $\mathcal{O}\left(\partial^{0}\right)$, similar to the number density. Since the spin density $\hat{S}^{\mu\nu}$ is transverse to the fluid velocity $u^\mu$, it is sufficient to assume that the spin chemical potential satisfies $\omega_{\mu\nu}u^\mu=0$. Note that the first-order dissipative current $\hat{\varpi}^{\lambda\mu\nu}$ is totally antisymmetric.

It is important to note that the equilibrium and bulk-viscous components of the pressure have not been explicitly separated in this analysis. The statistical average of the operator $\hat{p}$ yields the actual, isotropic pressure under nonequilibrium conditions. This pressure, generally deviates from the equilibrium pressure, $p\bigl(\langle\hat{\epsilon}\rangle,\langle\hat{n}_{a}\rangle,\langle\hat{S}^{\alpha\beta}\rangle\bigr)$, which is determined by averaging $\hat{p}$ over the local equilibrium distribution (formally evaluated at constant thermodynamic parameters). The bulk viscous pressure is consequently defined as the difference between these two averages.

The operators on the right-hand sides of Eqs.~\eqref{50}-\eqref{52} are defined by projections of $\hat{T}^{\mu\nu}$, $\hat{N}_{a}^{\mu}$, and $\hat{S}^{\lambda\mu\nu}$, respectively:
\begin{align} &\hat{\epsilon}=u_{\mu}u_{\nu}\hat{T}^{\mu\nu},\quad\hat{p}=-\frac{1}{3}\Delta_{\mu\nu}\hat{T}^{\mu\nu},\quad\hat{h}^{\mu}=\Delta_{(\alpha}^{\mu}u_{\beta)}\hat{T}^{\alpha\beta},\quad\hat{\pi}^{\mu\nu}=\Delta_{\alpha\beta}^{\mu\nu}\hat{T}^{\alpha\beta},\quad\hat{q}^{\mu}=\Delta_{[\alpha}^{\mu}u_{\beta]}\hat{T}^{\alpha\beta},\label{54}\\ &\hat{\phi}^{\mu\nu}=\mydelta_{\alpha\beta}^{\mu\nu}\hat{T}^{\alpha\beta},\quad\hat{n}_{a}=u_{\mu}\hat{N}_{a}^{\mu},\quad\hat{j}_{a}^{\nu}=\Delta_{\mu}^{\nu}\hat{N}_{a}^{\mu},\quad\hat{S}^{\mu\nu}=u_{\lambda}\hat{S}^{\lambda\mu\nu},\quad\hat{\varpi}^{\lambda\mu\nu}=\myDelta_{\rho\sigma\delta}^{\lambda\mu\nu}\hat{S}^{\rho\sigma\delta},\label{55}
\end{align}
where the following relations have been utilized
\begin{eqnarray}
u_{\mu}\Delta^{\mu\nu}=\Delta^{\mu\nu}u_{\nu}=0,\quad\Delta^{\mu\nu}\Delta_{\nu\lambda}=\Delta_{\lambda}^{\mu},\quad\Delta_{\mu}^{\mu}=3.
\label{56}
\end{eqnarray}
In Eqs.~\eqref{54} and \eqref{55} we also introduced the projectors orthogonal to $u^\mu$ via
\begin{align}
	\Delta_{\rho\sigma}^{\mu\nu}&=\frac{1}{2}\left(\Delta_{\rho}^{\mu}\Delta_{\sigma}^{\nu}+\Delta_{\sigma}^{\mu}\Delta_{\rho}^{\nu}\right)-\frac{1}{3}\Delta^{\mu\nu}\Delta_{\rho\sigma},\label{57}\\
	\mydelta_{\rho\sigma}^{\mu\nu}&=\frac{1}{2}\left(\Delta_{\rho}^{\mu}\Delta_{\sigma}^{\nu}-\Delta_{\sigma}^{\mu}\Delta_{\rho}^{\nu}\right),\label{58}\\
	\myDelta_{\alpha\beta\gamma}^{\lambda\mu\nu}&=\frac{1}{6}\left(\Delta_{\alpha}^{\lambda}\Delta_{\beta}^{\mu}\Delta_{\gamma}^{\nu}+\Delta_{\alpha}^{\mu}\Delta_{\beta}^{\nu}\Delta_{\gamma}^{\lambda}+\Delta_{\alpha}^{\nu}\Delta_{\beta}^{\lambda}\Delta_{\gamma}^{\mu}-\Delta_{\alpha}^{\mu}\Delta_{\beta}^{\lambda}\Delta_{\gamma}^{\nu}-\Delta_{\alpha}^{\nu}\Delta_{\beta}^{\mu}\Delta_{\gamma}^{\lambda}-\Delta_{\alpha}^{\lambda}\Delta_{\beta}^{\nu}\Delta_{\gamma}^{\mu}\right),\label{59}
\end{align}
which has the properties
\begin{align}
	\Delta_{\mu\nu\rho\sigma} & =\Delta_{\nu\mu\rho\sigma}=\Delta_{\rho\sigma\mu\nu},\quad u^{\mu}\Delta_{\mu\nu\rho\sigma}=0,\quad\Delta_{\alpha}^{\mu}\Delta_{\mu\nu\rho\sigma}=\Delta_{\alpha\nu\rho\sigma},\label{60}\\
	\Delta_{\mu\,\,\,\,\rho\sigma}^{\,\,\,\,\mu} & =0,\quad\Delta_{\nu\mu\,\,\,\,\sigma}^{\,\,\,\,\,\,\,\mu}=\frac{5}{3}\Delta_{\nu\sigma},\quad\Delta_{\mu\nu}^{\,\,\,\,\,\,\,\,\mu\nu}=5,\quad\Delta_{\mu\nu\rho\sigma}\Delta_{\alpha\beta}^{\rho\sigma}=\Delta_{\mu\nu\alpha\beta},\label{61}\\
	\mydelta_{\mu\nu\rho\sigma} & =-\mydelta_{\nu\mu\rho\sigma}=\mydelta_{\rho\sigma\mu\nu},\quad u^{\mu}\mydelta_{\mu\nu\rho\sigma}=0,\quad\Delta_{\alpha}^{\mu}\mydelta_{\mu\nu\rho\sigma}=\mydelta_{\alpha\nu\rho\sigma},\label{62}\\
	\mydelta_{\mu\,\,\,\,\rho\sigma}^{\,\,\,\,\mu} & =0,\quad\mydelta_{\nu\mu\,\,\,\,\sigma}^{\,\,\,\,\,\,\,\mu}=-\Delta_{\nu\sigma},\quad\mydelta_{\mu\nu}^{\,\,\,\,\,\,\,\,\mu\nu}=3,\quad\mydelta_{\mu\nu\rho\sigma}\mydelta_{\alpha\beta}^{\rho\sigma}=\mydelta_{\mu\nu\alpha\beta},\label{63}\\
	\myDelta_{\lambda\mu\nu\alpha\beta\gamma} & =-\myDelta_{\mu\lambda\nu\alpha\beta\gamma}=\myDelta_{\alpha\beta\gamma\lambda\mu\nu},\quad u^{\lambda}\myDelta_{\lambda\mu\nu\alpha\beta\gamma}=0,\quad\Delta_{\varepsilon}^{\lambda}\myDelta_{\lambda\mu\nu\alpha\beta\gamma}=\myDelta_{\varepsilon\mu\nu\alpha\beta\gamma},\label{64}\\
	\myDelta_{\lambda\,\,\,\,\nu\alpha\beta\gamma}^{\,\,\,\,\lambda} & =0,\quad\myDelta_{\lambda\mu\nu\,\,\,\,\beta\gamma}^{\,\,\,\,\,\,\,\,\,\,\,\nu}=\frac{1}{3}\mydelta_{\lambda\mu\beta\gamma},\quad\myDelta_{\lambda\mu\nu}^{\,\,\,\,\,\,\,\,\,\,\,\lambda\mu\nu}=1,\quad\myDelta_{\lambda\mu\nu\rho\sigma\delta}\myDelta_{\alpha\beta\gamma}^{\rho\sigma\delta}=\myDelta_{\lambda\mu\nu\alpha\beta\gamma},\label{65}
\end{align}

The equations of dissipative hydrodynamics are derived by averaging Eqs.~\eqref{50}-\eqref{52} with respect to the nonequilibrium statistical operator and then substituting the results into Eqs.~\eqref{1},~\eqref{2}, and~\eqref{3}:
\begin{align}
&Dn_{a}+n_{a}\theta+\partial_{\mu}j_{a}^{\mu}=0,\label{66}\\
&D\epsilon+(\epsilon+p+\Pi)\theta+\partial_{\mu}h^{\mu}-h^{\mu}Du_{\mu}-\pi^{\mu\nu}\sigma_{\mu\nu}+\partial_{\mu}q^{\mu}+q^{\mu}Du_{\mu}-\phi^{\mu\nu}\partial_{\mu}u_{\nu}=0,\label{67}\\
&(\epsilon+p+\Pi)Du_{\alpha}-\nabla_{\alpha}(p+\Pi)+\Delta_{\alpha\mu}Dh^{\mu}+h^{\mu}\partial_{\mu}u_{\alpha}+h_{\alpha}\theta+\Delta_{\alpha\nu}\partial_{\mu}\pi^{\mu\nu}+q^{\mu}\partial_{\mu}u_{\alpha}-q_{\alpha}\theta-\Delta_{\alpha\nu}Dq^{\nu}+\Delta_{\alpha\nu}\partial_{\mu}\phi^{\mu\nu}=0,\label{68}\\
&DS^{\mu\nu}+\theta S^{\mu\nu}+u^{\mu}\partial_{\lambda}S^{\nu\lambda}+S^{\nu\lambda}\partial_{\lambda}u^{\mu}+u^{\nu}\partial_{\lambda}S^{\lambda\mu}+S^{\lambda\mu}\partial_{\lambda}u^{\nu}+\partial_{\lambda}\varpi^{\lambda\mu\nu}+2q^{\mu}u^{\nu}-2q^{\nu}u^{\mu}+2\phi^{\mu\nu}=0,\label{69}
\end{align}
where $n_{a}\equiv\langle\hat{n}_{a}\rangle$, $j_{a}^{\mu}\equiv\langle\hat{j}_{a}^{\mu}\rangle$, $\epsilon\equiv\langle\hat{\epsilon}\rangle$, $h^{\mu}\equiv\langle\hat{h}^{\mu}\rangle$, $\pi^{\mu\nu}\equiv\langle\hat{\pi}^{\mu\nu}\rangle$, $q^{\mu}\equiv\langle\hat{q}^{\mu}\rangle$, $\phi^{\mu\nu}\equiv\langle\hat{\phi}^{\mu\nu}\rangle$, $S^{\mu\nu}\equiv\langle\hat{S}^{\mu\nu}\rangle$, and $\varpi^{\lambda\mu\nu}\equiv\langle\hat{\varpi}^{\lambda\mu\nu}\rangle$ are the statistical averages of the corresponding operators; $p_{eq}\equiv p\left(\epsilon,n_{a},S^{\alpha\beta}\right)$ is the local equilibrium pressure, given by the equation of state (EoS), whereas $\Pi$ is the nonequilibrium pressure. We employ the comoving derivative $D\equiv u^{\mu}\partial_{\mu}$, the covariant spatial derivative $\nabla_\alpha \equiv \Delta_{\alpha \beta} \partial^\beta$, the shear tensor $\sigma_{\mu \nu} \equiv \Delta_{\mu \nu}^{\alpha \beta} \partial_\alpha u_\beta$, and the expansion scalar $\theta\equiv\partial_{\mu}u^{\mu}$. The expansion scalar quantifies the rate of fluid expansion or contraction, depending on whether $\theta$ is positive or negative. The decomposition $\partial_{\mu}u_{\nu}=u_{\mu}Du_{\nu}+\frac{1}{3}\theta\Delta_{\mu\nu}+\sigma_{\mu\nu}+\nabla_{[\mu}u_{\nu]}$ is utilized. Equations ~\eqref{67} and \eqref{68} are derived by contracting Eq.~\eqref{2} with $u_\nu$ and $\Delta_{\nu\alpha}$, respectively.

In dissipative hydrodynamics, the selection of a suitable reference frame is crucial due to the inherent difficulties in defining flow velocity. The energy-momentum tensor $T^{\mu\nu}$ contains 16 independent components in four-dimensional spacetime. The components of dissipative hydrodynamics include $\epsilon,p,u^{\mu},h^{\mu},\pi^{\mu\nu},\Pi,q^{\mu}$, and $\phi^{\mu\nu}$. The equation of state establishes a relationship between $\epsilon$ and $p$, thereby reducing their degrees of freedom to one. The vectors $u^\mu,h^\mu$, and $q^\mu$ each possess three independent components, constrained by the conditions $u^{\mu}u_{\mu}=1,h^{\mu}u_{\mu}=0$, and $q^\mu u_\mu=0$. The symmetric and traceless tensor $\pi^{\mu\nu}$ has five degrees of freedom, while the antisymmetric tensor $\phi^{\mu\nu}$ contributes an additional three. The scalar $\Pi$ has one degree of freedom. Thus, $T^{\mu\nu}$ possesses 19 independent components, exceeding the expected 16. This overcounting necessitates the elimination of three degrees of freedom, which can be achieved through a choice of reference frame or a specific definition for $u^\mu$. In standard (spinless) hydrodynamics, the Landau frame—defined by $T^{\mu\nu}u_{\nu}=\epsilon u^{\mu}$—is a natural choice, resulting in a symmetric energy-momentum tensor with $h^\mu=0$. However, the presence of an antisymmetric component in the energy-momentum tensor of spin fluids introduces certain ambiguities. Two approaches can be considered: (i) applying the Landau frame condition solely to the symmetric part of $T^{\mu\nu}$, resulting in $h^\mu=0$; or (ii) extending the Landau frame condition to encompass the entire $T^{\mu\nu}$, leading to $h^\mu+q^\mu=0$. The latter approach permits nonzero values for $h^\mu$ and $q^\mu$ while still satisfying the Landau condition. In this work, we prioritize generality by refraining from adopting a specific frame choice unless explicitly necessitated.

The ensemble averages of the dissipative operators with respect to the local-equilibrium distribution are zero, as demonstrated in Ref.~\cite{zubarev1979derivation}:
\begin{equation}
\langle\hat{h}^{\mu}\rangle_{l}=0,\quad\langle\hat{\pi}^{\mu\nu}\rangle_{l}=0,\quad\langle\hat{q}^{\mu}\rangle_{l}=0,\quad\langle\hat{\phi}^{\mu\nu}\rangle_{l}=0,\quad\langle\hat{j}_{a}^{\mu}\rangle_{l}=0,\quad\langle\hat{\varpi}^{\lambda\mu\nu}\rangle_{l}=0.
\label{70}
\end{equation}
Indeed, the distributions defined in Eqs.~\eqref{16} and \eqref{17} depend exclusively on the scalar operators $\hat{\epsilon}$, $\hat{n}_a$, and the tensor operator $\hat{S}^{\alpha\beta}$. According to Curie's theorem~\cite{zubarev1979derivation,Hosoya:1983id,de2013non,de1963non}, these operators are uncorrelated with quantities of different rank and parity. Consequently, averaging Eqs.~\eqref{50}-\eqref{52} over these distributions and substituting the results into Eq.~\eqref{1}, \eqref{2}, and \eqref{3} yields the equations of ideal spin hydrodynamics, as shown below:
\begin{eqnarray}
\begin{aligned}
	&Dn_{a}+n_{a}\theta=0,\\
	&D\epsilon+\left(\epsilon+p\right)\theta=0,\\
	&DS^{\mu\nu}+\theta S^{\mu\nu}+u^{\mu}\partial_{\lambda}S^{\nu\lambda}+S^{\nu\lambda}\partial_{\lambda}u^{\mu}+u^{\nu}\partial_{\lambda}S^{\lambda\mu}+S^{\lambda\mu}\partial_{\lambda}u^{\nu}=0,\\
	&\left(\epsilon+p\right)Du_{\alpha}=\nabla_{\alpha}p.
\end{aligned}
\label{71}
\end{eqnarray}
The first three equations in Eq.~\eqref{71} represent the covariant formulations of charge, energy, and spin conservation, respectively. The fourth equation corresponds to the relativistic Euler equation. Notably, the rest-mass density is replaced by the enthalpy density $w$, which serves as the appropriate inertial measure for relativistic fluids.

\section{Second-order spin hydrodynamics with $\omega_{\mu\nu}\sim\mathcal{O}\left(\partial^{0}\right)$}
\label{section3}

To construct a spin hydrodynamic theory for dissipative fluids, we systematically expand the nonequilibrium statistical operator derived in the previous section using a perturbative approach. As previously shown, the zeroth-order approximation leads to the equations of ideal hydrodynamics as encapsulated in Eq.~\eqref{71}. In this section, we revisit the derivation of relativistic first-order and second-order spin hydrodynamics. Adopting the power counting scheme introduced in our previous work~\cite{She:2021lhe}, where $\omega_{\mu\nu}\sim\mathcal{O}\left(\partial^{0}\right)$, we employ a first-order and second-order expansion of the statistical operator in terms of thermodynamic forces to refine our analysis.

\subsection{Decomposition into different dissipative processes}

To facilitate subsequent calculations, we decompose the operator $\hat{C}$, as defined in Eq.~\eqref{42}, into its constituent dissipative quantities, which appear in Eqs.~\eqref{50}-\eqref{52}. By utilizing Eqs.~\eqref{53} and \eqref{56}, we derive
\begin{equation}
\begin{aligned}
	\hat{C}= & \hat{\epsilon}D\beta-\hat{p}\beta\theta-\sum_{a}\hat{n}_{a}D\alpha_{a}-\frac{1}{2}\hat{S}^{\alpha\beta}D\Omega_{\alpha\beta}+\hat{h}^{\mu}\left(\beta Du_{\mu}+\partial_{\mu}\beta\right)-\sum_{a}\hat{j}_{a}^{\mu}\partial_{\mu}\alpha_{a}\\
	& +\hat{q}^{\mu}\left(\partial_{\mu}\beta-\beta Du_{\mu}\right)+\beta\hat{\pi}^{\mu\nu}\partial_{\mu}u_{\nu}-u^{\alpha}\hat{S}^{\beta\lambda}\partial_{\lambda}\Omega_{\alpha\beta}+\hat{\phi}^{\mu\nu}\left(\beta\partial_{\mu}u_{\nu}+\Omega_{\mu\nu}\right)-\frac{1}{2}\hat{\varpi}^{\lambda\alpha\beta}\partial_{\lambda}\Omega_{\alpha\beta}.
\end{aligned}
\label{72}
\end{equation}
For convenience, the thermodynamic relations~\eqref{28},\eqref{31}, and \eqref{32} are expressed as
\begin{eqnarray}
	ds=\beta d\epsilon-\sum_{a}\alpha_{a}dn_{a}-\frac{1}{2}\Omega_{\alpha\beta}dS^{\alpha\beta},\quad\beta dp=-w d\beta+\sum_{a}n_{a}d\alpha_{a}+\frac{1}{2}S^{\alpha\beta}d\Omega_{\alpha\beta}.
	\label{73}
\end{eqnarray}
The first equation yields the Maxwell relations:
\begin{eqnarray}
	\begin{aligned}
		& \frac{\partial\beta}{\partial n_{a}}\bigg|_{\epsilon,n_{b}\neq n_{a},S^{\alpha\beta}}=-\frac{\partial\alpha_{a}}{\partial\epsilon}\bigg|_{n_{b},S^{\alpha\beta}},\quad\frac{\partial\alpha_{c}}{\partial n_{a}}\bigg|_{\epsilon,n_{b}\neq n_{a},S^{\alpha\beta}}=\frac{\partial\alpha_{a}}{\partial n_{c}}\bigg|_{\epsilon,n_{b}\neq n_{c},S^{\alpha\beta}},\\
		& \frac{\partial\beta}{\partial S^{\alpha\beta}}\bigg|_{\epsilon,n_{b}}=-\frac{1}{2}\frac{\partial\Omega_{\alpha\beta}}{\partial\epsilon}\bigg|_{n_{b},S^{\alpha\beta}},\quad\frac{1}{2}\frac{\partial\Omega_{\alpha\beta}}{\partial n_{a}}\bigg|_{\epsilon,n_{b}\neq n_{a},S^{\alpha\beta}}=\frac{\partial\alpha_{a}}{\partial S^{\alpha\beta}}\bigg|_{\epsilon,n_{b}},\\
		& \frac{\partial\Omega_{\alpha\beta}}{\partial S^{\lambda\rho}}\bigg|_{\epsilon,n_{b},S^{\alpha\beta}\neq S^{\lambda\rho}}=\frac{\partial\Omega_{\lambda\rho}}{\partial S^{\alpha\beta}}\bigg|_{\epsilon,n_{b},S^{\lambda\rho}\neq S^{\alpha\beta}}.
	\end{aligned}
	\label{74}
\end{eqnarray}
The second equation immediately yields
\begin{eqnarray}
	w=-\beta\frac{\partial p}{\partial\beta}\bigg|_{\alpha_{a},\Omega_{\alpha\beta}},\quad n_{a}=\beta\frac{\partial p}{\partial\alpha_{a}}\bigg|_{\beta,\alpha_{b}\neq\alpha_{a},\Omega_{\alpha\beta}},\quad S^{\alpha\beta}=2\beta\frac{\partial p}{\partial\Omega_{\alpha\beta}}\bigg|_{\beta,\alpha_{a}}.
	\label{75}
\end{eqnarray}

We now employ the full dissipative hydrodynamic equations~\eqref{66}-\eqref{69} to eliminate the terms $D\beta,D\alpha_{a}$, and $D\Omega_{\alpha\beta}$ in Eq.~\eqref{72}. By employing $\epsilon$, $n_a$, and $S^{\alpha\beta}$ as independent thermodynamic variables, we obtain
\begin{eqnarray}
\begin{aligned}
	D\beta= & \beta\theta\Gamma-\biggl(\Pi\theta+\partial_{\mu}h^{\mu}-h^{\mu}Du_{\mu}-\pi^{\mu\nu}\sigma_{\mu\nu}+\partial_{\mu}q^{\mu}+q^{\mu}Du_{\mu}-\phi^{\mu\nu}\partial_{\mu}u_{\nu}\biggr)\frac{\partial\beta}{\partial\epsilon}\bigg|_{n_{a},S^{\alpha\beta}}-\sum_{a}\partial_{\mu}j_{a}^{\mu}\frac{\partial\beta}{\partial n_{a}}\bigg|_{\epsilon,n_{b}\neq n_{a},S^{\alpha\beta}}\\
	&-\left(u^{\alpha}\partial_{\lambda}S^{\beta\lambda}+S^{\beta\lambda}\partial_{\lambda}u^{\alpha}+u^{\beta}\partial_{\lambda}S^{\lambda\alpha}+S^{\lambda\alpha}\partial_{\lambda}u^{\beta}+\partial_{\lambda}\varpi^{\lambda\alpha\beta}+2q^{\alpha}u^{\beta}-2q^{\beta}u^{\alpha}+2\phi^{\alpha\beta}\right)\frac{\partial\beta}{\partial S^{\alpha\beta}}\bigg|_{\epsilon,n_{a}},
\end{aligned}
\label{76}
\end{eqnarray}
\begin{eqnarray}
\begin{aligned}
	D\alpha_{c}= & -\beta\theta\delta_{c}-\biggl(\Pi\theta+\partial_{\mu}h^{\mu}-h^{\mu}Du_{\mu}-\pi^{\mu\nu}\sigma_{\mu\nu}+\partial_{\mu}q^{\mu}+q^{\mu}Du_{\mu}-\phi^{\mu\nu}\partial_{\mu}u_{\nu}\biggr)\frac{\partial\alpha_{c}}{\partial\epsilon}\bigg|_{n_{a},S^{\alpha\beta}}-\sum_{a}\partial_{\mu}j_{a}^{\mu}\frac{\partial\alpha_{c}}{\partial n_{a}}\bigg|_{\epsilon,n_{b}\neq n_{a},S^{\alpha\beta}}\\
	&-\left(u^{\alpha}\partial_{\lambda}S^{\beta\lambda}+S^{\beta\lambda}\partial_{\lambda}u^{\alpha}+u^{\beta}\partial_{\lambda}S^{\lambda\alpha}+S^{\lambda\alpha}\partial_{\lambda}u^{\beta}+\partial_{\lambda}\varpi^{\lambda\alpha\beta}+2q^{\alpha}u^{\beta}-2q^{\beta}u^{\alpha}+2\phi^{\alpha\beta}\right)\frac{\partial\alpha_{c}}{\partial S^{\alpha\beta}}\bigg|_{\epsilon,n_{a}},
\end{aligned}
\label{77}
\end{eqnarray}
and
\begin{eqnarray}
\begin{aligned}
	D\Omega_{\alpha\beta}= & -2\beta\theta\mathcal{K}_{\alpha\beta}-\biggl(\Pi\theta+\partial_{\mu}h^{\mu}-h^{\mu}Du_{\mu}-\pi^{\mu\nu}\sigma_{\mu\nu}+\partial_{\mu}q^{\mu}+q^{\mu}Du_{\mu}-\phi^{\mu\nu}\partial_{\mu}u_{\nu}\biggr)\frac{\partial\Omega_{\alpha\beta}}{\partial\epsilon}\bigg|_{n_{a},S^{\alpha\beta}}-\sum_{a}\partial_{\mu}j_{a}^{\mu}\frac{\partial\Omega_{\alpha\beta}}{\partial n_{a}}\bigg|_{\epsilon,n_{b}\neq n_{a},S^{\alpha\beta}}\\
	&-\left(u^{\lambda}\partial_{\delta}S^{\rho\delta}+S^{\rho\delta}\partial_{\delta}u^{\lambda}+u^{\rho}\partial_{\delta}S^{\delta\lambda}+S^{\delta\lambda}\partial_{\delta}u^{\rho}+\partial_{\delta}\varpi^{\delta\lambda\rho}+2q^{\lambda}u^{\rho}-2q^{\rho}u^{\lambda}+2\phi^{\lambda\rho}\right)\frac{\partial\Omega_{\alpha\beta}}{\partial S^{\lambda\rho}}\bigg|_{\epsilon,n_{a}},
\end{aligned}
\label{78}
\end{eqnarray}
where
\begin{eqnarray}
	\Gamma\equiv\frac{\partial p}{\partial\epsilon}\bigg|_{n_{a},S^{\alpha\beta}},\quad\delta_{c}\equiv\frac{\partial p}{\partial n_{c}}\bigg|_{\epsilon,n_{b}\neq n_{c},S^{\alpha\beta}},\quad\mathcal{K}_{\alpha\beta}\equiv\frac{\partial p}{\partial S^{\alpha\beta}}\bigg|_{\epsilon,n_{a}}.
	\label{79}
\end{eqnarray}

The first four terms in Eq.~\eqref{72} can be consolidated to yield a more compact expression:
\begin{eqnarray}
\begin{aligned} 
	&\hat{\epsilon}D\beta-\hat{p}\beta\theta-\sum_{c}\hat{n}_{c}D\alpha_{c}-\frac{1}{2}\hat{S}^{\alpha\beta}D\Omega_{\alpha\beta}\\
	=&-\beta\theta\hat{P}^{*}-\hat{\mathcal{A}}^{*}\left(\Pi\theta+\partial_{\mu}h^{\mu}-h^{\mu}Du_{\mu}-\pi^{\mu\nu}\sigma_{\mu\nu}+\partial_{\mu}q^{\mu}+q^{\mu}Du_{\mu}-\phi^{\mu\nu}\partial_{\mu}u_{\nu}\right)+\sum_{a}\hat{\mathcal{B}}_{a}^{*}\partial_{\mu}j_{a}^{\mu}\\
	&+\frac{1}{2}\left(u^{\alpha}\partial_{\lambda}S^{\beta\lambda}+S^{\beta\lambda}\partial_{\lambda}u^{\alpha}+u^{\beta}\partial_{\lambda}S^{\lambda\alpha}+S^{\lambda\alpha}\partial_{\lambda}u^{\beta}+\partial_{\lambda}\varpi^{\lambda\alpha\beta}+2q^{\alpha}u^{\beta}-2q^{\beta}u^{\alpha}+2\phi^{\alpha\beta}\right)\hat{\mathcal{C}}_{\alpha\beta}^{*},
\end{aligned}
\label{80}
\end{eqnarray}
where we define new operators
\begin{align} 
	&\hat{P}^{*}=\left(\hat{p}-\Gamma\hat{\epsilon}-\sum_{a}\delta_{a}\hat{n}_{a}\right)-\mathcal{K}_{\alpha\beta}\hat{S}^{\alpha\beta}=\hat{p}^{*}-\mathcal{K}_{\alpha\beta}\hat{S}^{\alpha\beta},\label{81}\\
	&\hat{\mathcal{A}}^{*}=\hat{\epsilon}\frac{\partial\beta}{\partial\epsilon}\bigg|_{n_{a},S^{\alpha\beta}}+\sum_{a}\hat{n}_{a}\frac{\partial\beta}{\partial n_{a}}\bigg|_{\epsilon,n_{b}\neq n_{a},S^{\alpha\beta}}+\hat{S}^{\alpha\beta}\frac{\partial\beta}{\partial S^{\alpha\beta}}\bigg|_{\epsilon,n_{b}}=\sum_{i}\hat{\mathfrak{D}}_{i}\partial_{\epsilon n}^{i}\beta+\hat{S}^{\alpha\beta}\mathcal{D}_{\alpha\beta},\label{82}\\
	&\hat{\mathcal{B}}_{a}^{*}=\hat{\epsilon}\frac{\partial\alpha_{a}}{\partial\epsilon}\bigg|_{n_{b},S^{\alpha\beta}}+\sum_{c}\hat{n}_{c}\frac{\partial\alpha_{a}}{\partial n_{c}}\bigg|_{\epsilon,n_{b}\neq n_{c},S^{\alpha\beta}}+\hat{S}^{\alpha\beta}\frac{\partial\alpha_{a}}{\partial S^{\alpha\beta}}\bigg|_{\epsilon,n_{b}}=\sum_{i}\hat{\mathfrak{D}}_{i}\partial_{\epsilon n}^{i}\alpha_{a}+\hat{S}^{\alpha\beta}\mathcal{E}_{\alpha\beta}^{a},\label{83}\\
	&\hat{\mathcal{C}}_{\alpha\beta}^{*}=\hat{\epsilon}\frac{\partial\Omega_{\alpha\beta}}{\partial\epsilon}\bigg|_{n_{b},S^{\alpha\beta}}+\sum_{c}\hat{n}_{c}\frac{\partial\Omega_{\alpha\beta}}{\partial n_{c}}\bigg|_{\epsilon,n_{b}\neq n_{c},S^{\alpha\beta}}+\hat{S}^{\delta\rho}\frac{\partial\Omega_{\alpha\beta}}{\partial S^{\delta\rho}}\bigg|_{\epsilon,n_{b},S^{\alpha\beta}\neq S^{\delta\rho}}=\sum_{i}\hat{\mathfrak{D}}_{i}\partial_{\epsilon n}^{i}\Omega_{\alpha\beta}+\hat{S}^{\delta\rho}\mathcal{F}_{\alpha\beta\delta\rho},\label{84}
\end{align}
and
\begin{eqnarray}
	\begin{aligned}
		\mathcal{D}_{\alpha\beta}=&\frac{\partial\beta}{\partial S^{\alpha\beta}}\bigg|_{\epsilon,n_{b}},\mathcal{E}_{\alpha\beta}^{a}=\frac{\partial\alpha_{a}}{\partial S^{\alpha\beta}}\bigg|_{\epsilon,n_{b}},\mathcal{F}_{\alpha\beta\delta\rho}=\frac{\partial\Omega_{\alpha\beta}}{\partial S^{\delta\rho}}\bigg|_{\epsilon,n_{b},S^{\alpha\beta}\neq S^{\delta\rho}},\\
		\hat{\mathfrak{D}}_{i}=&\left(\hat{\epsilon},\hat{n}_{1},\hat{n}_{2},\cdots,\hat{n}_{l}\right),\partial_{\epsilon n}^{i}=\left(\frac{\partial}{\partial\epsilon},\frac{\partial}{\partial n_{1}},\frac{\partial}{\partial n_{2}},\cdots,\frac{\partial}{\partial n_{l}}\right),\sum_{i}=\sum_{i=0}^{l}=\sum_{i=0}+\sum_{a=1}^{l}.
	\end{aligned}
\label{85}
\end{eqnarray}
In our notation, the index $i=0$ corresponds to the first element, denoted by $\hat{\mathfrak{D}}_{0}=\hat{\epsilon},\partial_{\epsilon n}^{0}=\frac{\partial}{\partial\epsilon}$, while other indices correspond to the subsequent elements in the sequence.

Next, we employ Eq.~\eqref{68} in conjunction with the modified pressure gradient derived from the second relation in Eq.~\eqref{73} to obtain the following:
\begin{eqnarray}
\begin{aligned}
wDu_{\sigma} =&-\Pi Du_{\sigma}-wT\nabla_{\sigma}\beta+T\sum_{a}n_{a}\nabla_{\sigma}\alpha_{a}+\frac{1}{2}TS^{\alpha\beta}\nabla_{\sigma}\Omega_{\alpha\beta}+\nabla_{\sigma}\Pi-\Delta_{\sigma\mu}Dh^{\mu}\\
&-h^{\mu}\partial_{\mu}u_{\sigma}-h_{\sigma}\theta-\Delta_{\sigma\nu}\partial_{\mu}\pi^{\mu\nu}-q^{\mu}\partial_{\mu}u_{\sigma}+q_{\sigma}\theta+\Delta_{\sigma\nu}Dq^{\nu}-\Delta_{\sigma\nu}\partial_{\mu}\phi^{\mu\nu},
\end{aligned}
\label{86}
\end{eqnarray}
which is subsequently utilized to modify the vector term containing $\hat{h}^\mu$ and $\hat{q}^\mu$ in Eq.~\eqref{72} using the identities $\hat{h}^{\sigma}\partial_{\sigma}=\hat{h}^{\sigma}\nabla_{\sigma}$ and $\hat{q}^{\sigma}\partial_{\sigma}=\hat{q}^{\sigma}\nabla_{\sigma}$,
\begin{eqnarray}
\begin{aligned}
\hat{h}^{\sigma}\left(\beta Du_{\sigma}+\partial_{\sigma}\beta\right)= & \sum_{a}n_{a}w^{-1}\hat{h}^{\sigma}\nabla_{\sigma}\alpha_{a}-\hat{h}^{\sigma}\beta w^{-1}\biggl(-\frac{1}{2}\beta^{-1}S^{\alpha\beta}\nabla_{\sigma}\Omega_{\alpha\beta}-\nabla_{\sigma}\Pi+\Pi Du_{\sigma}+\Delta_{\sigma\nu}Dh^{\nu}\\
&+h^{\mu}\partial_{\mu}u_{\sigma}+h_{\sigma}\theta+\Delta_{\sigma\nu}\partial_{\mu}\pi^{\mu\nu}+q^{\mu}\partial_{\mu}u_{\sigma}-q_{\sigma}\theta-\Delta_{\sigma\nu}Dq^{\nu}+\Delta_{\sigma\nu}\partial_{\mu}\phi^{\mu\nu}\biggr),
\end{aligned}
\label{87}
\end{eqnarray}
and
\begin{eqnarray}
\begin{aligned}
\hat{q}^{\mu}\left(\partial_{\mu}\beta-\beta Du_{\mu}\right)= & -\beta\hat{q}^{\mu}\left(w^{-1}\nabla_{\mu}p+\beta\nabla_{\mu}T\right)+\hat{q}^{\mu}\beta w^{-1}\Bigl(\Pi Du_{\mu}-\nabla_{\mu}\Pi+\Delta_{\mu\nu}Dh^{\nu}\\
 & +h^{\nu}\partial_{\nu}u_{\mu}+h_{\mu}\theta+\Delta_{\mu\nu}\partial_{\rho}\pi^{\rho\nu}+q^{\nu}\partial_{\nu}u^{\mu}-q_{\mu}\theta-\Delta_{\mu\nu}Dq^{\nu}+\Delta_{\mu\nu}\partial_{\rho}\phi^{\rho\nu}\Bigr).
\end{aligned}
\label{88}
\end{eqnarray}

Upon combining Eqs.~\eqref{72},\eqref{80},\eqref{87}, and~\eqref{88}, we arrive at the final expression for the operator $\hat{C}$ accurate to second order in gradients:
\begin{eqnarray}
	\hat{C}\left(x\right)=\hat{C}_{1}\left(x\right)+\hat{C}_{2}\left(x\right),
	\label{89}
\end{eqnarray}
where $\hat{C}_1$ and $\hat{C}_2$ represent the first- and second-order contributions, respectively:
\begin{align}
	\hat{C}_{1}\left(x\right)= & -\beta\theta\hat{p}^{*}+\beta\mathcal{R}_{\alpha\beta}\hat{S}^{\alpha\beta}+\beta\sum_{i}\left(\hat{\mathfrak{D}}_{i}\partial_{\epsilon n}^{i}\Omega_{\mu\nu}\right)\mathcal{Z}^{\mu\nu}-\sum_{a}\hat{\mathscr{J}}_{a}^{\sigma}\nabla_{\sigma}\alpha_{a}\nonumber\\
	&+\beta\hat{h}^{\sigma}N_{\sigma}+\beta\hat{q}^{\mu}M_{\mu}+\beta\hat{\pi}^{\mu\nu}\sigma_{\mu\nu}+\beta\hat{\phi}^{\mu\nu}\xi_{\mu\nu}+\hat{\varpi}^{\lambda\alpha\beta}\varXi_{\lambda\alpha\beta},\label{90}\\
	\hat{C}_{2}\left(x\right)= & \beta\sum_{i}\left[\left(\hat{\mathfrak{D}}_{i}\partial_{\epsilon n}^{i}\beta\right)\mathcal{X}+\sum_{a}\left(\hat{\mathfrak{D}}_{i}\partial_{\epsilon n}^{i}\alpha_{a}\right)\mathcal{Y}_{a}+\left(\hat{\mathfrak{D}}_{i}\partial_{\epsilon n}^{i}\Omega_{\alpha\beta}\right)\mathcal{G}^{\alpha\beta}\right]\nonumber\\
	&+\beta\hat{S}^{\alpha\beta}\mathcal{T}_{\alpha\beta}+\beta\hat{h}^{\sigma}\mathcal{H}_{\sigma}+\beta\hat{q}^{\mu}\mathcal{Q}_{\mu},\label{91}
\end{align}
where we define
\begin{eqnarray*}
	\begin{aligned}
		\hat{\mathscr{J}}_{a}^{\sigma}= & \hat{j}_{a}^{\sigma}-\frac{n_{a}}{\epsilon+p}\hat{h}^{\sigma},M_{\mu}=-\left(w^{-1}\nabla_{\mu}p+\beta\nabla_{\mu}T\right),\xi_{\mu\nu}=\mydelta_{\mu\nu}^{\lambda\delta}\left(\partial_{\lambda}u_{\delta}+\beta^{-1}\Omega_{\lambda\delta}\right),\\
		\mathcal{Z}^{\alpha\beta}= & \beta^{-1}\left(u^{[\alpha}\partial_{\lambda}S^{\beta]\lambda}+S^{[\beta\lambda}\partial_{\lambda}u^{\alpha]}+q^{\alpha}u^{\beta}-q^{\beta}u^{\alpha}+\phi^{\alpha\beta}\right),\mathcal{W}_{\beta\lambda}=-\beta^{-1}\mydelta_{\beta\lambda}^{\rho\sigma}u^{\alpha}\partial_{\sigma}\Omega_{\alpha\rho}\\
		N_{\sigma}= & \frac{1}{2}\beta^{-1}{w}^{-1}S^{\alpha\beta}\nabla_{\sigma}\Omega_{\alpha\beta},\varXi_{\lambda\alpha\beta}=-\frac{1}{2}\myDelta_{\lambda\alpha\beta\rho\sigma\delta}\partial^{\rho}\Omega^{\sigma\delta},\mathcal{Y}_{a}=\beta^{-1}\partial_{\mu}j_{a}^{\mu},\mathcal{G}^{\alpha\beta}=\frac{1}{2}\beta^{-1}\partial_{\lambda}\varpi^{\lambda\alpha\beta},\sigma_{\mu\nu}=\Delta_{\mu\nu}^{\lambda\delta}\partial_{\lambda}u_{\delta},\\
		\mathcal{X}= & -\beta^{-1}\left(\Pi\theta+\partial_{\mu}h^{\mu}-h^{\mu}Du_{\mu}-\pi^{\mu\nu}\sigma_{\mu\nu}+\partial_{\mu}q^{\mu}+q^{\mu}Du_{\mu}-\phi^{\mu\nu}\partial_{\mu}u_{\nu}\right),\\
		\mathcal{H}_{\sigma}= & -{w}^{-1}\left(-\nabla_{\sigma}\Pi+\Pi Du_{\sigma}+\Delta_{\sigma\nu}Dh^{\nu}+h^{\mu}\partial_{\mu}u_{\sigma}+h_{\sigma}\theta+\Delta_{\sigma\nu}\partial_{\mu}\pi^{\mu\nu}+q^{\mu}\partial_{\mu}u_{\sigma}-q_{\sigma}\theta-\Delta_{\sigma\nu}Dq^{\nu}+\Delta_{\sigma\nu}\partial_{\mu}\phi^{\mu\nu}\right),\\
		\mathcal{Q}_{\mu}= & {w}^{-1}\left(\Pi Du_{\mu}-\nabla_{\mu}\Pi+\Delta_{\mu\nu}Dh^{\nu}+h^{\nu}\partial_{\nu}u_{\mu}+h_{\mu}\theta+\Delta_{\mu\nu}\partial_{\rho}\pi^{\rho\nu}+q^{\nu}\partial_{\nu}u^{\mu}-q_{\mu}\theta-\Delta_{\mu\nu}Dq^{\nu}+\Delta_{\mu\nu}\partial_{\rho}\phi^{\rho\nu}\right),\\
		\mathcal{R}_{\alpha\beta}= & \theta\mathcal{K}_{\alpha\beta}+\mathcal{W}_{\alpha\beta}+\mathcal{Z}^{\rho\sigma}\mathcal{F}_{\rho\sigma\alpha\beta},\mathcal{T}_{\alpha\beta}=\mathcal{D}_{\alpha\beta}\mathcal{X}+\sum_{a}\mathcal{E}_{\alpha\beta}^{a}\mathcal{Y}_{a}+\mathcal{F}_{\rho\sigma\alpha\beta}\mathcal{G}^{\rho\sigma}.
	\end{aligned}
\end{eqnarray*}
Please note that, to avoid confusion in the subsequent analysis, we assume that the quantity $\xi_{\mu\nu}$ is of first order in both cases of power counting in this work. As indicated by Eq.~\eqref{90}, the operator $\hat{C}_1$ exhibits a linear dependence on the thermodynamic forces $\theta$, $\mathcal{R}^{\alpha\beta}$, $\mathcal{Z}^{\alpha\beta}$, $\nabla_{\sigma}\alpha_{a}$, $N_{\sigma}$, $M_{\mu}$, $\sigma_{\mu\nu}$, $\xi_{\mu\nu}$, and $\varXi_{\lambda\alpha\beta}$, which respectively correspond to bulk viscosity pressure, charge-diffusion currents, boost heat vector, shear stress tensor, rotational stress tensor, and spin tensor-related dissipative flux. The operator $\hat{C}_2$ comprises scalar, vector, and second-order antisymmetric tensor components, contributing to both bulk viscous pressure, diffusion currents, rotational stress tensor, and boost heat vector. This arises from the dissipative nature of the hydrodynamic equations [Eqs.~\eqref{66}-\eqref{69}] from which $\hat{C}_2$ is derived. The parenthetical expressions in Eq.~\eqref{91} can be naturally interpreted as generalized or extended thermodynamic forces, encompassing either spacetime derivatives of dissipative currents or their products with conventional thermodynamic forces. In the following, we derive the linear Navier-Stokes relations connecting these thermodynamic forces to the dissipative currents.

Leveraging Eq.~\eqref{46} for the statistical average of an arbitrary operator $\hat{X}\left(x\right)$, we expand the average up to second order as follows:
\begin{eqnarray}
	\langle\hat{X}\left(x\right)\rangle=\langle\hat{X}\left(x\right)\rangle_{l}+\langle\hat{X}\left(x\right)\rangle_{1}+\langle\hat{X}\left(x\right)\rangle_{2}.
	\label{92}
\end{eqnarray}
The first-order correction is given by
\begin{eqnarray}
	\langle \hat{X}\left(x\right)\rangle _{1}=\int d^{4}x_{1}\left(\hat{X}\left(x\right),\hat{C}_{1}\left(x_{1}\right)\right)\bigg|_{\text{loc}},
	\label{93}
\end{eqnarray}
where the subscript ``loc" denotes a local approximation of the thermodynamic forces within the integrand at point $x$, effectively neglecting nonlocal contributions.

The second-order correction, $\langle\hat{X}\left(x\right)\rangle_{2}$, can be decomposed into three contributions:
\begin{eqnarray}
	\langle\hat{X}\left(x\right)\rangle_{2}=\langle\hat{X}\left(x\right)\rangle_{2}^{1}+\langle\hat{X}\left(x\right)\rangle_{2}^{2}+\langle\hat{X}\left(x\right)\rangle_{2}^{3},
	\label{94}
\end{eqnarray}
where
\begin{align}
	& \langle\hat{X}\left(x\right)\rangle_{2}^{1}=\int d^{4}x_{1}\left(\hat{X}\left(x\right),\hat{C}_{1}\left(x_{1}\right)\right)-\langle\hat{X}\left(x\right)\rangle_{1},\label{95}\\
	& \langle\hat{X}\left(x\right)\rangle_{2}^{2}=\int d^{4}x_{1}\left(\hat{X}\left(x\right),\hat{C}_{2}\left(x_{1}\right)\right),\label{96}\\
	& \langle\hat{X}\left(x\right)\rangle_{2}^{3}=\int d^{4}x_{1}d^{4}x_{2}\left(\hat{X}\left(x\right),\hat{C}_{1}\left(x_{1}\right),\hat{C}_{1}\left(x_{2}\right)\right).\label{97}
\end{align}
The first term in Eq.~\eqref{94} accounts for corrections arising from the nonlocality of thermodynamic forces in the two-point correlators, reflecting the dependence of the dissipative currents and thermodynamic forces on the space-time separation between points $x$ and $x_1$. These corrections are inherently second order due to their dependence on this difference, as is evident in Eqs.~\eqref{90}, \eqref{93}, and \eqref{95}. The second term in Eq.~\eqref{94} incorporates corrections arising from generalized thermodynamic forces, which originate from the second-order corrections to be operator $\hat{C}$ in the two-point correlators. This third term in Eq.~\eqref{94} accounts for nonlinear (quadratic) contributions from the thermodynamic forces, arising from the three-point correlators that are quadratic in the operator $\hat{C}$.

\subsection{First-order spin hydrodynamics}

Given Curie's theorem, which states that there are no correlators between operators of different rank and spatial parity in isotropic systems~\cite{de1963non,de2013non}, we conclude from Eqs.~\eqref{90} and \eqref{93} that the shear-stress tensor, to leading order, is
\begin{equation}
\langle\hat{\pi}_{\mu\nu}\left(x\right)\rangle_{1}=\int d^{4}x_{1}\left(\hat{\pi}_{\mu\nu}\left(x\right),\hat{\pi}_{\rho\sigma}\left(x_{1}\right)\right)\beta\left(x_{1}\right)\sigma^{\rho\sigma}\left(x_{1}\right),
\label{98}
\end{equation}
where we used $\left(\hat{\pi}_{\mu\nu}\left(x\right),\hat{S}_{\rho\sigma}\left(x_{1}\right)\right)=0$ and $\left(\hat{\pi}_{\mu\nu}\left(x\right),\hat{\phi}_{\rho\sigma}\left(x_{1}\right)\right)=0$. The integrand in Eq.~\eqref{98} is dominated by contributions from the region $|x_{1}-x|\lesssim\lambda$, where $\lambda$ represents the characteristic microscopic length scale governing the decay of the shear-stress correlation function. This correlation length, equivalent to the mean free path between particle collisions, defines the interaction range. In contrast, thermodynamic parameters and fluid velocity vary over a macroscopic length scale, $L\gg\lambda$. The Knudsen number, $\text{Kn}=\lambda/L\ll1$, measures this difference. Corrections to ideal hydrodynamics can be systematically expanded in powers of Knudsen number (Kn). First-order dissipative effects are captured by terms linear in Kn.

Since the thermodynamic force $|\sigma^{\rho\sigma}|\simeq|u^{\rho}|/L$, while the integrand in Eq.~\eqref{98} is non-negligible only within a region of extent $\sim\lambda$, we approximate $\beta\sigma^{\rho\sigma}$ as constant and take it outside the integral. This approximation, justified by the integral mean value theorem, involves evaluating $\beta\sigma^{\rho\sigma}$ at the spacetime point $x$~\cite{zubarev1979derivation,Hosoya:1983id,Huang:2011dc}. This leads to a local, linear relationship between the shear stress tensor and the shear tensor
\begin{eqnarray}
\pi_{\mu\nu}\left(x\right)\equiv\langle\hat{\pi}_{\mu\nu}\left(x\right)\rangle_{1}=\beta\left(x\right)\sigma^{\rho\sigma}\left(x\right)\int d^{4}x_{1}\left(\hat{\pi}_{\mu\nu}\left(x\right),\hat{\pi}_{\rho\sigma}\left(x_{1}\right)\right).
\label{99}
\end{eqnarray}

The bulk viscous pressure, $\Pi$, is defined as the difference between the actual isotropic pressure, $\langle\hat{p}\rangle=\langle\hat{p}\rangle_{l}+\langle\hat{p}\rangle_{1}$, and its equilibrium value, $p\left(\epsilon,n_{a},S^{\alpha\beta}\right)$, as given by the equation of state. This deviation arises from fluid expansion or compression
\begin{eqnarray}
\Pi=\langle\hat{p}\rangle-p\left(\epsilon,n_{a},S^{\alpha\beta}\right)=\langle\hat{p}\rangle_{l}+\langle\hat{p}\rangle_{1}-p\left(\epsilon,n_{a},S^{\alpha\beta}\right).
\label{100}
\end{eqnarray}

We can decompose $\epsilon,n_a$, and $S^{\alpha\beta}$ as $\epsilon=\langle\hat{\epsilon}\rangle_{l}+\langle\hat{\epsilon}\rangle_{1},n_{a}=\langle\hat{n}_{a}\rangle_{l}+\langle\hat{n}_{a}\rangle_{1}$, and $S^{\alpha\beta}=\langle \hat{S}^{\alpha\beta}\rangle _{l}+\langle \hat{S}^{\alpha\beta}\rangle _{1}$, respectively. To first order in gradients,
\begin{eqnarray}
\begin{aligned}
\langle\hat{p}\rangle_{l}&= p\left(\langle\hat{\epsilon}\rangle_{l},\langle\hat{n}_{a}\rangle_{l},\langle\hat{S}^{\alpha\beta}\rangle_{l}\right)\\
&=p\left(\epsilon-\langle\hat{\epsilon}\rangle_{1},n_{a}-\langle\hat{n}_{a}\rangle_{1},S^{\alpha\beta}-\langle\hat{S}^{\alpha\beta}\rangle_{1}\right)\\
&=p\left(\epsilon,n_{a},S^{\alpha\beta}\right)-\Gamma\langle\hat{\epsilon}\rangle_{1}-\sum_{a}\delta_{a}\langle\hat{n}_{a}\rangle_{1}-\mathcal{K}_{\alpha\beta}\langle\hat{S}^{\alpha\beta}\rangle_{1},
\end{aligned}
\label{101}
\end{eqnarray}
where the coefficients $\Gamma,\delta_a$, and $\mathcal{K}_{\alpha\beta}$ are defined in Eq.\eqref{79}. Although the corrections $\langle\hat{\epsilon}\rangle_{1}$, $\langle\hat{n}_{a}\rangle_{1}$, and $\langle\hat{S}^{\alpha\beta}\rangle_{1}$ vanish when the matching conditions in Eq.~\eqref{13} are satisfied, we retain them for generality to ensure that the final expressions are independent of the specific matching conditions chosen. Substituting Eq.~\eqref{101} into Eq.\eqref{100} for the bulk viscous pressure, we obtain
\begin{eqnarray}
\Pi&=\langle\hat{p}-\Gamma\hat{\epsilon}-\Sigma_{a}\delta_{a}\hat{n}_{a}-\mathcal{K}_{\alpha\beta}\hat{S}^{\alpha\beta}\rangle_{1}=\langle\hat{P}^{*}\rangle_{1},
\label{102}
\end{eqnarray}
using the definition of $\hat{P}^{*}$ given in Eq.~\eqref{81}. Combining Eqs.~\eqref{93} and~\eqref{90}, and following a derivation similar to that of Eq.~\eqref{99}, we obtain the first-order correction to the bulk viscous pressure
\begin{equation}
\begin{aligned}
	\Pi\left(x\right)= & \langle\hat{P}^{*}\left(x\right)\rangle_{1}\\
	= & -\beta\left(x\right)\theta\left(x\right)\int d^{4}x_{1}\left(\hat{p}^{*}\left(x\right),\hat{p}^{*}\left(x_{1}\right)\right)+\sum_{i}\beta\left(x\right)\mathcal{Z}^{\mu\nu}\left(x\right)\partial_{\epsilon n}^{i}\Omega_{\mu\nu}\left(x\right)\int d^{4}x_{1}\left(\hat{p}^{*}\left(x\right),\hat{\mathfrak{D}}_{i}\left(x_{1}\right)\right)\\
	& -\beta\left(x\right)\mathcal{K}_{\alpha\beta}\left(x\right)\mathcal{R}_{\rho\sigma}\left(x\right)\int d^{4}x_{1}\left(\hat{S}^{\alpha\beta}\left(x\right),\hat{S}^{\rho\sigma}\left(x_{1}\right)\right)-\beta\left(x\right)\mathcal{K}_{\mu\nu}\left(x\right)\xi_{\rho\sigma}\left(x\right)\int d^{4}x_{1}\left(\hat{S}^{\mu\nu}\left(x\right),\hat{\phi}^{\rho\sigma}\left(x_{1}\right)\right).
\end{aligned}
\label{103}
\end{equation}

Applying Curie’s theorem, we can derive the remaining dissipative currents as follows:
\begin{eqnarray}
	\begin{aligned}\mathscr{J}_{a}^{\mu}\left(x\right)= & \langle\hat{\mathscr{J}}_{a}^{\mu}\left(x\right)\rangle_{1}\\
		= & -\sum_{b}\nabla_{\sigma}\alpha_{b}\left(x\right)\int d^{4}x_{1}\left(\hat{\mathscr{J}}_{a}^{\mu}\left(x\right),\hat{\mathscr{J}}_{b}^{\sigma}\left(x_{1}\right)\right)+\beta(x)N_{\sigma}\left(x\right)\int d^{4}x_{1}\left(\hat{\mathscr{J}}_{a}^{\mu}\left(x\right),\hat{h}^{\sigma}\left(x_{1}\right)\right)\\
		& +\beta\left(x\right)M_{\alpha}\left(x\right)\int d^{4}x_{1}\left(\hat{\mathscr{J}}_{a}^{\mu}\left(x\right),\hat{q}^{\alpha}\left(x_{1}\right)\right),
	\end{aligned}
	\label{104}
\end{eqnarray}
\begin{eqnarray}
	\begin{aligned}
		\phi^{\mu\nu}\left(x\right)= & \langle\hat{\phi}^{\mu\nu}\left(x\right)\rangle_{1}\\
		= & \beta\left(x\right)\mathcal{R}_{\alpha\beta}\left(x\right)\int d^{4}x_{1}\left(\hat{\phi}^{\mu\nu}\left(x\right),\hat{S}^{\alpha\beta}\left(x_{1}\right)\right)+\beta\left(x\right)\xi_{\rho\sigma}\left(x\right)\int d^{4}x_{1}\left(\hat{\phi}^{\mu\nu}\left(x\right),\hat{\phi}^{\rho\sigma}\left(x_{1}\right)\right),
	\end{aligned}
	\label{105}
\end{eqnarray}
\begin{eqnarray}
	\begin{aligned}
		q^{\mu}\left(x\right)= & \langle\hat{q}^{\mu}\left(x\right)\rangle_{1}\\
		= & -\sum_{a}\nabla_{\sigma}\alpha_{a}\left(x\right)\int d^{4}x_{1}\left(\hat{q}^{\mu}\left(x\right),\hat{\mathscr{J}}_{a}^{\sigma}\left(x_{1}\right)\right)+\beta(x)N_{\sigma}\left(x\right)\int d^{4}x_{1}\left(\hat{q}^{\mu}\left(x\right),\hat{h}^{\sigma}\left(x_{1}\right)\right)\\
		& +\beta\left(x\right)M_{\alpha}\left(x\right)\int d^{4}x_{1}\left(\hat{q}^{\mu}\left(x\right),\hat{q}^{\alpha}\left(x_{1}\right)\right),
	\end{aligned}
	\label{106}
\end{eqnarray}
and
\begin{eqnarray}
	\begin{aligned}
		\varpi^{\lambda\alpha\beta}\left(x\right)= & \langle\hat{\varpi}^{\lambda\alpha\beta}\left(x\right)\rangle_{1}\\
		= & \varXi_{abc}\left(x\right)\int d^{4}x_{1}\left(\hat{\varpi}^{\lambda\alpha\beta}\left(x\right),\hat{\varpi}^{abc}\left(x_{1}\right)\right).
	\end{aligned}
	\label{107}
\end{eqnarray}

Our chosen power counting scheme gives rise to a plethora of cross-correlations. This reveals the extent to which different physical quantities are intertwined within the system.

Due to the isotropic nature of the medium and the constraints imposed by Eq.~\eqref{56}, Ref.~\cite{Hosoya:1983id} implies that
\begin{align} 
	\biggl(\hat{\pi}_{\mu\nu}\left(x\right),\hat{\pi}_{\rho\sigma}\left(x_{1}\right)\biggr) &=\frac{1}{5}\Delta_{\mu\nu\rho\sigma}\left(x\right)\biggl(\hat{\pi}^{\lambda\eta}\left(x\right),\hat{\pi}_{\lambda\eta}\left(x_{1}\right)\biggr),\label{108}\\
	\left(\hat{S}^{\alpha\beta}\left(x\right),\hat{S}^{\rho\sigma}\left(x_{1}\right)\right) &=\frac{1}{3}\mydelta^{\alpha\beta\rho\sigma}\left(x\right)\left(\hat{S}^{\lambda\eta}\left(x\right),\hat{S}_{\lambda\eta}\left(x_{1}\right)\right),\label{109}\\
	\left(\hat{S}^{\alpha\beta}\left(x\right),\hat{\phi}^{\rho\sigma}\left(x_{1}\right)\right) &=\frac{1}{3}\mydelta^{\alpha\beta\rho\sigma}\left(x\right)\left(\hat{S}^{\lambda\eta}\left(x\right),\hat{\phi}_{\lambda\eta}\left(x_{1}\right)\right),\label{110}\\
	\biggl(\hat{\mathcal{\mathscr{J}}}_{a}^{\mu}\left(x\right),\hat{\mathscr{J}}_{b}^{\nu}\left(x_{1}\right)\biggr) &=\frac{1}{3}\Delta^{\mu\nu}\left(x\right)\biggl(\hat{\mathscr{J}}_{a}^{\lambda}\left(x\right),\hat{\mathscr{J}}_{b\lambda}\left(x_{1}\right)\biggr),\label{111}\\
	\left(\hat{\mathscr{J}}_{a}^{\mu}\left(x\right),\hat{h}^{\alpha}\left(x_{1}\right)\right) &=\frac{1}{3}\Delta^{\mu\alpha}\left(x\right)\left(\hat{\mathscr{J}}_{a}^{\lambda}\left(x\right),\hat{h}_{\lambda}\left(x_{1}\right)\right),\label{112}\\
	\left(\hat{\mathscr{J}}_{a}^{\mu}\left(x\right),\hat{q}^{\alpha}\left(x_{1}\right)\right) &=\frac{1}{3}\Delta^{\mu\alpha}\left(x\right)\left(\hat{\mathscr{J}}_{a}^{\lambda}\left(x\right),\hat{q}_{\lambda}\left(x_{1}\right)\right),\label{113}\\
	\left(\hat{\phi}^{\mu\nu}\left(x\right),\hat{S}^{\alpha\beta}\left(x_{1}\right)\right) &=\frac{1}{3}\mydelta^{\mu\nu\alpha\beta}\left(x\right)\left(\hat{\phi}^{\lambda\eta}\left(x\right),\hat{S}_{\lambda\eta}\left(x_{1}\right)\right),\label{114}\\
	\biggl(\hat{\phi}^{\mu\nu}\left(x\right),\hat{\phi}^{\rho\sigma}\left(x_{1}\right)\biggr) &=\frac{1}{3}\mydelta^{\mu\nu\rho\sigma}\left(x\right)\biggl(\hat{\phi}^{\lambda\eta}\left(x\right),\hat{\phi}_{\lambda\eta}\left(x_{1}\right)\biggr),\label{115}\\
	\left(\hat{q}^{\mu}\left(x\right),\hat{\mathscr{J}}_{a}^{\sigma}\left(x_{1}\right)\right) &=\frac{1}{3}\Delta^{\mu\sigma}\left(x\right)\left(\hat{q}^{\lambda}\left(x\right),\hat{\mathscr{J}}_{a\lambda}\left(x_{1}\right)\right),\label{116}\\
	\left(\hat{q}^{\mu}\left(x\right),\hat{h}^{\sigma}\left(x_{1}\right)\right) & =\frac{1}{3}\Delta^{\mu\sigma}\left(x\right)\left(\hat{q}^{\lambda}\left(x\right),\hat{h}_{\lambda}\left(x_{1}\right)\right),\label{117}\\
	\biggl(\hat{q}^{\mu}\left(x\right),\hat{q}^{\alpha}\left(x_{1}\right)\biggr) & =\frac{1}{3}\Delta^{\mu\alpha}\left(x\right)\biggl(\hat{q}^{\lambda}\left(x\right),\hat{q}_{\lambda}\left(x_{1}\right)\biggr),\label{118}\\
	\left(\hat{\varpi}^{\lambda\alpha\beta}\left(x\right),\hat{\varpi}^{abc}\left(x_{1}\right)\right) &=\myDelta^{\lambda\alpha\beta abc}\left(x\right)\left(\hat{\varpi}^{\rho\sigma\delta}\left(x\right),\hat{\varpi}_{\rho\sigma\delta}\left(x_{1}\right)\right).\label{119}
\end{align}

Using Eqs.~\eqref{99},\eqref{103}-\eqref{119}, we can derive
\begin{align}
	\pi_{\mu\nu}= & 2\eta\sigma_{\mu\nu},\label{120}\\
	\Pi= & -\zeta\theta+\sum_{i}\zeta_{p\mathfrak{D}_{i}}\mathcal{Z}^{\mu\nu}\partial_{\epsilon n}^{i}\Omega_{\mu\nu}+\zeta_{SS}\mathcal{K}_{\alpha\beta}\mathcal{R}^{\langle\alpha\rangle\langle\beta\rangle}+\zeta_{S\phi}\mathcal{K}_{\mu\nu}\xi^{\mu\nu},\label{121}\\
	\phi^{\mu\nu}= & 2\gamma\xi^{\mu\nu}+\gamma_{\phi S}\mathcal{R}^{\langle\mu\rangle\langle\nu\rangle},\label{122}\\
	\mathscr{J}_{a}^{\mu}= & \sum_{b}\chi_{ab}\nabla^{\mu}\alpha_{b}+\chi_{\mathscr{J}_{a}h}N^{\mu}+\chi_{\mathscr{J}_{a}q}M^{\mu},\label{123}\\
	q^{\mu}= & -\lambda M^{\mu}+\sum_{a}\lambda_{q\mathscr{J}_{a}}\nabla^{\mu}\alpha_{a}+\lambda_{qh}N^{\mu},\label{124}\\
	\varpi^{\lambda\alpha\beta} & =\varphi\varXi^{\lambda\alpha\beta},\label{125}
\end{align}
where the first-order transport coefficients are defined as
\begin{align}
	\eta= & \frac{\beta}{10}\int d^{4}x_{1}\biggl(\hat{\pi}_{\mu\nu}\left(x\right),\hat{\pi}^{\mu\nu}\left(x_{1}\right)\biggr)=-\frac{1}{10}\frac{d}{d\omega}\text{Im}G_{\hat{\pi}_{\mu\nu}\hat{\pi}^{\mu\nu}}^{R}\left(\omega\right)\bigg|_{\omega=0},\label{126}\\
	\zeta= & \beta\int d^{4}x_{1}\left(\hat{p}^{*}\left(x\right),\hat{p}^{*}\left(x_{1}\right)\right)=-\frac{d}{d\omega}\text{Im}G_{\hat{p}^{*}\hat{p}^{*}}^{R}\left(\omega\right)\bigg|_{\omega=0},\label{127}\\
	\zeta_{p\mathfrak{D}_{i}}= & \beta\int d^{4}x_{1}\left(\hat{p}^{*}\left(x\right),\hat{\mathfrak{D}}_{i}\left(x_{1}\right)\right)=-\frac{d}{d\omega}\text{Im}G_{\hat{p}^{*}\hat{\mathfrak{D}}_{i}}^{R}\left(\omega\right)\bigg|_{\omega=0},\label{128}\\
	\zeta_{SS}= & -\frac{1}{3}\beta\int d^{4}x_{1}\left(\hat{S}^{\lambda\eta}\left(x\right),\hat{S}_{\lambda\eta}\left(x_{1}\right)\right)=\frac{1}{3}\frac{d}{d\omega}\text{Im}G_{\hat{S}^{\lambda\eta}\hat{S}_{\lambda\eta}}^{R}\left(\omega\right)\bigg|_{\omega=0},\label{129}\\
	\zeta_{S\phi}= & -\frac{1}{3}\beta\int d^{4}x_{1}\left(\hat{S}^{\lambda\eta}\left(x\right),\hat{\phi}_{\lambda\eta}\left(x_{1}\right)\right)=\frac{1}{3}\frac{d}{d\omega}\text{Im}G_{\hat{S}^{\lambda\eta}\hat{\phi}_{\lambda\eta}}^{R}\left(\omega\right)\bigg|_{\omega=0},\label{130}\\
	\gamma= & \frac{1}{6}\beta\int d^{4}x_{1}\left(\hat{\phi}^{\lambda\eta}\left(x\right),\hat{\phi}_{\lambda\eta}\left(x_{1}\right)\right)=-\frac{1}{6}\frac{d}{d\omega}\text{Im}G_{\hat{\phi}^{\lambda\eta}\hat{\phi}_{\lambda\eta}}^{R}\left(\omega\right)\bigg|_{\omega=0},\label{131}\\
	\gamma_{\phi S}= & \frac{1}{3}\beta\int d^{4}x_{1}\left(\hat{\phi}^{\lambda\eta}\left(x\right),\hat{S}_{\lambda\eta}\left(x_{1}\right)\right)=-\frac{1}{3}\frac{d}{d\omega}\text{Im}G_{\hat{\phi}^{\lambda\eta}\hat{S}_{\lambda\eta}}^{R}\left(\omega\right)\bigg|_{\omega=0},\label{132}\\
	\chi_{ab}= & -\frac{1}{3}\int d^{4}x_{1}\left(\hat{\mathscr{J}}_{a}^{\lambda}\left(x\right),\hat{\mathscr{J}}_{b\lambda}\left(x_{1}\right)\right)=\frac{T}{3}\frac{d}{d\omega}\text{Im}G_{\hat{\mathscr{J}}_{a}^{\lambda}\hat{\mathscr{J}}_{b\lambda}}^{R}\left(\omega\right)\bigg|_{\omega=0},\label{133}\\
	\chi_{\mathscr{J}_{a}h}= & \frac{1}{3}\beta\int d^{4}x_{1}\left(\hat{\mathscr{J}}_{a}^{\lambda}\left(x\right),\hat{h}_{\lambda}\left(x_{1}\right)\right)=-\frac{1}{3}\frac{d}{d\omega}\text{Im}G_{\hat{\mathscr{J}}_{a}^{\lambda}\hat{h}_{\lambda}}^{R}\left(\omega\right)\bigg|_{\omega=0},\label{134}\\
	\chi_{\mathscr{Ja}_{a}q}= & \frac{1}{3}\beta\int d^{4}x_{1}\left(\hat{\mathscr{J}}_{a}^{\lambda}\left(x\right),\hat{q}_{\lambda}\left(x_{1}\right)\right)=-\frac{1}{3}\frac{d}{d\omega}\text{Im}G_{\hat{\mathscr{J}}_{a}^{\lambda}\hat{q}_{\lambda}}^{R}\left(\omega\right)\bigg|_{\omega=0},\label{135}\\
	\lambda= & -\frac{1}{3}\beta\int d^{4}x_{1}\left(\hat{q}^{\lambda}\left(x\right),\hat{q}_{\lambda}\left(x_{1}\right)\right)=\frac{1}{3}\frac{d}{d\omega}\text{Im}G_{\hat{q}^{\lambda}\hat{q}_{\lambda}}^{R}\left(\omega\right)\bigg|_{\omega=0},\label{136}\\
	\lambda_{q\mathscr{J}_{a}}= & -\frac{1}{3}\int d^{4}x_{1}\left(\hat{q}^{\lambda}\left(x\right),\hat{\mathscr{J}}_{a\lambda}\left(x_{1}\right)\right)=\frac{T}{3}\frac{d}{d\omega}\text{Im}G_{\hat{q}^{\lambda}\hat{\mathscr{J}}_{a\lambda}}^{R}\left(\omega\right)\bigg|_{\omega=0},\label{137}\\
	\lambda_{qh}= & \frac{1}{3}\beta\int d^{4}x_{1}\left(\hat{q}^{\lambda}\left(x\right),\hat{h}_{\lambda}\left(x_{1}\right)\right)=-\frac{1}{3}\frac{d}{d\omega}\text{Im}G_{\hat{q}^{\lambda}\hat{h}_{\lambda}}^{R}\left(\omega\right)\bigg|_{\omega=0},\label{138}\\
	\varphi= & \int d^{4}x_{1}\left(\hat{\varpi}^{\rho\sigma\delta}\left(x\right),\hat{\varpi}_{\rho\sigma\delta}\left(x_{1}\right)\right)=-T\frac{d}{d\omega}\text{Im}G_{\hat{\varpi}^{\rho\sigma\delta}\hat{\varpi}_{\rho\sigma\delta}}^{R}\left(\omega\right)\bigg|_{\omega=0},\label{139}
\end{align}
where
\begin{eqnarray}
	G_{\hat{X}\hat{Y}}^{R}\left(\omega\right)=-i\int_{0}^{\infty}dte^{i\omega t}\int d^{3}x\left\langle\left[\hat{X}\left(\boldsymbol{x},t\right),\hat{Y}\left(\boldsymbol{0},0\right)\right]\right\rangle_{l}
	\label{140}
\end{eqnarray}
is the zerowavenumber limit of the Fourier transform of the retarded two-point commutator and the square brackets denote the commutator. For details on how the two-point correlation functions can be expressed in terms of retarded two-point Green functions, please refer to Appendix C of Ref.~\cite{Harutyunyan:2021rmb}.

Equations~\eqref{120}-\eqref{139} establish the required linear relationships between the dissipative currents and thermodynamic forces, demonstrating that the nonequilibrium statistical operator accurately captures the Navier-Stokes limit of relativistic spin hydrodynamics.

\subsection{Second-order spin hydrodynamics}

This subsection presents a systematic calculation of all second-order corrections to the dissipative currents. These currents, along with thermodynamic forces, are considered first-order deviations from equilibrium. Second-order terms include those involving either spacetime derivatives of dissipative currents or products of dissipative currents and thermodynamic forces. These contributions arise from both three-point correlators in Eq.~\eqref{46}—quadratic in the operator $\hat{C}$—and two-point correlators with second-order corrections to $\hat{C}$. Furthermore, nonlocal effects in the two-point correlators, due to spacetime separations between dissipative currents and thermodynamic forces, introduce second-order gradient terms. As shown below, these nonlocalities lead to relaxation terms crucial for ensuring causality in the transport equations.

To accurately derive the nonlocal corrections in Eq.~\eqref{95} to the dissipative currents, the two-point correlators defined in Eqs.~\eqref{108}-\eqref{119} must be generalized, as they were originally formulated for first-order precision. The two-point correlators in Eqs.~\eqref{108}-\eqref{119} can be straightforwardly generalized to include nonlocal spatial projector effects as follows:
\begin{align} 
	\biggl(\hat{\pi}_{\mu\nu}\left(x\right),\hat{\pi}_{\rho\sigma}\left(x_{1}\right)\biggr) &=\frac{1}{5}\Delta_{\mu\nu\rho\sigma}\left(x,x_{1}\right)\biggl(\hat{\pi}^{\lambda\eta}\left(x\right),\hat{\pi}_{\lambda\eta}\left(x_{1}\right)\biggr),\label{141}\\
	\left(\hat{S}^{\alpha\beta}\left(x\right),\hat{S}^{\rho\sigma}\left(x_{1}\right)\right) &=\frac{1}{3}\mydelta^{\alpha\beta\rho\sigma}\left(x,x_{1}\right)\left(\hat{S}^{\lambda\eta}\left(x\right),\hat{S}_{\lambda\eta}\left(x_{1}\right)\right),\label{142}\\
	\left(\hat{S}^{\alpha\beta}\left(x\right),\hat{\phi}^{\rho\sigma}\left(x_{1}\right)\right) &=\frac{1}{3}\mydelta^{\alpha\beta\rho\sigma}\left(x,x_{1}\right)\left(\hat{S}^{\lambda\eta}\left(x\right),\hat{\phi}_{\lambda\eta}\left(x_{1}\right)\right),\label{143}\\	\biggl(\hat{\mathcal{\mathscr{J}}}_{a}^{\mu}\left(x\right),\hat{\mathscr{J}}_{b}^{\nu}\left(x_{1}\right)\biggr) &=\frac{1}{3}\Delta^{\mu\nu}\left(x,x_{1}\right)\biggl(\hat{\mathscr{J}}_{a}^{\lambda}\left(x\right),\hat{\mathscr{J}}_{b\lambda}\left(x_{1}\right)\biggr),\label{144}\\
	\left(\hat{\mathscr{J}}_{a}^{\mu}\left(x\right),\hat{h}^{\alpha}\left(x_{1}\right)\right) &=\frac{1}{3}\Delta^{\mu\alpha}\left(x,x_{1}\right)\left(\hat{\mathscr{J}}_{a}^{\lambda}\left(x\right),\hat{h}_{\lambda}\left(x_{1}\right)\right),\label{145}\\
	\left(\hat{\mathscr{J}}_{a}^{\mu}\left(x\right),\hat{q}^{\alpha}\left(x_{1}\right)\right) &=\frac{1}{3}\Delta^{\mu\alpha}\left(x,x_{1}\right)\left(\hat{\mathscr{J}}_{a}^{\lambda}\left(x\right),\hat{q}_{\lambda}\left(x_{1}\right)\right),\label{146}\\
	\left(\hat{\phi}^{\mu\nu}\left(x\right),\hat{S}^{\alpha\beta}\left(x_{1}\right)\right) &=\frac{1}{3}\mydelta^{\mu\nu\alpha\beta}\left(x,x_{1}\right)\left(\hat{\phi}^{\lambda\eta}\left(x\right),\hat{S}_{\lambda\eta}\left(x_{1}\right)\right),\label{147}\\
	\biggl(\hat{\phi}^{\mu\nu}\left(x\right),\hat{\phi}^{\rho\sigma}\left(x_{1}\right)\biggr) &=\frac{1}{3}\mydelta^{\mu\nu\rho\sigma}\left(x,x_{1}\right)\biggl(\hat{\phi}^{\lambda\eta}\left(x\right),\hat{\phi}_{\lambda\eta}\left(x_{1}\right)\biggr),\label{148}\\
	\left(\hat{q}^{\mu}\left(x\right),\hat{\mathscr{J}}_{a}^{\sigma}\left(x_{1}\right)\right) &=\frac{1}{3}\Delta^{\mu\sigma}\left(x,x_{1}\right)\left(\hat{q}^{\lambda}\left(x\right),\hat{\mathscr{J}}_{a\lambda}\left(x_{1}\right)\right),\label{149}\\
	\left(\hat{q}^{\mu}\left(x\right),\hat{h}^{\sigma}\left(x_{1}\right)\right)
	&=\frac{1}{3}\Delta^{\mu\sigma}\left(x,x_{1}\right)\left(\hat{q}^{\lambda}\left(x\right),\hat{h}_{\lambda}\left(x_{1}\right)\right),\label{150}\\
	\biggl(\hat{q}^{\mu}\left(x\right),\hat{q}^{\alpha}\left(x_{1}\right)\biggr)
	&=\frac{1}{3}\Delta^{\mu\alpha}\left(x,x_{1}\right)\biggl(\hat{q}^{\lambda}\left(x\right),\hat{q}_{\lambda}\left(x_{1}\right)\biggr),\label{151}\\
	\left(\hat{\varpi}^{\lambda\alpha\beta}\left(x\right),\hat{\varpi}^{abc}\left(x_{1}\right)\right) &=\myDelta^{\lambda\alpha\beta abc}\left(x,x_{1}\right)\left(\hat{\varpi}^{\rho\sigma\delta}\left(x\right),\hat{\varpi}_{\rho\sigma\delta}\left(x_{1}\right)\right).\label{152}
\end{align}
Here the new nonlocal projection operators are defined as follows:
\begin{align}
	\Delta_{\mu\nu}\left(x,x_{1}\right) & =\Delta_{\mu\lambda}(x)\Delta_{\nu}^{\lambda}\left(x_{1}\right),\label{153}\\
	\Delta_{\mu\nu\rho\sigma}\left(x,x_{1}\right) & =\Delta_{\mu\nu\alpha\beta}(x)\Delta_{\rho\sigma}^{\alpha\beta}\left(x_{1}\right),\label{154}\\
	\mydelta^{\mu\nu\rho\sigma}\left(x,x_{1}\right) & =\mydelta^{\mu\nu\alpha\beta}\left(x\right)\mydelta_{\alpha\beta}^{\rho\sigma}\left(x_{1}\right),\label{155}\\
	\myDelta^{\lambda\mu\nu\alpha\beta\gamma}\left(x,x_{1}\right) & =\myDelta^{\lambda\mu\nu\rho\sigma\delta}\left(x\right)\myDelta_{\rho\sigma\delta}^{\alpha\beta\gamma}\left(x_{1}\right).\label{156}
\end{align}
These operators are natural nonlocal generalizations of their local counterparts: the second-rank $\Delta_{\mu \nu}$, the fourth-rank symmetric traceless $\Delta_{\mu \nu \rho \sigma}$, the fourth-rank antisymmetric $\mydelta_{\mu\nu\rho\sigma}$, and the sixth-rank totally antisymmetric $\myDelta_{\lambda\mu\nu\alpha\beta\gamma}$ projectors. The normalization of the right-hand sides of Eqs.~\eqref{141}-\eqref{152} is performed at leading order in velocity. The nonlocal structure of the projectors in Eqs.~\eqref{153}-\eqref{156} ensures that the orthogonality conditions in Eq.~\eqref{53} are satisfied by the correlation functions defined in Eqs.~\eqref{141}-\eqref{152} at both spacetime point $x$ and $x_1$.

To improve computational efficiency, we truncate the expansion of the nonlocal projectors around $x_1$ linear order in the difference $x_1-x$,
\begin{align}
\frac{\partial}{\partial x_{1}^{\alpha}}\Delta_{\mu\nu\rho\sigma}\left(x,x_{1}\right)\bigg|_{x_{1}=x}= & -\biggl[\Delta_{\mu\nu\rho\beta}\left(x\right)u_{\sigma}\left(x\right)+\Delta_{\mu\nu\sigma\beta}\left(x\right)u_{\rho}\left(x\right)\biggr]\frac{\partial u^{\beta}\left(x_{1}\right)}{\partial x_{1}^{\alpha}}\bigg|_{x_{1}=x},\label{157}\\
\frac{\partial}{\partial x_{1}^{\alpha}}\mydelta_{\mu\nu\rho\sigma}\left(x,x_{1}\right)\bigg|_{x_{1}=x}= & -\biggl[\mydelta_{\mu\nu\rho\beta}\left(x\right)u_{\sigma}\left(x\right)-\mydelta_{\mu\nu\sigma\beta}\left(x\right)u_{\rho}\left(x\right)\biggr]\frac{\partial u^{\beta}\left(x_{1}\right)}{\partial x_{1}^{\alpha}}\bigg|_{x_{1}=x},\label{158}\\
\frac{\partial}{\partial x_{1}^{\alpha}}\myDelta_{\lambda\mu\nu\varepsilon\beta\gamma}\left(x,x_{1}\right)\bigg|_{x_{1}=x}= & -\biggl[\myDelta_{\lambda\mu\nu\beta\gamma\rho}\left(x\right)u_{\varepsilon}\left(x\right)+\myDelta_{\lambda\mu\nu\gamma\varepsilon\rho}\left(x\right)u_{\beta}\left(x\right)+\myDelta_{\lambda\mu\nu\varepsilon\beta\rho}\left(x\right)u_{\gamma}\left(x\right)\biggr]\frac{\partial u^{\rho}\left(x_{1}\right)}{\partial x_{1}^{\alpha}}\bigg|_{x_{1}=x},\label{159}
\end{align}
which are subsequently used in our analysis. Furthermore, we assume that Curie's theorem holds in this approximation, implying that two-point correlations between tensors of different rank vanish.

\subsubsection{Second-order corrections to the shear stress tensor}

Substituting the decomposition in Eq.~\eqref{90} into Eq.~\eqref{95} and applying Curie's theorem, we find
\begin{eqnarray}
\left\langle \hat{\pi}_{\mu\nu}\left(x\right)\right\rangle _{2}^{1}=\int d^{4}x_{1}\left(\hat{\pi}_{\mu\nu}\left(x\right),\hat{\pi}_{\rho\sigma}\left(x_{1}\right)\right)\beta\left(x_{1}\right)\sigma^{\rho\sigma}\left(x_{1}\right)-2\eta\left(x\right)\sigma_{\mu\nu}\left(x\right),
\label{160}
\end{eqnarray}
where we used the first-order relation $\langle\hat{\pi}_{\mu\nu}\left(x\right)\rangle_{1}$ from Eq.~\eqref{120}. The thermodynamic force $\beta\left(x_{1}\right)\sigma^{\rho\sigma}\left(x_{1}\right)$ cannot be factored out of the integral with its value at $x$ and must be expanded to first order in $x_1-x$.

Substituting the two-point correlation functions from Eqs.~\eqref{141} and~\eqref{154} into Eq.~\eqref{160} and using the identity $\Delta_{\rho\sigma}^{\alpha\beta}\sigma^{\rho\sigma}\equiv\sigma^{\alpha\beta}$, we find
\begin{eqnarray}
\langle\hat{\pi}_{\mu\nu}\left(x\right)\rangle_{2}^{1}=\frac{1}{5}\Delta_{\mu\nu\rho\sigma}\left(x\right)\int d^{4}x_{1}\left(\hat{\pi}^{\lambda\eta}\left(x\right),\hat{\pi}_{\lambda\eta}\left(x_{1}\right)\right)\beta\left(x_{1}\right)\sigma^{\rho\sigma}\left(x_{1}\right)-2\eta\left(x\right)\sigma_{\mu\nu}\left(x\right).
\label{161}
\end{eqnarray}

To capture all nonlocal second-order corrections, the nonlocality of the fluid velocity $u^\lambda$ in $\hat{\pi}_{\lambda\eta}\left(x_{1}\right)=\Delta_{\lambda\eta\gamma\delta}\left(x_{1}\right)\hat{T}^{\gamma\delta}\left(x_{1}\right)$ must also be considered. Unlike the hydrodynamic quantity $\Delta_{\lambda\eta\gamma\delta}\left(x_{1}\right)$, the microscopic energy-momentum tensor $\hat{T}^{\gamma\delta}\left(x_{1}\right)$ does not require an expansion. Performing a Taylor expansion of the hydrodynamic quantities around $x_1=x$ and retaining linear terms, we obtain
\begin{eqnarray} \hat{\pi}_{\lambda\eta}\left(x_{1}\right)=\hat{\pi}_{\lambda\eta}\left(x_{1}\right)\bigg|_{x_{1}=x}+\left(x_{1}-x\right)^{\tau}\frac{\partial\hat{\pi}_{\lambda\eta}\left(x_{1}\right)}{\partial x_{1}^{\tau}}\bigg|_{x_{1}=x},
	\label{162}
\end{eqnarray}
where
\begin{eqnarray}
\hat{\pi}_{\lambda\eta}\left(x_{1}\right)\bigg|_{x_{1}=x}=\Delta_{\lambda\eta\mu\nu}\left(x\right)\hat{T}^{\mu\nu}\left(x_{1}\right),\qquad\frac{\partial\hat{\pi}_{\lambda\eta}\left(x_{1}\right)}{\partial x_{1}^{\tau}}\bigg|_{x_{1}=x}=-2\Delta_{\lambda\eta\mu\beta}\left(x\right)\hat{h}^{\mu}\left(x_{1}\right)\frac{\partial u^{\beta}\left(x_{1}\right)}{\partial x_{1}^{\tau}}\bigg|_{x_{1}=x},
\label{163}
\end{eqnarray}
where the approximation $u^{\mu}\left(x\right)\simeq u^{\mu}\left(x_{1}\right)$ has been used. All operators in Eq.~\eqref{163} are evaluated at $x_1$, while hydrodynamic quantities are evaluated at $x$.

Substituting Eq.~\eqref{162} into Eq.~\eqref{161} and expanding the thermodynamic force $\beta\sigma^{\rho\sigma}$ around $x_1=x$, we obtain up to the second order in gradients
\begin{eqnarray}
	\begin{aligned}
\langle\hat{\pi}_{\mu\nu}\left(x\right)\rangle_{2}^{1}= & \frac{1}{5}\Delta_{\mu\nu\rho\sigma}\left(x\right)\int d^{4}x_{1}\left(\hat{\pi}^{\lambda\eta}\left(x\right),\frac{\partial\hat{\pi}_{\lambda\eta}\left(x_{1}\right)}{\partial x_{1}^{\tau}}\bigg|_{x_{1}=x}\right)\left(x_{1}-x\right)^{\tau}\beta\left(x\right)\sigma^{\rho\sigma}\left(x\right)\\
 & +\frac{1}{5}\Delta_{\mu\nu\rho\sigma}\left(x\right)\int d^{4}x_{1}\left(\hat{\pi}^{\lambda\eta}\left(x\right),\hat{\pi}_{\lambda\eta}\left(x_{1}\right)\bigg|_{x_{1}=x}\right)\left(x_{1}-x\right)^{\tau}\frac{\partial}{\partial x_{1}^{\tau}}\left[\beta\left(x_{1}\right)\sigma^{\rho\sigma}\left(x_{1}\right)\right]\bigg|_{x_{1}=x},
\end{aligned}
    \label{164}
\end{eqnarray}
where we used the following relation to eliminate the first-order term
\begin{eqnarray}
	\eta=\frac{\beta}{10}\int d^{4}x_{1}\left(\hat{\pi}_{\mu\nu}\left(x\right),\hat{\pi}^{\mu\nu}\left(x_{1}\right)\bigg|_{x_{1}=x}\right).
	\label{165}
\end{eqnarray}

Substituting Eq.~\eqref{163} into Eq.~\eqref{164}, applying Curie’s theorem, and utilizing the approximation $D\beta\simeq\beta\theta\Gamma-2\beta\mathcal{Z}^{\alpha\beta}\mathcal{D}_{\alpha\beta}$, we obtain the nonlocal corrections from the two-point correlation function to shear stress tensor
\begin{eqnarray} \langle\hat{\pi}_{\mu\nu}\rangle_{2}^{1}=2\widetilde{\eta}\Delta_{\mu\nu\rho\sigma}\beta^{-1}D\left(\beta\sigma^{\rho\sigma}\right)=2\widetilde{\eta}\Delta_{\mu\nu\rho\sigma}D\sigma^{\rho\sigma}+2\widetilde{\eta}\left(\theta\Gamma-2\mathcal{Z}^{\alpha\beta}\mathcal{D}_{\alpha\beta}\right)\sigma_{\mu\nu},
\label{166}
\end{eqnarray}
where we define
\begin{eqnarray} \widetilde{\eta}=i\frac{d}{d\omega}\eta\left(\omega\right)\bigg|_{\omega=0}=-\frac{1}{20}\frac{d^{2}}{d\omega^{2}}\text{Re}G_{\hat{\pi}_{\mu\nu}\hat{\pi}^{\mu\nu}}^{R}\left(\omega\right)\bigg|_{\omega=0},
\label{167}
\end{eqnarray}
with the frequency-dependent transport coefficient $\eta(\omega)$ is defined as
\begin{eqnarray}
	\eta\left(\omega\right)=\frac{\beta}{10}\int d^{4}x_{1}\int_{-\infty}^{t}e^{i\omega\left(t-t_{1}\right)}\left(\hat{\pi}_{\mu\nu}\left(\boldsymbol{x},t\right),\hat{\pi}^{\mu\nu}\left(\boldsymbol{x}_{1},t_{1}\right)\right).
	\label{168}
\end{eqnarray}

Since the operator $\hat{C}_2$ given by Eq.~\eqref{91} does not contain any term involving  symmetric second-rank operator, hence all the correlations vanishes by using Curie's theorem. Specifically, we find the corrections from extended thermodynamic forces
\begin{eqnarray}
\left\langle \hat{\pi}_{\mu\nu}\left(x\right)\right\rangle _{2}^{2}=\int d^{4}x_{1}\left(\hat{\pi}_{\mu\nu}\left(x\right),\hat{C}_{2}\left(x_{1}\right)\right)=0.
\label{169}
\end{eqnarray}

Combining Eqs.~\eqref{90} and \eqref{97}, the correction to shear stress tensor from the three-point correlation function is given by
\begin{eqnarray}
\begin{aligned} & \langle\hat{\pi}_{\mu\nu}(x)\rangle_{2}^{3}\\
	= & \int d^{4}x_{1}d^{4}x_{2}\biggl(\hat{\pi}_{\mu\nu}(x),\biggl[-\beta\theta\hat{p}^{*}+\beta\mathcal{R}_{\alpha\beta}\hat{S}^{\alpha\beta}+\beta\sum_{i}\left(\hat{\mathfrak{D}}_{i}\partial_{\epsilon n}^{i}\Omega_{\mu\nu}\right)\mathcal{Z}^{\mu\nu}-\sum_{a}\hat{\mathscr{J}}_{a}^{\sigma}\nabla_{\sigma}\alpha_{a}\\
	& +\beta\hat{h}^{\sigma}N_{\sigma}+\beta\hat{q}^{\mu}M_{\mu}+\beta\hat{\pi}^{\mu\nu}\sigma_{\mu\nu}+\beta\hat{\phi}^{\mu\nu}\xi_{\mu\nu}+\hat{\varpi}^{\lambda\alpha\beta}\varXi_{\lambda\alpha\beta}\biggr]_{x_{1}},\biggl[-\beta\theta\hat{p}^{*}+\beta\mathcal{R}_{\alpha\beta}\hat{S}^{\alpha\beta}+\beta\sum_{i}\left(\hat{\mathfrak{D}}_{i}\partial_{\epsilon n}^{i}\Omega_{\mu\nu}\right)\mathcal{Z}^{\mu\nu}\\
	& -\sum_{a}\hat{\mathscr{J}}_{a}^{\sigma}\nabla_{\sigma}\alpha_{a}+\beta\hat{h}^{\sigma}N_{\sigma}+\beta\hat{q}^{\mu}M_{\mu}+\beta\hat{\pi}^{\mu\nu}\sigma_{\mu\nu}+\beta\hat{\phi}^{\mu\nu}\xi_{\mu\nu}+\hat{\varpi}^{\lambda\alpha\beta}\varXi_{\lambda\alpha\beta}\biggr]_{x_{2}}\biggr).
\end{aligned}
\label{170}
\end{eqnarray}
Unlike two-point correlators, which exclusively couple operators of the same rank, three-point correlators can exhibit mixing between operators of different ranks. The nonvanishing mixed three-point correlators derived from Eq.~\eqref{170} are
\begin{align}
	\biggl(\hat{\pi}_{\mu\nu}\left(x\right),\hat{p}^{*}\left(x_{1}\right),\hat{\pi}_{\alpha\beta}\left(x_{2}\right)\biggr)= & \frac{1}{5}\Delta_{\mu\nu\alpha\beta}\biggl(\hat{\pi}_{\gamma\delta}\left(x\right),\hat{p}^{*}\left(x_{1}\right),\hat{\pi}^{\gamma\delta}\left(x_{2}\right)\biggr),\label{171}\\
	\left(\hat{\pi}_{\mu\nu}\left(x\right),\hat{S}_{\rho\sigma}\left(x_{1}\right),\hat{S}_{\alpha\beta}\left(x_{2}\right)\right)= & -\frac{1}{5}\left(\Delta_{\rho\alpha}\Delta_{\mu\nu\sigma\beta}-\Delta_{\rho\beta}\Delta_{\mu\nu\sigma\alpha}-\Delta_{\sigma\alpha}\Delta_{\mu\nu\rho\beta}+\Delta_{\sigma\beta}\Delta_{\mu\nu\rho\alpha}\right)\left(\hat{\pi}_{\gamma}^{\,\,\,\delta}\left(x\right),\hat{S}_{\delta}^{\,\,\,\lambda}\left(x_{1}\right),\hat{S}_{\lambda}^{\,\,\,\gamma}\left(x_{2}\right)\right),\label{172}\\
	\left(\hat{\pi}_{\mu\nu}\left(x\right),\hat{S}_{\rho\sigma}\left(x_{1}\right),\hat{\pi}_{\alpha\beta}\left(x_{2}\right)\right)= & \frac{1}{15}\left(-\Delta_{\rho\alpha}\Delta_{\mu\nu\sigma\beta}-\Delta_{\rho\beta}\Delta_{\mu\nu\sigma\alpha}+\Delta_{\sigma\alpha}\Delta_{\mu\nu\rho\beta}+\Delta_{\sigma\beta}\Delta_{\mu\nu\rho\alpha}\right)\left(\hat{\pi}_{\gamma}^{\,\,\,\delta}\left(x\right),\hat{S}_{\delta}^{\,\,\,\lambda}\left(x_{1}\right),\hat{\pi}_{\lambda}^{\,\,\,\gamma}\left(x_{2}\right)\right),\label{173}\\
	\left(\hat{\pi}_{\mu\nu}\left(x\right),\hat{S}_{\rho\sigma}\left(x_{1}\right),\hat{\phi}_{\alpha\beta}\left(x_{2}\right)\right)= & -\frac{1}{5}\left(\Delta_{\rho\alpha}\Delta_{\mu\nu\sigma\beta}-\Delta_{\rho\beta}\Delta_{\mu\nu\sigma\alpha}-\Delta_{\sigma\alpha}\Delta_{\mu\nu\rho\beta}+\Delta_{\sigma\beta}\Delta_{\mu\nu\rho\alpha}\right)\left(\hat{\pi}_{\gamma}^{\,\,\,\delta}\left(x\right),\hat{S}_{\delta}^{\,\,\,\lambda}\left(x_{1}\right),\hat{\phi}_{\lambda}^{\,\,\,\gamma}\left(x_{2}\right)\right),\label{174}\\
	\left(\hat{\pi}_{\mu\nu}\left(x\right),\hat{\mathfrak{D}}_{i}\left(x_{1}\right),\hat{\pi}_{\alpha\beta}\left(x_{2}\right)\right)= & \frac{1}{5}\Delta_{\mu\nu\alpha\beta}\left(\hat{\pi}_{\lambda\eta}\left(x\right),\hat{\mathfrak{D}}_{i}\left(x_{1}\right),\hat{\pi}^{\lambda\eta}\left(x_{1}\right)\right),\label{175}\\
	\biggl(\hat{\pi}_{\mu\nu}\left(x\right),\hat{\mathcal{\mathscr{J}}}_{a\sigma}\left(x_{1}\right),\hat{\mathcal{\mathscr{J}}}_{b\alpha}\left(x_{2}\right)\biggr)= & \frac{1}{5}\Delta_{\mu\nu\sigma\alpha}\biggl(\hat{\pi}_{\gamma\delta}\left(x\right),\hat{\mathcal{\mathscr{J}}}_{a}^{\gamma}\left(x_{1}\right),\hat{\mathcal{\mathscr{J}}}_{b}^{\delta}\left(x_{2}\right)\biggr),\label{176}\\
	\biggl(\hat{\pi}_{\mu\nu}\left(x\right),\hat{\mathcal{\mathscr{J}}}_{a\sigma}\left(x_{1}\right),\hat{h}_{\alpha}\left(x_{2}\right)\biggr)= & \frac{1}{5}\Delta_{\mu\nu\sigma\alpha}\biggl(\hat{\pi}_{\gamma\delta}\left(x\right),\hat{\mathcal{\mathscr{J}}}_{a}^{\gamma}\left(x_{1}\right),\hat{h}^{\delta}\left(x_{2}\right)\biggr),\label{177}\\
	\biggl(\hat{\pi}_{\mu\nu}\left(x\right),\hat{\mathcal{\mathscr{J}}}_{a\sigma}\left(x_{1}\right),\hat{q}_{\alpha}\left(x_{2}\right)\biggr)= & \frac{1}{5}\Delta_{\mu\nu\sigma\alpha}\biggl(\hat{\pi}_{\gamma\delta}\left(x\right),\hat{\mathcal{\mathscr{J}}}_{a}^{\gamma}\left(x_{1}\right),\hat{q}^{\delta}\left(x_{2}\right)\biggr),\label{178}\\
	\biggl(\hat{\pi}_{\mu\nu}\left(x\right),\hat{h}_{\sigma}\left(x_{1}\right),\hat{h}_{\alpha}\left(x_{2}\right)\biggr)= & \frac{1}{5}\Delta_{\mu\nu\sigma\alpha}\biggl(\hat{\pi}_{\gamma\delta}\left(x\right),\hat{h}^{\gamma}\left(x_{1}\right),\hat{h}^{\delta}\left(x_{2}\right)\biggr),\label{179}\\
	\biggl(\hat{\pi}_{\mu\nu}\left(x\right),\hat{h}_{\sigma}\left(x_{1}\right),\hat{q}_{\alpha}\left(x_{2}\right)\biggr)= & \frac{1}{5}\Delta_{\mu\nu\sigma\alpha}\biggl(\hat{\pi}_{\gamma\delta}\left(x\right),\hat{h}^{\gamma}\left(x_{1}\right),\hat{q}^{\delta}\left(x_{2}\right)\biggr),\label{180}\\
	\biggl(\hat{\pi}_{\mu\nu}\left(x\right),\hat{q}_{\sigma}\left(x_{1}\right),\hat{q}_{\alpha}\left(x_{2}\right)\biggr)= & \frac{1}{5}\Delta_{\mu\nu\sigma\alpha}\biggl(\hat{\pi}_{\gamma\delta}\left(x\right),\hat{q}^{\gamma}\left(x_{1}\right),\hat{q}^{\delta}\left(x_{2}\right)\biggr),\label{181}\\
	\left(\hat{\pi}_{\mu\nu}\left(x\right),\hat{\pi}_{\rho\sigma}\left(x_{1}\right),\hat{\pi}_{\alpha\beta}\left(x_{2}\right)\right)= & \frac{1}{35}\biggl[3\left(\Delta_{\rho\alpha}\Delta_{\mu\nu\sigma\beta}+\Delta_{\rho\beta}\Delta_{\mu\nu\sigma\alpha}+\Delta_{\sigma\alpha}\Delta_{\mu\nu\rho\beta}+\Delta_{\sigma\beta}\Delta_{\mu\nu\rho\alpha}\right)\nonumber\\
	&-4\left(\Delta_{\rho\sigma}\Delta_{\mu\nu\alpha\beta}+\Delta_{\alpha\beta}\Delta_{\mu\nu\rho\sigma}\right)\biggr]\left(\hat{\pi}_{\gamma}^{\delta}\left(x\right),\hat{\pi}_{\delta}^{\lambda}\left(x_{1}\right),\hat{\pi}_{\lambda}^{\gamma}\left(x_{2}\right)\right),\label{182}\\
	\left(\hat{\pi}_{\mu\nu}\left(x\right),\hat{\pi}_{\rho\sigma}\left(x_{1}\right),\hat{\phi}_{\alpha\beta}\left(x_{2}\right)\right)= & -\frac{1}{15}\left(-\Delta_{\rho\alpha}\Delta_{\mu\nu\sigma\beta}+\Delta_{\rho\beta}\Delta_{\mu\nu\sigma\alpha}-\Delta_{\sigma\alpha}\Delta_{\mu\nu\rho\beta}+\Delta_{\sigma\beta}\Delta_{\mu\nu\rho\alpha}\right)\left(\hat{\pi}_{\gamma}^{\,\,\,\delta}\left(x\right),\hat{\pi}_{\delta}^{\,\,\,\lambda}\left(x_{1}\right),\hat{\phi}_{\lambda}^{\,\,\,\gamma}\left(x_{2}\right)\right),\label{183}\\
	\left(\hat{\pi}_{\mu\nu}\left(x\right),\hat{\phi}_{\rho\sigma}\left(x_{1}\right),\hat{\phi}_{\alpha\beta}\left(x_{2}\right)\right)= & -\frac{1}{5}\left(\Delta_{\rho\alpha}\Delta_{\mu\nu\sigma\beta}-\Delta_{\rho\beta}\Delta_{\mu\nu\sigma\alpha}-\Delta_{\sigma\alpha}\Delta_{\mu\nu\rho\beta}+\Delta_{\sigma\beta}\Delta_{\mu\nu\rho\alpha}\right)\left(\hat{\pi}_{\gamma}^{\,\,\,\delta}\left(x\right),\hat{\phi}_{\delta}^{\,\,\,\lambda}\left(x_{1}\right),\hat{\phi}_{\lambda}^{\,\,\,\gamma}\left(x_{2}\right)\right),\label{184}
\end{align}
where we utilized the inherent symmetries of the operators and the three-point correlators [cf. Eq.~\eqref{49}].  Since the correlators in Eqs.~\eqref{171}-\eqref{184} are associated with second-order thermodynamic force terms, evaluating all $\Delta$ projectors at point $x$ suffices.

By substituting the correlation functions defined in Eqs.~\eqref{171}-\eqref{184} into Eq.~\eqref{170}, factoring out the thermodynamic forces evaluated at point $x$, and applying the symmetry property from Eq.~\eqref{49}, we introduce the transport coefficients
\begin{align}
	\eta_{\pi p\pi} & =-\frac{1}{5}\beta^{2}\int d^{4}x_{1}d^{4}x_{2}\left(\hat{\pi}_{\gamma\delta}(x),\hat{p}^{*}\left(x_{1}\right),\hat{\pi}^{\gamma\delta}\left(x_{2}\right)\right),\eta_{\pi SS}=\frac{4}{5}\beta^{2}\int d^{4}x_{1}d^{4}x_{2}\left(\hat{\pi}_{\gamma}^{\,\,\,\delta}\left(x\right),\hat{S}_{\delta}^{\,\,\,\lambda}\left(x_{1}\right),\hat{S}_{\lambda}^{\,\,\,\gamma}\left(x_{2}\right)\right),\label{185}\\
	\eta_{\pi S\pi} & =-\frac{4}{15}\beta^{2}\int d^{4}x_{1}d^{4}x_{2}\left(\hat{\pi}_{\gamma}^{\,\,\,\delta}\left(x\right),\hat{S}_{\delta}^{\,\,\,\lambda}\left(x_{1}\right),\hat{\pi}_{\lambda}^{\,\,\,\gamma}\left(x_{2}\right)\right),\eta_{\pi S\phi}=\frac{4}{5}\beta^{2}\int d^{4}x_{1}d^{4}x_{2}\left(\hat{\pi}_{\gamma}^{\,\,\,\delta}\left(x\right),\hat{S}_{\delta}^{\,\,\,\lambda}\left(x_{1}\right),\hat{\phi}_{\lambda}^{\,\,\,\gamma}\left(x_{2}\right)\right),\label{186}\\
	\eta_{\pi\mathfrak{D}_{i}\pi} & =\frac{1}{5}\beta^{2}\int d^{4}x_{1}d^{4}x_{2}\left(\hat{\pi}_{\lambda\eta}\left(x\right),\hat{\mathfrak{D}}_{i}\left(x_{1}\right),\hat{\pi}^{\lambda\eta}\left(x_{2}\right)\right),\eta_{\pi\mathscr{J}_{a}\mathscr{J}_{b}}=\frac{1}{5}\int d^{4}x_{1}d^{4}x_{2}\left(\hat{\pi}_{\gamma\delta}\left(x\right),\hat{\mathscr{J}}_{a}^{\gamma}\left(x_{1}\right),\hat{\mathscr{J}}_{b}^{\delta}\left(x_{2}\right)\right),\label{187}\\
	\eta_{\pi\mathscr{J}_{a}h} & =-\frac{1}{5}\beta\int d^{4}x_{1}d^{4}x_{2}\left(\hat{\pi}_{\gamma\delta}\left(x\right),\hat{\mathscr{J}}_{a}^{\gamma}\left(x_{1}\right),\hat{h}^{\delta}\left(x_{2}\right)\right),\eta_{\pi\mathscr{J}_{a}q}=-\frac{1}{5}\beta\int d^{4}x_{1}d^{4}x_{2}\left(\hat{\pi}_{\gamma\delta}\left(x\right),\hat{\mathscr{J}}_{a}^{\gamma}\left(x_{1}\right),\hat{q}^{\delta}\left(x_{2}\right)\right),\label{188}\\
	\eta_{\pi hh} & =\frac{1}{5}\beta^{2}\int d^{4}x_{1}d^{4}x_{2}\left(\hat{\pi}_{\gamma\delta}\left(x\right),\hat{h}^{\gamma}\left(x_{1}\right),\hat{h}^{\delta}\left(x_{2}\right)\right),\eta_{\pi hq}=\frac{1}{5}\beta^{2}\int d^{4}x_{1}d^{4}x_{2}\left(\hat{\pi}_{\gamma\delta}\left(x\right),\hat{h}^{\gamma}\left(x_{1}\right),\hat{q}^{\delta}\left(x_{2}\right)\right),\label{189}\\
	\eta_{\pi qq} & =\frac{1}{5}\beta^{2}\int d^{4}x_{1}d^{4}x_{2}\left(\hat{\pi}_{\gamma\delta}\left(x\right),\hat{q}^{\gamma}\left(x_{1}\right),\hat{q}^{\delta}\left(x_{2}\right)\right),\eta_{\pi\pi\pi}=\frac{12}{35}\beta^{2}\int d^{4}x_{1}d^{4}x_{2}\left(\hat{\pi}_{\gamma}^{\,\,\,\delta}\left(x\right),\hat{\pi}_{\delta}^{\,\,\,\lambda}\left(x_{1}\right),\hat{\pi}_{\lambda}^{\,\,\,\gamma}\left(x_{2}\right)\right),\label{190}\\
	\eta_{\pi\pi\phi} & =-\frac{4}{15}\beta^{2}\int d^{4}x_{1}d^{4}x_{2}\left(\hat{\pi}_{\gamma}^{\,\,\,\delta}\left(x\right),\hat{\pi}_{\delta}^{\,\,\,\lambda}\left(x_{1}\right),\hat{\phi}_{\lambda}^{\,\,\,\gamma}\left(x_{2}\right)\right),\eta_{\pi\phi\phi}=\frac{4}{5}\beta^{2}\int d^{4}x_{1}d^{4}x_{2}\left(\hat{\pi}_{\gamma}^{\,\,\,\delta}\left(x\right),\hat{\phi}_{\delta}^{\,\,\,\lambda}\left(x_{1}\right),\hat{\phi}_{\lambda}^{\,\,\,\gamma}\left(x_{2}\right)\right),\label{191}
\end{align}
Consequently, the second-order corrections to $\langle\hat{\pi}_{\mu\nu}\rangle$ from the three-point correlation function contribution are expressed as
\begin{eqnarray}
\begin{aligned}
	\langle\hat{\pi}_{\mu\nu}\rangle_{2}^{3}= & 2\eta_{\pi p\pi}\theta\sigma_{\mu\nu}+\eta_{\pi SS}\mathcal{R}_{\langle\alpha\rangle\langle\mu}\mathcal{R}_{\nu\rangle}^{\,\,\,\,\alpha}+2\eta_{\pi S\pi}\mathcal{R}_{\langle\alpha\rangle\langle\mu}\sigma_{\nu\rangle}^{\,\,\,\,\alpha}+2\eta_{\pi S\phi}\mathcal{R}_{\langle\alpha\rangle\langle\mu}\xi_{\nu\rangle}^{\,\,\,\,\alpha}+2\sum_{i}\eta_{\pi\mathfrak{D}_{i}\pi}\mathcal{Z}^{\rho\sigma}\left(\partial_{\epsilon n}^{i}\Omega_{\rho\sigma}\right)\sigma_{\mu\nu}\\
	&+\sum_{ab}\eta_{\pi\mathscr{J}_{a}\mathscr{J}_{b}}\nabla_{\langle\mu}\alpha_{a}\nabla_{\nu\rangle}\alpha_{b}+2\sum_{a}\eta_{\pi\mathscr{J}_{a}h}\nabla_{\langle\mu}\alpha_{a}N_{\nu\rangle}+2\sum_{a}\eta_{\pi\mathscr{J}_{a}q}\nabla_{\langle\mu}\alpha_{a}M_{\nu\rangle}+\eta_{\pi hh}N_{\langle\mu}N_{\nu\rangle}\\
	& +2\eta_{\pi hq}N_{\langle\mu}M_{\nu\rangle}+\eta_{\pi qq}M_{\langle\mu}M_{\nu\rangle}+\eta_{\pi\pi\pi}\sigma_{\alpha\langle\mu}\sigma_{\nu\rangle}^{\,\,\,\,\alpha}+2\eta_{\pi\pi\phi}\sigma_{\alpha\langle\mu}\xi_{\nu\rangle}^{\,\,\,\,\alpha}+\eta_{\pi\phi\phi}\xi_{\alpha\langle\mu}\xi_{\nu\rangle}^{\,\,\,\,\alpha},
\end{aligned}
\label{192}
\end{eqnarray}
where we have introduced the notation
\begin{eqnarray}
A_{\langle\mu\nu\rangle}\equiv\Delta_{\mu\nu}^{\alpha\beta}A_{\alpha\beta},
\label{193}
\end{eqnarray}
and utilized the identities $\sigma^{\langle\alpha\beta\rangle}=\sigma^{\alpha\beta},\sigma^{\alpha\beta}\Delta_{\beta\lambda}=\sigma_{\lambda}^{\alpha}$, and $\sigma_{\alpha}^{\alpha}=0$.

Combining the corrections from Eqs.~\eqref{120}, \eqref{166}, \eqref{169}, and \eqref{192} with the results from Eqs.~\eqref{70}, \eqref{92}, and \eqref{94}, we derive the complete second-order expression for the shear stress tensor:
\begin{equation}
\begin{aligned}
\pi_{\mu\nu}	=&2\eta\sigma_{\mu\nu}+2\widetilde{\eta}\left(\Delta_{\mu\nu\rho\sigma}D\sigma^{\rho\sigma}+\theta\Gamma\sigma_{\mu\nu}-2\mathcal{Z}^{\alpha\beta}\mathcal{D}_{\alpha\beta}\sigma_{\mu\nu}\right)\\
&+2\eta_{\pi p\pi}\theta\sigma_{\mu\nu}+\eta_{\pi SS}\mathcal{R}_{\langle\alpha\rangle\langle\mu}\mathcal{R}_{\nu\rangle}^{\,\,\,\,\alpha}+2\eta_{\pi S\pi}\mathcal{R}_{\langle\alpha\rangle\langle\mu}\sigma_{\nu\rangle}^{\,\,\,\,\alpha}+2\eta_{\pi S\phi}\mathcal{R}_{\langle\alpha\rangle\langle\mu}\xi_{\nu\rangle}^{\,\,\,\,\alpha}+2\sum_{i}\eta_{\pi\mathfrak{D}_{i}\pi}\mathcal{Z}^{\rho\sigma}\left(\partial_{\epsilon n}^{i}\Omega_{\rho\sigma}\right)\sigma_{\mu\nu}\\
&+\sum_{ab}\eta_{\pi\mathscr{J}_{a}\mathscr{J}_{b}}\nabla_{\langle\mu}\alpha_{a}\nabla_{\nu\rangle}\alpha_{b}+2\sum_{a}\eta_{\pi\mathscr{J}_{a}h}\nabla_{\langle\mu}\alpha_{a}N_{\nu\rangle}+2\sum_{a}\eta_{\pi\mathscr{J}_{a}q}\nabla_{\langle\mu}\alpha_{a}M_{\nu\rangle}+\eta_{\pi hh}N_{\langle\mu}N_{\nu\rangle}\\
&+2\eta_{\pi hq}N_{\langle\mu}M_{\nu\rangle}+\eta_{\pi qq}M_{\langle\mu}M_{\nu\rangle}+\eta_{\pi\pi\pi}\sigma_{\alpha\langle\mu}\sigma_{\nu\rangle}^{\,\,\,\,\alpha}+2\eta_{\pi\pi\phi}\sigma_{\alpha\langle\mu}\xi_{\nu\rangle}^{\,\,\,\,\alpha}+\eta_{\pi\phi\phi}\xi_{\alpha\langle\mu}\xi_{\nu\rangle}^{\,\,\,\,\alpha},
\end{aligned}
\label{194}
\end{equation}
where the second term on the right-hand side represents the nonlocal corrections, while the other second-order terms capture nonlinear effects arising from three-point correlations.

To derive a relaxation-type equation for $\pi_{\mu\nu}$ from Eq.~\eqref{194}, we substitute $2\sigma^{\rho\sigma}$ with $\eta^{-1}\pi^{\rho\sigma}$ in the second term on the right-hand side of Eq.~\eqref{194}. This approximation is justified since the term is already second order in spacetime gradients, rendering higher-order corrections negligible. We then have
\begin{eqnarray}
2\widetilde{\eta}\Delta_{\mu\nu\rho\sigma}D\sigma^{\rho\sigma}\simeq  \widetilde{\eta}\eta^{-1}\Delta_{\mu\nu\rho\sigma}D\pi^{\rho\sigma}-\widetilde{\eta}\eta^{-2}\beta\left[\left(\frac{\partial\eta}{\partial\beta}\Gamma-\sum_{a}\frac{\partial\eta}{\partial\alpha_{a}}\delta_{a}-2\frac{\partial\eta}{\partial\Omega_{\alpha\beta}}\mathcal{K}_{\alpha\beta}\right)-2\theta^{-1}\mathcal{Z}^{\alpha\beta}\frac{\partial\eta}{\partial S^{\alpha\beta}}\right]\theta\pi_{\mu\nu},
\label{195}
\end{eqnarray}
where Eqs.~\eqref{76}-\eqref{78} have been used. Introducing the coefficients
\begin{align} 
	&\tau_{\pi}=-\widetilde{\eta}\eta^{-1},\label{196}\\
	&\widetilde{\eta}_{\pi}=\tau_{\pi}\eta^{-1}\beta\left[\left(\frac{\partial\eta}{\partial\beta}\Gamma-\sum_{a}\frac{\partial\eta}{\partial\alpha_{a}}\delta_{a}-2\frac{\partial\eta}{\partial\Omega_{\alpha\beta}}\mathcal{K}_{\alpha\beta}\right)-2\theta^{-1}\mathcal{Z}^{\alpha\beta}\frac{\partial\eta}{\partial S^{\alpha\beta}}\right],\label{197}
\end{align}
and combining Eqs.~\eqref{194} and \eqref{195}, we arrive at the following relaxation equation for the shear-stress tensor:
\begin{eqnarray}
\begin{aligned}
	\pi_{\mu\nu}+\tau_{\pi}\dot{\pi}_{\mu\nu}=& 2\eta\sigma_{\mu\nu}+\widetilde{\eta}_{\pi}\theta\pi_{\mu\nu}+2\widetilde{\eta}\theta\Gamma\sigma_{\mu\nu}-4\widetilde{\eta}\mathcal{Z}^{\alpha\beta}\mathcal{D}_{\alpha\beta}\sigma_{\mu\nu}\\
	&+2\eta_{\pi p\pi}\theta\sigma_{\mu\nu}+\eta_{\pi SS}\mathcal{R}_{\langle\alpha\rangle\langle\mu}\mathcal{R}_{\nu\rangle}^{\,\,\,\,\alpha}+2\eta_{\pi S\pi}\mathcal{R}_{\langle\alpha\rangle\langle\mu}\sigma_{\nu\rangle}^{\,\,\,\,\alpha}+2\eta_{\pi S\phi}\mathcal{R}_{\langle\alpha\rangle\langle\mu}\xi_{\nu\rangle}^{\,\,\,\,\alpha}+2\sum_{i}\eta_{\pi\mathfrak{D}_{i}\pi}\mathcal{Z}^{\rho\sigma}\left(\partial_{\epsilon n}^{i}\Omega_{\rho\sigma}\right)\sigma_{\mu\nu}\\
	&+\sum_{ab}\eta_{\pi\mathscr{J}_{a}\mathscr{J}_{b}}\nabla_{\langle\mu}\alpha_{a}\nabla_{\nu\rangle}\alpha_{b}+2\sum_{a}\eta_{\pi\mathscr{J}_{a}h}\nabla_{\langle\mu}\alpha_{a}N_{\nu\rangle}+2\sum_{a}\eta_{\pi\mathscr{J}_{a}q}\nabla_{\langle\mu}\alpha_{a}M_{\nu\rangle}+\eta_{\pi hh}N_{\langle\mu}N_{\nu\rangle}\\
	&+2\eta_{\pi hq}N_{\langle\mu}M_{\nu\rangle}+\eta_{\pi qq}M_{\langle\mu}M_{\nu\rangle}+\eta_{\pi\pi\pi}\sigma_{\alpha\langle\mu}\sigma_{\nu\rangle}^{\,\,\,\,\alpha}+2\eta_{\pi\pi\phi}\sigma_{\alpha\langle\mu}\xi_{\nu\rangle}^{\,\,\,\,\alpha}+\eta_{\pi\phi\phi}\xi_{\alpha\langle\mu}\xi_{\nu\rangle}^{\,\,\,\,\alpha},
\end{aligned}
\label{198}
\end{eqnarray}
where we define $\dot{\pi}_{\mu\nu}=\Delta_{\mu\nu\rho\sigma}D\pi^{\rho\sigma}$.

\subsubsection{Second-order corrections to the bulk viscous pressure}

To accurately evaluate the bulk viscous pressure to second order, it is necessary to incorporate corresponding corrections to Eqs.~\eqref{101} and \eqref{102}. Defining $\Delta\epsilon=\langle\hat{\epsilon}\rangle_{1}+\langle\hat{\epsilon}\rangle_{2}$, $\Delta n_{a}=\langle\hat{n}_{a}\rangle_{1}+\langle\hat{n}_{a}\rangle_{2}$, and $\Delta S^{\alpha\beta}=\langle\hat{S}^{\alpha\beta}\rangle_{1}+\langle\hat{S}^{\alpha\beta}\rangle_{2}$, we obtain
\begin{eqnarray}
\begin{aligned}
	\langle\hat{p}\rangle_{l}\equiv & p\left(\langle\hat{\epsilon}\rangle_{l},\langle\hat{n}_{a}\rangle_{l},\langle\hat{S}^{\alpha\beta}\rangle_{l}\right)\\
	= & p\left(\epsilon-\Delta\epsilon,n_{a}-\Delta n_{a},S^{\alpha\beta}-\Delta S^{\alpha\beta}\right)\\
	= & p\left(\epsilon,n_{a},S^{\alpha\beta}\right)-\Gamma\Delta\epsilon-\sum_{a}\delta_{a}\Delta n_{a}-\mathcal{K}_{\alpha\beta}\Delta S^{\alpha\beta}+\frac{1}{2}\frac{\partial^{2}p}{\partial\epsilon^{2}}\left(\Delta\epsilon\right)^{2}+\frac{1}{2}\times2\sum_{a}\frac{\partial^{2}p}{\partial\epsilon\partial n_{a}}\Delta\epsilon\Delta n_{a}\\
	& +\frac{1}{2}\sum_{ab}\frac{\partial^{2}p}{\partial n_{a}\partial n_{b}}\Delta n_{a}\Delta n_{b}+\frac{1}{2}\times2\sum_{a}\frac{\partial^{2}p}{\partial n_{a}\partial S^{\alpha\beta}}\Delta n_{a}\Delta S^{\alpha\beta}+\frac{1}{2}\frac{\partial^{2}p}{\partial S^{\alpha\beta}\partial S^{\rho\sigma}}\Delta S^{\alpha\beta}\Delta S^{\rho\sigma}+\frac{1}{2}\times2\frac{\partial^{2}p}{\partial\epsilon\partial S^{\alpha\beta}}\Delta\epsilon\Delta S^{\alpha\beta}.
\end{aligned}
\label{199}
\end{eqnarray}

Then, the bulk-viscous pressure can be expressed as
\begin{eqnarray}
\begin{aligned}
\Pi=&\langle\hat{p}\rangle_{l}+\langle\hat{p}\rangle_{1}+\langle\hat{p}\rangle_{2}-p\left(\epsilon,n_{a},S^{\alpha\beta}\right)\\
    =&\langle\hat{p}\rangle_{1}+\langle\hat{p}\rangle_{2}-\Gamma\Delta\epsilon-\sum_{a}\delta_{a}\Delta n_{a}-\mathcal{K}_{\alpha\beta}\Delta S^{\alpha\beta}+\frac{1}{2}\frac{\partial^{2}p}{\partial\epsilon^{2}}\left(\Delta\epsilon\right)^{2}+\sum_{a}\frac{\partial^{2}p}{\partial\epsilon\partial n_{a}}\Delta\epsilon\Delta n_{a}\\
	&+\frac{1}{2}\sum_{ab}\frac{\partial^{2}p}{\partial n_{a}\partial n_{b}}\Delta n_{a}\Delta n_{b}+\sum_{a}\frac{\partial^{2}p}{\partial n_{a}\partial S^{\alpha\beta}}\Delta n_{a}\Delta S^{\alpha\beta}+\frac{1}{2}\frac{\partial^{2}p}{\partial S^{\alpha\beta}\partial S^{\rho\sigma}}\Delta S^{\alpha\beta}\Delta S^{\rho\sigma}+\frac{\partial^{2}p}{\partial\epsilon\partial S^{\alpha\beta}}\Delta\epsilon\Delta S^{\alpha\beta}.
\end{aligned}
\label{200}
\end{eqnarray}

Substituting $\Delta \epsilon$, $\Delta n_a$, and $\Delta S^{\alpha\beta}$ while neglecting higher-order terms, we obtain
\begin{eqnarray}
\begin{aligned}
	\Pi= & \langle\hat{P}^{*}\rangle_{1}+\langle\hat{P}^{*}\rangle_{2}+\frac{1}{2}\frac{\partial^{2}p}{\partial\epsilon^{2}}\langle\hat{\epsilon}\rangle_{1}^{2}+\sum_{a}\frac{\partial^{2}p}{\partial\epsilon\partial n_{a}}\langle\hat{\epsilon}\rangle_{1}\langle\hat{n}_{a}\rangle_{1}+\frac{1}{2}\sum_{ab}\frac{\partial^{2}p}{\partial n_{a}\partial n_{b}}\langle\hat{n}_{a}\rangle_{1}\langle\hat{n}_{b}\rangle_{1}\\
	& +\sum_{a}\frac{\partial^{2}p}{\partial n_{a}\partial S^{\alpha\beta}}\langle\hat{n}_{a}\rangle_{1}\langle\hat{S}^{\alpha\beta}\rangle_{1}+\frac{1}{2}\frac{\partial^{2}p}{\partial S^{\alpha\beta}\partial S^{\rho\sigma}}\langle\hat{S}^{\alpha\beta}\rangle_{1}\langle\hat{S}^{\rho\sigma}\rangle_{1}+\frac{\partial^{2}p}{\partial\epsilon\partial S^{\alpha\beta}}\langle\hat{\epsilon}\rangle_{1}\langle\hat{S}^{\alpha\beta}\rangle_{1},
\end{aligned}
\label{201}
\end{eqnarray}
where we have used the definition of $\hat{P}^*$ from Eq.~\eqref{81}.

Introducing the new coefficients
\begin{align}
	\zeta_{\epsilon p} & =\beta\int d^{4}x_{1}\left(\hat{\epsilon}\left(x\right),\hat{p}^{*}\left(x_{1}\right)\right)=-\frac{d}{d\omega}\text{Im}G_{\hat{\epsilon}\hat{p}^{*}}^{R}\left(\omega\right)\bigg|_{\omega=0},\label{202}\\
	\zeta_{\epsilon\mathfrak{D}_{i}} & =\beta\int d^{4}x_{1}\left(\hat{\epsilon}\left(x\right),\hat{\mathfrak{D}}_{i}\left(x_{1}\right)\right)=-\frac{d}{d\omega}\text{Im}G_{\hat{\epsilon}\hat{\mathfrak{D}}_{i}}^{R}\left(\omega\right)\bigg|_{\omega=0},\label{203}\\
	\zeta_{n_{a}p} & =\beta\int d^{4}x_{1}\left(\hat{n}_{a}\left(x\right),\hat{p}^{*}\left(x_{1}\right)\right)=-\frac{d}{d\omega}\text{Im}G_{\hat{n}_{a}\hat{p}^{*}}^{R}\left(\omega\right)\bigg|_{\omega=0},\label{204}\\
	\zeta_{n_{a}\mathfrak{D}_{i}} & =\beta\int d^{4}x_{1}\left(\hat{n}_{a}\left(x\right),\hat{\mathfrak{D}}_{i}\left(x_{1}\right)\right)=-\frac{d}{d\omega}\text{Im}G_{\hat{n}_{a}\hat{\mathfrak{D}}_{i}}^{R}\left(\omega\right)\bigg|_{\omega=0},\label{205}
\end{align}
and utilizing Eqs.~\eqref{90} and \eqref{93}, the first-order averages $\langle\hat{\epsilon}\rangle_{1},\langle\hat{n}_{a}\rangle_{1}$, and $\langle\hat{S}^{\alpha\beta}\rangle_{1}$ can be expressed as
\begin{align}
	\langle\hat{\epsilon}\rangle_{1}= & -\zeta_{\epsilon p}\theta+\sum_{i}\zeta_{\epsilon\mathfrak{D}_{i}}\mathcal{Z}^{\mu\nu}\partial_{\epsilon n}^{i}\Omega_{\mu\nu},\label{206}\\
	\langle\hat{n}_{a}\rangle_{1}= & -\zeta_{n_{a}p}\theta+\sum_{i}\zeta_{n_{a}\mathfrak{D}_{i}}\mathcal{Z}^{\mu\nu}\partial_{\epsilon n}^{i}\Omega_{\mu\nu},\label{207}\\
	\langle\hat{S}^{\alpha\beta}\rangle_{1}= & -\zeta_{SS}\mathcal{R}^{\langle\alpha\rangle\langle\beta\rangle}-\zeta_{S\phi}\xi^{\alpha\beta}.\label{208}
\end{align}
Then we obtain from Eqs.~\eqref{121}, \eqref{201}, \eqref{206}, \eqref{207}, and \eqref{208}
\begin{eqnarray}
\begin{aligned}
	\Pi= & -\zeta\theta+\sum_{i}\zeta_{p\mathfrak{D}_{i}}\mathcal{Z}^{\mu\nu}\partial_{\epsilon n}^{i}\Omega_{\mu\nu}+\zeta_{SS}\mathcal{K}_{\alpha\beta}\mathcal{R}^{\langle\alpha\rangle\langle\beta\rangle}+\zeta_{S\phi}\mathcal{K}_{\mu\nu}\xi^{\mu\nu}+\frac{1}{2}\frac{\partial^{2}p}{\partial\epsilon^{2}}\zeta_{\epsilon p}^{2}\theta^{2}\\
	&+\frac{1}{2}\frac{\partial^{2}p}{\partial\epsilon^{2}}\sum_{i}\sum_{j}\zeta_{\epsilon\mathfrak{D}_{i}}\zeta_{\epsilon\mathfrak{D}_{j}}\mathcal{Z}^{\mu\nu}\mathcal{Z}^{\alpha\beta}\partial_{\epsilon n}^{i}\Omega_{\mu\nu}\partial_{\epsilon n}^{j}\Omega_{\alpha\beta}-\frac{\partial^{2}p}{\partial\epsilon^{2}}\zeta_{\epsilon p}\theta\sum_{i}\zeta_{\epsilon\mathfrak{D}_{i}}\mathcal{Z}^{\mu\nu}\partial_{\epsilon n}^{i}\Omega_{\mu\nu}+\sum_{a}\frac{\partial^{2}p}{\partial\epsilon\partial n_{a}}\zeta_{\epsilon p}\zeta_{n_{a}p}\theta^{2}\\
	& -\sum_{a}\frac{\partial^{2}p}{\partial\epsilon\partial n_{a}}\zeta_{\epsilon p}\theta\sum_{j}\zeta_{n_{a}\mathfrak{D}_{j}}\mathcal{Z}^{\rho\sigma}\partial_{\epsilon n}^{j}\Omega_{\rho\sigma}-\sum_{a}\frac{\partial^{2}p}{\partial\epsilon\partial n_{a}}\zeta_{n_{a}p}\theta\sum_{i}\zeta_{\epsilon\mathfrak{D}_{i}}\mathcal{Z}^{\alpha\beta}\partial_{\epsilon n}^{i}\Omega_{\alpha\beta}\\
	& +\sum_{a}\frac{\partial^{2}p}{\partial\epsilon\partial n_{a}}\sum_{i}\sum_{j}\zeta_{\epsilon\mathfrak{D}_{i}}\mathcal{Z}^{\alpha\beta}\partial_{\epsilon n}^{i}\Omega_{\alpha\beta}\zeta_{n_{a}\mathfrak{D}_{j}}\mathcal{Z}^{\rho\sigma}\partial_{\epsilon n}^{j}\Omega_{\rho\sigma}+\frac{1}{2}\sum_{ab}\frac{\partial^{2}p}{\partial n_{a}\partial n_{b}}\zeta_{n_{a}p}\zeta_{n_{b}p}\theta^{2}\\
	& -\frac{1}{2}\sum_{ab}\frac{\partial^{2}p}{\partial n_{a}\partial n_{b}}\zeta_{n_{a}p}\theta\sum_{j}\zeta_{n_{b}\mathfrak{D}_{j}}\mathcal{Z}^{\rho\sigma}\partial_{\epsilon n}^{j}\Omega_{\rho\sigma}-\frac{1}{2}\sum_{ab}\frac{\partial^{2}p}{\partial n_{a}\partial n_{b}}\zeta_{n_{b}p}\theta\sum_{j}\zeta_{n_{a}\mathfrak{D}_{j}}\mathcal{Z}^{\alpha\beta}\partial_{\epsilon n}^{j}\Omega_{\alpha\beta}\\
	& +\frac{1}{2}\sum_{ab}\frac{\partial^{2}p}{\partial n_{a}\partial n_{b}}\sum_{i}\zeta_{n_{a}\mathfrak{D}_{i}}\mathcal{Z}^{\alpha\beta}\partial_{\epsilon n}^{i}\Omega_{\alpha\beta}\sum_{j}\zeta_{n_{b}\mathfrak{D}_{j}}\mathcal{Z}^{\rho\sigma}\partial_{\epsilon n}^{j}\Omega_{\rho\sigma}+\sum_{a}\frac{\partial^{2}p}{\partial n_{a}\partial S^{\alpha\beta}}\zeta_{n_{a}p}\theta\zeta_{SS}\mathcal{R}^{\langle\alpha\rangle\langle\beta\rangle}\\
	& +\sum_{a}\frac{\partial^{2}p}{\partial n_{a}\partial S^{\alpha\beta}}\zeta_{n_{a}p}\theta\zeta_{S\phi}\xi^{\alpha\beta}-\sum_{a}\frac{\partial^{2}p}{\partial n_{a}\partial S^{\alpha\beta}}\zeta_{SS}\mathcal{R}^{\langle\alpha\rangle\langle\beta\rangle}\sum_{j}\zeta_{n_{a}\mathfrak{D}_{j}}\mathcal{Z}^{\rho\sigma}\partial_{\epsilon n}^{j}\Omega_{\rho\sigma}\\
	& -\sum_{a}\frac{\partial^{2}p}{\partial n_{a}\partial S^{\alpha\beta}}\zeta_{S\phi}\xi^{\alpha\beta}\sum_{j}\zeta_{n_{a}\mathfrak{D}_{j}}\mathcal{Z}^{\rho\sigma}\partial_{\epsilon n}^{j}\Omega_{\rho\sigma}+\frac{1}{2}\frac{\partial^{2}p}{\partial S^{\alpha\beta}\partial S^{\rho\sigma}}\zeta_{SS}\mathcal{R}^{\langle\alpha\rangle\langle\beta\rangle}\zeta_{SS}\mathcal{R}^{\langle\rho\rangle\langle\sigma\rangle}+\frac{1}{2}\frac{\partial^{2}p}{\partial S^{\alpha\beta}\partial S^{\rho\sigma}}\zeta_{SS}\mathcal{R}^{\langle\alpha\rangle\langle\beta\rangle}\zeta_{S\phi}\xi^{\rho\sigma}\\
	& +\frac{1}{2}\frac{\partial^{2}p}{\partial S^{\alpha\beta}\partial S^{\rho\sigma}}\zeta_{S\phi}\xi^{\alpha\beta}\zeta_{SS}\mathcal{R}^{\langle\rho\rangle\langle\sigma\rangle}+\frac{1}{2}\frac{\partial^{2}p}{\partial S^{\alpha\beta}\partial S^{\rho\sigma}}\zeta_{S\phi}\xi^{\alpha\beta}\zeta_{S\phi}\xi^{\rho\sigma}+\frac{\partial^{2}p}{\partial\epsilon\partial S^{\alpha\beta}}\zeta_{\epsilon p}\theta\zeta_{SS}\mathcal{R}^{\langle\alpha\rangle\langle\beta\rangle}+\frac{\partial^{2}p}{\partial\epsilon\partial S^{\alpha\beta}}\zeta_{\epsilon p}\theta\zeta_{S\phi}\xi^{\alpha\beta}\\
	& -\frac{\partial^{2}p}{\partial\epsilon\partial S^{\alpha\beta}}\zeta_{SS}\mathcal{R}^{\langle\alpha\rangle\langle\beta\rangle}\sum_{i}\zeta_{\epsilon\mathfrak{D}_{i}}\mathcal{Z}^{\mu\nu}\partial_{\epsilon n}^{i}\Omega_{\mu\nu}-\frac{\partial^{2}p}{\partial\epsilon\partial S^{\alpha\beta}}\zeta_{S\phi}\xi^{\alpha\beta}\sum_{i}\zeta_{\epsilon\mathfrak{D}_{i}}\mathcal{Z}^{\mu\nu}\partial_{\epsilon n}^{i}\Omega_{\mu\nu}+\langle\hat{P}^{*}\rangle_{2}.
\end{aligned}
\label{209}
\end{eqnarray}

Substituting Eq.~\eqref{90} into Eq.~\eqref{95} and applying Curie's theorem, we find
\begin{equation}
\begin{aligned}
	\langle\hat{P}^{*}\left(x\right)\rangle_{2}^{1}= & -\int d^{4}x_{1}\left(\hat{p}^{*}\left(x\right),\hat{p}^{*}\left(x_{1}\right)\right)\beta\left(x_{1}\right)\theta\left(x_{1}\right)+\sum_{i}\int d^{4}x_{1}\left(\hat{p}^{*}\left(x\right),\hat{\mathfrak{D}}_{i}\left(x_{1}\right)\right)\beta\left(x_{1}\right)\mathcal{Z}^{\mu\nu}\left(x_{1}\right)\partial_{\epsilon n}^{i}\Omega_{\mu\nu}\left(x_{1}\right)\\
	&-\mathcal{K}_{\alpha\beta}\left(x\right)\int d^{4}x_{1}\left(\hat{S}^{\alpha\beta}\left(x\right),\hat{S}^{\rho\sigma}\left(x_{1}\right)\right)\beta\left(x_{1}\right)\mathcal{R}_{\rho\sigma}\left(x_{1}\right)-\mathcal{K}_{\mu\nu}\left(x\right)\int d^{4}x_{1}\left(\hat{S}^{\mu\nu}\left(x\right),\hat{\phi}^{\rho\sigma}\left(x_{1}\right)\right)\beta\left(x_{1}\right)\xi_{\rho\sigma}\left(x_{1}\right)\\
	&+\zeta\left(x\right)\theta\left(x\right)-\sum_{i}\zeta_{p\mathfrak{D}_{i}}\left(x\right)\mathcal{Z}^{\mu\nu}\left(x\right)\partial_{\epsilon n}^{i}\Omega_{\mu\nu}\left(x\right)-\zeta_{SS}\left(x\right)\mathcal{K}_{\alpha\beta}\left(x\right)\mathcal{R}^{\langle\alpha\rangle\langle\beta\rangle}\left(x\right)-\zeta_{S\phi}\left(x\right)\mathcal{K}_{\mu\nu}\left(x\right)\xi^{\mu\nu}\left(x\right),
\end{aligned}
\label{210}
\end{equation}
where we used the first-order relation $\langle\hat{P}^{*}\rangle_{1}$ from Eq.~\eqref{121}. Substituting the two-point correlation functions from Eqs.~\eqref{142},\eqref{143}, and \eqref{155} into Eq.~\eqref{210}, we obtain
\begin{eqnarray}
\begin{aligned}
	&\langle\hat{P}^{*}\left(x\right)\rangle_{2}^{1}\\
	=&-\int d^{4}x_{1}\left(\hat{p}^{*}\left(x\right),\hat{p}^{*}\left(x_{1}\right)\right)\beta\left(x_{1}\right)\theta\left(x_{1}\right)+\sum_{i}\int d^{4}x_{1}\left(\hat{p}^{*}(x),\hat{\mathfrak{D}}_{i}(x_{1})\right)\beta(x_{1})\mathcal{Z}^{\mu\nu}(x_{1})\partial_{\epsilon n}^{i}\Omega_{\mu\nu}(x_{1})\\
	& -\frac{1}{3}\mathcal{K}^{\langle\mu\rangle\langle\nu\rangle}\left(x\right)\int d^{4}x_{1}\left(\hat{S}^{\lambda\eta}\left(x\right),\hat{S}_{\lambda\eta}\left(x_{1}\right)\right)\beta\left(x_{1}\right)\mathcal{R}_{\langle\mu\rangle\langle\nu\rangle}\left(x_{1}\right)-\frac{1}{3}\mathcal{K}^{\langle a\rangle\langle b\rangle}\left(x\right)\int d^{4}x_{1}\left(\hat{S}^{\lambda\eta}\left(x\right),\hat{\phi}_{\lambda\eta}\left(x_{1}\right)\right)\beta\left(x_{1}\right)\xi_{ab}\left(x_{1}\right)\\
	&+\zeta\left(x\right)\theta\left(x\right)-\sum_{i}\zeta_{p\mathfrak{D}_{i}}\left(x\right)\mathcal{Z}^{\mu\nu}\left(x\right)\partial_{\epsilon n}^{i}\Omega_{\mu\nu}\left(x\right)-\zeta_{SS}\left(x\right)\mathcal{K}_{\alpha\beta}\left(x\right)\mathcal{R}^{\langle\alpha\rangle\langle\beta\rangle}\left(x\right)-\zeta_{S\phi}\left(x\right)\mathcal{K}_{\mu\nu}\left(x\right)\xi^{\mu\nu}\left(x\right).
\end{aligned}
\label{211}
\end{eqnarray}
Performing a Taylor expansion of the hydrodynamic quantities around $x_1=x$ and retaining linear terms, we obtain
\begin{align}
	\hat{p}^{*}\left(x_{1}\right)= & \hat{p}^{*}\left(x_{1}\right)\bigg|_{x_{1}=x}+\left(x_{1}-x\right)^{\tau}\frac{\partial\hat{p}^{*}\left(x_{1}\right)}{\partial x_{1}^{\tau}}\bigg|_{x_{1}=x},\label{212}\\
	\sum_{i}\hat{\mathfrak{D}}_{i}\left(x_{1}\right)\partial_{\epsilon n}^{i}\Omega_{\alpha\beta}(x_{1})= & \sum_{i}\hat{\mathfrak{D}}_{i}\left(x_{1}\right)\partial_{\epsilon n}^{i}\Omega_{\alpha\beta}(x_{1})\bigg|_{x_{1}=x}+\left(x_{1}-x\right)^{\tau}\frac{\partial\left(\sum_{i}\hat{\mathfrak{D}}_{i}\partial_{\epsilon n}^{i}\Omega_{\alpha\beta}\right)}{\partial x_{1}^{\tau}}\bigg|_{x_{1}=x},\label{213}\\
	\hat{S}_{\lambda\eta}\left(x_{1}\right)= & \hat{S}_{\lambda\eta}\left(x_{1}\right)\bigg|_{x_{1}=x}+\left(x_{1}-x\right)^{\tau}\frac{\partial\hat{S}_{\lambda\eta}\left(x_{1}\right)}{\partial x_{1}^{\tau}}\bigg|_{x_{1}=x},\label{214}\\
	\hat{\phi}_{\lambda\eta}\left(x_{1}\right)= & \hat{\phi}_{\lambda\eta}\left(x_{1}\right)\bigg|_{x_{1}=x}+\left(x_{1}-x\right)^{\tau}\frac{\partial\hat{\phi}_{\lambda\eta}\left(x_{1}\right)}{\partial x_{1}^{\tau}}\bigg|_{x_{1}=x},\label{215}
\end{align}
where
\begin{align}
	\hat{p}^{*}\left(x_{1}\right)\bigg|_{x_{1}=x}= & -\frac{1}{3}\Delta_{\mu\nu}\left(x\right)\hat{T}^{\mu\nu}\left(x_{1}\right)-\Gamma\left(x\right)u_{\mu}\left(x\right)u_{\nu}\left(x\right)\hat{T}^{\mu\nu}\left(x_{1}\right)-\sum_{a}\delta_{a}\left(x\right)u_{\mu}\left(x\right)\hat{N}_{a}^{\mu}\left(x_{1}\right),\label{216}\\
	\sum_{i}\hat{\mathfrak{D}}_{i}(x_{1})\partial_{\epsilon n}^{i}\Omega_{\alpha\beta}(x_{1})\bigg|_{x_{1}=x}= & \hat{\epsilon}\left(x_{1}\right)\frac{\partial\Omega_{\alpha\beta}}{\partial\epsilon}\left(x\right)+\sum_{c}\hat{n}_{c}\left(x_{1}\right)\frac{\partial\Omega_{\alpha\beta}}{\partial n_{c}}\left(x\right),\label{217}\\
	\hat{S}_{\lambda\eta}\left(x_{1}\right)\bigg|_{x_{1}=x}= & u^{\mu}\left(x\right)\hat{S}_{\mu\lambda\eta}\left(x_{1}\right),\label{218}\\
	\hat{\phi}_{\lambda\eta}\left(x_{1}\right)\bigg|_{x_{1}=x}= & \mydelta_{\lambda\eta\rho\sigma}\left(x\right)\hat{T}^{\rho\sigma}\left(x_{1}\right),\label{219}\\
	\frac{\partial\hat{p}^{*}\left(x_{1}\right)}{\partial x_{1}^{\tau}}\bigg|_{x_{1}=x}= & 2\left[\frac{1}{3}-\Gamma\left(x\right)\right]\hat{h}^{\mu}\left(x_{1}\right)\frac{\partial u_{\mu}\left(x_{1}\right)}{\partial x_{1}^{\tau}}\bigg|_{x_{1}=x}-\hat{\epsilon}\left(x_{1}\right)\biggl[\frac{\partial\Gamma\left(x\right)}{\partial\epsilon\left(x\right)}\frac{\partial\epsilon\left(x_{1}\right)}{\partial x_{1}^{\tau}}\bigg|_{x_{1}=x}+\sum_{a}\frac{\partial\Gamma\left(x\right)}{\partial n_{a}\left(x\right)}\frac{\partial n_{a}\left(x_{1}\right)}{\partial x_{1}^{\tau}}\bigg|_{x_{1}=x}\nonumber\\
	& +\frac{\partial\Gamma\left(x\right)}{\partial S^{\alpha\beta}\left(x\right)}\frac{\partial S^{\alpha\beta}\left(x_{1}\right)}{\partial x_{1}^{\tau}}\bigg|_{x_{1}=x}\biggr]-\sum_{a}\hat{n}_{a}\left(x\right)\biggl[\frac{\partial\delta_{a}\left(x\right)}{\partial\epsilon\left(x\right)}\frac{\partial\epsilon\left(x_{1}\right)}{\partial x_{1}^{\tau}}\bigg|_{x_{1}=x}+\sum_{b}\frac{\partial\delta_{a}\left(x\right)}{\partial n_{b}\left(x\right)}\frac{\partial n_{b}\left(x_{1}\right)}{\partial x_{1}}\bigg|_{x_{1}=x}\nonumber\\
	& +\frac{\partial\delta_{a}\left(x\right)}{\partial S^{\alpha\beta}\left(x\right)}\frac{\partial S^{\alpha\beta}\left(x_{1}\right)}{\partial x_{1}^{\tau}}\bigg|_{x_{1}=x}\biggr]-\sum_{a}\delta_{a}\left(x\right)\hat{j}_{a}^{\mu}\left(x_{1}\right)\frac{\partial u_{\mu}\left(x_{1}\right)}{\partial x_{1}^{\tau}}\bigg|_{x_{1}=x},\label{220}\\
	\frac{\partial\left(\sum_{i}\hat{\mathfrak{D}}_{i}\partial_{\epsilon n}^{i}\Omega_{\alpha\beta}\right)\left(x_{1}\right)}{\partial x_{1}^{\tau}}\bigg|_{x_{1}=x}= & 2\hat{h}^{\mu}\left(x_{1}\right)\frac{\partial\Omega_{\alpha\beta}\left(x\right)}{\partial\epsilon\left(x\right)}\frac{\partial u_{\mu}\left(x_{1}\right)}{\partial x_{1}^{\tau}}\bigg|_{x_{1}=x}+\hat{\epsilon}\left(x_{1}\right)\biggl[\frac{\partial^{2}\Omega_{\alpha\beta}\left(x\right)}{\partial\epsilon\left(x\right)^{2}}\frac{\partial\epsilon\left(x_{1}\right)}{\partial x_{1}^{\tau}}\bigg|_{x_{1}=x}\nonumber\\
	& +\sum_{a}\frac{\partial^{2}\Omega_{\alpha\beta}\left(x\right)}{\partial n_{a}\left(x\right)\partial\epsilon\left(x\right)}\frac{\partial n_{a}\left(x_{1}\right)}{\partial x_{1}^{\tau}}\bigg|_{x_{1}=x}+\frac{\partial^{2}\Omega_{\alpha\beta}\left(x\right)}{\partial S^{\rho\sigma}\left(x\right)\partial\epsilon\left(x\right)}\frac{\partial S^{\rho\sigma}\left(x_{1}\right)}{\partial x_{1}^{\tau}}\bigg|_{x_{1}=x}\biggr]\nonumber\\
	& +\sum_{c}\hat{n}_{c}\left(x_{1}\right)\biggl[\frac{\partial^{2}\Omega_{\alpha\beta}\left(x\right)}{\partial\epsilon\left(x\right)\partial n_{c}\left(x\right)}\frac{\partial\epsilon\left(x_{1}\right)}{\partial x_{1}^{\tau}}\bigg|_{x_{1}=x}+\sum_{a}\frac{\partial^{2}\Omega_{\alpha\beta}\left(x\right)}{\partial n_{a}\left(x\right)\partial n_{c}\left(x\right)}\frac{\partial n_{a}\left(x_{1}\right)}{\partial x_{1}^{\tau}}\bigg|_{x_{1}=x}\nonumber\\
	& +\frac{\partial^{2}\Omega_{\alpha\beta}\left(x\right)}{\partial S^{\rho\sigma}\left(x\right)\partial n_{c}\left(x\right)}\frac{\partial S^{\rho\sigma}\left(x_{1}\right)}{\partial x_{1}^{\tau}}\bigg|_{x_{1}=x}\biggr]+\sum_{c}\hat{j}_{c}^{\mu}\left(x_{1}\right)\frac{\partial\Omega_{\alpha\beta}\left(x\right)}{\partial n_{c}\left(x\right)}\frac{\partial u_{\mu}\left(x_{1}\right)}{\partial x_{1}^{\tau}}\bigg|_{x_{1}=x},\label{221}\\
	\frac{\partial\hat{S}_{\lambda\eta}\left(x_{1}\right)}{\partial x_{1}^{\tau}}\bigg|_{x_{1}=x}= & u_{\lambda}\left(x\right)\hat{S}_{\eta\mu}\left(x_{1}\right)\frac{\partial u^{\mu}\left(x_{1}\right)}{\partial x_{1}^{\tau}}\bigg|_{x_{1}=x}+u_{\eta}\left(x\right)\hat{S}_{\mu\lambda}\left(x_{1}\right)\frac{\partial u^{\mu}\left(x_{1}\right)}{\partial x_{1}^{\tau}}\bigg|_{x_{1}=x}+\hat{\varpi}_{\mu\lambda\eta}\left(x_{1}\right)\frac{\partial u^{\mu}\left(x_{1}\right)}{\partial x_{1}^{\tau}}\bigg|_{x_{1}=x},\label{222}\\
	\frac{\partial\hat{\phi}_{\lambda\eta}\left(x_{1}\right)}{\partial x_{1}^{\tau}}\bigg|_{x_{1}=x}= & -2\mydelta_{\lambda\eta\rho\beta}\left(x\right)\hat{q}^{\rho}\left(x_{1}\right)\frac{\partial u^{\beta}\left(x_{1}\right)}{\partial x_{1}}\bigg|_{x_{1}=x},\label{223}
\end{align}
where we have used Eqs.~\eqref{54}, \eqref{55}, and \eqref{79}, with $u_{\mu}\left(x\right)\approx u_{\mu}\left(x_{1}\right)$ for a higher-order approximation. All operators in Eqs.\eqref{220}-\eqref{223} are evaluated at $x_1$, while hydrodynamic quantities are evaluated at $x$.

Substituting Eqs.\eqref{212}-\eqref{215} into Eq.\eqref{211} and expanding the thermodynamic forces around $x_1=x$, we obtain up to the second order in gradients
\begin{eqnarray}
\begin{aligned}
	\langle\hat{P}^{*}\left(x\right)\rangle_{2}^{1}= & -\beta\left(x\right)\theta\left(x\right)\int d^{4}x_{1}\left(\hat{p}^{*}(x),\frac{\partial\hat{p}^{*}\left(x_{1}\right)}{\partial x_{1}^{\tau}}\bigg|_{x_{1}=x}\right)\left(x_{1}-x\right)^{\tau}\\
	& -\frac{\partial}{\partial x_{1}^{\tau}}\left[\beta\left(x_{1}\right)\theta\left(x_{1}\right)\right]\bigg|_{x_{1}=x}\int d^{4}x_{1}\left(\hat{p}^{*}(x),\hat{p}^{*}\left(x_{1}\right)\bigg|_{x_{1}=x}\right)\left(x_{1}-x\right)^{\tau}\\
	& +\beta\left(x\right)\mathcal{Z}^{\alpha\beta}\left(x\right)\int d^{4}x_{1}\left(\hat{p}^{*}\left(x\right),\frac{\partial\sum_{i}\hat{\mathfrak{D}_{i}}\partial_{\epsilon n}^{i}\Omega_{\alpha\beta}\left(x_{1}\right)}{\partial x_{1}^{\tau}}\bigg|_{x_{1}=x}\right)\left(x_{1}-x\right)^{\tau}\\
	& +\frac{\partial}{\partial x_{1}^{\tau}}\left[\beta\left(x_{1}\right)\mathcal{Z}^{\alpha\beta}\left(x_{1}\right)\right]\bigg|_{x_{1}=x}\int d^{4}x_{1}\left(\hat{p}^{*}\left(x\right),\sum_{i}\hat{\mathfrak{D}_{i}}\partial_{\epsilon n}^{i}\Omega_{\alpha\beta}\left(x_{1}\right)\bigg|_{x_{1}=x}\right)\left(x_{1}-x\right)^{\tau}\\
	&-\frac{1}{3}\mathcal{K}^{\langle\mu\rangle\langle\nu\rangle}\left(x\right)\beta\left(x\right)\mathcal{R}_{\langle\mu\rangle\langle\nu\rangle}\left(x\right)\int d^{4}x_{1}\left(\hat{S}^{\lambda\eta}\left(x\right),\frac{\partial\hat{S}_{\lambda\eta}\left(x_{1}\right)}{\partial x_{1}^{\tau}}\bigg|_{x_{1}=x}\right)\left(x_{1}-x\right)^{\tau}\\
	&-\frac{1}{3}\mathcal{K}^{\langle\mu\rangle\langle\nu\rangle}\left(x\right)\frac{\partial}{\partial x_{1}^{\tau}}\left[\beta\left(x_{1}\right)\mathcal{R}_{\langle\mu\rangle\langle\nu\rangle}\left(x_{1}\right)\right]\bigg|_{x_{1}=x}\int d^{4}x_{1}\left(\hat{S}^{\lambda\eta}\left(x\right),\hat{S}_{\lambda\eta}\left(x_{1}\right)\bigg|_{x_{1}=x}\right)\left(x_{1}-x\right)^{\tau}\\
	&-\frac{1}{3}\mathcal{K}^{\langle\mu\rangle\langle\nu\rangle}\left(x\right)\beta\left(x\right)\xi_{\mu\nu}\left(x\right)\int d^{4}x_{1}\left(\hat{S}^{\lambda\eta}\left(x\right),\frac{\partial\hat{\phi}_{\lambda\eta}\left(x_{1}\right)}{\partial x_{1}^{\tau}}\bigg|_{x_{1}=x}\right)\left(x_{1}-x\right)^{\tau}\\
	&-\frac{1}{3}\mathcal{K}^{\langle\mu\rangle\langle\nu\rangle}\left(x\right)\frac{\partial}{\partial x_{1}^{\tau}}\left[\beta\left(x_{1}\right)\xi_{\mu\nu}\left(x_{1}\right)\right]\bigg|_{x_{1}=x}\int d^{4}x_{1}\left(\hat{S}^{\lambda\eta}\left(x\right),\hat{\phi}_{\lambda\eta}\left(x_{1}\right)\bigg|_{x_{1}=x}\right)\left(x_{1}-x\right)^{\tau}
\end{aligned}
\label{224}
\end{eqnarray}
where we used the following relation to eliminate the first-order terms:
\begin{align}
	\zeta & =\beta\int d^{4}x_{1}\biggl(\hat{p}^{*}\left(x\right),\hat{p}^{*}\left(x_{1}\right)\bigg|_{x_{1}=x}\biggr),\label{225}\\
	\zeta_{p\mathfrak{D}_{i}} & =\beta\int d^{4}x_{1}\left(\hat{p}^{*}\left(x\right),\hat{\mathfrak{D}}_{i}\left(x_{1}\right)\bigg|_{x_{1}=x}\right),\label{226}\\
	\zeta_{SS} & =-\frac{\beta}{3}\int d^{4}x_{1}\left(\hat{S}^{\lambda\eta}\left(x\right),\hat{S}_{\lambda\eta}\left(x_{1}\right)\bigg|_{x_{1}=x}\right),\label{227}\\
	\zeta_{S\phi} & =-\frac{\beta}{3}\int d^{4}x_{1}\left(\hat{S}^{\lambda\eta}\left(x\right),\hat{\phi}_{\lambda\eta}\left(x_{1}\right)\bigg|_{x_{1}=x}\right).\label{228}
\end{align}
Substituting Eqs.~\eqref{220}-\eqref{223} into Eq.~\eqref{224}, applying Curie's theorem, and imposing the orthogonality condition $u^{\mu}\hat{S}_{\mu\nu}=0$, we obtain
\begin{eqnarray}
\begin{aligned}
	\langle\hat{P}^{*}\rangle_{2}^{1}= & \widetilde{\zeta}_{p\epsilon}\theta\left(\frac{\partial\Gamma}{\partial\epsilon}D\epsilon+\sum_{a}\frac{\partial\Gamma}{\partial n_{a}}Dn_{a}+\frac{\partial\Gamma}{\partial S^{\alpha\beta}}DS^{\alpha\beta}\right)\\
	&+\sum_{a}\widetilde{\zeta}_{pn_{a}}\theta\left(\frac{\partial\delta_{a}}{\partial\epsilon}D\epsilon+\sum_{b}\frac{\partial\delta_{a}}{\partial n_{b}}Dn_{b}+\frac{\partial\delta_{a}}{\partial S^{\alpha\beta}}DS^{\alpha\beta}\right)-\widetilde{\zeta}\beta^{-1}D\left(\beta\theta\right)\\
	&+\widetilde{\zeta}_{p\epsilon}\mathcal{Z}^{\alpha\beta}\left(\frac{\partial^{2}\Omega_{\alpha\beta}}{\partial\epsilon^{2}}D\epsilon+\sum_{a}\frac{\partial^{2}\Omega_{\alpha\beta}}{\partial n_{a}\partial\epsilon}Dn_{a}+\frac{\partial^{2}\Omega_{\alpha\beta}}{\partial S^{\rho\sigma}\partial\epsilon}DS^{\rho\sigma}\right)\\
	&+\sum_{c}\widetilde{\zeta}_{pn_{c}}\mathcal{Z}^{\alpha\beta}\left(\frac{\partial^{2}\Omega_{\alpha\beta}}{\partial\epsilon\partial n_{c}}D\epsilon+\sum_{a}\frac{\partial^{2}\Omega_{\alpha\beta}}{\partial n_{a}\partial n_{c}}Dn_{a}+\frac{\partial^{2}\Omega_{\alpha\beta}}{\partial S^{\rho\sigma}\partial n_{c}}DS^{\rho\sigma}\right)\\
	&+\sum_{i}\widetilde{\zeta}_{p\mathfrak{D}_{i}}\beta^{-1}D\left(\beta\mathcal{Z}^{\alpha\beta}\right)\partial_{\epsilon n}^{i}\Omega_{\alpha\beta}+\widetilde{\zeta}_{SS}\beta^{-1}\mathcal{K}^{\langle\mu\rangle\langle\nu\rangle}D\left(\beta\mathcal{R}_{\langle\mu\rangle\langle\nu\rangle}\right)+\widetilde{\zeta}_{S\phi}\mathcal{K}^{\langle\mu\rangle\langle\nu\rangle}\beta^{-1}D\left(\beta\xi_{\mu\nu}\right),
\end{aligned}
\label{229}
\end{eqnarray}
where we define
\begin{align} 
	&\widetilde{\zeta}=i\frac{d}{d\omega}\zeta(\omega)\bigg|_{\omega=0}=-\frac{1}{2}\frac{d^{2}}{d\omega^{2}}\mathrm{Re}G_{\hat{p}^{*}\hat{p}^{*}}^{R}(\omega)\bigg|_{\omega=0},\label{230}\\
	&\widetilde{\zeta}_{p\epsilon}=i\frac{d}{d\omega}\zeta_{p\epsilon}(\omega)\bigg|_{\omega=0}=-\frac{1}{2}\frac{d^{2}}{d\omega^{2}}\mathrm{Re}G_{\hat{p}^{*}\hat{\epsilon}}^{R}(\omega)\bigg|_{\omega=0},\label{231}\\
	&\widetilde{\zeta}_{pn_{a}}=i\frac{d}{d\omega}\zeta_{pn_{a}}(\omega)\bigg|_{\omega=0}=-\frac{1}{2}\frac{d^{2}}{d\omega^{2}}\mathrm{Re}G_{\hat{p}^{*}\hat{n}_{a}}^{R}(\omega)\bigg|_{\omega=0},\label{232}\\
	&\widetilde{\zeta}_{p\mathfrak{D}_{i}}=i\frac{d}{d\omega}\zeta_{p\mathfrak{D}_{i}}\left(\omega\right)\bigg|_{\omega=0}=-\frac{1}{2}\frac{d^{2}}{d\omega^{2}}\text{Re}G_{\hat{p}^{*}\mathfrak{D}_{i}}^{R}\left(\omega\right)\bigg|_{\omega=0},\label{233}\\
	&\widetilde{\zeta}_{SS}=i\frac{d}{d\omega}\zeta_{SS}\left(\omega\right)\bigg|_{\omega=0}=\frac{1}{6}\frac{d^{2}}{d\omega^{2}}\text{Re}G_{\hat{S}^{\lambda\eta}\hat{S}_{\lambda\eta}}^{R}\left(\omega\right)\bigg|_{\omega=0},\label{234}\\
	&\widetilde{\zeta}_{S\phi}=i\frac{d}{d\omega}\zeta_{S\phi}\left(\omega\right)\bigg|_{\omega=0}=\frac{1}{6}\frac{d^{2}}{d\omega^{2}}\text{Re}G_{\hat{S}^{\lambda\eta}\hat{\phi}_{\lambda\eta}}^{R}\left(\omega\right)\bigg|_{\omega=0},\label{235}
\end{align}
with the frequency-dependent transport coefficients $\zeta\left(\omega\right)$,$\zeta_{p\epsilon}\left(\omega\right)$,$\zeta_{pn_{a}}\left(\omega\right)$,$\zeta_{p\mathfrak{D}_{i}}\left(\omega\right)$,$\zeta_{SS}\left(\omega\right)$, and $\zeta_{S\phi}\left(\omega\right)$ are expressed as
\begin{align}
	\zeta\left(\omega\right) & =\beta\int d^{4}x_{1}\int_{-\infty}^{t}e^{i\omega\left(t-t_{1}\right)}\biggl(\hat{p}^{*}\left(\boldsymbol{x},t\right),\hat{p}^{*}\left(\boldsymbol{x}_{1},t_{1}\right)\biggr),\label{236}\\
	\zeta_{p\epsilon}\left(\omega\right) & =\beta\int d^{4}x_{1}\int_{-\infty}^{t}e^{i\omega\left(t-t_{1}\right)}\biggl(\hat{p}^{*}\left(\boldsymbol{x},t\right),\hat{\epsilon}\left(\boldsymbol{x}_{1},t_{1}\right)\biggr),\label{237}\\
	\zeta_{pn_{a}}\left(\omega\right) & =\beta\int d^{4}x_{1}\int_{-\infty}^{t}e^{i\omega\left(t-t_{1}\right)}\biggl(\hat{p}^{*}\left(\boldsymbol{x},t\right),\hat{n}_{a}\left(\boldsymbol{x}_{1},t_{1}\right)\biggr),\label{238}\\
	\zeta_{p\mathfrak{D}_{i}}\left(\omega\right) & =\beta\int d^{4}x_{1}\int_{-\infty}^{t}e^{i\omega\left(t-t_{1}\right)}\left(\hat{p}^{*}\left(x\right),\hat{\mathfrak{D}}_{i}\left(x_{1}\right)\right),\label{239}\\
	\zeta_{SS}\left(\omega\right) & =-\frac{1}{3}\beta\int d^{4}x_{1}\int_{-\infty}^{t}e^{i\omega\left(t-t_{1}\right)}\left(\hat{S}^{\lambda\eta}\left(\boldsymbol{x},t\right),\hat{S}_{\lambda\eta}\left(\boldsymbol{x}_{1},t_{1}\right)\right),\label{240}\\
	\zeta_{S\phi}= & -\frac{\beta}{3}\int d^{4}x_{1}\int_{-\infty}^{t}e^{i\omega\left(t-t_{1}\right)}\left(\hat{S}^{\lambda\eta}\left(\boldsymbol{x},t\right),\hat{\phi}_{\lambda\eta}\left(\boldsymbol{x}_{1},t_{1}\right)\right).\label{241}
\end{align}
Employing Eqs.~\eqref{71},\eqref{76}-\eqref{78} to eliminate the derivatives $D\beta,D\epsilon,Dn_{a}$, and $DS^{\alpha\beta}$ in Eq.~\eqref{229}, denoting
\begin{align}
	\widetilde{\Gamma} & =\frac{\partial\Gamma}{\partial\epsilon}{w}+\sum_{a}\frac{\partial\Gamma}{\partial n_{a}}n_{a}+\frac{\partial\Gamma}{\partial S^{\alpha\beta}}\left[S^{\alpha\beta}+\theta^{-1}\left(u^{\alpha}\partial_{\lambda}S^{\beta\lambda}+S^{\beta\lambda}\partial_{\lambda}u^{\alpha}+u^{\beta}\partial_{\lambda}S^{\lambda\alpha}+S^{\lambda\alpha}\partial_{\lambda}u^{\beta}\right)\right],\label{242}\\
	\widetilde{\delta}_{a} & =\frac{\partial\delta_{a}}{\partial\epsilon}{w}+\sum_{b}\frac{\partial\delta_{a}}{\partial n_{b}}n_{b}+\frac{\partial\delta_{a}}{\partial S^{\alpha\beta}}\left[S^{\alpha\beta}+\theta^{-1}\left(u^{\alpha}\partial_{\lambda}S^{\beta\lambda}+S^{\beta\lambda}\partial_{\lambda}u^{\alpha}+u^{\beta}\partial_{\lambda}S^{\lambda\alpha}+S^{\lambda\alpha}\partial_{\lambda}u^{\beta}\right)\right],\label{243}\\
	\zeta_{p\epsilon}^{*} & =\frac{\partial^{2}\Omega_{\alpha\beta}}{\partial\epsilon^{2}}{w}+\sum_{a}\frac{\partial^{2}\Omega_{\alpha\beta}}{\partial n_{a}\partial\epsilon}n_{a}+\frac{\partial^{2}\Omega_{\alpha\beta}}{\partial S^{\rho\sigma}\partial\epsilon}\left[S^{\rho\sigma}+\theta^{-1}\left(u^{\rho}\partial_{\lambda}S^{\sigma\lambda}+S^{\sigma\lambda}\partial_{\lambda}u^{\rho}+u^{\sigma}\partial_{\lambda}S^{\lambda\rho}+S^{\lambda\rho}\partial_{\lambda}u^{\sigma}\right)\right],\label{244}\\
	\zeta_{pn_{c}}^{*} & =\frac{\partial^{2}\Omega_{\alpha\beta}}{\partial\epsilon\partial n_{c}}{w}+\sum_{a}\frac{\partial^{2}\Omega_{\alpha\beta}}{\partial n_{a}\partial n_{c}}n_{a}+\frac{\partial^{2}\Omega_{\alpha\beta}}{\partial S^{\rho\sigma}\partial n_{c}}\left[S^{\rho\sigma}+\theta^{-1}\left(u^{\rho}\partial_{\lambda}S^{\sigma\lambda}+S^{\sigma\lambda}\partial_{\lambda}u^{\rho}+u^{\sigma}\partial_{\lambda}S^{\lambda\rho}+S^{\lambda\rho}\partial_{\lambda}u^{\sigma}\right)\right],\label{245}
\end{align}
we can obtain the nonlocal corrections from the two-point correlation function to bulk viscous pressure as derived from Eqs.~\eqref{229}-\eqref{245}
\begin{eqnarray}
\begin{aligned}
	\langle\hat{P}^{*}\left(x\right)\rangle_{2}^{1}=&-\widetilde{\zeta}_{p\epsilon}\theta^{2}\widetilde{\Gamma}-\sum_{a}\widetilde{\zeta}_{pn_{a}}\theta^{2}\widetilde{\delta}_{a}-\widetilde{\zeta}\theta\left(\theta\Gamma-2\mathcal{Z}^{\alpha\beta}\mathcal{D}_{\alpha\beta}\right)-\widetilde{\zeta}D\theta+\sum_{i}\widetilde{\zeta}_{p\mathfrak{D}_{i}}\mathcal{Z}^{\alpha\beta}\partial_{\epsilon n}^{i}\Omega_{\alpha\beta}\left(\theta\Gamma-2\mathcal{Z}_{\rho\sigma}\mathcal{D}^{\rho\sigma}\right)\\
	&+\sum_{i}\widetilde{\zeta}_{p\mathfrak{D}_{i}}\partial_{\epsilon n}^{i}\Omega_{\alpha\beta}D\mathcal{Z}^{\alpha\beta}-\widetilde{\zeta}_{p\epsilon}\zeta_{p\epsilon}^{*}\theta\mathcal{Z}^{\alpha\beta}-\sum_{c}\widetilde{\zeta}_{pn_{c}}\zeta_{pn_{c}}^{*}\theta\mathcal{Z}^{\alpha\beta}+\widetilde{\zeta}_{SS}\mathcal{K}^{\langle\mu\rangle\langle\nu\rangle}\mathcal{R}_{\langle\mu\rangle\langle\nu\rangle}\left(\theta\Gamma-2\mathcal{Z}_{\rho\sigma}\mathcal{D}^{\rho\sigma}\right)\\
	&+\widetilde{\zeta}_{SS}\mathcal{K}^{\langle\mu\rangle\langle\nu\rangle}D\mathcal{R}_{\langle\mu\rangle\langle\nu\rangle}+\widetilde{\zeta}_{S\phi}\mathcal{K}^{\langle\mu\rangle\langle\nu\rangle}\xi_{\mu\nu}\left(\theta\Gamma-2\mathcal{Z}^{\alpha\beta}\mathcal{D}_{\alpha\beta}\right)+\widetilde{\zeta}_{S\phi}\mathcal{K}^{\langle\mu\rangle\langle\nu\rangle}D\xi_{\mu\nu}.
\end{aligned}
\label{246}
\end{eqnarray}
Substituting Eq.~\eqref{91} into Eq.\eqref{96} and applying Curie's theorem, we obtain corrections from extended thermodynamic forces to bulk viscous pressure
\begin{eqnarray}
\langle\hat{P}^{*}\rangle_{2}^{2}=\sum_{i}\zeta_{p\mathfrak{D}_{i}}\Bigl[\left(\partial_{\epsilon n}^{i}\beta\right)\mathcal{X}+\sum_{a}\left(\partial_{\epsilon n}^{i}\alpha_{a}\right)\mathcal{Y}_{a}+\left(\partial_{\epsilon n}^{i}\Omega_{\alpha\beta}\right)\mathcal{Z}^{\alpha\beta}\Bigr]+\zeta_{SS}\mathcal{K}_{\mu\nu}\mathcal{T}^{\langle\mu\rangle\langle\nu\rangle},
\label{247}
\end{eqnarray}
where we extracted the thermodynamic forces at the point $x$ from the integral and used the definitions \eqref{128} and \eqref{129}.

From Eqs.~\eqref{90} and \eqref{97}, the correction to bulk viscous pressure from the three-point correlation function is given by
\begin{eqnarray}
\begin{aligned}
	&\left\langle \hat{P}^{*}(x)\right\rangle _{2}^{3}\\
	= & \int d^{4}x_{1}d^{4}x_{2}\biggl(\left[\hat{p}^{*}-\mathcal{K}_{\alpha\beta}\hat{S}^{\alpha\beta}\right]_{x},\biggl[-\beta\theta\hat{p}^{*}+\beta\mathcal{R}_{\alpha\beta}\hat{S}^{\alpha\beta}+\beta\sum_{i}\left(\hat{\mathfrak{D}}_{i}\partial_{\epsilon n}^{i}\Omega_{\mu\nu}\right)\mathcal{Z}^{\mu\nu}-\sum_{a}\hat{\mathscr{J}}_{a}^{\sigma}\nabla_{\sigma}\alpha_{a}\\
	& +\beta\hat{h}^{\sigma}N_{\sigma}+\beta\hat{q}^{\mu}M_{\mu}+\beta\hat{\pi}^{\mu\nu}\sigma_{\mu\nu}+\beta\hat{\phi}^{\mu\nu}\xi_{\mu\nu}+\hat{\varpi}^{\lambda\alpha\beta}\varXi_{\lambda\alpha\beta}\biggr]_{x_{1}},\biggl[-\beta\theta\hat{p}^{*}+\beta\mathcal{R}_{\alpha\beta}\hat{S}^{\alpha\beta}+\beta\sum_{i}\left(\hat{\mathfrak{D}}_{i}\partial_{\epsilon n}^{i}\Omega_{\mu\nu}\right)\mathcal{Z}^{\mu\nu}\\
	& -\sum_{a}\hat{\mathscr{J}}_{a}^{\sigma}\nabla_{\sigma}\alpha_{a}+\beta\hat{h}^{\sigma}N_{\sigma}+\beta\hat{q}^{\mu}M_{\mu}+\beta\hat{\pi}^{\mu\nu}\sigma_{\mu\nu}+\beta\hat{\phi}^{\mu\nu}\xi_{\mu\nu}+\hat{\varpi}^{\lambda\alpha\beta}\varXi_{\lambda\alpha\beta}\biggr]_{x_{2}}\biggr).
\end{aligned}
\label{248}
\end{eqnarray}
Among the correlators in Eq.~\eqref{248}, the nonvanishing ones are $\biggl(\hat{p}^{*}\left(x\right),\hat{p}^{*}\left(x_{1}\right),\hat{p}^{*}\left(x_{2}\right)\biggr)$,$\left(\hat{p}^{*}\left(x\right),\hat{p}^{*}\left(x_{1}\right),\hat{\mathfrak{D}}_{j}\left(x_{2}\right)\right)$,\\
$\left(\hat{p}^{*}\left(x\right),\hat{\mathfrak{D}}_{i}\left(x_{1}\right),\hat{\mathfrak{D}}_{j}\left(x_{2}\right)\right)$, and
\begin{align} 
	\left(\hat{p}^{*}\left(x\right),\hat{S}^{\mu\nu}\left(x_{1}\right),\hat{S}^{\alpha\beta}\left(x_{2}\right)\right)=  &  \frac{1}{3}\mydelta^{\mu\nu\alpha\beta}\left(\hat{p}^{*}\left(x\right),\hat{S}^{\lambda\eta}\left(x_{1}\right),\hat{S}_{\lambda\eta}\left(x_{2}\right)\right),\label{249}\\
	\left(\hat{p}^{*}\left(x\right),\hat{S}^{\mu\nu}\left(x_{1}\right),\hat{\phi}^{\alpha\beta}\left(x_{2}\right)\right)=  &  \frac{1}{3}\mydelta^{\mu\nu\alpha\beta}\left(\hat{p}^{*}\left(x\right),\hat{S}^{\lambda\eta}\left(x_{1}\right),\hat{\phi}_{\lambda\eta}\left(x_{2}\right)\right),\label{250}\\
	\left(\hat{p}^{*}\left(x\right),\hat{\mathscr{J}}_{a\sigma}\left(x_{1}\right),\hat{\mathscr{J}}_{b\alpha}\left(x_{2}\right)\right)=  &  \frac{1}{3}\Delta_{\sigma\alpha}\left(\hat{p}^{*}\left(x\right),\hat{\mathscr{J}}_{a\gamma}\left(x_{1}\right),\hat{\mathscr{J}}_{b}^{\gamma}\left(x_{2}\right)\right),\label{251}\\
	\left(\hat{p}^{*}\left(x\right),\hat{\mathscr{J}}_{a\sigma}\left(x_{1}\right),\hat{h}_{\alpha}\left(x_{2}\right)\right)=  &  \frac{1}{3}\Delta_{\sigma\alpha}\left(\hat{p}^{*}\left(x\right),\hat{\mathscr{J}}_{a\gamma}\left(x_{1}\right),\hat{h}^{\gamma}\left(x_{2}\right)\right),\label{252}\\
	\left(\hat{p}^{*}\left(x\right),\hat{\mathscr{J}}_{a\sigma}\left(x_{1}\right),\hat{q}_{\alpha}\left(x_{2}\right)\right)=  &  \frac{1}{3}\Delta_{\sigma\alpha}\left(\hat{p}^{*}\left(x\right),\hat{\mathscr{J}}_{a\gamma}\left(x_{1}\right),\hat{q}^{\gamma}\left(x_{2}\right)\right),\label{253}\\
	\left(\hat{p}^{*}\left(x\right),\hat{h}_{\sigma}\left(x_{1}\right),\hat{h}_{\alpha}\left(x_{2}\right)\right)=  &  \frac{1}{3}\Delta_{\sigma\alpha}\left(\hat{p}^{*}\left(x\right),\hat{h}_{\gamma}\left(x_{1}\right),\hat{h}^{\gamma}\left(x_{2}\right)\right),\label{254}\\
	\left(\hat{p}^{*}\left(x\right),\hat{h}_{\sigma}\left(x_{1}\right),\hat{q}_{\alpha}\left(x_{2}\right)\right)=  &  \frac{1}{3}\Delta_{\sigma\alpha}\left(\hat{p}^{*}\left(x\right),\hat{h}_{\gamma}\left(x_{1}\right),\hat{q}^{\gamma}\left(x_{2}\right)\right),\label{255}\\
	\left(\hat{p}^{*}\left(x\right),\hat{q}_{\sigma}\left(x_{1}\right),\hat{q}_{\alpha}\left(x_{2}\right)\right)=  &  \frac{1}{3}\Delta_{\sigma\alpha}\left(\hat{p}^{*}\left(x\right),\hat{q}_{\gamma}\left(x_{1}\right),\hat{q}^{\gamma}\left(x_{2}\right)\right),\label{256}\\
	\left(\hat{p}^{*}\left(x\right),\hat{\pi}_{\rho\sigma}\left(x_{1}\right),\hat{\pi}_{\alpha\beta}\left(x_{2}\right)\right)=  &  \frac{1}{5}\Delta_{\rho\sigma\alpha\beta}\left(\hat{p}^{*}\left(x\right),\hat{\pi}_{\gamma\delta}\left(x_{1}\right),\hat{\pi}^{\gamma\delta}\left(x_{2}\right)\right),\label{257}\\
	\left(\hat{p}^{*}\left(x\right),\hat{\phi}_{\rho\sigma}\left(x_{1}\right),\hat{\phi}_{\alpha\beta}\left(x_{2}\right)\right)=  &  \frac{1}{3}\mydelta_{\rho\sigma\alpha\beta}\left(\hat{p}^{*}\left(x\right),\hat{\phi}_{\gamma\delta}\left(x_{1}\right),\hat{\phi}^{\gamma\delta}\left(x_{2}\right)\right),\label{258}\\
	\left(\hat{p}^{*}\left(x\right),\hat{\varpi}^{\mu\nu\sigma}\left(x_{1}\right),\hat{\varpi}^{\lambda\alpha\beta}\left(x_{2}\right)\right)=  &  \myDelta^{\mu\nu\sigma\lambda\alpha\beta}\left(\hat{p}^{*}\left(x\right),\hat{\varpi}^{\rho\gamma\delta}\left(x_{1}\right),\hat{\varpi}_{\rho\gamma\delta}\left(x_{2}\right)\right),\label{259}\\
	\left(\hat{S}^{\mu\nu}\left(x\right),\hat{p}^{*}\left(x_{1}\right),\hat{S}^{\alpha\beta}\left(x_{2}\right)\right)=  &  \frac{1}{3}\mydelta^{\mu\nu\alpha\beta}\left(\hat{S}^{\lambda\eta}\left(x\right),\hat{p}^{*}\left(x_{1}\right),\hat{S}_{\lambda\eta}\left(x_{2}\right)\right),\label{260}\\
	\left(\hat{S}^{\mu\nu}\left(x\right),\hat{p}^{*}\left(x_{1}\right),\hat{\phi}^{\alpha\beta}\left(x_{2}\right)\right)=  &  \frac{1}{3}\mydelta^{\mu\nu\alpha\beta}\left(\hat{S}^{\lambda\eta}\left(x\right),\hat{p}^{*}\left(x_{1}\right),\hat{\phi}_{\lambda\eta}\left(x_{2}\right)\right),\label{261}\\
	\left(\hat{S}^{\mu\nu}\left(x\right),\hat{S}^{\rho\sigma}\left(x_{1}\right),\hat{S}^{\alpha\beta}\left(x_{2}\right)\right)=  &  \frac{1}{3}\left(\Delta^{\rho\alpha}\mydelta^{\mu\nu\sigma\beta}-\Delta^{\rho\beta}\mydelta^{\mu\nu\sigma\alpha}-\Delta^{\sigma\alpha}\mydelta^{\mu\nu\rho\beta}+\Delta^{\sigma\beta}\mydelta^{\mu\nu\rho\alpha}\right)\left(\hat{S}_{\lambda}^{\,\,\,\,\delta}\left(x\right),\hat{S}_{\delta}^{\,\,\,\,\eta}\left(x_{1}\right),\hat{S}_{\eta}^{\,\,\,\,\lambda}\left(x_{2}\right)\right),\label{262}\\
	\left(\hat{S}^{\mu\nu}\left(x\right),\hat{S}^{\rho\sigma}\left(x_{1}\right),\hat{\mathfrak{D}}_{j}\left(x_{2}\right)\right)=  &\frac{1}{3}\mydelta^{\mu\nu\rho\sigma}\left(\hat{S}^{\lambda\eta}\left(x\right),\hat{S}_{\lambda\eta}\left(x_{1}\right),\hat{\mathfrak{D}}_{j}\left(x_{2}\right)\right),\label{263}\\
	\left(\hat{S}^{\mu\nu}\left(x\right),\hat{S}^{\rho\sigma}\left(x_{1}\right),\hat{\pi}^{\alpha\beta}\left(x_{2}\right)\right)=  &  -\frac{1}{5}\left(-\Delta^{\rho\alpha}\mydelta^{\mu\nu\sigma\beta}-\Delta^{\rho\beta}\mydelta^{\mu\nu\sigma\alpha}+\Delta^{\sigma\alpha}\mydelta^{\mu\nu\rho\beta}+\Delta^{\sigma\beta}\mydelta^{\mu\nu\rho\alpha}-\frac{4}{3}\Delta^{\alpha\beta}\mydelta^{\mu\nu\rho\sigma}\right)\nonumber\\
	&  \times\left(\hat{S}_{\lambda}^{\,\,\,\,\delta}\left(x\right),\hat{S}_{\delta}^{\,\,\,\,\eta}\left(x_{1}\right),\hat{\pi}_{\eta}^{\,\,\,\,\lambda}\left(x_{2}\right)\right),\label{264}\\
	\left(\hat{S}^{\mu\nu}\left(x\right),\hat{S}^{\rho\sigma}\left(x_{1}\right),\hat{\phi}^{\alpha\beta}\left(x_{2}\right)\right)=  &  \frac{1}{3}\left(\Delta^{\rho\alpha}\mydelta^{\mu\nu\sigma\beta}-\Delta^{\rho\beta}\mydelta^{\mu\nu\sigma\alpha}-\Delta^{\sigma\alpha}\mydelta^{\mu\nu\rho\beta}+\Delta^{\sigma\beta}\mydelta^{\mu\nu\rho\alpha}\right)\left(\hat{S}_{\lambda}^{\,\,\,\,\delta}\left(x\right),\hat{S}_{\delta}^{\,\,\,\,\eta}\left(x_{1}\right),\hat{\phi}_{\eta}^{\,\,\,\,\lambda}\left(x_{2}\right)\right),\label{265}\\
	\left(\hat{S}^{\mu\nu}\left(x\right),\hat{\mathfrak{D}}_{i}\left(x_{1}\right),\hat{\phi}^{\alpha\beta}\left(x_{2}\right)\right)=&  \frac{1}{3}\mydelta^{\mu\nu\alpha\beta}\left(\hat{S}^{\lambda\eta}\left(x\right),\hat{\mathfrak{D}}_{i}\left(x_{1}\right),\hat{\phi}_{\lambda\eta}\left(x_{2}\right)\right),\label{266}\\
	\left(\hat{S}^{\mu\nu}\left(x\right),\hat{\mathscr{J}}_{a}^{\alpha}\left(x_{1}\right),\hat{\mathscr{J}}_{b}^{\beta}\left(x_{1}\right)\right)=  &  \frac{1}{3}\mydelta^{\mu\nu\alpha\beta}\left(\hat{S}^{\lambda\eta}\left(x\right),\hat{\mathscr{J}}_{a\lambda}\left(x_{1}\right),\hat{\mathscr{J}}_{b\eta}\left(x_{2}\right)\right),\label{267}\\
	\left(\hat{S}^{\mu\nu}\left(x\right),\hat{\mathscr{J}}_{a}^{\alpha}\left(x_{1}\right),\hat{h}^{\beta}\left(x_{1}\right)\right)=  &  \frac{1}{3}\mydelta^{\mu\nu\alpha\beta}\left(\hat{S}^{\lambda\eta}\left(x\right),\hat{\mathscr{J}}_{a\lambda}\left(x_{1}\right),\hat{h}_{\eta}\left(x_{2}\right)\right),\label{268}\\
	\left(\hat{S}^{\mu\nu}\left(x\right),\hat{\mathscr{J}}_{a}^{\alpha}\left(x_{1}\right),\hat{q}^{\beta}\left(x_{1}\right)\right)=  &  \frac{1}{3}\mydelta^{\mu\nu\alpha\beta}\left(\hat{S}^{\lambda\eta}\left(x\right),\hat{\mathscr{J}}_{a\lambda}\left(x_{1}\right),\hat{q}_{\eta}\left(x_{2}\right)\right),\label{269}\\
	\left(\hat{S}^{\mu\nu}\left(x\right),\hat{\mathscr{J}}_{a}^{\alpha}\left(x_{1}\right),\hat{\varpi}^{\rho\sigma\delta}\left(x_{1}\right)\right)=  &  \myDelta^{\mu\nu\alpha\rho\sigma\delta}\left(\hat{S}^{\lambda\eta}\left(x\right),\hat{\mathscr{J}}_{a}^{\beta}\left(x_{1}\right),\hat{\varpi}_{\lambda\eta\beta}\left(x_{2}\right)\right),\label{270}\\
	\left(\hat{S}^{\mu\nu}\left(x\right),\hat{h}^{\alpha}\left(x_{1}\right),\hat{h}^{\beta}\left(x_{1}\right)\right)=  &  \frac{1}{3}\mydelta^{\mu\nu\alpha\beta}\left(\hat{S}^{\lambda\eta}\left(x\right),\hat{h}_{\lambda}\left(x_{1}\right),\hat{h}_{\eta}\left(x_{2}\right)\right),\label{271}\\
	\left(\hat{S}^{\mu\nu}\left(x\right),\hat{h}^{\alpha}\left(x_{1}\right),\hat{q}^{\beta}\left(x_{1}\right)\right)=  &  \frac{1}{3}\mydelta^{\mu\nu\alpha\beta}\left(\hat{S}^{\lambda\eta}\left(x\right),\hat{h}_{\lambda}\left(x_{1}\right),\hat{q}_{\eta}\left(x_{2}\right)\right),\label{272} 
\end{align}
\begin{align}
	\left(\hat{S}^{\mu\nu}\left(x\right),\hat{h}^{\alpha}\left(x_{1}\right),\hat{\varpi}^{\rho\sigma\delta}\left(x_{1}\right)\right)= & \myDelta^{\mu\nu\alpha\rho\sigma\delta}\left(\hat{S}^{\lambda\eta}\left(x\right),\hat{h}^{\beta}\left(x_{1}\right),\hat{\varpi}_{\lambda\eta\beta}\left(x_{2}\right)\right),\label{273}\\
	\left(\hat{S}^{\mu\nu}\left(x\right),\hat{q}^{\alpha}\left(x_{1}\right),\hat{q}^{\beta}\left(x_{2}\right)\right)= & \frac{1}{3}\mydelta^{\mu\nu\alpha\beta}\left(\hat{S}^{\lambda\eta}\left(x\right),\hat{q}_{\lambda}\left(x_{1}\right),\hat{q}_{\eta}\left(x_{2}\right)\right),\label{274}\\
	\left(\hat{S}^{\mu\nu}\left(x\right),\hat{q}^{\alpha}\left(x_{1}\right),\hat{\varpi}^{\rho\sigma\delta}\left(x_{1}\right)\right)= & \myDelta^{\mu\nu\alpha\rho\sigma\delta}\left(\hat{S}^{\lambda\eta}\left(x\right),\hat{q}^{\beta}\left(x_{1}\right),\hat{\varpi}_{\lambda\eta\beta}\left(x_{2}\right)\right),\label{275}\\
	\left(\hat{S}^{\mu\nu}\left(x\right),\hat{\pi}^{\rho\sigma}\left(x_{1}\right),\hat{\pi}^{\alpha\beta}\left(x_{2}\right)\right)= & -\frac{1}{15}\left(\Delta^{\rho\alpha}\mydelta^{\mu\nu\sigma\beta}+\Delta^{\rho\beta}\mydelta^{\mu\nu\sigma\alpha}+\Delta^{\sigma\alpha}\mydelta^{\mu\nu\rho\beta}+\Delta^{\sigma\beta}\mydelta^{\mu\nu\rho\alpha}\right)\left(\hat{S}_{\lambda}^{\,\,\,\,\delta}\left(x\right),\hat{\pi}_{\delta}^{\,\,\,\,\eta}\left(x_{1}\right),\hat{\pi}_{\eta}^{\,\,\,\,\lambda}\left(x_{2}\right)\right),\label{276}\\
	\left(\hat{S}^{\mu\nu}\left(x\right),\hat{\pi}^{\rho\sigma}\left(x_{1}\right),\hat{\phi}^{\alpha\beta}\left(x_{2}\right)\right)= & \frac{1}{5}\left(-\Delta^{\rho\alpha}\mydelta^{\mu\nu\sigma\beta}+\Delta^{\rho\beta}\mydelta^{\mu\nu\sigma\alpha}-\Delta^{\sigma\alpha}\mydelta^{\mu\nu\rho\beta}+\Delta^{\sigma\beta}\mydelta^{\mu\nu\rho\alpha}+\frac{4}{3}\Delta^{\rho\sigma}\mydelta^{\mu\nu\alpha\beta}\right)\nonumber\\
	&\times\left(\hat{S}_{\lambda}^{\,\,\,\,\delta}\left(x\right),\hat{\pi}_{\delta}^{\,\,\,\,\eta}\left(x_{1}\right),\hat{\phi}_{\eta}^{\,\,\,\,\lambda}\left(x_{2}\right)\right),\label{277}\\
	\left(\hat{S}^{\mu\nu}\left(x\right),\hat{\phi}^{\rho\sigma}\left(x_{1}\right),\hat{\phi}^{\alpha\beta}\left(x_{2}\right)\right)= & \frac{1}{3}\left(\Delta^{\rho\alpha}\mydelta^{\mu\nu\sigma\beta}-\Delta^{\rho\beta}\mydelta^{\mu\nu\sigma\alpha}-\Delta^{\sigma\alpha}\mydelta^{\mu\nu\rho\beta}+\Delta^{\sigma\beta}\mydelta^{\mu\nu\rho\alpha}\right)\left(\hat{S}_{\lambda}^{\,\,\,\,\delta}\left(x\right),\hat{\phi}_{\delta}^{\,\,\,\,\eta}\left(x_{1}\right),\hat{\phi}_{\eta}^{\,\,\,\,\lambda}\left(x_{2}\right)\right),\label{278}
\end{align}
Substituting these expressions into Eq.~\eqref{248}, we obtain
\begin{eqnarray}
\begin{aligned}
	\langle\hat{P}^{*}\left(x\right)\rangle_{2}^{3}= & \zeta_{ppp}\theta^{2}+2\theta\sum_{i}\zeta_{pp\mathfrak{D}_{i}}\mathcal{Z}^{\mu\nu}\partial_{\epsilon n}^{i}\Omega_{\mu\nu}+\zeta_{pSS}\mathcal{R}_{\mu\nu}\mathcal{R}^{\langle\mu\rangle\langle\nu\rangle}+2\zeta_{pS\phi}\mathcal{R}_{\mu\nu}\xi^{\mu\nu}+\sum_{i}\sum_{j}\zeta_{p\mathfrak{D}_{i}\mathfrak{D}_{j}}\mathcal{Z}^{\mu\nu}\partial_{\epsilon n}^{i}\Omega_{\mu\nu}\mathcal{Z}^{\alpha\beta}\partial_{\epsilon n}^{i}\Omega_{\alpha\beta}\\
	&+\sum_{ab}\zeta_{p\mathscr{J}_{a}\mathscr{J}_{b}}\nabla^{\sigma}\alpha_{a}\nabla_{\sigma}\alpha_{b}+2\sum_{a}\zeta_{p\mathscr{J}_{a}h}\nabla^{\sigma}\alpha_{a}N_{\sigma}+2\sum_{a}\zeta_{p\mathscr{J}_{a}q}\nabla^{\sigma}\alpha_{a}M_{\sigma}+\zeta_{phh}N^{\sigma}N_{\sigma}+2\zeta_{phq}N^{\sigma}M_{\sigma}\\
	&+\zeta_{pqq}M^{\sigma}M_{\sigma}+\zeta_{p\pi\pi}\sigma^{\rho\sigma}\sigma_{\rho\sigma}+\zeta_{p\phi\phi}\xi^{\rho\sigma}\xi_{\rho\sigma}+\zeta_{p\varpi\varpi}\varXi_{\mu\nu\sigma}\varXi^{\mu\nu\sigma}+2\zeta_{SpS}\mathcal{K}_{\mu\nu}\theta\mathcal{R}^{\langle\mu\rangle\langle\nu\rangle}+2\zeta_{Sp\phi}\mathcal{K}_{\mu\nu}\theta\xi^{\mu\nu}\\
	&+\zeta_{SSS}\mathcal{K}^{\langle\sigma\rangle\langle\beta\rangle}\mathcal{R}_{\,\,\,\,\sigma}^{\langle\alpha\rangle}\mathcal{R}_{\alpha\beta}+2\sum_{i}\zeta_{SS\mathfrak{D}_{i}}\mathcal{K}_{\mu\nu}\mathcal{R}^{\langle\mu\rangle\langle\nu\rangle}\mathcal{Z}^{\alpha\beta}\partial_{\epsilon n}^{i}\Omega_{\alpha\beta}+2\zeta_{SS\pi}\mathcal{K}^{\langle\sigma\rangle\langle\beta\rangle}\mathcal{R}_{\,\,\,\,\sigma}^{\langle\alpha\rangle}\sigma_{\alpha\beta}\\
	&+2\zeta_{SS\phi}\mathcal{K}^{\langle\sigma\rangle\langle\beta\rangle}\mathcal{R}_{\,\,\,\,\sigma}^{\langle\alpha\rangle}\xi_{\alpha\beta}+2\sum_{i}\zeta_{S\mathfrak{D}_{i}\phi}\mathcal{K}^{\langle\alpha\rangle\langle\beta\rangle}\mathcal{Z}_{\rho\sigma}\partial_{\epsilon n}^{i}\Omega^{\rho\sigma}\xi_{\alpha\beta}+\sum_{ab}\zeta_{S\mathscr{J}_{a}\mathscr{J}_{b}}\mathcal{K}^{\langle\alpha\rangle\langle\beta\rangle}\nabla_{\alpha}\alpha_{a}\nabla_{\beta}\alpha_{b}\\
	&+2\sum_{a}\zeta_{S\mathscr{J}_{a}h}\mathcal{K}^{\langle\alpha\rangle\langle\beta\rangle}\nabla_{\alpha}\alpha_{a}N_{\beta}+2\sum_{a}\zeta_{S\mathscr{J}_{a}q}\mathcal{K}^{\langle\alpha\rangle\langle\beta\rangle}\nabla_{\alpha}\alpha_{a}M_{\beta}+2\sum_{a}\zeta_{S\mathscr{J}_{a}\varpi}\mathcal{K}_{\mu\nu}\nabla_{\alpha}\alpha_{a}\varXi^{\mu\nu\alpha}\\
	&+\zeta_{Shh}\mathcal{K}^{\langle\alpha\rangle\langle\beta\rangle}N_{\alpha}N_{\beta}+2\zeta_{Shq}\mathcal{K}^{\langle\alpha\rangle\langle\beta\rangle}N_{\alpha}M_{\beta}+2\zeta_{Sh\varpi}\mathcal{K}_{\mu\nu}N_{\alpha}\varXi^{\mu\nu\alpha}+\zeta_{Sqq}\mathcal{K}^{\langle\alpha\rangle\langle\beta\rangle}M_{\alpha}M_{\beta}\\
	&+2\zeta_{Sq\varpi}\mathcal{K}_{\mu\nu}M_{\alpha}\varXi^{\mu\nu\alpha}+\zeta_{S\pi\pi}\mathcal{K}^{\langle\sigma\rangle\langle\beta\rangle}\sigma_{\,\,\,\,\sigma}^{\alpha}\sigma_{\alpha\beta}+2\zeta_{S\pi\phi}\mathcal{K}^{\langle\sigma\rangle\langle\beta\rangle}\sigma_{\,\,\,\,\sigma}^{\alpha}\xi_{\alpha\beta}+\zeta_{S\phi\phi}\mathcal{K}^{\langle\sigma\rangle\langle\beta\rangle}\xi_{\,\,\,\,\sigma}^{\alpha}\xi_{\alpha\beta},
\end{aligned}
\label{279}
\end{eqnarray}
where the coefficients are defined as
\begin{align}
	\zeta_{ppp} & =\beta^{2}\int d^{4}x_{1}d^{4}x_{2}\left(\hat{p}^{*}\left(x\right),\hat{p}^{*}\left(x_{1}\right),\hat{p}^{*}\left(x_{2}\right)\right),\zeta_{pp\mathfrak{D}_{i}}=-\beta^{2}\int d^{4}x_{1}d^{4}x_{2}\left(\hat{p}^{*}\left(x\right),\hat{p}^{*}\left(x_{1}\right),\hat{\mathfrak{D}}_{i}\left(x_{2}\right)\right),\label{280}\\
	\zeta_{pSS} & =\frac{1}{3}\beta^{2}\int d^{4}x_{1}\left(\hat{p}^{*}\left(x\right),\hat{S}^{\lambda\eta}\left(x_{1}\right),\hat{S}_{\lambda\eta}\left(x_{2}\right)\right),\zeta_{pS\phi}=\frac{1}{3}\beta^{2}\int d^{4}x_{1}d^{4}x_{2}\left(\hat{p}^{*}\left(x\right),\hat{S}^{\lambda\eta}\left(x_{1}\right),\hat{\phi}_{\lambda\eta}\left(x_{2}\right)\right),\label{281}\\
	\zeta_{p\mathfrak{D}_{i}\mathfrak{D}_{j}} & =\beta^{2}\int d^{4}x_{1}d^{4}x_{2}\left(\hat{p}^{*}\left(x\right),\hat{\mathfrak{D}}_{i}\left(x_{1}\right),\hat{\mathfrak{D}}_{j}\left(x_{2}\right)\right),\zeta_{p\mathscr{J}_{a}\mathscr{J}_{b}}=\frac{1}{3}\int d^{4}x_{1}d^{4}x_{2}\left(\hat{p}^{*}\left(x\right),\hat{\mathscr{J}}_{a\gamma}\left(x_{1}\right),\hat{\mathscr{J}}_{b}^{\gamma}\left(x_{2}\right)\right),\label{282}\\
	\zeta_{p\mathscr{J}_{a}h} & =-\frac{1}{3}\beta\int d^{4}x_{1}d^{4}x_{2}\left(\hat{p}^{*}\left(x\right),\hat{\mathscr{J}}_{a\gamma}\left(x_{1}\right),\hat{h}^{\gamma}\left(x_{2}\right)\right),\zeta_{p\mathscr{J}_{a}q}=-\frac{1}{3}\beta\int d^{4}x_{1}d^{4}x_{2}\left(\hat{p}^{*}\left(x\right),\hat{\mathscr{J}}_{a\gamma}\left(x_{1}\right),\hat{q}^{\gamma}\left(x_{2}\right)\right),\label{283}\\
	\zeta_{phh} & =\frac{1}{3}\beta^{2}\int d^{4}x_{1}d^{4}x_{2}\left(\hat{p}^{*}(x),\hat{h}_{\gamma}(x_{1}),\hat{h}^{\gamma}(x_{2})\right),\zeta_{phq}=\frac{1}{3}\beta^{2}\int d^{4}x_{1}d^{4}x_{2}\left(\hat{p}^{*}\left(x\right),\hat{h}_{\gamma}\left(x_{1}\right),\hat{q}^{\gamma}\left(x_{2}\right)\right),\label{284}\\
	\zeta_{pqq} & =\frac{1}{3}\beta^{2}\int d^{4}x_{1}d^{4}x_{2}\left(\hat{p}^{*}(x),\hat{q}_{\gamma}(x_{1}),\hat{q}^{\gamma}(x_{2})\right),\zeta_{p\pi\pi}=\frac{1}{5}\beta^{2}\int d^{4}x_{1}d^{4}x_{2}\left(\hat{p}^{*}\left(x\right),\hat{\pi}_{\gamma\delta}\left(x_{1}\right),\hat{\pi}^{\gamma\delta}\left(x_{2}\right)\right),\label{285}\\
	\zeta_{p\phi\phi} & =\frac{1}{3}\beta^{2}\int d^{4}x_{1}d^{4}x_{2}\left(\hat{p}^{*}\left(x\right),\hat{\phi}_{\gamma\delta}\left(x_{1}\right),\hat{\phi}^{\gamma\delta}\left(x_{2}\right)\right),\zeta_{p\varpi\varpi}=\int d^{4}x_{1}d^{4}x_{2}\left(\hat{p}^{*}\left(x\right),\hat{\varpi}^{\rho\gamma\delta}\left(x_{1}\right),\hat{\varpi}_{\rho\gamma\delta}\left(x_{2}\right)\right),\label{286}\\
	\zeta_{SpS} & =\frac{1}{3}\beta^{2}\int d^{4}x_{1}d^{4}x_{2}\left(\hat{S}^{\lambda\eta}\left(x\right),\hat{p}^{*}\left(x_{1}\right),\hat{S}_{\lambda\eta}\left(x_{2}\right)\right),\zeta_{Sp\phi}=\frac{1}{3}\beta^{2}\int d^{4}x_{1}d^{4}x_{2}\left(\hat{S}^{\lambda\eta}\left(x\right),\hat{p}^{*}\left(x_{1}\right),\hat{\phi}_{\lambda\eta}\left(x_{2}\right)\right),\label{287}\\
	\zeta_{SSS} & =-\frac{4}{3}\beta^{2}\int d^{4}x_{1}d^{4}x_{2}\left(\hat{S_{\lambda}}^{\delta}\left(x\right),\hat{S_{\delta}}^{\eta}\left(x_{1}\right),\hat{S_{\eta}}^{\lambda}\left(x_{2}\right)\right),\zeta_{SS\mathfrak{D}_{i}}=-\frac{1}{3}\beta^{2}\int d^{4}x_{1}d^{4}x_{2}\left(\hat{S}^{\lambda\eta}\left(x\right),\hat{S}_{\lambda\eta}\left(x_{1}\right),\hat{\mathfrak{D}}_{i}\left(x_{2}\right)\right),\label{288}\\
	\zeta_{SS\pi} & =-\frac{4}{5}\beta^{2}\int d^{4}x_{1}d^{4}x_{2}\left(\hat{S_{\lambda}}^{\delta}\left(x\right),\hat{S_{\delta}}^{\eta}\left(x_{1}\right),\hat{\pi_{\eta}}^{\lambda}\left(x_{2}\right)\right),\zeta_{SS\phi}=-\frac{4}{3}\beta^{2}\int d^{4}x_{1}d^{4}x_{2}\left(\hat{S_{\lambda}}^{\delta}\left(x\right),\hat{S_{\delta}}^{\eta}\left(x_{1}\right),\hat{\phi_{\eta}}^{\lambda}\left(x_{2}\right)\right),\label{289}\\
	\zeta_{S\mathfrak{D}_{i}\phi} & =-\frac{1}{3}\beta^{2}\int d^{4}x_{1}d^{4}x_{2}\left(\hat{S}^{\lambda\eta}\left(x\right),\hat{\mathfrak{D}}_{i}\left(x_{1}\right),\hat{\phi}_{\lambda\eta}\left(x_{2}\right)\right),\zeta_{S\mathscr{J}_{a}\mathscr{J}_{b}}=-\frac{1}{3}\int d^{4}x_{1}d^{4}x_{2}\left(\hat{S}^{\lambda\eta}\left(x\right),\hat{\mathscr{J}}_{a\lambda}\left(x_{1}\right),\hat{\mathscr{J}}_{b\eta}\left(x_{2}\right)\right),\label{290}\\
	\zeta_{S\mathscr{J}_{a}h} & =\frac{1}{3}\beta\int d^{4}x_{1}d^{4}x_{2}\left(\hat{S}^{\lambda\eta}\left(x\right),\hat{\mathscr{J}}_{a\lambda}\left(x_{1}\right),\hat{h}_{\eta}\left(x_{2}\right)\right),\zeta_{S\mathscr{J}_{a}q}=\frac{1}{3}\beta\int d^{4}x_{1}d^{4}x_{2}\left(\hat{S}^{\lambda\eta}\left(x\right),\hat{\mathscr{J}}_{a\lambda}\left(x_{1}\right),\hat{q}_{\eta}\left(x_{2}\right)\right),\label{291}\\
	\zeta_{S\mathscr{J}_{a}\varpi} & =\int d^{4}x_{1}d^{4}x_{2}\left(\hat{S}^{\lambda\eta}\left(x\right),\hat{\mathscr{J}}_{a}^{\beta}\left(x_{1}\right),\hat{\varpi}_{\lambda\eta\beta}\left(x_{2}\right)\right),\zeta_{Shh}=-\frac{1}{3}\beta^{2}\int d^{4}x_{1}d^{4}x_{2}\left(\hat{S}^{\lambda\eta}\left(x\right),\hat{h}_{\lambda}\left(x_{1}\right),\hat{h}_{\eta}\left(x_{2}\right)\right),\label{292}\\
	\zeta_{Shq} & =-\frac{1}{3}\beta^{2}\int d^{4}x_{1}d^{4}x_{2}\left(\hat{S}^{\lambda\eta}\left(x\right),\hat{h}_{\lambda}\left(x_{1}\right),\hat{q}_{\eta}\left(x_{2}\right)\right),\zeta_{Sh\varpi}=-\beta\int d^{4}x_{1}d^{4}x_{2}\left(\hat{S}^{\lambda\eta}\left(x\right),\hat{h}^{\beta}\left(x_{1}\right),\hat{\varpi}_{\lambda\eta\beta}\left(x_{2}\right)\right),\label{293}\\
	\zeta_{Sqq} & =-\frac{1}{3}\beta^{2}\int d^{4}x_{1}d^{4}x_{2}\left(\hat{S}^{\lambda\eta}\left(x\right),\hat{q}_{\lambda}\left(x_{1}\right),\hat{q}_{\eta}\left(x_{2}\right)\right),\zeta_{Sq\varpi}=-\beta\int d^{4}x_{1}d^{4}x_{2}\left(\hat{S}^{\lambda\eta}\left(x\right),\hat{q}^{\beta}\left(x_{1}\right),\hat{\varpi}_{\lambda\eta\beta}\left(x_{2}\right)\right),\label{294}\\
	\zeta_{S\pi\pi} & =\frac{4}{15}\beta^{2}\int d^{4}x_{1}d^{4}x_{2}\left(\hat{S}_{\lambda}^{\delta}\left(x\right),\hat{\pi}_{\delta}^{\eta}\left(x_{1}\right),\hat{\pi}_{\eta}^{\lambda}\left(x_{2}\right)\right),\zeta_{S\pi\phi}=\frac{4}{5}\beta^{2}\int d^{4}x_{1}d^{4}x_{2}\left(\hat{S}_{\lambda}^{\delta}\left(x\right),\hat{\pi}_{\delta}^{\eta}\left(x_{1}\right),\hat{\phi}_{\eta}^{\lambda}\left(x_{2}\right)\right),\label{295}\\
	\zeta_{S\phi\phi} & =-\frac{4}{3}\beta^{2}\int d^{4}x_{1}d^{4}x_{2}\left(\hat{S}_{\lambda}^{\delta}\left(x\right),\hat{\phi}_{\delta}^{\eta}\left(x_{1}\right),\hat{\phi}_{\eta}^{\lambda}\left(x_{2}\right)\right).\label{296}
\end{align}
Combining the contributions from Eqs.~\eqref{209}, \eqref{246}, \eqref{247}, and \eqref{279}, we obtain the complete second-order expression for the bulk viscous pressure:
\begin{eqnarray}
\begin{aligned}
	\Pi=&-\zeta\theta+\sum_{i}\zeta_{p\mathfrak{D}_{i}}\mathcal{Z}^{\mu\nu}\partial_{\epsilon n}^{i}\Omega_{\mu\nu}+\zeta_{SS}\mathcal{K}_{\alpha\beta}\mathcal{R}^{\langle\alpha\rangle\langle\beta\rangle}+\zeta_{S\phi}\mathcal{K}_{\mu\nu}\xi^{\mu\nu}+\frac{1}{2}\frac{\partial^{2}p}{\partial\epsilon^{2}}\zeta_{\epsilon p}^{2}\theta^{2}+\frac{1}{2}\frac{\partial^{2}p}{\partial\epsilon^{2}}\sum_{i}\sum_{j}\zeta_{\epsilon\mathfrak{D}_{i}}\zeta_{\epsilon\mathfrak{D}_{j}}\mathcal{Z}^{\mu\nu}\mathcal{Z}^{\alpha\beta}\partial_{\epsilon n}^{i}\Omega_{\mu\nu}\partial_{\epsilon n}^{j}\Omega_{\alpha\beta}\\
	&-\frac{\partial^{2}p}{\partial\epsilon^{2}}\zeta_{\epsilon p}\theta\sum_{i}\zeta_{\epsilon\mathfrak{D}_{i}}\mathcal{Z}^{\mu\nu}\partial_{\epsilon n}^{i}\Omega_{\mu\nu}+\sum_{a}\frac{\partial^{2}p}{\partial\epsilon\partial n_{a}}\zeta_{\epsilon p}\zeta_{n_{a}p}\theta^{2}-\sum_{a}\frac{\partial^{2}p}{\partial\epsilon\partial n_{a}}\zeta_{\epsilon p}\theta\sum_{j}\zeta_{n_{a}\mathfrak{D}_{j}}\mathcal{Z}^{\rho\sigma}\partial_{\epsilon n}^{j}\Omega_{\rho\sigma}\\
	&-\sum_{a}\frac{\partial^{2}p}{\partial\epsilon\partial n_{a}}\zeta_{n_{a}p}\theta\sum_{i}\zeta_{\epsilon\mathfrak{D}_{i}}\mathcal{Z}^{\alpha\beta}\partial_{\epsilon n}^{i}\Omega_{\alpha\beta}+\sum_{a}\frac{\partial^{2}p}{\partial\epsilon\partial n_{a}}\sum_{i}\sum_{j}\zeta_{\epsilon\mathfrak{D}_{i}}\mathcal{Z}^{\alpha\beta}\partial_{\epsilon n}^{i}\Omega_{\alpha\beta}\zeta_{n_{a}\mathfrak{D}_{j}}\mathcal{Z}^{\rho\sigma}\partial_{\epsilon n}^{j}\Omega_{\rho\sigma}+\frac{1}{2}\sum_{ab}\frac{\partial^{2}p}{\partial n_{a}\partial n_{b}}\zeta_{n_{a}p}\zeta_{n_{b}p}\theta^{2}\\
	&-\frac{1}{2}\sum_{ab}\frac{\partial^{2}p}{\partial n_{a}\partial n_{b}}\zeta_{n_{a}p}\theta\sum_{j}\zeta_{n_{b}\mathfrak{D}_{j}}\mathcal{Z}^{\rho\sigma}\partial_{\epsilon n}^{j}\Omega_{\rho\sigma}-\frac{1}{2}\sum_{ab}\frac{\partial^{2}p}{\partial n_{a}\partial n_{b}}\zeta_{n_{b}p}\theta\sum_{j}\zeta_{n_{a}\mathfrak{D}_{j}}\mathcal{Z}^{\alpha\beta}\partial_{\epsilon n}^{j}\Omega_{\alpha\beta}\\
	&+\frac{1}{2}\sum_{ab}\frac{\partial^{2}p}{\partial n_{a}\partial n_{b}}\sum_{i}\zeta_{n_{a}\mathfrak{D}_{i}}\mathcal{Z}^{\alpha\beta}\partial_{\epsilon n}^{i}\Omega_{\alpha\beta}\sum_{j}\zeta_{n_{b}\mathfrak{D}_{j}}\mathcal{Z}^{\rho\sigma}\partial_{\epsilon n}^{j}\Omega_{\rho\sigma}+\sum_{a}\frac{\partial^{2}p}{\partial n_{a}\partial S^{\alpha\beta}}\zeta_{n_{a}p}\theta\zeta_{SS}\mathcal{R}^{\langle\alpha\rangle\langle\beta\rangle}+\sum_{a}\frac{\partial^{2}p}{\partial n_{a}\partial S^{\alpha\beta}}\zeta_{n_{a}p}\theta\zeta_{S\phi}\xi^{\alpha\beta}\\
	&-\sum_{a}\frac{\partial^{2}p}{\partial n_{a}\partial S^{\alpha\beta}}\zeta_{SS}\mathcal{R}^{\langle\alpha\rangle\langle\beta\rangle}\sum_{j}\zeta_{n_{a}\mathfrak{D}_{j}}\mathcal{Z}^{\rho\sigma}\partial_{\epsilon n}^{j}\Omega_{\rho\sigma}-\sum_{a}\frac{\partial^{2}p}{\partial n_{a}\partial S^{\alpha\beta}}\zeta_{S\phi}\xi^{\alpha\beta}\sum_{j}\zeta_{n_{a}\mathfrak{D}_{j}}\mathcal{Z}^{\rho\sigma}\partial_{\epsilon n}^{j}\Omega_{\rho\sigma}\\
	&+\frac{1}{2}\frac{\partial^{2}p}{\partial S^{\alpha\beta}\partial S^{\rho\sigma}}\zeta_{SS}\mathcal{R}^{\langle\alpha\rangle\langle\beta\rangle}\zeta_{SS}\mathcal{R}^{\langle\rho\rangle\langle\sigma\rangle}+\frac{1}{2}\frac{\partial^{2}p}{\partial S^{\alpha\beta}\partial S^{\rho\sigma}}\zeta_{SS}\mathcal{R}^{\langle\alpha\rangle\langle\beta\rangle}\zeta_{S\phi}\xi^{\rho\sigma}+\frac{1}{2}\frac{\partial^{2}p}{\partial S^{\alpha\beta}\partial S^{\rho\sigma}}\zeta_{S\phi}\xi^{\alpha\beta}\zeta_{SS}\mathcal{R}^{\langle\rho\rangle\langle\sigma\rangle}\\
	&+\frac{1}{2}\frac{\partial^{2}p}{\partial S^{\alpha\beta}\partial S^{\rho\sigma}}\zeta_{S\phi}\xi^{\alpha\beta}\zeta_{S\phi}\xi^{\rho\sigma}+\frac{\partial^{2}p}{\partial\epsilon\partial S^{\alpha\beta}}\zeta_{\epsilon p}\theta\zeta_{SS}\mathcal{R}^{\langle\alpha\rangle\langle\beta\rangle}+\frac{\partial^{2}p}{\partial\epsilon\partial S^{\alpha\beta}}\zeta_{\epsilon p}\theta\zeta_{S\phi}\xi^{\alpha\beta}-\frac{\partial^{2}p}{\partial\epsilon\partial S^{\alpha\beta}}\zeta_{SS}\mathcal{R}^{\langle\alpha\rangle\langle\beta\rangle}\sum_{i}\zeta_{\epsilon\mathfrak{D}_{i}}\mathcal{Z}^{\mu\nu}\partial_{\epsilon n}^{i}\Omega_{\mu\nu}\\
	&-\frac{\partial^{2}p}{\partial\epsilon\partial S^{\alpha\beta}}\zeta_{S\phi}\xi^{\alpha\beta}\sum_{i}\zeta_{\epsilon\mathfrak{D}_{i}}\mathcal{Z}^{\mu\nu}\partial_{\epsilon n}^{i}\Omega_{\mu\nu}-\widetilde{\zeta}_{p\epsilon}\theta^{2}\widetilde{\Gamma}-\sum_{a}\widetilde{\zeta}_{pn_{a}}\theta^{2}\widetilde{\delta}_{a}-\widetilde{\zeta}\theta\left(\theta\Gamma-2\mathcal{Z}^{\alpha\beta}\mathcal{D}_{\alpha\beta}\right)-\widetilde{\zeta}D\theta\\
	&+\sum_{i}\widetilde{\zeta}_{p\mathfrak{D}_{i}}\mathcal{Z}^{\alpha\beta}\partial_{\epsilon n}^{i}\Omega_{\alpha\beta}\left(\theta\Gamma-2\mathcal{Z}_{\rho\sigma}\mathcal{D}^{\rho\sigma}\right)+\sum_{i}\widetilde{\zeta}_{p\mathfrak{D}_{i}}\partial_{\epsilon n}^{i}\Omega_{\alpha\beta}D\mathcal{Z}^{\alpha\beta}-\widetilde{\zeta}_{p\epsilon}\zeta_{p\epsilon}^{*}\theta\mathcal{Z}^{\alpha\beta}-\sum_{c}\widetilde{\zeta}_{pn_{c}}\zeta_{pn_{c}}^{*}\theta\mathcal{Z}^{\alpha\beta}\\
	&+\widetilde{\zeta}_{SS}\mathcal{K}^{\langle\mu\rangle\langle\nu\rangle}\mathcal{R}_{\langle\mu\rangle\langle\nu\rangle}\left(\theta\Gamma-2\mathcal{Z}_{\rho\sigma}\mathcal{D}^{\rho\sigma}\right)+\widetilde{\zeta}_{SS}\mathcal{K}^{\langle\mu\rangle\langle\nu\rangle}D\mathcal{R}_{\langle\mu\rangle\langle\nu\rangle}+\widetilde{\zeta}_{S\phi}\mathcal{K}^{\langle\mu\rangle\langle\nu\rangle}\xi_{\mu\nu}\left(\theta\Gamma-2\mathcal{Z}^{\alpha\beta}\mathcal{D}_{\alpha\beta}\right)+\widetilde{\zeta}_{S\phi}\mathcal{K}^{\langle\mu\rangle\langle\nu\rangle}D\xi_{\mu\nu}\\
	&+\sum_{i}\zeta_{p\mathfrak{D}_{i}}\Bigl[\left(\partial_{\epsilon n}^{i}\beta\right)\mathcal{X}+\sum_{a}\left(\partial_{\epsilon n}^{i}\alpha_{a}\right)\mathcal{Y}_{a}+\left(\partial_{\epsilon n}^{i}\Omega_{\alpha\beta}\right)\mathcal{Z}^{\alpha\beta}\Bigr]+\zeta_{SS}\mathcal{K}_{\mu\nu}\mathcal{T}^{\langle\mu\rangle\langle\nu\rangle}+\zeta_{ppp}\theta^{2}+2\theta\sum_{i}\zeta_{pp\mathfrak{D}_{i}}\mathcal{Z}^{\mu\nu}\partial_{\epsilon n}^{i}\Omega_{\mu\nu}\\
	&+\zeta_{pSS}\mathcal{R}_{\mu\nu}\mathcal{R}^{\langle\mu\rangle\langle\nu\rangle}+2\zeta_{pS\phi}\mathcal{R}_{\mu\nu}\xi^{\mu\nu}+\sum_{i}\sum_{j}\zeta_{p\mathfrak{D}_{i}\mathfrak{D}_{j}}\mathcal{Z}^{\mu\nu}\partial_{\epsilon n}^{i}\Omega_{\mu\nu}\mathcal{Z}^{\alpha\beta}\partial_{\epsilon n}^{i}\Omega_{\alpha\beta}+\sum_{ab}\zeta_{p\mathscr{J}_{a}\mathscr{J}_{b}}\nabla^{\sigma}\alpha_{a}\nabla_{\sigma}\alpha_{b}\\
	&+2\sum_{a}\zeta_{p\mathscr{J}_{a}h}\nabla^{\sigma}\alpha_{a}N_{\sigma}+2\sum_{a}\zeta_{p\mathscr{J}_{a}q}\nabla^{\sigma}\alpha_{a}M_{\sigma}+\zeta_{phh}N^{\sigma}N_{\sigma}+2\zeta_{phq}N^{\sigma}M_{\sigma}+\zeta_{pqq}M^{\sigma}M_{\sigma}+\zeta_{p\pi\pi}\sigma^{\rho\sigma}\sigma_{\rho\sigma}+\zeta_{p\phi\phi}\xi^{\rho\sigma}\xi_{\rho\sigma}\\
	&+\zeta_{p\varpi\varpi}\varXi_{\mu\nu\sigma}\varXi^{\mu\nu\sigma}+2\zeta_{SpS}\mathcal{K}_{\mu\nu}\theta\mathcal{R}^{\langle\mu\rangle\langle\nu\rangle}+2\zeta_{Sp\phi}\mathcal{K}_{\mu\nu}\theta\xi^{\mu\nu}+\zeta_{SSS}\mathcal{K}^{\langle\sigma\rangle\langle\beta\rangle}\mathcal{R}_{\,\,\,\,\sigma}^{\langle\alpha\rangle}\mathcal{R}_{\alpha\beta}+2\sum_{i}\zeta_{SS\mathfrak{D}_{i}}\mathcal{K}_{\mu\nu}\mathcal{R}^{\langle\mu\rangle\langle\nu\rangle}\mathcal{Z}^{\alpha\beta}\partial_{\epsilon n}^{i}\Omega_{\alpha\beta}\\
	&+2\zeta_{SS\pi}\mathcal{K}^{\langle\sigma\rangle\langle\beta\rangle}\mathcal{R}_{\,\,\,\,\sigma}^{\langle\alpha\rangle}\sigma_{\alpha\beta}+2\zeta_{SS\phi}\mathcal{K}^{\langle\sigma\rangle\langle\beta\rangle}\mathcal{R}_{\,\,\,\,\sigma}^{\langle\alpha\rangle}\xi_{\alpha\beta}+2\sum_{i}\zeta_{S\mathfrak{D}_{i}\phi}\mathcal{K}^{\langle\alpha\rangle\langle\beta\rangle}\mathcal{Z}_{\rho\sigma}\partial_{\epsilon n}^{i}\Omega^{\rho\sigma}\xi_{\alpha\beta}+\sum_{ab}\zeta_{S\mathscr{J}_{a}\mathscr{J}_{b}}\mathcal{K}^{\langle\alpha\rangle\langle\beta\rangle}\nabla_{\alpha}\alpha_{a}\nabla_{\beta}\alpha_{b}\\
	&+2\sum_{a}\zeta_{S\mathscr{J}_{a}h}\mathcal{K}^{\langle\alpha\rangle\langle\beta\rangle}\nabla_{\alpha}\alpha_{a}N_{\beta}+2\sum_{a}\zeta_{S\mathscr{J}_{a}q}\mathcal{K}^{\langle\alpha\rangle\langle\beta\rangle}\nabla_{\alpha}\alpha_{a}M_{\beta}+2\sum_{a}\zeta_{S\mathscr{J}_{a}\varpi}\mathcal{K}_{\mu\nu}\nabla_{\alpha}\alpha_{a}\varXi^{\mu\nu\alpha}+\zeta_{Shh}\mathcal{K}^{\langle\alpha\rangle\langle\beta\rangle}N_{\alpha}N_{\beta}\\
	&+2\zeta_{Shq}\mathcal{K}^{\langle\alpha\rangle\langle\beta\rangle}N_{\alpha}M_{\beta}+2\zeta_{Sh\varpi}\mathcal{K}_{\mu\nu}N_{\alpha}\varXi^{\mu\nu\alpha}+\zeta_{Sqq}\mathcal{K}^{\langle\alpha\rangle\langle\beta\rangle}M_{\alpha}M_{\beta}+2\zeta_{Sq\varpi}\mathcal{K}_{\mu\nu}M_{\alpha}\varXi^{\mu\nu\alpha}+\zeta_{S\pi\pi}\mathcal{K}^{\langle\sigma\rangle\langle\beta\rangle}\sigma_{\,\,\,\,\sigma}^{\alpha}\sigma_{\alpha\beta}\\
	&+2\zeta_{S\pi\phi}\mathcal{K}^{\langle\sigma\rangle\langle\beta\rangle}\sigma_{\,\,\,\,\sigma}^{\alpha}\xi_{\alpha\beta}+\zeta_{S\phi\phi}\mathcal{K}^{\langle\sigma\rangle\langle\beta\rangle}\xi_{\,\,\,\,\sigma}^{\alpha}\xi_{\alpha\beta}
\end{aligned}
\label{297}
\end{eqnarray}
The principles of symmetry necessitate that the terms $\zeta_{Shh}\mathcal{K}^{\langle\alpha\rangle\langle\beta\rangle}N_{\alpha}N_{\beta}$ and $\zeta_{Sqq}\mathcal{K}^{\langle\alpha\rangle\langle\beta\rangle}M_{\alpha}M_{\beta}$ evaluate to zero.

To derive a relaxation-type equation for $\Pi$ from Eq.~\eqref{297}, we use the first-order approximation
\begin{eqnarray}
	\theta\simeq-\zeta^{-1}\left(\Pi-\sum_{i}\zeta_{p\mathfrak{D}_{i}}\mathcal{Z}^{\mu\nu}\partial_{\epsilon n}^{i}\Omega_{\mu\nu}-\zeta_{SS}\mathcal{K}_{\alpha\beta}\mathcal{R}^{\langle\alpha\rangle\langle\beta\rangle}-\zeta_{S\phi}\mathcal{K}_{\mu\nu}\xi^{\mu\nu}\right),
	\label{298}
\end{eqnarray}
in the term $-\widetilde{\zeta}D\theta$ on the right-hand side of Eq.~\eqref{297}. We then have 
\begin{equation}
	\begin{aligned}-\widetilde{\zeta}D\theta\simeq & -\widetilde{\zeta}\zeta^{-2}\beta\theta\left[\left(\frac{\partial\zeta}{\partial\beta}\Gamma-\sum_{d}\frac{\partial\zeta}{\partial\alpha_{d}}\delta_{d}-2\frac{\partial\zeta}{\partial\Omega_{\alpha\beta}}\mathcal{K}_{\alpha\beta}\right)-2\theta^{-1}\mathcal{Z}^{\alpha\beta}\frac{\partial\zeta}{\partial S^{\alpha\beta}}\right]\\
		& \times\left[\Pi-\sum_{i}\zeta_{p\mathfrak{D}_{i}}\mathcal{Z}^{\mu\nu}\partial_{\epsilon n}^{i}\Omega_{\mu\nu}-\zeta_{SS}\mathcal{K}_{\alpha\beta}\mathcal{R}^{\langle\alpha\rangle\langle\beta\rangle}-\zeta_{S\phi}\mathcal{K}_{\mu\nu}\xi^{\mu\nu}\right]+\widetilde{\zeta}\zeta^{-1}D\Pi\\
		& -\widetilde{\zeta}\zeta^{-1}\beta\theta\sum_{i}\left[\left(\frac{\partial\zeta_{p\mathfrak{D}_{i}}}{\partial\beta}\Gamma-\sum_{d}\frac{\partial\zeta_{p\mathfrak{D}_{i}}}{\partial\alpha_{d}}\delta_{d}-2\frac{\partial\zeta_{p\mathfrak{D}_{i}}}{\partial\Omega_{\alpha\beta}}\mathcal{K}_{\alpha\beta}\right)-2\theta^{-1}\mathcal{Z}^{\alpha\beta}\frac{\partial\zeta_{p\mathfrak{D}_{i}}}{\partial S^{\alpha\beta}}\right]\mathcal{Z}^{\mu\nu}\partial_{\epsilon n}^{i}\Omega_{\mu\nu}-\widetilde{\zeta}\zeta^{-1}\sum_{i}\zeta_{p\mathfrak{D}_{i}}D\left(\mathcal{Z}^{\mu\nu}\partial_{\epsilon n}^{i}\Omega_{\mu\nu}\right)\\
		& -\widetilde{\zeta}\zeta^{-1}\beta\theta\left[\left(\frac{\partial\zeta_{SS}}{\partial\beta}\Gamma-\sum_{d}\frac{\partial\zeta_{SS}}{\partial\alpha_{d}}\delta_{d}-2\frac{\partial\zeta_{SS}}{\partial\Omega_{\alpha\beta}}\mathcal{K}_{\alpha\beta}\right)-2\theta^{-1}\mathcal{Z}^{\alpha\beta}\frac{\partial\zeta_{SS}}{\partial S^{\alpha\beta}}\right]\mathcal{K}_{\alpha\beta}\mathcal{R}^{\langle\alpha\rangle\langle\beta\rangle}-\widetilde{\zeta}\zeta^{-1}\zeta_{SS}D\left(\mathcal{K}_{\alpha\beta}\mathcal{R}^{\langle\alpha\rangle\langle\beta\rangle}\right)\\
		& -\widetilde{\zeta}\zeta^{-1}\beta\theta\left[\left(\frac{\partial\zeta_{S\phi}}{\partial\beta}\Gamma-\sum_{d}\frac{\partial\zeta_{S\phi}}{\partial\alpha_{d}}\delta_{d}-2\frac{\partial\zeta_{S\phi}}{\partial\Omega_{\alpha\beta}}\mathcal{K}_{\alpha\beta}\right)-2\theta^{-1}\mathcal{Z}^{\alpha\beta}\frac{\partial\zeta_{S\phi}}{\partial S^{\alpha\beta}}\right]\mathcal{K}_{\mu\nu}\xi^{\mu\nu}-\widetilde{\zeta}\zeta^{-1}\zeta_{S\phi}D\left(\mathcal{K}_{\mu\nu}\xi^{\mu\nu}\right),
	\end{aligned}
	\label{299}
\end{equation}
where Eqs.~\eqref{76}-\eqref{78} have been used. Introducing the coefficients
\begin{align}
	\tau_{\Pi}= & -\widetilde{\zeta}\zeta^{-1},\label{300}\\
	\widetilde{\zeta}_{\Pi}= & \tau_{\Pi}\zeta^{-1}\beta\left[\left(\frac{\partial\zeta}{\partial\beta}\Gamma-\sum_{d}\frac{\partial\zeta}{\partial\alpha_{d}}\delta_{d}-2\frac{\partial\zeta}{\partial\Omega_{\alpha\beta}}\mathcal{K}_{\alpha\beta}\right)-2\theta^{-1}\mathcal{Z}^{\alpha\beta}\frac{\partial\zeta}{\partial S^{\alpha\beta}}\right],\label{301}\\
	\overline{\zeta}_{p\mathfrak{D}_{i}} =&\tau_{\Pi}\beta\left[\left(\frac{\partial\zeta_{p\mathfrak{D}_{i}}}{\partial\beta}\Gamma-\sum_{d}\frac{\partial\zeta_{p\mathfrak{D}_{i}}}{\partial\alpha_{d}}\delta_{d}-2\frac{\partial\zeta_{p\mathfrak{D}_{i}}}{\partial\Omega_{\alpha\beta}}\mathcal{K}_{\alpha\beta}\right)-2\theta^{-1}\mathcal{Z}^{\alpha\beta}\frac{\partial\zeta_{p\mathfrak{D}_{i}}}{\partial S^{\alpha\beta}}\right],\label{302}\\
	\overline{\zeta}_{SS} = &\tau_{\Pi}\beta\left[\left(\frac{\partial\zeta_{SS}}{\partial\beta}\Gamma-\sum_{d}\frac{\partial\zeta_{SS}}{\partial\alpha_{d}}\delta_{d}-2\frac{\partial\zeta_{SS}}{\partial\Omega_{\alpha\beta}}\mathcal{K}_{\alpha\beta}\right)-2\theta^{-1}\mathcal{Z}^{\alpha\beta}\frac{\partial\zeta_{SS}}{\partial S^{\alpha\beta}}\right],\label{303}\\
	\overline{\zeta}_{S\phi} = &\tau_{\Pi}\beta\left[\left(\frac{\partial\zeta_{S\phi}}{\partial\beta}\Gamma-\sum_{d}\frac{\partial\zeta_{S\phi}}{\partial\alpha_{d}}\delta_{d}-2\frac{\partial\zeta_{S\phi}}{\partial\Omega_{\alpha\beta}}\mathcal{K}_{\alpha\beta}\right)-2\theta^{-1}\mathcal{Z}^{\alpha\beta}\frac{\partial\zeta_{S\phi}}{\partial S^{\alpha\beta}}\right],\label{304}
\end{align}
and combining Eqs.~\eqref{297} and \eqref{299}, we arrive at the following relaxation equation for the bulk viscous pressure:
\begin{eqnarray}
	\begin{small}
	\begin{aligned}
		\Pi+\tau_{\Pi}\dot{\Pi}= & -\zeta\theta+\sum_{i}\zeta_{p\mathfrak{D}_{i}}\mathcal{Z}^{\mu\nu}\partial_{\epsilon n}^{i}\Omega_{\mu\nu}+\zeta_{SS}\mathcal{K}_{\alpha\beta}\mathcal{R}^{\langle\alpha\rangle\langle\beta\rangle}+\zeta_{S\phi}\mathcal{K}_{\mu\nu}\xi^{\mu\nu}+\frac{1}{2}\frac{\partial^{2}p}{\partial\epsilon^{2}}\zeta_{\epsilon p}^{2}\theta^{2}+\frac{1}{2}\frac{\partial^{2}p}{\partial\epsilon^{2}}\sum_{i}\sum_{j}\zeta_{\epsilon\mathfrak{D}_{i}}\zeta_{\epsilon\mathfrak{D}_{j}}\mathcal{Z}^{\mu\nu}\mathcal{Z}^{\alpha\beta}\partial_{\epsilon n}^{i}\Omega_{\mu\nu}\partial_{\epsilon n}^{j}\Omega_{\alpha\beta}\\
		& -\frac{\partial^{2}p}{\partial\epsilon^{2}}\zeta_{\epsilon p}\theta\sum_{i}\zeta_{\epsilon\mathfrak{D}_{i}}\mathcal{Z}^{\mu\nu}\partial_{\epsilon n}^{i}\Omega_{\mu\nu}+\sum_{a}\frac{\partial^{2}p}{\partial\epsilon\partial n_{a}}\zeta_{\epsilon p}\zeta_{n_{a}p}\theta^{2}-\sum_{a}\frac{\partial^{2}p}{\partial\epsilon\partial n_{a}}\zeta_{\epsilon p}\theta\sum_{j}\zeta_{n_{a}\mathfrak{D}_{j}}\mathcal{Z}^{\rho\sigma}\partial_{\epsilon n}^{j}\Omega_{\rho\sigma}\\
		& -\sum_{a}\frac{\partial^{2}p}{\partial\epsilon\partial n_{a}}\zeta_{n_{a}p}\theta\sum_{i}\zeta_{\epsilon\mathfrak{D}_{i}}\mathcal{Z}^{\alpha\beta}\partial_{\epsilon n}^{i}\Omega_{\alpha\beta}+\sum_{a}\frac{\partial^{2}p}{\partial\epsilon\partial n_{a}}\sum_{i}\sum_{j}\zeta_{\epsilon\mathfrak{D}_{i}}\mathcal{Z}^{\alpha\beta}\partial_{\epsilon n}^{i}\Omega_{\alpha\beta}\zeta_{n_{a}\mathfrak{D}_{j}}\mathcal{Z}^{\rho\sigma}\partial_{\epsilon n}^{j}\Omega_{\rho\sigma}+\frac{1}{2}\sum_{ab}\frac{\partial^{2}p}{\partial n_{a}\partial n_{b}}\zeta_{n_{a}p}\zeta_{n_{b}p}\theta^{2}\\
		& -\frac{1}{2}\sum_{ab}\frac{\partial^{2}p}{\partial n_{a}\partial n_{b}}\zeta_{n_{a}p}\theta\sum_{j}\zeta_{n_{b}\mathfrak{D}_{j}}\mathcal{Z}^{\rho\sigma}\partial_{\epsilon n}^{j}\Omega_{\rho\sigma}-\frac{1}{2}\sum_{ab}\frac{\partial^{2}p}{\partial n_{a}\partial n_{b}}\zeta_{n_{b}p}\theta\sum_{j}\zeta_{n_{a}\mathfrak{D}_{j}}\mathcal{Z}^{\alpha\beta}\partial_{\epsilon n}^{j}\Omega_{\alpha\beta}\\
		& +\frac{1}{2}\sum_{ab}\frac{\partial^{2}p}{\partial n_{a}\partial n_{b}}\sum_{i}\zeta_{n_{a}\mathfrak{D}_{i}}\mathcal{Z}^{\alpha\beta}\partial_{\epsilon n}^{i}\Omega_{\alpha\beta}\sum_{j}\zeta_{n_{b}\mathfrak{D}_{j}}\mathcal{Z}^{\rho\sigma}\partial_{\epsilon n}^{j}\Omega_{\rho\sigma}+\sum_{a}\frac{\partial^{2}p}{\partial n_{a}\partial S^{\alpha\beta}}\zeta_{n_{a}p}\theta\zeta_{SS}\mathcal{R}^{\langle\alpha\rangle\langle\beta\rangle}+\sum_{a}\frac{\partial^{2}p}{\partial n_{a}\partial S^{\alpha\beta}}\zeta_{n_{a}p}\theta\zeta_{S\phi}\xi^{\alpha\beta}\\
		& -\sum_{a}\frac{\partial^{2}p}{\partial n_{a}\partial S^{\alpha\beta}}\zeta_{SS}\mathcal{R}^{\langle\alpha\rangle\langle\beta\rangle}\sum_{j}\zeta_{n_{a}\mathfrak{D}_{j}}\mathcal{Z}^{\rho\sigma}\partial_{\epsilon n}^{j}\Omega_{\rho\sigma}-\sum_{a}\frac{\partial^{2}p}{\partial n_{a}\partial S^{\alpha\beta}}\zeta_{S\phi}\xi^{\alpha\beta}\sum_{j}\zeta_{n_{a}\mathfrak{D}_{j}}\mathcal{Z}^{\rho\sigma}\partial_{\epsilon n}^{j}\Omega_{\rho\sigma}\\
		& +\frac{1}{2}\frac{\partial^{2}p}{\partial S^{\alpha\beta}\partial S^{\rho\sigma}}\zeta_{SS}\mathcal{R}^{\langle\alpha\rangle\langle\beta\rangle}\zeta_{SS}\mathcal{R}^{\langle\rho\rangle\langle\sigma\rangle}+\frac{1}{2}\frac{\partial^{2}p}{\partial S^{\alpha\beta}\partial S^{\rho\sigma}}\zeta_{SS}\mathcal{R}^{\langle\alpha\rangle\langle\beta\rangle}\zeta_{S\phi}\xi^{\rho\sigma}+\frac{1}{2}\frac{\partial^{2}p}{\partial S^{\alpha\beta}\partial S^{\rho\sigma}}\zeta_{S\phi}\xi^{\alpha\beta}\zeta_{SS}\mathcal{R}^{\langle\rho\rangle\langle\sigma\rangle}\\
		& +\frac{1}{2}\frac{\partial^{2}p}{\partial S^{\alpha\beta}\partial S^{\rho\sigma}}\zeta_{S\phi}\xi^{\alpha\beta}\zeta_{S\phi}\xi^{\rho\sigma}+\frac{\partial^{2}p}{\partial\epsilon\partial S^{\alpha\beta}}\zeta_{\epsilon p}\theta\zeta_{SS}\mathcal{R}^{\langle\alpha\rangle\langle\beta\rangle}+\frac{\partial^{2}p}{\partial\epsilon\partial S^{\alpha\beta}}\zeta_{\epsilon p}\theta\zeta_{S\phi}\xi^{\alpha\beta}-\frac{\partial^{2}p}{\partial\epsilon\partial S^{\alpha\beta}}\zeta_{SS}\mathcal{R}^{\langle\alpha\rangle\langle\beta\rangle}\sum_{i}\zeta_{\epsilon\mathfrak{D}_{i}}\mathcal{Z}^{\mu\nu}\partial_{\epsilon n}^{i}\Omega_{\mu\nu}\\
		& -\frac{\partial^{2}p}{\partial\epsilon\partial S^{\alpha\beta}}\zeta_{S\phi}\xi^{\alpha\beta}\sum_{i}\zeta_{\epsilon\mathfrak{D}_{i}}\mathcal{Z}^{\mu\nu}\partial_{\epsilon n}^{i}\Omega_{\mu\nu}-\widetilde{\zeta}_{p\epsilon}\theta^{2}\widetilde{\Gamma}-\sum_{a}\widetilde{\zeta}_{pn_{a}}\theta^{2}\widetilde{\delta}_{a}-\widetilde{\zeta}\theta\left(\theta\Gamma-2\mathcal{Z}^{\alpha\beta}\mathcal{D}_{\alpha\beta}\right)\\
		& +\widetilde{\zeta}_{\Pi}\theta\left(\Pi-\sum_{i}\zeta_{p\mathfrak{D}_{i}}\mathcal{Z}^{\mu\nu}\partial_{\epsilon n}^{i}\Omega_{\mu\nu}-\zeta_{SS}\mathcal{K}_{\alpha\beta}\mathcal{R}^{\langle\alpha\rangle\langle\beta\rangle}-\zeta_{S\phi}\mathcal{K}_{\mu\nu}\xi^{\mu\nu}\right)+\sum_{i}\overline{\zeta}_{p\mathfrak{D}_{i}}\theta\mathcal{Z}^{\mu\nu}\partial_{\epsilon n}^{i}\Omega_{\mu\nu}\\
		& +\tau_{\Pi}\sum_{i}\zeta_{p\mathfrak{D}_{i}}D\left(\mathcal{Z}^{\mu\nu}\partial_{\epsilon n}^{i}\Omega_{\mu\nu}\right)+\overline{\zeta}_{SS}\theta\mathcal{K}_{\alpha\beta}\mathcal{R}^{\langle\alpha\rangle\langle\beta\rangle}+\tau_{\Pi}\zeta_{SS}D\left(\mathcal{K}_{\alpha\beta}\mathcal{R}^{\langle\alpha\rangle\langle\beta\rangle}\right)+\overline{\zeta}_{S\phi}\theta\mathcal{K}_{\mu\nu}\xi^{\mu\nu}+\tau_{\Pi}\zeta_{S\phi}D\left(\mathcal{K}_{\mu\nu}\xi^{\mu\nu}\right)\\
		& +\sum_{i}\widetilde{\zeta}_{p\mathfrak{D}_{i}}\mathcal{Z}^{\alpha\beta}\partial_{\epsilon n}^{i}\Omega_{\alpha\beta}\left(\theta\Gamma-2\mathcal{Z}_{\rho\sigma}\mathcal{D}^{\rho\sigma}\right)+\sum_{i}\widetilde{\zeta}_{p\mathfrak{D}_{i}}\partial_{\epsilon n}^{i}\Omega_{\alpha\beta}D\mathcal{Z}^{\alpha\beta}-\widetilde{\zeta}_{p\epsilon}\zeta_{p\epsilon}^{*}\theta\mathcal{Z}^{\alpha\beta}-\sum_{c}\widetilde{\zeta}_{pn_{c}}\zeta_{pn_{c}}^{*}\theta\mathcal{Z}^{\alpha\beta}\\
		& +\widetilde{\zeta}_{SS}\mathcal{K}^{\langle\mu\rangle\langle\nu\rangle}\mathcal{R}_{\langle\mu\rangle\langle\nu\rangle}\left(\theta\Gamma-2\mathcal{Z}_{\rho\sigma}\mathcal{D}^{\rho\sigma}\right)+\widetilde{\zeta}_{SS}\mathcal{K}^{\langle\mu\rangle\langle\nu\rangle}D\mathcal{R}_{\langle\mu\rangle\langle\nu\rangle}+\widetilde{\zeta}_{S\phi}\mathcal{K}^{\langle\mu\rangle\langle\nu\rangle}\xi_{\mu\nu}\left(\theta\Gamma-2\mathcal{Z}^{\alpha\beta}\mathcal{D}_{\alpha\beta}\right)+\widetilde{\zeta}_{S\phi}\mathcal{K}^{\langle\mu\rangle\langle\nu\rangle}D\xi_{\mu\nu}\\
		& +\sum_{i}\zeta_{p\mathfrak{D}_{i}}\Bigl[\left(\partial_{\epsilon n}^{i}\beta\right)\mathcal{X}+\sum_{a}\left(\partial_{\epsilon n}^{i}\alpha_{a}\right)\mathcal{Y}_{a}+\left(\partial_{\epsilon n}^{i}\Omega_{\alpha\beta}\right)\mathcal{Z}^{\alpha\beta}\Bigr]+\zeta_{SS}\mathcal{K}_{\mu\nu}\mathcal{T}^{\langle\mu\rangle\langle\nu\rangle}+\zeta_{ppp}\theta^{2}+2\theta\sum_{i}\zeta_{pp\mathfrak{D}_{i}}\mathcal{Z}^{\mu\nu}\partial_{\epsilon n}^{i}\Omega_{\mu\nu}\\
		& +\zeta_{pSS}\mathcal{R}_{\mu\nu}\mathcal{R}^{\langle\mu\rangle\langle\nu\rangle}+2\zeta_{pS\phi}\mathcal{R}_{\mu\nu}\xi^{\mu\nu}+\sum_{i}\sum_{j}\zeta_{p\mathfrak{D}_{i}\mathfrak{D}_{j}}\mathcal{Z}^{\mu\nu}\partial_{\epsilon n}^{i}\Omega_{\mu\nu}\mathcal{Z}^{\alpha\beta}\partial_{\epsilon n}^{i}\Omega_{\alpha\beta}+\sum_{ab}\zeta_{p\mathscr{J}_{a}\mathscr{J}_{b}}\nabla^{\sigma}\alpha_{a}\nabla_{\sigma}\alpha_{b}\\
		& +2\sum_{a}\zeta_{p\mathscr{J}_{a}h}\nabla^{\sigma}\alpha_{a}N_{\sigma}+2\sum_{a}\zeta_{p\mathscr{J}_{a}q}\nabla^{\sigma}\alpha_{a}M_{\sigma}+\zeta_{phh}N^{\sigma}N_{\sigma}+2\zeta_{phq}N^{\sigma}M_{\sigma}+\zeta_{pqq}M^{\sigma}M_{\sigma}+\zeta_{p\pi\pi}\sigma^{\rho\sigma}\sigma_{\rho\sigma}+\zeta_{p\phi\phi}\xi^{\rho\sigma}\xi_{\rho\sigma}\\
		& +\zeta_{p\varpi\varpi}\varXi_{\mu\nu\sigma}\varXi^{\mu\nu\sigma}+2\zeta_{SpS}\mathcal{K}_{\mu\nu}\theta\mathcal{R}^{\langle\mu\rangle\langle\nu\rangle}+2\zeta_{Sp\phi}\mathcal{K}_{\mu\nu}\theta\xi^{\mu\nu}+\zeta_{SSS}\mathcal{K}^{\langle\sigma\rangle\langle\beta\rangle}\mathcal{R}_{\,\,\,\,\sigma}^{\langle\alpha\rangle}\mathcal{R}_{\alpha\beta}+2\sum_{i}\zeta_{SS\mathfrak{D}_{i}}\mathcal{K}_{\mu\nu}\mathcal{R}^{\langle\mu\rangle\langle\nu\rangle}\mathcal{Z}^{\alpha\beta}\partial_{\epsilon n}^{i}\Omega_{\alpha\beta}\\
		& +2\zeta_{SS\pi}\mathcal{K}^{\langle\sigma\rangle\langle\beta\rangle}\mathcal{R}_{\,\,\,\,\sigma}^{\langle\alpha\rangle}\sigma_{\alpha\beta}+2\zeta_{SS\phi}\mathcal{K}^{\langle\sigma\rangle\langle\beta\rangle}\mathcal{R}_{\,\,\,\,\sigma}^{\langle\alpha\rangle}\xi_{\alpha\beta}+2\sum_{i}\zeta_{S\mathfrak{D}_{i}\phi}\mathcal{K}^{\langle\alpha\rangle\langle\beta\rangle}\mathcal{Z}_{\rho\sigma}\partial_{\epsilon n}^{i}\Omega^{\rho\sigma}\xi_{\alpha\beta}+\sum_{ab}\zeta_{S\mathscr{J}_{a}\mathscr{J}_{b}}\mathcal{K}^{\langle\alpha\rangle\langle\beta\rangle}\nabla_{\alpha}\alpha_{a}\nabla_{\beta}\alpha_{b}\\
		& +2\sum_{a}\zeta_{S\mathscr{J}_{a}h}\mathcal{K}^{\langle\alpha\rangle\langle\beta\rangle}\nabla_{\alpha}\alpha_{a}N_{\beta}+2\sum_{a}\zeta_{S\mathscr{J}_{a}q}\mathcal{K}^{\langle\alpha\rangle\langle\beta\rangle}\nabla_{\alpha}\alpha_{a}M_{\beta}+2\sum_{a}\zeta_{S\mathscr{J}_{a}\varpi}\mathcal{K}_{\mu\nu}\nabla_{\alpha}\alpha_{a}\varXi^{\mu\nu\alpha}+\zeta_{Shh}\mathcal{K}^{\langle\alpha\rangle\langle\beta\rangle}N_{\alpha}N_{\beta}\\
		& +2\zeta_{Shq}\mathcal{K}^{\langle\alpha\rangle\langle\beta\rangle}N_{\alpha}M_{\beta}+2\zeta_{Sh\varpi}\mathcal{K}_{\mu\nu}N_{\alpha}\varXi^{\mu\nu\alpha}+\zeta_{Sqq}\mathcal{K}^{\langle\alpha\rangle\langle\beta\rangle}M_{\alpha}M_{\beta}+2\zeta_{Sq\varpi}\mathcal{K}_{\mu\nu}M_{\alpha}\varXi^{\mu\nu\alpha}+\zeta_{S\pi\pi}\mathcal{K}^{\langle\sigma\rangle\langle\beta\rangle}\sigma_{\,\,\,\,\sigma}^{\alpha}\sigma_{\alpha\beta}\\
		& +2\zeta_{S\pi\phi}\mathcal{K}^{\langle\sigma\rangle\langle\beta\rangle}\sigma_{\,\,\,\,\sigma}^{\alpha}\xi_{\alpha\beta}+\zeta_{S\phi\phi}\mathcal{K}^{\langle\sigma\rangle\langle\beta\rangle}\xi_{\,\,\,\,\sigma}^{\alpha}\xi_{\alpha\beta}
	\end{aligned}
	\end{small}
	\label{305}
\end{eqnarray}
where we define $\dot{\Pi}=D\Pi$.

\subsubsection{Second-order corrections to the charge-diffusion currents}

Utilizing Eqs.~\eqref{90}, \eqref{95}, and \eqref{123} in conjunction with Curie’s theorem leads to the following expression:
\begin{eqnarray}
\begin{aligned}
	\langle\hat{\mathscr{J}}_{c\mu}\left(x\right)\rangle_{2}^{1}= & -\sum_{a}\int d^{4}x_{1}\left(\hat{\mathscr{J}}_{c\mu}\left(x\right),\hat{\mathscr{J}}_{a\sigma}\left(x_{1}\right)\right)\nabla_{x_{1}}^{\sigma}\alpha_{a}\left(x_{1}\right)+\int d^{4}x_{1}\left(\hat{\mathscr{J}}_{c\mu}\left(x\right),\hat{h}_{\sigma}\left(x_{1}\right)\right)\beta\left(x_{1}\right)N^{\sigma}\left(x_{1}\right)\\
	&+\int d^{4}x_{1}\left(\hat{\mathscr{J}}_{c\mu}\left(x\right),\hat{q}_{\sigma}\left(x_{1}\right)\right)\beta\left(x_{1}\right)M^{\sigma}\left(x_{1}\right)-\sum_{a}\chi_{ca}\left(x\right)\nabla_{\mu}\alpha_{a}\left(x\right)-\chi_{\mathscr{J}_{c}h}\left(x\right)N_{\mu}\left(x\right)-\chi_{\mathscr{J}_{c}q}\left(x\right)M_{\mu}\left(x\right).
\end{aligned}
\label{306}
\end{eqnarray}
Substituting the correlation functions defined in Eqs.~\eqref{144}-\eqref{146} and \eqref{153} into Eq.~\eqref{306}, we obtain the following expression:
\begin{eqnarray}
\begin{aligned}
	\langle\hat{\mathscr{J}}_{c\mu}\left(x\right)\rangle_{2}^{1}= & -\frac{1}{3}\sum_{a}\Delta_{\mu\beta}\left(x\right)\int d^{4}x_{1}\left(\hat{\mathscr{J}}_{c}^{\lambda}\left(x\right),\hat{\mathscr{J}}_{a\lambda}\left(x_{1}\right)\right)\nabla_{x_{1}}^{\beta}\alpha_{a}\left(x_{1}\right)+\frac{1}{3}\Delta_{\mu\beta}\left(x\right)\int d^{4}x_{1}\left(\hat{\mathscr{J}}_{c}^{\lambda}\left(x\right),\hat{h}_{\lambda}\left(x_{1}\right)\right)\beta\left(x_{1}\right)N^{\beta}\left(x_{1}\right)\\
	& +\frac{1}{3}\Delta_{\mu\beta}\left(x\right)\int d^{4}x_{1}\left(\hat{\mathscr{J}}_{c}^{\lambda}\left(x\right),\hat{q}_{\lambda}\left(x_{1}\right)\right)\beta\left(x_{1}\right)M^{\beta}\left(x_{1}\right)-\sum_{a}\chi_{ca}\left(x\right)\nabla_{\mu}\alpha_{a}\left(x\right)-\chi_{\mathscr{J}_{c}h}\left(x\right)N_{\mu}\left(x\right)-\chi_{\mathscr{J}_{c}q}\left(x\right)M_{\mu}\left(x\right).
\end{aligned}
\label{307}
\end{eqnarray}
Performing a Taylor expansion of the hydrodynamic quantities around $x_1=x$ and retaining linear terms, we obtain
\begin{align}
\hat{\mathscr{J}}_{a\lambda}\left(x_{1}\right) & =\hat{\mathscr{J}}_{a\lambda}\left(x_{1}\right)\bigg|_{x_{1}=x}+\left(x_{1}-x\right)^{\tau}\frac{\partial\hat{\mathscr{J}}_{a\lambda}\left(x_{1}\right)}{\partial x_{1}^{\tau}}\bigg|_{x_{1}=x},\label{308}\\
\hat{h}_{\lambda}\left(x_{1}\right) & =\hat{h}_{\lambda}\left(x_{1}\right)\bigg|_{x_{1}=x}+\left(x_{1}-x\right)^{\tau}\frac{\partial\hat{h}_{\lambda}\left(x_{1}\right)}{\partial x_{1}^{\tau}}\bigg|_{x_{1}=x},\label{309}\\
\hat{q}_{\lambda}\left(x_{1}\right) & =\hat{q}_{\lambda}\left(x_{1}\right)\bigg|_{x_{1}=x}+\left(x_{1}-x\right)^{\tau}\frac{\partial\hat{q}_{\lambda}\left(x_{1}\right)}{\partial x_{1}^{\tau}}\bigg|_{x_{1}=x},\label{310}
\end{align}
where
\begin{align}
	\hat{\mathscr{J}}_{a\lambda}\left(x_{1}\right)\bigg|_{x_{1}=x}=&\Delta_{\lambda\mu}\left(x\right)\hat{N}_{a}^{\mu}\left(x_{1}\right)-\frac{n_{a}\left(x\right)}{h\left(x\right)}\Delta_{\lambda\mu}\left(x\right)u_{\nu}\left(x\right)\hat{T}^{\mu\nu}\left(x_{1}\right),\label{311}\\
	\hat{h}_{\lambda}\left(x_{1}\right)\bigg|_{x_{1}=x}=&\Delta_{\lambda(\mu}\left(x\right)u_{\nu)}\left(x\right)\hat{T}^{\mu\nu}\left(x_{1}\right),\label{312}\\
	\hat{q}_{\lambda}\left(x_{1}\right)\bigg|_{x_{1}=x}=&\Delta_{\lambda[\mu}\left(x\right)u_{\nu]}\left(x\right)\hat{T}^{\mu\nu}\left(x_{1}\right),\label{313}\\
	\frac{\partial\hat{\mathscr{J}}_{a\lambda}\left(x_{1}\right)}{\partial x_{1}^{\tau}}\bigg|_{x_{1}=x}= & -\hat{\mathscr{J}}_{a}^{\mu}\left(x_{1}\right)u_{\lambda}\left(x\right)\frac{\partial u_{\mu}\left(x_{1}\right)}{\partial x_{1}^{\tau}}\bigg|_{x_{1}=x}-\hat{n}_{a}\left(x_{1}\right)\frac{\partial u_{\lambda}\left(x_{1}\right)}{\partial x_{1}^{\tau}}\bigg|_{x_{1}=x}-\hat{h}_{\lambda}\left(x_{1}\right)\frac{\partial}{\partial x_{1}^{\tau}}\left[n_{a}\left(x_{1}\right){w}^{-1}\left(x_{1}\right)\right]\bigg|_{x_{1}=x}\nonumber\\
	& +n_{a}\left(x\right){w}^{-1}\left(x\right)\left[\hat{\epsilon}\left(x_{1}\right)\frac{\partial u_{\lambda}\left(x_{1}\right)}{\partial x_{1}^{\tau}}\bigg|_{x_{1}=x}+\hat{p}\left(x_{1}\right)\frac{\partial u_{\lambda}\left(x_{1}\right)}{\partial x_{1}^{\tau}}\bigg|_{x_{1}=x}-\hat{\pi}_{\lambda\mu}\left(x_{1}\right)\frac{\partial u^{\mu}\left(x_{1}\right)}{\partial x_{1}^{\tau}}\bigg|_{x_{1}=x}\right],\label{314}\\
	\frac{\partial\hat{h}_{\lambda}\left(x_{1}\right)}{\partial x_{1}^{\tau}}\bigg|_{x_{1}=x}= & -\left[\hat{\epsilon}\left(x_{1}\right)\frac{\partial u_{\lambda}\left(x_{1}\right)}{\partial x_{1}^{\tau}}\bigg|_{x_{1}=x}+\hat{h}^{\mu}\left(x_{1}\right)u_{\lambda}\left(x\right)\frac{\partial u_{\mu}\left(x_{1}\right)}{\partial x_{1}^{\tau}}\bigg|_{x_{1}=x}+\hat{p}\left(x_{1}\right)\frac{\partial u_{\lambda}\left(x_{1}\right)}{\partial x_{1}^{\tau}}\bigg|_{x_{1}=x}-\hat{\pi}_{\lambda\mu}\left(x_{1}\right)\frac{\partial u^{\mu}\left(x_{1}\right)}{\partial x_{1}^{\tau}}\bigg|_{x_{1}=x}\right],\label{315}\\
	\frac{\partial\hat{q}_{\lambda}\left(x_{1}\right)}{\partial x_{1}^{\tau}}\bigg|_{x_{1}=x}= & -\left[\hat{q}^{\mu}\left(x_{1}\right)u_{\lambda}\left(x\right)\frac{\partial u_{\mu}\left(x_{1}\right)}{\partial x_{1}^{\tau}}\bigg|_{x_{1}=x}-\hat{\phi}_{\lambda\nu}\left(x_{1}\right)\frac{\partial u^{\nu}\left(x_{1}\right)}{\partial x_{1}^{\tau}}\bigg|_{x_{1}=x}\right].\label{316}
\end{align}
Substituting Eqs.~\eqref{308}-\eqref{310} into Eq.~\eqref{307} and expanding the thermodynamic forces around $x_1=x$, we obtain up to the second order in gradients
\begin{eqnarray}
\begin{aligned}\langle\hat{\mathscr{J}}_{c\mu}(x)\rangle_{2}^{1}= & -\frac{1}{3}\sum_{a}\nabla_{\mu}\alpha_{a}(x)\int d^{4}x_{1}\left(\hat{\mathscr{J}}_{c}^{\lambda}(x),\frac{\partial\hat{\mathscr{J}}_{a\lambda}\left(x_{1}\right)}{\partial x_{1}^{\tau}}\bigg|_{x_{1}=x}\right)\left(x_{1}-x\right)^{\tau}\\
	& -\frac{1}{3}\sum_{a}\Delta_{\mu\beta}(x)\frac{\partial}{\partial x_{1}^{\tau}}\left[\nabla^{\beta}\alpha_{a}\left(x_{1}\right)\right]\bigg|_{x_{1}=x}\int d^{4}x_{1}\left(\hat{\mathscr{J}}_{c}^{\lambda}(x),\hat{\mathscr{J}}_{a\lambda}\left(x_{1}\right)\bigg|_{x_{1}=x}\right)\left(x_{1}-x\right)^{\tau}\\
	& +\frac{1}{3}\beta(x)N_{\mu}\left(x\right)\int d^{4}x_{1}\left(\hat{\mathscr{J}}_{c}^{\lambda}\left(x\right),\frac{\partial\hat{h}_{\lambda}\left(x_{1}\right)}{\partial x_{1}^{\tau}}\bigg|_{x_{1}=x}\right)\left(x_{1}-x\right)^{\tau}\\
	& +\frac{1}{3}\Delta_{\mu\beta}\left(x\right)\frac{\partial}{\partial x_{1}^{\tau}}\left[\beta\left(x_{1}\right)N^{\beta}\left(x_{1}\right)\right]\bigg|_{x_{1}=x}\int d^{4}x_{1}\left(\hat{\mathscr{J}}_{c}^{\lambda}\left(x\right),\hat{h}_{\lambda}\left(x_{1}\right)\bigg|_{x_{1}=x}\right)\left(x_{1}-x\right)^{\tau}\\
	& +\frac{1}{3}\beta(x)M_{\mu}(x)\int d^{4}x_{1}\left(\hat{\mathscr{J}}_{c}^{\lambda}\left(x\right),\frac{\partial\hat{q}_{\lambda}\left(x_{1}\right)}{\partial x_{1}^{\tau}}\bigg|_{x_{1}=x}\right)\left(x_{1}-x\right)^{\tau}\\
	& +\frac{1}{3}\Delta_{\mu\beta}\left(x\right)\frac{\partial}{\partial x_{1}^{\tau}}\left[\beta\left(x_{1}\right)M^{\beta}\left(x_{1}\right)\right]\bigg|_{x_{1}=x}\int d^{4}x_{1}\left(\hat{\mathscr{J}}_{c}^{\lambda}\left(x\right),\hat{q}_{\lambda}\left(x_{1}\right)\bigg|_{x_{1}=x}\right)\left(x_{1}-x\right)^{\tau}
\end{aligned}
\label{317}
\end{eqnarray}
where we used the following relation to eliminate the first-order terms
\begin{align}
	\chi_{ab} & =-\frac{1}{3}\int d^{4}x_{1}\left(\hat{\mathscr{J}}_{a}^{\lambda}\left(x\right),\hat{\mathscr{J}}_{b\lambda}\left(x_{1}\right)\bigg|_{x_{1}=x}\right),\label{318}\\
	\chi_{\mathscr{J}_{a}h} & =\frac{1}{3}\beta\int d^{4}x_{1}\left(\hat{\mathscr{J}}_{a}^{\lambda}\left(x\right),\hat{h}_{\lambda}\left(x_{1}\right)\bigg|_{x_{1}=x}\right),\label{319}\\
	\chi_{\mathscr{J}_{a}q} & =\frac{1}{3}\beta\int d^{4}x_{1}\left(\hat{\mathscr{J}}_{a}^{\lambda}\left(x\right),\hat{q}_{\lambda}\left(x_{1}\right)\bigg|_{x_{1}=x}\right).\label{320}
\end{align}
Substituting Eqs.~\eqref{314}-\eqref{316} into Eq.~\eqref{317}, applying Curie's theorem, using the first-order approximation $D\beta\simeq\beta\theta\Gamma-2\beta\mathcal{Z}^{\alpha\beta}\mathcal{D}_{\alpha\beta}$, and imposing the orthogonality condition $u^{\lambda}\hat{\mathscr{J}}_{c\lambda}=0$, we derive the nonlocal corrections to the charge-diffusion currents from the two-point correlation function
\begin{eqnarray}
\begin{aligned}
	\langle\hat{\mathscr{J}}_{c\mu}(x)\rangle_{2}^{1}= & \widetilde{\chi}_{\mathscr{J}_{c}h}\sum_{a}D\left(n_{a}{w}^{-1}\right)\nabla_{\mu}\alpha_{a}+\sum_{a}\widetilde{\chi}_{ca}\Delta_{\mu\beta}D\left(\nabla^{\beta}\alpha_{a}\right)+\widetilde{\chi}_{\mathscr{J}_{c}h}N_{\mu}\beta^{-1}\left(\beta\theta\Gamma-2\beta\mathcal{Z}^{\alpha\beta}\mathcal{D}_{\alpha\beta}\right)\\
	&+\widetilde{\chi}_{\mathscr{J}_{c}h}\Delta_{\mu\beta}DN^{\beta}+\widetilde{\chi}_{\mathscr{J}_{c}q}M_{\mu}\beta^{-1}\left(\beta\theta\Gamma-2\beta\mathcal{Z}^{\alpha\beta}\mathcal{D}_{\alpha\beta}\right)+\widetilde{\chi}_{\mathscr{J}_{c}q}\Delta_{\mu\beta}DM^{\beta},
\end{aligned}
\label{321}
\end{eqnarray}
where we define
\begin{align} 
	&\widetilde{\chi}_{ac}=i\frac{d}{d\omega}\chi_{ac}(\omega)\bigg|_{\omega=0}=\frac{T}{6}\frac{d^{2}}{d\omega^{2}}\mathrm{Re}G_{\hat{\mathscr{J}}_{a}^{\lambda}\hat{\mathscr{J}}_{c\lambda}}^{R}(\omega)\bigg|_{\omega=0},\label{322}\\
	&\widetilde{\chi}_{\mathscr{J}_{c}h}=i\frac{d}{d\omega}\chi_{\mathscr{J}_{c}h}(\omega)\bigg|_{\omega=0}=-\frac{1}{6}\frac{d^{2}}{d\omega^{2}}\mathrm{Re}G_{\mathcal{\mathscr{\hat{J}}}_{c}^{\lambda}\hat{h}_{\lambda}}^{R}(\omega)\bigg|_{\omega=0},\label{323}\\
	&\widetilde{\chi}_{\mathscr{J}_{c}q}=i\frac{d}{d\omega}\chi_{\mathscr{J}_{c}q}\left(\omega\right)\bigg|_{\omega=0}=-\frac{1}{6}\frac{d^{2}}{d\omega^{2}}\text{Re}G_{\hat{\mathscr{J}}_{c}^{\lambda}\hat{q}_{\lambda}}^{R}\left(\omega\right)\bigg|_{\omega=0},\label{324}
\end{align}
with the frequency-dependent transport coefficients $\chi_{ac}\left(\omega\right),\chi_{ch}\left(\omega\right)$, and $\chi_{cq}\left(\omega\right)$ are expressed as
\begin{align}
	\chi_{ac}\left(\omega\right) & =-\frac{1}{3}\int d^{4}x_{1}\int_{-\infty}^{t}e^{i\omega\left(t-t_{1}\right)}\left(\hat{\mathscr{J}}_{c}^{\lambda}\left(\boldsymbol{x},t\right),\hat{\mathscr{J}}_{a\lambda}\left(\boldsymbol{x}_{1},t_{1}\right)\right),\label{325}\\
	\chi_{\mathscr{J}_{c}h}\left(\omega\right) & =\frac{1}{3}\beta\int d^{4}x_{1}\int_{-\infty}^{t}e^{i\omega\left(t-t_{1}\right)}\left(\hat{\mathscr{J}}_{c}^{\lambda}\left(\boldsymbol{x},t\right),\hat{h}_{\lambda}\left(\boldsymbol{x}_{1},t_{1}\right)\right),\label{326}\\
	\chi_{\mathscr{J}_{c}q}\left(\omega\right) & =\frac{1}{3}\beta\int d^{4}x_{1}\int_{-\infty}^{t}e^{i\omega\left(t-t_{1}\right)}\left(\hat{\mathscr{J}}_{c}^{\lambda}\left(\boldsymbol{x},t\right),\hat{q}_{\lambda}\left(\boldsymbol{x}_{1},t_{1}\right)\right).\label{327}
\end{align}
Since Eq.~\eqref{321} is already of second order, we can utilize Eq.~\eqref{71} to replace $D\left(n_{a}{w}^{-1}\right)=-n_{a}{w}^{-2}Dp$. By applying Eqs.~\eqref{71} and \eqref{79}, we arrive at the following expression for $Dp$:
\begin{eqnarray}
\begin{aligned}
	Dp & =\Gamma D\epsilon+\sum_{a}\delta_{a}Dn_{a}+\mathcal{K}_{\alpha\beta}DS^{\alpha\beta}\\
	& =-\left[\Gamma {w}\theta+\sum_{a}\delta_{a}n_{a}\theta+\mathcal{K}_{\alpha\beta}\left(\theta S^{\alpha\beta}+u^{\alpha}\partial_{\lambda}S^{\beta\lambda}+S^{\beta\lambda}\partial_{\lambda}u^{\alpha}+u^{\beta}\partial_{\lambda}S^{\lambda\alpha}+S^{\lambda\alpha}\partial_{\lambda}u^{\beta}\right)\right].
\end{aligned}
\label{328}
\end{eqnarray}
Substituting these results into Eq.~\eqref{321}, we obtain the nonlocal corrections from the two-point correlation function to charge-diffusion currents
\begin{eqnarray}
\begin{aligned}
	\langle\hat{\mathscr{J}}_{c\mu}(x)\rangle_{2}^{1}= & \widetilde{\chi}_{\mathscr{J}_{c}h}\sum_{a}n_{a}{w}^{-2}\left[\Gamma {w}\theta+\sum_{c}\delta_{c}n_{c}\theta+\mathcal{K}_{\alpha\beta}\left(\theta S^{\alpha\beta}+u^{\alpha}\partial_{\lambda}S^{\beta\lambda}+S^{\beta\lambda}\partial_{\lambda}u^{\alpha}+u^{\beta}\partial_{\lambda}S^{\lambda\alpha}+S^{\lambda\alpha}\partial_{\lambda}u^{\beta}\right)\right]\nabla_{\mu}\alpha_{a}\\
	&+\sum_{a}\widetilde{\chi}_{ca}\Delta_{\mu\beta}D\left(\nabla^{\beta}\alpha_{a}\right)+\widetilde{\chi}_{\mathscr{J}_{c}h}N_{\mu}\left(\theta\Gamma-2\mathcal{Z}^{\alpha\beta}\mathcal{D}_{\alpha\beta}\right)+\widetilde{\chi}_{\mathscr{J}_{c}h}\Delta_{\mu\beta}DN^{\beta}+\widetilde{\chi}_{\mathscr{J}_{c}q}M_{\mu}\left(\theta\Gamma-2\mathcal{Z}^{\alpha\beta}\mathcal{D}_{\alpha\beta}\right)\\
	&+\widetilde{\chi}_{\mathscr{J}_{c}q}\Delta_{\mu\beta}DM^{\beta}.
\end{aligned}
\label{329}
\end{eqnarray}
Applying Eqs.~\eqref{91} and \eqref{96} and considering Curie’s theorem, we obtain corrections from extended thermodynamic forces to charge-diffusion currents
\begin{eqnarray}
\langle\hat{\mathscr{J}}_{c\mu}\rangle_{2}^{2}=\chi_{\mathscr{J}_{c}h}\mathcal{H}_{\mu}+\chi_{\mathscr{J}_{c}q}\mathcal{Q}_{\mu}.
\label{330}
\end{eqnarray}
Substituting Eq.~\eqref{90} into Eq.~\eqref{97}, we obtain the corrections from the three-point correlation function to charge-diffusion currents
\begin{equation}
\begin{aligned}
	\langle\hat{\mathscr{J}}_{c\mu}\left(x\right)\rangle_{2}^{3}= & \int d^{4}x_{1}d^{4}x_{2}\Bigl(\hat{\mathcal{\mathscr{J}}}_{c\mu}\left(x\right),\biggl[-\beta\theta\hat{p}^{*}+\beta\mathcal{R}_{\alpha\beta}\hat{S}^{\alpha\beta}+\beta\sum_{i}\left(\hat{\mathfrak{D}}_{i}\partial_{\epsilon n}^{i}\Omega_{\mu\nu}\right)\mathcal{Z}^{\mu\nu}-\sum_{a}\hat{\mathscr{J}}_{a}^{\sigma}\nabla_{\sigma}\alpha_{a}\\
	&+\beta\hat{h}^{\sigma}N_{\sigma}+\beta\hat{q}^{\mu}M_{\mu}+\beta\hat{\pi}^{\mu\nu}\sigma_{\mu\nu}+\beta\hat{\phi}^{\mu\nu}\xi_{\mu\nu}+\hat{\varpi}^{\lambda\alpha\beta}\varXi_{\lambda\alpha\beta}\biggr]_{x_{1}},\biggl[-\beta\theta\hat{p}^{*}+\beta\mathcal{R}_{\alpha\beta}\hat{S}^{\alpha\beta}+\beta\sum_{i}\left(\hat{\mathfrak{D}}_{i}\partial_{\epsilon n}^{i}\Omega_{\mu\nu}\right)\mathcal{Z}^{\mu\nu}\\
	&-\sum_{a}\hat{\mathscr{J}}_{a}^{\sigma}\nabla_{\sigma}\alpha_{a}+\beta\hat{h}^{\sigma}N_{\sigma}+\beta\hat{q}^{\mu}M_{\mu}+\beta\hat{\pi}^{\mu\nu}\sigma_{\mu\nu}+\beta\hat{\phi}^{\mu\nu}\xi_{\mu\nu}+\hat{\varpi}^{\lambda\alpha\beta}\varXi_{\lambda\alpha\beta}\biggr]_{x_{2}}\biggr).
\end{aligned}
\label{331}
\end{equation}
The nonvanishing correlators are given by:
\begin{align}
	\left(\hat{\mathscr{J}}_{c\mu}\left(x\right),\hat{p}^{*}\left(x_{1}\right),\hat{\mathscr{J}}_{a\sigma}\left(x_{2}\right)\right) & =\frac{1}{3}\Delta_{\mu\sigma}\left(x\right)\left(\hat{\mathscr{J}}_{c\beta}\left(x\right),\hat{p}^{*}\left(x_{1}\right),\hat{\mathscr{J}}_{a}^{\beta}\left(x_{2}\right)\right),\label{332}\\
	\left(\hat{\mathscr{J}}_{c\mu}\left(x\right),\hat{p}^{*}\left(x_{1}\right),\hat{h}_{\sigma}\left(x_{2}\right)\right) & =\frac{1}{3}\Delta_{\mu\sigma}\left(x\right)\left(\hat{\mathscr{J}}_{c\beta}\left(x\right),\hat{p}^{*}\left(x_{1}\right),\hat{h}^{\beta}\left(x_{2}\right)\right),\label{333}\\
	\left(\hat{\mathscr{J}}_{c\mu}\left(x\right),\hat{p}^{*}\left(x_{1}\right),\hat{q}_{\sigma}\left(x_{2}\right)\right) & =\frac{1}{3}\Delta_{\mu\sigma}\left(x\right)\left(\hat{\mathscr{J}}_{c\beta}\left(x\right),\hat{p}^{*}\left(x_{1}\right),\hat{q}^{\beta}\left(x_{2}\right)\right),\label{334}\\
	\left(\hat{\mathscr{J}}_{c\mu}\left(x\right),\hat{S}_{\rho\sigma}\left(x_{1}\right),\hat{\mathscr{J}}_{a\alpha}\left(x_{2}\right)\right) & =\frac{1}{3}\mydelta_{\mu\alpha\rho\sigma}\left(x\right)\left(\hat{\mathscr{J}}_{c\gamma}\left(x\right),\hat{S}^{\gamma\delta}\left(x_{1}\right),\hat{\mathscr{J}}_{a\delta}\left(x_{2}\right)\right),\label{335}\\
	\left(\hat{\mathscr{J}}_{c\mu}\left(x\right),\hat{S}_{\rho\sigma}\left(x_{1}\right),\hat{h}_{\alpha}\left(x_{2}\right)\right) & =\frac{1}{3}\mydelta_{\mu\alpha\rho\sigma}\left(x\right)\left(\hat{\mathscr{J}}_{c\gamma}\left(x\right),\hat{S}^{\gamma\delta}\left(x_{1}\right),\hat{h}_{\delta}\left(x_{2}\right)\right),\label{336}\\
	\left(\hat{\mathscr{J}}_{c\mu}\left(x\right),\hat{S}_{\rho\sigma}\left(x_{1}\right),\hat{q}_{\alpha}\left(x_{2}\right)\right) & =\frac{1}{3}\mydelta_{\mu\alpha\rho\sigma}\left(x\right)\left(\hat{\mathscr{J}}_{c\gamma}\left(x\right),\hat{S}^{\gamma\delta}\left(x_{1}\right),\hat{q}_{\delta}\left(x_{2}\right)\right),\label{337}\\
	\left(\hat{\mathscr{J}}_{c\mu}\left(x\right),\hat{S}_{\rho\sigma}\left(x_{1}\right),\hat{\varpi}_{\lambda\alpha\beta}\left(x_{2}\right)\right) & =\myDelta_{\mu\rho\sigma\lambda\alpha\beta}\left(x\right)\left(\hat{\mathscr{J}}_{c\nu}\left(x\right),\hat{S}_{\gamma\delta}\left(x_{1}\right),\hat{\varpi}^{\nu\gamma\delta}\left(x_{2}\right)\right),\label{338}\\
	\left(\hat{\mathscr{J}}_{c\mu}\left(x\right),\hat{\mathfrak{D}}_{i}\left(x_{1}\right),\hat{\mathscr{J}}_{a\alpha}\left(x_{2}\right)\right) & =\frac{1}{3}\Delta_{\mu\alpha}\left(x\right)\left(\hat{\mathscr{J}}_{c\lambda}\left(x\right),\hat{\mathfrak{D}}_{i}\left(x_{1}\right),\hat{\mathscr{J}}_{a}^{\lambda}\left(x_{2}\right)\right),\label{339}\\
	\left(\hat{\mathscr{J}}_{c\mu}\left(x\right),\hat{\mathfrak{D}}_{i}\left(x_{1}\right),\hat{h}_{\alpha}\left(x_{2}\right)\right) & =\frac{1}{3}\Delta_{\mu\alpha}\left(x\right)\left(\hat{\mathscr{J}}_{c\lambda}\left(x\right),\hat{\mathfrak{D}}_{i}\left(x_{1}\right),\hat{h}^{\lambda}\left(x_{2}\right)\right),\label{340}\\
	\left(\hat{\mathscr{J}}_{c\mu}\left(x\right),\hat{\mathfrak{D}}_{i}\left(x_{1}\right),\hat{q}_{\alpha}\left(x_{2}\right)\right)& =\frac{1}{3}\Delta_{\mu\alpha}\left(x\right)\left(\hat{\mathscr{J}}_{c\lambda}\left(x\right),\hat{\mathfrak{D}}_{i}\left(x_{1}\right),\hat{q}^{\lambda}\left(x_{2}\right)\right),\label{341}\\
	\left(\hat{\mathscr{J}}_{c\mu}\left(x\right),\hat{\mathscr{J}}_{a\nu}\left(x_{1}\right),\hat{\pi}_{\alpha\beta}\left(x_{2}\right)\right) & =\frac{1}{5}\Delta_{\mu\nu\alpha\beta}\left(x\right)\left(\hat{\mathscr{J}}_{c\lambda}\left(x\right),\hat{\mathscr{J}}_{a\delta}\left(x_{1}\right),\hat{\pi}^{\lambda\delta}\left(x_{2}\right)\right),\label{342}\\
	\left(\hat{\mathscr{J}}_{c\mu}\left(x\right),\hat{\mathscr{J}}_{a\nu}\left(x_{1}\right),\hat{\phi}_{\alpha\beta}\left(x_{2}\right)\right) & =\frac{1}{3}\mydelta_{\mu\nu\alpha\beta}\left(x\right)\left(\hat{\mathscr{J}}_{c\lambda}\left(x\right),\hat{\mathscr{J}}_{a\delta}\left(x_{1}\right),\hat{\phi}^{\lambda\delta}\left(x_{2}\right)\right),\label{343}\\
	\left(\hat{\mathscr{J}}_{c\mu}\left(x\right),\hat{h}_{\nu}\left(x_{1}\right),\hat{\pi}_{\alpha\beta}\left(x_{2}\right)\right) & =\frac{1}{5}\Delta_{\mu\nu\alpha\beta}\left(x\right)\left(\hat{\mathscr{J}}_{c\lambda}\left(x\right),\hat{h}_{\delta}\left(x_{1}\right),\hat{\pi}^{\lambda\delta}\left(x_{2}\right)\right),\label{344}\\
	\left(\hat{\mathscr{J}}_{c\mu}\left(x\right),\hat{h}_{\nu}\left(x_{1}\right),\hat{\phi}_{\alpha\beta}\left(x_{2}\right)\right) & =\frac{1}{3}\mydelta_{\mu\nu\alpha\beta}\left(x\right)\left(\hat{\mathscr{J}}_{c\lambda}\left(x\right),\hat{h}_{\delta}\left(x_{1}\right),\hat{\phi}^{\lambda\delta}\left(x_{2}\right)\right),\label{345}\\
	\left(\hat{\mathscr{J}}_{c\mu}\left(x\right),\hat{q}_{\nu}\left(x_{1}\right),\hat{\pi}_{\alpha\beta}\left(x_{2}\right)\right) & =\frac{1}{5}\Delta_{\mu\nu\alpha\beta}\left(x\right)\left(\hat{\mathscr{J}}_{c\lambda}\left(x\right),\hat{q}_{\delta}\left(x_{1}\right),\hat{\pi}^{\lambda\delta}\left(x_{2}\right)\right),\label{346}\\
	\left(\hat{\mathscr{J}}_{c\mu}\left(x\right),\hat{q}_{\nu}\left(x_{1}\right),\hat{\phi}_{\alpha\beta}\left(x_{2}\right)\right) & =\frac{1}{3}\mydelta_{\mu\nu\alpha\beta}\left(x\right)\left(\hat{\mathscr{J}}_{c\lambda}\left(x\right),\hat{q}_{\delta}\left(x_{1}\right),\hat{\phi}^{\lambda\delta}\left(x_{2}\right)\right),\label{347}\\
	\left(\hat{\mathscr{J}}_{c\mu}\left(x\right),\hat{\phi}_{\nu\sigma}\left(x_{1}\right),\hat{\varpi}_{\lambda\alpha\beta}\left(x_{2}\right)\right) & =\myDelta_{\mu\nu\sigma\lambda\alpha\beta}\left(x\right)\left(\hat{\mathscr{J}}_{c\rho}\left(x\right),\hat{\phi}_{\gamma\delta}\left(x_{1}\right),\hat{\varpi}^{\rho\gamma\delta}\left(x_{2}\right)\right).\label{348}
\end{align}
We now define the following coefficients:
\begin{align}
	\chi_{\mathscr{J}_{c}p\mathscr{J}_{a}} & =\frac{1}{3}\beta\int d^{4}x_{1}d^{4}x_{2}\left(\hat{\mathscr{J}}_{c\beta}\left(x\right),\hat{p}^{*}\left(x_{1}\right),\hat{\mathscr{J}}_{a}^{\beta}\left(x_{2}\right)\right),\chi_{\mathscr{J}_{c}ph}=-\frac{1}{3}\beta^{2}\int d^{4}x_{1}d^{4}x_{2}\left(\hat{\mathscr{J}}_{c\beta}\left(x\right),\hat{p}^{*}\left(x_{1}\right),\hat{h}^{\beta}\left(x_{2}\right)\right),\label{349}\\
	\chi_{\mathscr{J}_{c}pq} & =-\frac{1}{3}\beta^{2}\int d^{4}x_{1}d^{4}x_{2}\left(\hat{\mathscr{J}}_{c\beta}\left(x\right),\hat{p}^{*}\left(x_{1}\right),\hat{q}^{\beta}\left(x_{2}\right)\right),\chi_{\mathscr{J}_{c}S\mathscr{J}_{a}}=-\frac{1}{3}\beta\int d^{4}x_{1}d^{4}x_{2}\left(\hat{\mathscr{J}}_{c\gamma}\left(x\right),\hat{S}^{\gamma\delta}\left(x_{1}\right),\hat{\mathscr{J}}_{a\delta}\left(x_{2}\right)\right),\label{350}\\
	\chi_{\mathscr{J}_{c}Sh} & =\frac{1}{3}\beta^{2}\int d^{4}x_{1}d^{4}x_{2}\left(\hat{\mathscr{J}}_{c\gamma}\left(x\right),\hat{S}^{\gamma\delta}\left(x_{1}\right),\hat{h}_{\delta}\left(x_{2}\right)\right),\chi_{\mathscr{J}_{c}Sq}=\frac{1}{3}\beta^{2}\int d^{4}x_{1}d^{4}x_{2}\left(\hat{\mathscr{J}}_{c\gamma}\left(x\right),\hat{S}^{\gamma\delta}\left(x_{1}\right),\hat{q}_{\delta}\left(x_{2}\right)\right),\label{351}\\
	\chi_{\mathscr{J}_{c}S\varpi} & =\beta\int d^{4}x_{1}d^{4}x_{2}\left(\hat{\mathscr{J}}_{c\nu}\left(x\right),\hat{S}_{\gamma\delta}\left(x_{1}\right),\hat{\varpi}^{\nu\gamma\delta}\left(x_{2}\right)\right),\chi_{\mathscr{J}_{c}\mathfrak{D}_{i}\mathscr{J}_{a}}=-\frac{1}{3}\beta\int d^{4}x_{1}d^{4}x_{2}\left(\hat{\mathscr{J}}_{c\lambda}\left(x\right),\hat{\mathfrak{D}}_{i}\left(x_{1}\right),\hat{\mathscr{J}}_{a}^{\lambda}\left(x_{2}\right)\right),\label{352}\\
	\chi_{\mathscr{J}_{c}\mathfrak{D}_{i}h} & =\frac{1}{3}\beta^{2}\int d^{4}x_{1}d^{4}x_{2}\left(\hat{\mathscr{J}}_{c\lambda}\left(x\right),\hat{\mathfrak{D}}_{i}\left(x_{1}\right),\hat{h}^{\lambda}\left(x_{2}\right)\right),\chi_{\mathscr{J}_{c}\mathfrak{D}_{i}q}=\frac{1}{3}\beta^{2}\int d^{4}x_{1}d^{4}x_{2}\left(\hat{\mathscr{J}}_{c\lambda}\left(x\right),\hat{\mathfrak{D}}_{i}\left(x_{1}\right),\hat{q}^{\lambda}\left(x_{2}\right)\right),\label{353}\\
	\chi_{\mathscr{J}_{c}\mathscr{J}_{a}\pi} & =-\frac{1}{5}\beta\int d^{4}x_{1}d^{4}x_{2}\left(\hat{\mathscr{J}}_{c\lambda}\left(x\right),\hat{\mathscr{J}}_{a\delta}\left(x_{1}\right),\hat{\pi}^{\lambda\delta}\left(x_{2}\right)\right),\chi_{\mathscr{J}_{c}\mathscr{J}_{a}\phi}=-\frac{1}{3}\beta\int d^{4}x_{1}d^{4}x_{2}\left(\hat{\mathscr{J}}_{c\lambda}\left(x\right),\hat{\mathscr{J}}_{a\delta}\left(x_{1}\right),\hat{\phi}^{\lambda\delta}\left(x_{2}\right)\right),\label{354}\\
	\chi_{\mathscr{J}_{c}h\pi} & =\frac{1}{5}\beta^{2}\int d^{4}x_{1}d^{4}x_{2}\left(\hat{\mathscr{J}}_{c\lambda}\left(x\right),\hat{h}_{\delta}\left(x_{1}\right),\hat{\pi}^{\lambda\delta}\left(x_{2}\right)\right),\chi_{\mathscr{J}_{c}h\phi}=\frac{1}{3}\beta^{2}\int d^{4}x_{1}d^{4}x_{2}\left(\hat{\mathscr{J}}_{c\lambda}\left(x\right),\hat{h}_{\delta}\left(x_{1}\right),\hat{\phi}^{\lambda\delta}\left(x_{2}\right)\right),\label{355}\\
	\chi_{\mathscr{J}_{c}q\pi} & =\frac{1}{5}\beta^{2}\int d^{4}x_{1}d^{4}x_{2}\left(\hat{\mathscr{J}}_{c\lambda}\left(x\right),\hat{q}_{\delta}\left(x_{1}\right),\hat{\pi}^{\lambda\delta}\left(x_{2}\right)\right),\chi_{\mathscr{J}_{c}q\phi}=\frac{1}{3}\beta^{2}\int d^{4}x_{1}d^{4}x_{2}\left(\hat{\mathscr{J}}_{c\lambda}\left(x\right),\hat{q}_{\delta}\left(x_{1}\right),\hat{\phi}^{\lambda\delta}\left(x_{2}\right)\right),\label{356}\\
	\chi_{\mathscr{J}_{c}\phi\varpi} & =\beta\int d^{4}x_{1}d^{4}x_{2}\left(\hat{\mathscr{J}}_{c\rho}\left(x\right),\hat{\phi}_{\gamma\delta}\left(x_{1}\right),\hat{\varpi}^{\rho\gamma\delta}\left(x_{2}\right)\right).\label{357}
\end{align}
Combining these definitions with Eqs.~\eqref{331}–\eqref{357} and the symmetry property in Eq.~\eqref{49}, we obtain
\begin{equation}
\begin{aligned}
	\langle\hat{\mathscr{J}}_{c\mu}\rangle_{2}^{3}= & 2\sum_{a}\chi_{\mathscr{J}_{c}p\mathscr{J}_{a}}\theta\nabla_{\mu}\alpha_{a}+2\chi_{\mathscr{J}_{c}ph}\theta N_{\mu}+2\chi_{\mathscr{J}_{c}pq}\theta M_{\mu}+2\sum_{a}\chi_{\mathscr{J}_{c}S\mathscr{J}_{a}}\mathcal{R}_{\langle\mu\rangle\langle\alpha\rangle}\nabla^{\alpha}\alpha_{a}+2\chi_{\mathscr{J}_{c}Sh}\mathcal{R}_{\langle\mu\rangle\langle\alpha\rangle}N^{\alpha}\\
	&+2\chi_{\mathscr{J}_{c}Sq}\mathcal{R}_{\langle\mu\rangle\langle\alpha\rangle}M^{\alpha}+2\chi_{\mathscr{J}_{c}S\varpi}\mathcal{R}^{\rho\sigma}\varXi_{\mu\rho\sigma}+2\sum_{i}\mathcal{Z}^{\rho\sigma}\partial_{\epsilon n}^{i}\Omega_{\rho\sigma}\sum_{a}\chi_{\mathscr{J}_{c}\mathfrak{D}_{i}\mathscr{J}_{a}}\nabla_{\mu}\alpha_{a}+2\sum_{i}\mathcal{Z}^{\rho\sigma}\partial_{\epsilon n}^{i}\Omega_{\rho\sigma}\chi_{\mathscr{J}_{c}\mathfrak{D}_{i}h}N_{\mu}\\
	&+2\sum_{i}\chi_{\mathscr{J}_{c}\mathfrak{D}_{i}q}\mathcal{Z}^{\rho\sigma}\partial_{\epsilon n}^{i}\Omega_{\rho\sigma}M_{\mu}+2\sum_{a}\chi_{\mathscr{J}_{c}\mathscr{J}_{a}\pi}\nabla^{\nu}\alpha_{a}\sigma_{\mu\nu}+2\sum_{a}\chi_{\mathscr{J}_{c}\mathscr{J}_{a}\phi}\nabla^{\nu}\alpha_{a}\xi_{\mu\nu}\\
	&+2\chi_{\mathscr{J}_{c}h\pi}N^{\nu}\sigma_{\mu\nu}+2\chi_{\mathscr{J}_{c}h\phi}N^{\nu}\xi_{\mu\nu}+2\chi_{\mathscr{J}_{c}q\pi}M^{\nu}\sigma_{\mu\nu}+2\chi_{\mathscr{J}_{c}q\phi}M^{\nu}\xi_{\mu\nu}+2\chi_{\mathscr{J}_{c}\phi\varpi}\xi^{\nu\sigma}\varXi_{\mu\nu\sigma}.
\end{aligned}
\label{358}
\end{equation}
Combining the results from Eqs.~\eqref{70}, \eqref{92}, \eqref{94}, \eqref{123}, \eqref{329}, \eqref{330}, and \eqref{358}, we obtain the complete second-order expression for the charge-diffusion currents:
\begin{equation}
\begin{aligned}
	\mathscr{J}_{c\mu}= & \sum_{b}\chi_{cb}\nabla_{\mu}\alpha_{b}+\chi_{\mathscr{J}_{c}h}N_{\mu}+\chi_{\mathscr{J}_{c}q}M_{\mu}+\widetilde{\chi}_{\mathscr{J}_{c}h}\sum_{a}n_{a}{w}^{-2}\Bigl[\Gamma {w}\theta+\sum_{c}\delta_{c}n_{c}\theta+\mathcal{K}_{\alpha\beta}\Bigl(\theta S^{\alpha\beta}+u^{\alpha}\partial_{\lambda}S^{\beta\lambda}\\
	&+S^{\beta\lambda}\partial_{\lambda}u^{\alpha}+u^{\beta}\partial_{\lambda}S^{\lambda\alpha}+S^{\lambda\alpha}\partial_{\lambda}u^{\beta}\Bigr)\Bigr]\nabla_{\mu}\alpha_{a}+\sum_{a}\widetilde{\chi}_{ca}\Delta_{\mu\beta}D\left(\nabla^{\beta}\alpha_{a}\right)+\widetilde{\chi}_{\mathscr{J}_{c}h}N_{\mu}\left(\theta\Gamma-2\mathcal{Z}^{\alpha\beta}\mathcal{D}_{\alpha\beta}\right)\\
	&+\widetilde{\chi}_{\mathscr{J}_{c}h}\Delta_{\mu\beta}DN^{\beta}+\widetilde{\chi}_{\mathscr{J}_{c}q}M_{\mu}\left(\theta\Gamma-2\mathcal{Z}^{\alpha\beta}\mathcal{D}_{\alpha\beta}\right)+\widetilde{\chi}_{\mathscr{J}_{c}q}\Delta_{\mu\beta}DM^{\beta}+\chi_{\mathscr{J}_{c}h}\mathcal{H}_{\mu}+\chi_{\mathscr{J}_{c}q}\mathcal{Q}_{\mu}\\
	&+2\sum_{a}\chi_{\mathscr{J}_{c}p\mathscr{J}_{a}}\theta\nabla_{\mu}\alpha_{a}+2\chi_{\mathscr{J}_{c}ph}\theta N_{\mu}+2\chi_{\mathscr{J}_{c}pq}\theta M_{\mu}+2\sum_{a}\chi_{\mathscr{J}_{c}S\mathscr{J}_{a}}\mathcal{R}_{\langle\mu\rangle\langle\alpha\rangle}\nabla^{\alpha}\alpha_{a}+2\chi_{\mathscr{J}_{c}Sh}\mathcal{R}_{\langle\mu\rangle\langle\alpha\rangle}N^{\alpha}\\
	&+2\chi_{\mathscr{J}_{c}Sq}\mathcal{R}_{\langle\mu\rangle\langle\alpha\rangle}M^{\alpha}+2\chi_{\mathscr{J}_{c}S\varpi}\mathcal{R}^{\rho\sigma}\varXi_{\mu\rho\sigma}+2\sum_{i}\mathcal{Z}^{\rho\sigma}\partial_{\epsilon n}^{i}\Omega_{\rho\sigma}\sum_{a}\chi_{\mathscr{J}_{c}\mathfrak{D}_{i}\mathscr{J}_{a}}\nabla_{\mu}\alpha_{a}\\
	&+2\sum_{i}\mathcal{Z}^{\rho\sigma}\partial_{\epsilon n}^{i}\Omega_{\rho\sigma}\chi_{\mathscr{J}_{c}\mathfrak{D}_{i}h}N_{\mu}+2\sum_{i}\chi_{\mathscr{J}_{c}\mathfrak{D}_{i}q}\mathcal{Z}^{\rho\sigma}\partial_{\epsilon n}^{i}\Omega_{\rho\sigma}M_{\mu}+2\sum_{a}\chi_{\mathscr{J}_{c}\mathscr{J}_{a}\pi}\nabla^{\nu}\alpha_{a}\sigma_{\mu\nu}\\
	&+2\sum_{a}\chi_{\mathscr{J}_{c}\mathscr{J}_{a}\phi}\nabla^{\nu}\alpha_{a}\xi_{\mu\nu}+2\chi_{\mathscr{J}_{c}h\pi}N^{\nu}\sigma_{\mu\nu}+2\chi_{\mathscr{J}_{c}h\phi}N^{\nu}\xi_{\mu\nu}+2\chi_{\mathscr{J}_{c}q\pi}M^{\nu}\sigma_{\mu\nu}+2\chi_{\mathscr{J}_{c}q\phi}M^{\nu}\xi_{\mu\nu}\\
	&+2\chi_{\mathscr{J}_{c}\phi\varpi}\xi^{\nu\sigma}\varXi_{\mu\nu\sigma}.
\end{aligned}
\label{359}
\end{equation}
To derive a relaxation-type equation for $\mathscr{J}_{c\mu}$ from \eqref{359}, we use the first-order approximation
\begin{eqnarray}
	\nabla^{\beta}\alpha_{a}\simeq\sum_{b}\left(\chi^{-1}\right)_{ab}\left(\mathscr{J}_{b}^{\beta}-\chi_{\mathscr{J}_{b}q}M^{\beta}-\chi_{\mathscr{J}_{b}h}N^{\beta}\right),
	\label{360}
\end{eqnarray}
in the term $\sum_{a}\widetilde{\chi}_{ca}\Delta_{\mu\beta}D\left(\nabla^{\beta}\alpha_{a}\right)$ on the right-hand side of Eq.~\eqref{359}. We then have
\begin{equation}
	\begin{aligned} 
		&\sum_{a}\widetilde{\chi}_{ca}\Delta_{\mu\beta}D\left(\nabla^{\beta}\alpha_{a}\right)\\
		\simeq&\sum_{b}\left(\widetilde{\chi}\chi^{-1}\right)_{cb}\Delta_{\mu\beta}D\mathscr{J}_{b}^{\beta}-\sum_{b}\left(\widetilde{\chi}\chi^{-1}\right)_{cb}\beta\theta M_{\mu}\biggl[\left(\Gamma\frac{\partial\chi_{\mathscr{J}_{b}q}}{\partial\beta}-\sum_{d}\delta_{d}\frac{\partial\chi_{\mathscr{J}_{b}q}}{\partial\alpha_{d}}-2\frac{\partial\chi_{\mathscr{J}_{b}q}}{\partial\Omega_{\alpha\beta}}\mathcal{K}_{\alpha\beta}\right)-2\theta^{-1}\mathcal{Z}^{\alpha\beta}\frac{\partial\chi_{\mathscr{J}_{b}q}}{\partial S^{\alpha\beta}}\biggr]\\
		&-\sum_{b}\left(\widetilde{\chi}\chi^{-1}\right)_{cb}\Delta_{\mu\beta}\chi_{\mathscr{J}_{b}q}DM^{\beta}-\sum_{b}\left(\widetilde{\chi}\chi^{-1}\right)_{cb}\beta\theta N_{\mu}\biggl[\left(\Gamma\frac{\partial\chi_{\mathscr{J}_{b}h}}{\partial\beta}-\sum_{d}\delta_{d}\frac{\partial\chi_{\mathscr{J}_{b}h}}{\partial\alpha_{d}}-2\frac{\partial\chi_{\mathscr{J}_{b}h}}{\partial\Omega_{\alpha\beta}}\mathcal{K}_{\alpha\beta}\right)-2\theta^{-1}\mathcal{Z}^{\alpha\beta}\frac{\partial\chi_{\mathscr{J}_{b}h}}{\partial S^{\alpha\beta}}\biggr]\\
		&-\sum_{b}\left(\widetilde{\chi}\chi^{-1}\right)_{cb}\Delta_{\mu\beta}\chi_{\mathscr{J}_{b}h}DN^{\beta}+\sum_{ab}\widetilde{\chi}_{ca}\left(\mathscr{J}_{b\mu}-\chi_{\mathscr{J}_{b}q}M_{\mu}-\chi_{\mathscr{J}_{b}h}N_{\mu}\right)\beta\theta\\
		&\times\left[\left(\Gamma\frac{\partial\left(\chi^{-1}\right)_{ab}}{\partial\beta}-\sum_{d}\delta_{d}\frac{\partial\left(\chi^{-1}\right)_{ab}}{\partial\alpha_{d}}-2\mathcal{K}_{\alpha\beta}\frac{\partial\left(\chi^{-1}\right)_{ab}}{\partial\Omega_{\alpha\beta}}\right)-2\theta^{-1}\mathcal{Z}^{\alpha\beta}\frac{\partial\left(\chi^{-1}\right)_{ab}}{\partial\Omega_{\alpha\beta}}\right],
	\end{aligned}
	\label{361}
\end{equation}
where Eqs.~\eqref{76}-\eqref{78} have been used. Introducing the coefficients
\begin{align}
	\tau_{\mathscr{J}}^{cb}= & -\left(\widetilde{\chi}\chi^{-1}\right)_{cb}=-\sum_{a}\widetilde{\chi}_{ca}\left(\chi^{-1}\right)_{ab},\label{362}\\
	\widetilde{\chi}_{\mathscr{J}}^{cb}= & \beta\sum_{a}\widetilde{\chi}_{ca}\left[\left(\Gamma\frac{\partial\left(\chi^{-1}\right)_{ab}}{\partial\beta}-\sum_{d}\delta_{d}\frac{\partial\left(\chi^{-1}\right)_{ab}}{\partial\alpha_{d}}-2\mathcal{K}_{\alpha\beta}\frac{\partial\left(\chi^{-1}\right)_{ab}}{\partial\Omega_{\alpha\beta}}\right)-2\theta^{-1}\mathcal{Z}^{\alpha\beta}\frac{\partial\left(\chi^{-1}\right)_{ab}}{\partial\Omega_{\alpha\beta}}\right],\label{363}\\
	\overline{\chi}^{c}= & \beta\sum_{b}\tau_{\mathscr{J}}^{cb}\left[\left(\Gamma\frac{\partial\chi_{\mathscr{J}_{b}q}}{\partial\beta}-\sum_{d}\delta_{d}\frac{\partial\chi_{\mathscr{J}_{b}q}}{\partial\alpha_{d}}-2\frac{\partial\chi_{\mathscr{J}_{b}q}}{\partial\Omega_{\alpha\beta}}\mathcal{K}_{\alpha\beta}\right)-2\theta^{-1}\mathcal{Z}^{\alpha\beta}\frac{\partial\chi_{\mathscr{J}_{b}q}}{\partial S^{\alpha\beta}}\right],\label{364}\\
	\widehat{\chi}^{c}= & \beta\sum_{b}\tau_{\mathscr{J}}^{cb}\left[\left(\Gamma\frac{\partial\chi_{\mathscr{J}_{b}h}}{\partial\beta}-\sum_{d}\delta_{d}\frac{\partial\chi_{\mathscr{J}_{b}h}}{\partial\alpha_{d}}-2\frac{\partial\chi_{\mathscr{J}_{b}h}}{\partial\Omega_{\alpha\beta}}\mathcal{K}_{\alpha\beta}\right)-2\theta^{-1}\mathcal{Z}^{\alpha\beta}\frac{\partial\chi_{bh}}{\partial S^{\alpha\beta}}\right],\label{365}
\end{align}
and combining Eqs.~\eqref{359} and~\eqref{361}, we arrive at the following relaxation equation for the charge-diffusion currents:
\begin{equation}
\begin{aligned}
	\mathscr{J}_{c\mu}+\sum_{b}\tau_{\mathscr{J}}^{cb}\dot{\mathscr{J}}_{b\mu}= & \sum_{b}\chi_{cb}\nabla_{\mu}\alpha_{b}+\chi_{\mathscr{J}_{c}h}N_{\mu}+\chi_{\mathscr{J}_{c}q}M_{\mu}+\widetilde{\chi}_{\mathscr{J}_{c}h}\sum_{a}n_{a}{w}^{-2}\Bigl[\Gamma {w}\theta+\sum_{c}\delta_{c}n_{c}\theta+\mathcal{K}_{\alpha\beta}\Bigl(\theta S^{\alpha\beta}\\
	&+u^{\alpha}\partial_{\lambda}S^{\beta\lambda}+S^{\beta\lambda}\partial_{\lambda}u^{\alpha}+u^{\beta}\partial_{\lambda}S^{\lambda\alpha}+S^{\lambda\alpha}\partial_{\lambda}u^{\beta}\Bigr)\Bigr]\nabla_{\mu}\alpha_{a}+\overline{\chi}^{c}\theta M_{\mu}+\sum_{b}\tau_{\mathscr{J}}^{cb}\Delta_{\mu\beta}\chi_{\mathscr{J}_{b}q}DM^{\beta}\\
	& +\widehat{\chi}^{c}\theta N_{\mu}+\sum_{b}\tau_{\mathscr{J}}^{cb}\Delta_{\mu\beta}\chi_{\mathscr{J}_{b}h}DN^{\beta}+\sum_{b}\widetilde{\chi}_{\mathscr{J}}^{cb}\theta\left(\mathscr{J}_{b\mu}-\chi_{\mathscr{J}_{b}q}M_{\mu}-\chi_{\mathscr{J}_{b}h}N_{\mu}\right)\\
	& +\widetilde{\chi}_{\mathscr{J}_{c}h}N_{\mu}\left(\theta\Gamma-2\mathcal{Z}^{\alpha\beta}\mathcal{D}_{\alpha\beta}\right)+\widetilde{\chi}_{\mathscr{J}_{c}h}\Delta_{\mu\beta}DN^{\beta}+\widetilde{\chi}_{\mathscr{J}_{c}q}M_{\mu}\left(\theta\Gamma-2\mathcal{Z}^{\alpha\beta}\mathcal{D}_{\alpha\beta}\right)\\
	& +\widetilde{\chi}_{\mathscr{J}_{c}q}\Delta_{\mu\beta}DM^{\beta}+\chi_{\mathscr{J}_{c}h}\mathcal{H}_{\mu}+\chi_{\mathscr{J}_{c}q}\mathcal{Q}_{\mu}+2\sum_{a}\chi_{\mathscr{J}_{c}p\mathscr{J}_{a}}\theta\nabla_{\mu}\alpha_{a}+2\chi_{\mathscr{J}_{c}ph}\theta N_{\mu}\\
	& +2\chi_{\mathscr{J}_{c}pq}\theta M_{\mu}+2\sum_{a}\chi_{\mathscr{J}_{c}S\mathscr{J}_{a}}\mathcal{R}_{\langle\mu\rangle\langle\alpha\rangle}\nabla^{\alpha}\alpha_{a}+2\chi_{\mathscr{J}_{c}Sh}\mathcal{R}_{\langle\mu\rangle\langle\alpha\rangle}N^{\alpha}+2\chi_{\mathscr{J}_{c}Sq}\mathcal{R}_{\langle\mu\rangle\langle\alpha\rangle}M^{\alpha}\\
	& +2\chi_{\mathscr{J}_{c}S\varpi}\mathcal{R}^{\rho\sigma}\varXi_{\mu\rho\sigma}+2\sum_{i}\mathcal{Z}^{\rho\sigma}\partial_{\epsilon n}^{i}\Omega_{\rho\sigma}\sum_{a}\chi_{\mathscr{J}_{c}\mathfrak{D}_{i}\mathscr{J}_{a}}\nabla_{\mu}\alpha_{a}+2\sum_{i}\mathcal{Z}^{\rho\sigma}\partial_{\epsilon n}^{i}\Omega_{\rho\sigma}\chi_{\mathscr{J}_{c}\mathfrak{D}_{i}h}N_{\mu}\\
	& +2\sum_{i}\chi_{\mathscr{J}_{c}\mathfrak{D}_{i}q}\mathcal{Z}^{\rho\sigma}\partial_{\epsilon n}^{i}\Omega_{\rho\sigma}M_{\mu}+2\sum_{a}\chi_{\mathscr{J}_{c}\mathscr{J}_{a}\pi}\nabla^{\nu}\alpha_{a}\sigma_{\mu\nu}+2\sum_{a}\chi_{\mathscr{J}_{c}\mathscr{J}_{a}\phi}\nabla^{\nu}\alpha_{a}\xi_{\mu\nu}\\
	& +2\chi_{\mathscr{J}_{c}h\pi}N^{\nu}\sigma_{\mu\nu}+2\chi_{\mathscr{J}_{c}h\phi}N^{\nu}\xi_{\mu\nu}+2\chi_{\mathscr{J}_{c}q\pi}M^{\nu}\sigma_{\mu\nu}+2\chi_{\mathscr{J}_{c}q\phi}M^{\nu}\xi_{\mu\nu}+2\chi_{\mathscr{J}_{c}\phi\varpi}\xi^{\nu\sigma}\varXi_{\mu\nu\sigma}.
\end{aligned}
	\label{366}
\end{equation}
where we define $\dot{\mathscr{J}}_{b\mu}=\Delta_{\mu\beta}D\mathscr{J}_{b}^{\beta}$.

\subsubsection{Second-order corrections to the rotational stress tensor}

Substituting Eq.~\eqref{90} into Eq.~\eqref{95} and applying Curie's theorem, we obtain
\begin{eqnarray}
\begin{aligned}\langle\hat{\phi}_{\mu\nu}\left(x\right)\rangle_{2}^{1}= & \int d^{4}x_{1}\left(\hat{\phi}_{\mu\nu}\left(x\right),\hat{S}_{\alpha\beta}\left(x_{1}\right)\right)\beta\left(x_{1}\right)\mathcal{R}^{\alpha\beta}\left(x_{1}\right)\\
 & +\int d^{4}x_{1}\left(\hat{\phi}_{\mu\nu}\left(x\right),\hat{\phi}_{\alpha\beta}\left(x_{1}\right)\right)\beta\left(x_{1}\right)\xi^{\alpha\beta}\left(x_{1}\right)-\gamma_{\phi S}\left(x\right)\mathcal{R}_{\langle\mu\rangle\langle\nu\rangle}\left(x\right)-2\gamma\left(x\right)\xi_{\mu\nu}\left(x\right),
\end{aligned}
\label{367}
\end{eqnarray}
where we used the first-order relation $\langle\hat{\phi}_{\mu\nu}\left(x\right)\rangle_{1}$ from Eq.~\eqref{122}. Substituting the two-point correlation functions from Eqs.~\eqref{147},\eqref{148}, and~\eqref{155} into Eq.~\eqref{367}, we obtain
\begin{eqnarray}
\begin{aligned}
\langle\hat{\phi}_{\mu\nu}\left(x\right)\rangle_{2}^{1}= & \frac{1}{3}\mydelta_{\mu\nu\rho\sigma}\left(x\right)\int d^{4}x_{1}\left(\hat{\phi}^{\lambda\eta}\left(x\right),\hat{S}_{\lambda\eta}\left(x_{1}\right)\right)\beta\left(x_{1}\right)\mathcal{R}^{\langle\rho\rangle\langle\sigma\rangle}\left(x_{1}\right)\\
 & +\frac{1}{3}\mydelta_{\mu\nu\rho\sigma}\left(x\right)\int d^{4}x_{1}\left(\hat{\phi}^{\lambda\eta}\left(x\right),\hat{\phi}_{\lambda\eta}\left(x_{1}\right)\right)\beta\left(x_{1}\right)\xi^{\rho\sigma}\left(x_{1}\right)\\
 & -\gamma_{\phi S}\left(x\right)\mathcal{R}_{\langle\mu\rangle\langle\nu\rangle}\left(x\right)-2\gamma\left(x\right)\xi_{\mu\nu}\left(x\right).
\end{aligned}
\label{368}
\end{eqnarray}
Substituting Eq.~\eqref{214} and \eqref{215} into Eq.~\eqref{368} and expanding the thermodynamic forces around $x_1=x$, we obtain up to the second order in gradients
\begin{eqnarray}
\begin{aligned}
\langle\hat{\phi}_{\mu\nu}\left(x\right)\rangle_{2}^{1}= & \frac{1}{3}\mydelta_{\mu\nu\rho\sigma}\left(x\right)\beta\left(x\right)\mathcal{R}^{\langle\rho\rangle\langle\sigma\rangle}\left(x\right)\int d^{4}x_{1}\left(\hat{\phi}^{\lambda\eta}\left(x\right),\frac{\partial\hat{S}_{\lambda\eta}\left(x_{1}\right)}{\partial x_{1}^{\tau}}\bigg|_{x_{1}=x}\right)\left(x_{1}-x\right)^{\tau}\\
 & +\frac{1}{3}\mydelta_{\mu\nu\rho\sigma}\left(x\right)\frac{\partial}{\partial x_{1}^{\tau}}\left[\beta\left(x_{1}\right)\mathcal{R}^{\langle\rho\rangle\langle\sigma\rangle}\left(x_{1}\right)\right]\bigg|_{x_{1}=x}\int d^{4}x_{1}\left(\hat{\phi}^{\lambda\eta}\left(x\right),\hat{S}_{\lambda\eta}\left(x_{1}\right)\bigg|_{x_{1}=x}\right)\left(x_{1}-x\right)^{\tau}\\
 & +\frac{1}{3}\mydelta_{\mu\nu\rho\sigma}\left(x\right)\beta\left(x\right)\xi^{\rho\sigma}\left(x\right)\int d^{4}x_{1}\left(\hat{\phi}^{\lambda\eta}\left(x\right),\frac{\partial\hat{\phi}_{\lambda\eta}\left(x_{1}\right)}{\partial x_{1}^{\tau}}\bigg|_{x_{1}=x}\right)\left(x_{1}-x\right)^{\tau}\\
 & +\frac{1}{3}\mydelta_{\mu\nu\rho\sigma}\left(x\right)\frac{\partial}{\partial x_{1}^{\tau}}\left[\beta\left(x_{1}\right)\xi^{\rho\sigma}\left(x_{1}\right)\right]\bigg|_{x_{1}=x}\int d^{4}x_{1}\left(\hat{\phi}^{\lambda\eta}\left(x\right),\hat{\phi}_{\lambda\eta}\left(x_{1}\right)\bigg|_{x_{1}=x}\right)\left(x_{1}-x\right)^{\tau},
\end{aligned}
\label{369}
\end{eqnarray}
where we used the following relation to eliminate the first-order terms
\begin{align}
	\gamma= & \frac{\beta}{6}\int d^{4}x_{1}\left(\hat{\phi}_{\mu\nu}\left(x\right),\hat{\phi}^{\mu\nu}\left(x_{1}\right)\bigg|_{x_{1}=x}\right),\label{370}\\
	\gamma_{\phi S}= & \frac{\beta}{3}\int d^{4}x_{1}\left(\hat{\phi}_{\mu\nu}\left(x\right),\hat{S}^{\mu\nu}\left(x_{1}\right)\bigg|_{x_{1}=x}\right).\label{371}
\end{align}
Substituting Eqs.~\eqref{222} and \eqref{223} into Eq.~\eqref{369}, applying Curie's theorem, and imposing the orthogonality condition $\hat{\phi}^{\lambda\eta}u_{\lambda}=0$, we obtain
\begin{eqnarray}
\langle\hat{\phi}_{\mu\nu}\rangle_{2}^{1}=\widetilde{\gamma}_{\phi S}\mydelta_{\mu\nu\rho\sigma}\beta^{-1}D\left(\beta\mathcal{R}^{\langle\rho\rangle\langle\sigma\rangle}\right)+2\widetilde{\gamma}\mydelta_{\mu\nu\rho\sigma}\beta^{-1}D\left(\beta\xi^{\rho\sigma}\right),
\label{372}
\end{eqnarray}
where we define
\begin{align}
\widetilde{\gamma}_{\phi S}&=i\frac{d}{d\omega}\gamma_{\phi S}\left(\omega\right)\bigg|_{\omega=0}=-\frac{1}{6}\frac{d^{2}}{d\omega^{2}}\text{Re}G_{\hat{\phi}^{\lambda\eta}\hat{S}_{\lambda\eta}}^{R}\left(\omega\right)\bigg|_{\omega=0},\label{373}\\
\widetilde{\gamma}&=	i\frac{d}{d\omega}\gamma\left(\omega\right)\bigg|_{\omega=0}=-\frac{1}{12}\frac{d^{2}}{d\omega^{2}}\text{Re}G_{\hat{\phi}_{\mu\nu}\hat{\phi}^{\mu\nu}}^{R}\left(\omega\right)\bigg|_{\omega=0},\label{374}
\end{align}
with the frequency-dependent transport coefficients $\gamma_{\phi S}\left(\omega\right)$ and $\gamma\left(\omega\right)$ are denoted by
\begin{align}
	\gamma_{\phi S}\left(\omega\right)&=\frac{1}{3}\beta\int d^{4}x_{1}\int_{-\infty}^{t}e^{i\omega\left(t-t_{1}\right)}\left(\hat{\phi}^{\lambda\eta}\left(x\right),\hat{S}_{\lambda\eta}\left(x_{1}\right)\right),\label{375}\\
	\gamma\left(\omega\right)=&\frac{\beta}{6}\int d^{4}x_{1}\int_{-\infty}^{t}e^{i\omega\left(t-t_{1}\right)}\left(\hat{\phi}_{\mu\nu}\left(\boldsymbol{x},t\right),\hat{\phi}^{\mu\nu}\left(\boldsymbol{x}_{1},t_{1}\right)\right).\label{376}
\end{align}
Employing the first-order approximation $D\beta\simeq\beta\theta\Gamma-2\beta\mathcal{Z}^{\alpha\beta}\mathcal{D}_{\alpha\beta}$, we obtain the nonlocal corrections from the two-point correlation function to the rotational stress tensor
\begin{eqnarray}
\begin{aligned} \langle\hat{\phi}_{\mu\nu}\left(x\right)\rangle_{2}^{1}=2\widetilde{\gamma}\mydelta_{\mu\nu\rho\sigma}D\xi^{\rho\sigma}+2\widetilde{\gamma}\xi_{\mu\nu}\left(\theta\Gamma-2\mathcal{Z}^{\alpha\beta}\mathcal{D}_{\alpha\beta}\right)+\widetilde{\gamma}_{\phi S}\mathcal{R}_{\langle\mu\rangle\langle\nu\rangle}\left(\theta\Gamma-2\mathcal{Z}^{\alpha\beta}\mathcal{D}_{\alpha\beta}\right)+\widetilde{\gamma}_{\phi S}\mydelta_{\mu\nu\rho\sigma}D\mathcal{R}^{\langle\rho\rangle\langle\sigma\rangle}.
\end{aligned}
\label{377}
\end{eqnarray}
Applying Eqs.~\eqref{91} and~\eqref{96} in conjunction with Curie's theorem, the corrections of $\hat{\phi}_{\mu\nu}$ from extended thermodynamic forces are expressed as
\begin{eqnarray}
\langle\hat{\phi}_{\mu\nu}\rangle_{2}^{2}=\gamma_{\phi S}\mathcal{T}_{\langle\mu\rangle\langle\nu\rangle}.
\label{378}
\end{eqnarray}
Substituting Eq.~\eqref{90} into Eq.~\eqref{97}, we obtain the corrections from the three-point correlation function to rotational stress tensor
\begin{equation}
\begin{aligned}
	\langle\hat{\phi}_{\mu\nu}\left(x\right)\rangle_{2}^{3}= & \int d^{4}x_{1}d^{4}x_{2}\biggl(\hat{\phi}_{\mu\nu}\left(x\right),\biggl[-\beta\theta\hat{p}^{*}+\beta\mathcal{R}_{\alpha\beta}\hat{S}^{\alpha\beta}+\beta\sum_{i}\left(\hat{\mathfrak{D}}_{i}\partial_{\epsilon n}^{i}\Omega_{\mu\nu}\right)\mathcal{Z}^{\mu\nu}-\sum_{a}\hat{\mathscr{J}}_{a}^{\sigma}\nabla_{\sigma}\alpha_{a}\\
	&+\beta\hat{h}^{\sigma}N_{\sigma}+\beta\hat{q}^{\mu}M_{\mu}+\beta\hat{\pi}^{\mu\nu}\sigma_{\mu\nu}+\beta\hat{\phi}^{\mu\nu}\xi_{\mu\nu}+\hat{\varpi}^{\lambda\alpha\beta}\varXi_{\lambda\alpha\beta}\biggr]_{x_{1}},\biggl[-\beta\theta\hat{p}^{*}+\beta\mathcal{R}_{\alpha\beta}\hat{S}^{\alpha\beta}+\beta\sum_{i}\left(\hat{\mathfrak{D}}_{i}\partial_{\epsilon n}^{i}\Omega_{\mu\nu}\right)\mathcal{Z}^{\mu\nu}\\
	&-\sum_{a}\hat{\mathscr{J}}_{a}^{\sigma}\nabla_{\sigma}\alpha_{a}+\beta\hat{h}^{\sigma}N_{\sigma}+\beta\hat{q}^{\mu}M_{\mu}+\beta\hat{\pi}^{\mu\nu}\sigma_{\mu\nu}+\beta\hat{\phi}^{\mu\nu}\xi_{\mu\nu}+\hat{\varpi}^{\lambda\alpha\beta}\varXi_{\lambda\alpha\beta}\biggr]_{x_{2}}\biggr).
\end{aligned}
\label{379}
\end{equation}
The nonvanishing three-point correlators in Eq.~\eqref{379} are given by
\begin{align}
	\left(\hat{\phi}_{\mu\nu}\left(x\right),\hat{p}^{*}\left(x_{1}\right),\hat{S}_{\alpha\beta}\left(x_{2}\right)\right) & =\frac{1}{3}\mydelta_{\mu\nu\alpha\beta}\left(\hat{\phi}_{\gamma\delta}\left(x\right),\hat{p}^{*}\left(x_{1}\right),\hat{S}^{\gamma\delta}\left(x_{2}\right)\right),\label{380}\\
	\left(\hat{\phi}_{\mu\nu}\left(x\right),\hat{p}^{*}\left(x_{1}\right),\hat{\phi}_{\alpha\beta}\left(x_{2}\right)\right) & =\frac{1}{3}\mydelta_{\mu\nu\alpha\beta}\left(\hat{\phi}_{\gamma\delta}\left(x\right),\hat{p}^{*}\left(x_{1}\right),\hat{\phi}^{\gamma\delta}\left(x_{2}\right)\right),\label{381}\\
	\left(\hat{\phi}_{\mu\nu}\left(x\right),\hat{S}_{\rho\sigma}\left(x_{1}\right),\hat{S}_{\alpha\beta}\left(x_{2}\right)\right) & =\frac{1}{3}\left[\Delta_{\rho\alpha}\mydelta_{\mu\nu\sigma\beta}-\Delta_{\rho\beta}\mydelta_{\mu\nu\sigma\alpha}-\Delta_{\sigma\alpha}\mydelta_{\mu\nu\rho\beta}+\Delta_{\sigma\beta}\mydelta_{\mu\nu\rho\alpha}\right]\nonumber\\
	\times & \left(\hat{\phi}_{\lambda}^{\,\,\,\,\delta}\left(x\right),\hat{S}_{\delta}^{\,\,\,\,\eta}\left(x_{1}\right),\hat{S}_{\eta}^{\,\,\,\,\lambda}\left(x_{2}\right)\right),\label{382}\\
	\left(\hat{\phi}_{\mu\nu}\left(x\right),\hat{S}_{\rho\sigma}\left(x_{1}\right),\hat{\mathfrak{D}}_{i}\left(x_{2}\right)\right) &=\frac{1}{3}\mydelta_{\mu\nu\rho\sigma}\left(\hat{\phi}_{\gamma\delta}\left(x\right),\hat{S}^{\gamma\delta}\left(x_{1}\right),\hat{\mathfrak{D}}_{i}\left(x_{2}\right)\right),\label{383}\\
	\left(\hat{\phi}_{\mu\nu}\left(x\right),\hat{S}_{\rho\sigma}\left(x_{1}\right),\hat{\pi}_{\alpha\beta}\left(x_{2}\right)\right) & =-\frac{1}{5}\left[-\Delta_{\rho\alpha}\mydelta_{\mu\nu\sigma\beta}-\Delta_{\rho\beta}\mydelta_{\mu\nu\sigma\alpha}+\Delta_{\sigma\alpha}\mydelta_{\mu\nu\rho\beta}+\Delta_{\sigma\beta}\mydelta_{\mu\nu\rho\alpha}-\frac{4}{3}\Delta_{\alpha\beta}\mydelta_{\mu\nu\rho\sigma}\right]\nonumber\\
	\times & \left(\hat{\phi}_{\lambda}^{\,\,\,\,\delta}\left(x\right),\hat{S}_{\delta}^{\,\,\,\,\eta}\left(x_{1}\right),\hat{\pi}_{\eta}^{\,\,\,\,\lambda}\left(x_{2}\right)\right),\label{384}\\
	\left(\hat{\phi}_{\mu\nu}\left(x\right),\hat{S}_{\rho\sigma}\left(x_{1}\right),\hat{\phi}_{\alpha\beta}\left(x_{2}\right)\right) & =\frac{1}{3}\left[\Delta_{\rho\alpha}\mydelta_{\mu\nu\sigma\beta}-\Delta_{\rho\beta}\mydelta_{\mu\nu\sigma\alpha}-\Delta_{\sigma\alpha}\mydelta_{\mu\nu\rho\beta}+\Delta_{\sigma\beta}\mydelta_{\mu\nu\rho\alpha}\right]\nonumber\\
	\times & \left(\hat{\phi}_{\lambda}^{\,\,\,\,\delta}\left(x\right),\hat{S}_{\delta}^{\,\,\,\,\eta}\left(x_{1}\right),\hat{\phi}_{\eta}^{\,\,\,\,\lambda}\left(x_{2}\right)\right),\label{385}\\
	\left(\hat{\phi}_{\mu\nu}\left(x\right),\hat{\mathfrak{D}}_{i}\left(x_{1}\right),\hat{\phi}_{\alpha\beta}\left(x_{2}\right)\right) & =\frac{1}{3}\mydelta_{\mu\nu\alpha\beta}\left(\hat{\phi}_{\gamma\delta}\left(x\right),\hat{\mathfrak{D}}_{i}\left(x_{1}\right),\hat{\phi}^{\gamma\delta}\left(x_{2}\right)\right),\label{386}\\
	\left(\hat{\phi}_{\mu\nu}\left(x\right),\hat{\mathscr{J}}_{a\alpha}\left(x_{1}\right),\hat{\mathscr{J}}_{b\beta}\left(x_{2}\right)\right) & =\frac{1}{3}\mydelta_{\mu\nu\alpha\beta}\left(\hat{\phi}_{\gamma\delta}\left(x\right),\hat{\mathscr{J}}_{a}^{\gamma}\left(x_{1}\right),\hat{\mathscr{J}}_{b}^{\delta}\left(x_{2}\right)\right),\label{387}\\
	\left(\hat{\phi}_{\mu\nu}\left(x\right),\hat{\mathscr{J}}_{a\alpha}\left(x_{1}\right),\hat{h}_{\beta}\left(x_{2}\right)\right) & =\frac{1}{3}\mydelta_{\mu\nu\alpha\beta}\left(\hat{\phi}_{\gamma\delta}\left(x\right),\hat{\mathscr{J}}_{a}^{\gamma}\left(x_{1}\right),\hat{h}^{\delta}\left(x_{2}\right)\right),\label{388}\\
	\left(\hat{\phi}_{\mu\nu}\left(x\right),\hat{\mathscr{J}}_{a\alpha}\left(x_{1}\right),\hat{q}_{\beta}\left(x_{2}\right)\right) & =\frac{1}{3}\mydelta_{\mu\nu\alpha\beta}\left(\hat{\phi}_{\gamma\delta}\left(x\right),\hat{\mathscr{J}}_{a}^{\gamma}\left(x_{1}\right),\hat{q}^{\delta}\left(x_{2}\right)\right),\label{389}\\
	\left(\hat{\phi}_{\mu\nu}\left(x\right),\hat{\mathscr{J}}_{a\alpha}\left(x_{1}\right),\hat{\varpi}_{\rho\sigma\delta}\left(x_{2}\right)\right) & =\myDelta_{\mu\nu\alpha\rho\sigma\delta}\left(\hat{\phi}_{\gamma\eta}\left(x\right),\hat{\mathscr{J}}_{a\varepsilon}\left(x_{1}\right),\hat{\varpi}^{\gamma\eta\varepsilon}\left(x_{2}\right)\right),\label{390}\\
	\left(\hat{\phi}_{\mu\nu}\left(x\right),\hat{h}_{\alpha}\left(x_{1}\right),\hat{h}_{\beta}\left(x_{2}\right)\right) & =\frac{1}{3}\mydelta_{\mu\nu\alpha\beta}\left(\hat{\phi}_{\gamma\delta}\left(x\right),\hat{h}^{\gamma}\left(x_{1}\right),\hat{h}^{\delta}\left(x_{2}\right)\right),\label{391}\\
	\left(\hat{\phi}_{\mu\nu}\left(x\right),\hat{h}_{\alpha}\left(x_{1}\right),\hat{q}_{\beta}\left(x_{2}\right)\right) & =\frac{1}{3}\mydelta_{\mu\nu\alpha\beta}\left(\hat{\phi}_{\gamma\delta}\left(x\right),\hat{h}^{\gamma}\left(x_{1}\right),\hat{q}^{\delta}\left(x_{2}\right)\right),\label{392}\\
	\left(\hat{\phi}_{\mu\nu}\left(x\right),\hat{h}_{\alpha}\left(x_{1}\right),\hat{\varpi}_{\rho\sigma\delta}\left(x_{2}\right)\right) & =\myDelta_{\mu\nu\alpha\rho\sigma\delta}\left(\hat{\phi}_{\gamma\eta}\left(x\right),\hat{h}_{\varepsilon}\left(x_{1}\right),\hat{\varpi}^{\gamma\eta\varepsilon}\left(x_{2}\right)\right),\label{393}\\
	\left(\hat{\phi}_{\mu\nu}\left(x\right),\hat{q}_{\alpha}\left(x_{1}\right),\hat{q}_{\beta}\left(x_{2}\right)\right) & =\frac{1}{3}\mydelta_{\mu\nu\alpha\beta}\left(\hat{\phi}_{\gamma\delta}\left(x\right),\hat{q}^{\gamma}\left(x_{1}\right),\hat{q}^{\delta}\left(x_{2}\right)\right),\label{394}\\
	\left(\hat{\phi}_{\mu\nu}\left(x\right),\hat{q}_{\alpha}\left(x_{1}\right),\hat{\varpi}_{\rho\sigma\delta}\left(x_{2}\right)\right) & =\myDelta_{\mu\nu\alpha\rho\sigma\delta}\left(\hat{\phi}_{\gamma\eta}\left(x\right),\hat{q}_{\varepsilon}\left(x_{1}\right),\hat{\varpi}^{\gamma\eta\varepsilon}\left(x_{2}\right)\right),\label{395}\\
	\left(\hat{\phi}_{\mu\nu}\left(x\right),\hat{\pi}_{\rho\sigma}\left(x_{1}\right),\hat{\pi}_{\alpha\beta}\left(x_{2}\right)\right) & =-\frac{1}{15}\left[\Delta_{\rho\alpha}\mydelta_{\mu\nu\sigma\beta}+\Delta_{\rho\beta}\mydelta_{\mu\nu\sigma\alpha}+\Delta_{\sigma\alpha}\mydelta_{\mu\nu\rho\beta}+\Delta_{\sigma\beta}\mydelta_{\mu\nu\rho\alpha}\right]\nonumber\\
	\times & \left(\hat{\phi}_{\lambda}^{\,\,\,\,\delta}\left(x\right),\hat{\pi}_{\delta}^{\,\,\,\,\eta}\left(x_{1}\right),\hat{\pi}_{\eta}^{\,\,\,\,\lambda}\left(x_{2}\right)\right),\label{396}\\
	\left(\hat{\phi}_{\mu\nu}\left(x\right),\hat{\pi}_{\rho\sigma}\left(x_{1}\right),\hat{\phi}_{\alpha\beta}\left(x_{2}\right)\right) & =\frac{1}{5}\left[-\Delta_{\rho\alpha}\mydelta_{\mu\nu\sigma\beta}+\Delta_{\rho\beta}\mydelta_{\mu\nu\sigma\alpha}-\Delta_{\sigma\alpha}\mydelta_{\mu\nu\rho\beta}+\Delta_{\sigma\beta}\mydelta_{\mu\nu\rho\alpha}+\frac{4}{3}\Delta_{\rho\sigma}\mydelta_{\mu\nu\alpha\beta}\right]\nonumber\\
	\times & \left(\hat{\phi}_{\lambda}^{\,\,\,\,\delta}\left(x\right),\hat{\pi}_{\delta}^{\,\,\,\,\eta}\left(x_{1}\right),\hat{\phi}_{\eta}^{\,\,\,\,\lambda}\left(x_{2}\right)\right),\label{397}\\
	\left(\hat{\phi}_{\mu\nu}\left(x\right),\hat{\phi}_{\rho\sigma}\left(x_{1}\right),\hat{\phi}_{\alpha\beta}\left(x_{2}\right)\right) & =\frac{1}{3}\left[\Delta_{\rho\alpha}\mydelta_{\mu\nu\sigma\beta}-\Delta_{\rho\beta}\mydelta_{\mu\nu\sigma\alpha}-\Delta_{\sigma\alpha}\mydelta_{\mu\nu\rho\beta}+\Delta_{\sigma\beta}\mydelta_{\mu\nu\rho\alpha}\right]\nonumber\\
	\times & \left(\hat{\phi}_{\lambda}^{\,\,\,\,\delta}\left(x\right),\hat{\phi}_{\delta}^{\,\,\,\,\eta}\left(x_{1}\right),\hat{\phi}_{\eta}^{\,\,\,\,\lambda}\left(x_{2}\right)\right),\label{398}
\end{align}
Substituting the correlation functions from Eqs.~\eqref{380}-\eqref{398} into Eq.~\eqref{379} and factoring out the thermodynamic forces evaluated at point $x$, we define a set of transport coefficients
\begin{align}
	\gamma_{\phi pS} & =-\frac{1}{3}\beta^{2}\int d^{4}x_{1}d^{4}x_{2}\left(\hat{\phi}_{\gamma\delta}\left(x\right),\hat{p}^{*}\left(x_{1}\right),\hat{S}^{\gamma\delta}\left(x_{2}\right)\right),\gamma_{\phi p\phi}=-\frac{1}{3}\beta^{2}\int d^{4}x_{1}d^{4}x_{2}\left(\hat{\phi}_{\gamma\delta}\left(x\right),\hat{p}^{*}\left(x_{1}\right),\hat{\phi}^{\gamma\delta}\left(x_{2}\right)\right),\label{399}\\
	\gamma_{\phi SS} & =-\frac{4}{3}\beta^{2}\int d^{4}x_{1}d^{4}x_{2}\left(\hat{\phi}_{\lambda}^{\,\,\,\,\delta}\left(x\right),\hat{S}_{\delta}^{\,\,\,\,\eta}\left(x_{1}\right),\hat{S}_{\eta}^{\,\,\,\,\lambda}\left(x_{2}\right)\right),\gamma_{\phi S\mathfrak{D}_{i}}=\frac{1}{3}\beta^{2}\int d^{4}x_{1}d^{4}x_{2}\left(\hat{\phi}_{\gamma\delta}\left(x\right),\hat{S}^{\gamma\delta}\left(x_{1}\right),\hat{\mathfrak{D}}_{i}\left(x_{2}\right)\right),\label{400}\\
	\gamma_{\phi S\pi} & =\frac{4}{5}\beta^{2}\int d^{4}x_{1}d^{4}x_{2}\left(\hat{\phi}_{\lambda}^{\,\,\,\,\delta}\left(x\right),\hat{S}_{\delta}^{\,\,\,\,\eta}\left(x_{1}\right),\hat{\pi}_{\eta}^{\,\,\,\,\lambda}\left(x_{2}\right)\right),\gamma_{\phi S\phi}=-\frac{4}{3}\beta^{2}\int d^{4}x_{1}d^{4}x_{2}\left(\hat{\phi}_{\lambda}^{\,\,\,\,\delta}\left(x\right),\hat{S}_{\delta}^{\,\,\,\,\eta}\left(x_{1}\right),\hat{\phi}_{\eta}^{\,\,\,\,\lambda}\left(x_{2}\right)\right),\label{401}\\
	\gamma_{\phi\mathfrak{D}_{i}\phi} & =\frac{1}{3}\beta^{2}\int d^{4}x_{1}d^{4}x_{2}\left(\hat{\phi}_{\gamma\delta}\left(x\right),\hat{\mathfrak{D}}_{i}\left(x_{1}\right),\hat{\phi}^{\gamma\delta}\left(x_{2}\right)\right),\gamma_{\phi\mathscr{J}_{a}\mathscr{J}_{b}}=\frac{1}{3}\int d^{4}x_{1}d^{4}x_{2}\left(\hat{\phi}_{\gamma\delta}\left(x\right),\hat{\mathscr{J}}_{a}^{\gamma}\left(x_{1}\right),\hat{\mathscr{J}}_{b}^{\delta}\left(x_{2}\right)\right),\label{402}\\
	\gamma_{\phi\mathscr{J}_{a}h} & =-\frac{1}{3}\beta\int d^{4}x_{1}d^{4}x_{2}\left(\hat{\phi}_{\gamma\delta}\left(x\right),\hat{\mathscr{J}}_{a}^{\gamma}\left(x_{1}\right),\hat{h}^{\delta}\left(x_{2}\right)\right),\gamma_{\phi\mathscr{J}_{a}q}=-\frac{1}{3}\beta\int d^{4}x_{1}d^{4}x_{2}\left(\hat{\phi}_{\gamma\delta}\left(x\right),\hat{\mathscr{J}}_{a}^{\gamma}\left(x_{1}\right),\hat{q}^{\delta}\left(x_{2}\right)\right),\label{403}\\
	\gamma_{\phi\mathscr{J}_{a}\varpi} & =-\int d^{4}x_{1}d^{4}x_{2}\left(\hat{\phi}_{\gamma\eta}\left(x\right),\hat{\mathscr{J}}_{a\varepsilon}\left(x_{1}\right),\hat{\varpi}^{\gamma\eta\varepsilon}\left(x_{2}\right)\right),\gamma_{\phi hh}=\frac{1}{3}\beta^{2}\int d^{4}x_{1}d^{4}x_{2}\left(\hat{\phi}_{\gamma\delta}\left(x\right),\hat{h}^{\gamma}\left(x_{1}\right),\hat{h}^{\delta}\left(x_{2}\right)\right),\label{404}\\
	\gamma_{\phi hq} & =\frac{1}{3}\beta^{2}\int d^{4}x_{1}d^{4}x_{2}\left(\hat{\phi}_{\gamma\delta}\left(x\right),\hat{h}^{\gamma}\left(x_{1}\right),\hat{q}^{\delta}\left(x_{2}\right)\right),\gamma_{\phi h\varpi}=\beta\int d^{4}x_{1}d^{4}x_{2}\left(\hat{\phi}_{\gamma\eta}\left(x\right),\hat{h}_{\varepsilon}\left(x_{1}\right),\hat{\varpi}^{\gamma\eta\varepsilon}\left(x_{2}\right)\right),\label{405}\\
	\gamma_{\phi qq} & =\frac{1}{3}\beta^{2}\int d^{4}x_{1}d^{4}x_{2}\left(\hat{\phi}_{\gamma\delta}\left(x\right),\hat{q}^{\gamma}\left(x_{1}\right),\hat{q}^{\delta}\left(x_{2}\right)\right),\gamma_{\phi q\varpi}=\beta\int d^{4}x_{1}d^{4}x_{2}\left(\hat{\phi}_{\gamma\eta}\left(x\right),\hat{q}_{\varepsilon}\left(x_{1}\right),\hat{\varpi}^{\gamma\eta\varepsilon}\left(x_{2}\right)\right),\label{406}\\
	\gamma_{\phi\pi\pi} & =-\frac{4}{15}\beta^{2}\int d^{4}x_{1}d^{4}x_{2}\left(\hat{\phi}_{\lambda}^{\,\,\,\,\delta}\left(x\right),\hat{\pi}_{\delta}^{\,\,\,\,\eta}\left(x_{1}\right),\hat{\pi}_{\eta}^{\,\,\,\,\lambda}\left(x_{2}\right)\right),\gamma_{\phi\pi\phi}=\frac{4}{5}\beta^{2}\int d^{4}x_{1}d^{4}x_{2}\left(\hat{\phi}_{\lambda}^{\,\,\,\,\delta}\left(x\right),\hat{\pi}_{\delta}^{\,\,\,\,\eta}\left(x_{1}\right),\hat{\phi}_{\eta}^{\,\,\,\,\lambda}\left(x_{2}\right)\right),\label{407}\\
	\gamma_{\phi\phi\phi} & =-\frac{4}{3}\beta^{2}\int d^{4}x_{1}d^{4}x_{2}\left(\hat{\phi}_{\lambda}^{\,\,\,\,\delta}\left(x\right),\hat{\phi}_{\delta}^{\,\,\,\,\eta}\left(x_{1}\right),\hat{\phi}_{\eta}^{\,\,\,\,\lambda}\left(x_{2}\right)\right).\label{408}
\end{align}
Combining these definitions with Eqs.~\eqref{379}-\eqref{408} and the symmetry property in Eq.~\eqref{49}, we obtain
\begin{eqnarray}
\begin{aligned}
\langle\hat{\phi}_{\mu\nu}\rangle_{2}^{3}= & 2\gamma_{\phi pS}\theta\mathcal{R}_{\langle\mu\rangle\langle\nu\rangle}+2\gamma_{\phi p\phi}\theta\xi_{\mu\nu}+\gamma_{\phi SS}\mathcal{R}_{\langle\alpha\rangle[\langle\mu\rangle}\mathcal{R}_{\langle\nu\rangle]}^{\,\,\,\,\,\,\alpha}+2\mathcal{R}_{\langle\mu\rangle\langle\nu\rangle}\sum_{i}\gamma_{\phi S\mathfrak{D}_{i}}\mathcal{Z}^{\alpha\beta}\partial_{\epsilon n}^{i}\Omega_{\alpha\beta}+2\gamma_{\phi S\pi}\mathcal{R}_{\langle\alpha\rangle[\langle\mu\rangle}\sigma_{\nu]}^{\,\,\,\,\,\,\alpha}\\
&+2\gamma_{\phi S\phi}\mathcal{R}_{\langle\alpha\rangle[\langle\mu\rangle}\xi_{\nu]}^{\,\,\,\,\,\alpha}+2\sum_{i}\gamma_{\phi\mathfrak{D}_{i}\phi}\mathcal{Z}^{\rho\sigma}\partial_{\epsilon n}^{i}\Omega_{\rho\sigma}\xi_{\mu\nu}+\sum_{ab}\gamma_{\phi\mathscr{J}_{a}\mathscr{J}_{b}}\nabla_{[\mu}\alpha_{a}\nabla_{\nu]}\alpha_{b}+2\sum_{a}\gamma_{\phi\mathscr{J}_{a}h}\nabla_{[\mu}\alpha_{a}N_{\nu]}\\
&+2\sum_{a}\gamma_{\phi\mathscr{J}_{a}q}\nabla_{[\mu}\alpha_{a}M_{\nu]}+2\sum_{a}\gamma_{\phi\mathscr{J}_{a}\varpi}\nabla^{\alpha}\alpha_{a}\varXi_{\mu\nu\alpha}+\gamma_{\phi hh}N_{[\mu}N_{\nu]}+2\gamma_{\phi hq}N_{[\mu}M_{\nu]}+2\gamma_{\phi h\varpi}N^{\alpha}\varXi_{\mu\nu\alpha}\\
&+\gamma_{\phi qq}M_{[\mu}M_{\nu]}++2\gamma_{\phi q\varpi}M^{\alpha}\varXi_{\mu\nu\alpha}+\gamma_{\phi\pi\pi}\sigma_{\alpha[\mu}\sigma_{\nu]}^{\,\,\,\,\alpha}+2\gamma_{\phi\pi\phi}\sigma_{\alpha[\mu}\xi_{\nu]}^{\,\,\,\,\alpha}+\gamma_{\phi\phi\phi}\xi_{\alpha[\mu}\xi_{\nu]}^{\,\,\,\,\alpha}.
\end{aligned}
\label{409}
\end{eqnarray}
Combining the contributions from Eqs.~\eqref{70},~\eqref{122},~\eqref{377},~\eqref{378}, and~\eqref{409}, and employing Eqs.~\eqref{92}, and~\eqref{94}, we obtain the complete second-order expression for the rotational stress tensor:
\begin{eqnarray}
\begin{aligned}
\phi_{\mu\nu}= & 2\gamma\xi_{\mu\nu}+\gamma_{\phi S}\mathcal{R}_{\langle\mu\rangle\langle\nu\rangle}+2\widetilde{\gamma}\mydelta_{\mu\nu\rho\sigma}D\xi^{\rho\sigma}+2\widetilde{\gamma}\xi_{\mu\nu}\left(\theta\Gamma-2\mathcal{Z}^{\alpha\beta}\mathcal{D}_{\alpha\beta}\right)+\widetilde{\gamma}_{\phi S}\mathcal{R}_{\langle\mu\rangle\langle\nu\rangle}\left(\theta\Gamma-2\mathcal{Z}^{\alpha\beta}\mathcal{D}_{\alpha\beta}\right)\\
 & +\widetilde{\gamma}_{\phi S}\mydelta_{\mu\nu\rho\sigma}D\mathcal{R}^{\langle\rho\rangle\langle\sigma\rangle}+\gamma_{\phi S}\mathcal{T}_{\langle\mu\rangle\langle\nu\rangle}+2\gamma_{\phi pS}\theta\mathcal{R}_{\langle\mu\rangle\langle\nu\rangle}+2\gamma_{\phi p\phi}\theta\xi_{\mu\nu}+\gamma_{\phi SS}\mathcal{R}_{\langle\alpha\rangle[\langle\mu\rangle}\mathcal{R}_{\langle\nu\rangle]}^{\,\,\,\,\,\,\alpha}\\
 & +2\mathcal{R}_{\langle\mu\rangle\langle\nu\rangle}\sum_{i}\gamma_{\phi S\mathfrak{D}_{i}}\mathcal{Z}^{\alpha\beta}\partial_{\epsilon n}^{i}\Omega_{\alpha\beta}+2\gamma_{\phi S\pi}\mathcal{R}_{\langle\alpha\rangle[\langle\mu\rangle}\sigma_{\nu]}^{\,\,\,\,\,\,\alpha}+2\gamma_{\phi S\phi}\mathcal{R}_{\langle\alpha\rangle[\langle\mu\rangle}\xi_{\nu]}^{\,\,\,\,\,\alpha}+2\sum_{i}\gamma_{\phi\mathfrak{D}_{i}\phi}\mathcal{Z}^{\rho\sigma}\partial_{\epsilon n}^{i}\Omega_{\rho\sigma}\xi_{\mu\nu}\\
 & +\sum_{ab}\gamma_{\phi\mathscr{J}_{a}\mathscr{J}_{b}}\nabla_{[\mu}\alpha_{a}\nabla_{\nu]}\alpha_{b}+2\sum_{a}\gamma_{\phi\mathscr{J}_{a}h}\nabla_{[\mu}\alpha_{a}N_{\nu]}+2\sum_{a}\gamma_{\phi\mathscr{J}_{a}q}\nabla_{[\mu}\alpha_{a}M_{\nu]}+2\sum_{a}\gamma_{\phi\mathscr{J}_{a}\varpi}\nabla^{\alpha}\alpha_{a}\varXi_{\mu\nu\alpha}\\
 & +\gamma_{\phi hh}N_{[\mu}N_{\nu]}+2\gamma_{\phi hq}N_{[\mu}M_{\nu]}+2\gamma_{\phi h\varpi}N^{\alpha}\varXi_{\mu\nu\alpha}+\gamma_{\phi qq}M_{[\mu}M_{\nu]}+2\gamma_{\phi q\varpi}M^{\alpha}\varXi_{\mu\nu\alpha}+\gamma_{\phi\pi\pi}\sigma_{\alpha[\mu}\sigma_{\nu]}^{\,\,\,\,\alpha}\\
 &+2\gamma_{\phi\pi\phi}\sigma_{\alpha[\mu}\xi_{\nu]}^{\,\,\,\,\alpha}+\gamma_{\phi\phi\phi}\xi_{\alpha[\mu}\xi_{\nu]}^{\,\,\,\,\alpha}.
\end{aligned}
\label{410}
\end{eqnarray}
Note that symmetry arguments dictate that terms $\gamma_{\phi hh}N_{[\mu}N_{\nu]}$ and $\gamma_{\phi qq}M_{[\mu}M_{\nu]}$ are zero.

To derive a relaxation-type equation for $\phi_{\mu\nu}$ from \eqref{410}, we use the first-order approximation
\begin{eqnarray}
	\xi^{\rho\sigma}\simeq\frac{1}{2\gamma}\left(\phi^{\rho\sigma}-\gamma_{\phi S}\mathcal{R}^{\langle\rho\rangle\langle\sigma\rangle}\right),
	\label{411}
\end{eqnarray}
in the term $2\widetilde{\gamma}\mydelta_{\mu\nu\rho\sigma}D\xi^{\rho\sigma}$ on the right-hand side of Eq.~\eqref{410}. We then have
\begin{eqnarray}
\begin{aligned} 
2\widetilde{\gamma}\mydelta_{\mu\nu\rho\sigma}D\xi^{\rho\sigma}
\simeq &\widetilde{\gamma}\gamma^{-1}\mydelta_{\mu\nu\rho\sigma}D\phi^{\rho\sigma}-\widetilde{\gamma}\gamma^{-1}\gamma_{\phi S}\mydelta_{\mu\nu\rho\sigma}D\mathcal{R}^{\langle\rho\rangle\langle\sigma\rangle}\\
&-\widetilde{\gamma}\gamma^{-1}\mathcal{R}_{\langle\mu\rangle\langle\nu\rangle}\beta\theta\left[\left(\frac{\partial\gamma_{\phi S}}{\partial\beta}\Gamma-\sum_{a}\frac{\partial\gamma_{\phi S}}{\partial\alpha_{a}}\delta_{a}-2\frac{\partial\gamma_{\phi S}}{\partial\Omega_{\alpha\beta}}\mathcal{K}_{\alpha\beta}\right)-2\theta^{-1}\mathcal{Z}^{\alpha\beta}\frac{\partial\gamma_{\phi S}}{\partial S^{\alpha\beta}}\right]\\
&-\widetilde{\gamma}\gamma^{-2}\left(\phi_{\mu\nu}-\gamma_{\phi S}\mathcal{R}_{\langle\mu\rangle\langle\nu\rangle}\right)\beta\theta\left[\left(\frac{\partial\gamma}{\partial\beta}\Gamma-\sum_{a}\frac{\partial\gamma}{\partial\alpha_{a}}\delta_{a}-2\frac{\partial\gamma}{\partial\Omega_{\alpha\beta}}\mathcal{K}_{\alpha\beta}\right)-2\theta^{-1}\mathcal{Z}^{\alpha\beta}\frac{\partial\gamma}{\partial S^{\alpha\beta}}\right].
\end{aligned}
\label{412}
\end{eqnarray}
where Eqs.~\eqref{76}-\eqref{78} have been used. Introducing the coefficients
\begin{align}
	\tau_{\phi}= & -\widetilde{\gamma}\gamma^{-1},\label{413}\\
	\widetilde{\gamma}_{\phi}= & \tau_{\phi}\gamma^{-1}\beta\left[\left(\frac{\partial\gamma}{\partial\beta}\Gamma-\sum_{a}\frac{\partial\gamma}{\partial\alpha_{a}}\delta_{a}-2\frac{\partial\gamma}{\partial\Omega_{\alpha\beta}}\mathcal{K}_{\alpha\beta}\right)-2\theta^{-1}\mathcal{Z}^{\alpha\beta}\frac{\partial\gamma}{\partial S^{\alpha\beta}}\right],\label{414}\\
	\overline{\gamma}= & \tau_{\phi}\beta\left[\left(\frac{\partial\gamma_{\phi S}}{\partial\beta}\Gamma-\sum_{a}\frac{\partial\gamma_{\phi S}}{\partial\alpha_{a}}\delta_{a}-2\frac{\partial\gamma_{\phi S}}{\partial\Omega_{\alpha\beta}}\mathcal{K}_{\alpha\beta}\right)-2\theta^{-1}\mathcal{Z}^{\alpha\beta}\frac{\partial\gamma_{\phi S}}{\partial S^{\alpha\beta}}\right],\label{415}
\end{align}
and combining Eqs.~\eqref{410} and \eqref{412}, we arrive at the following relaxation equation for the  rotational stress tensor:
\begin{eqnarray}
\begin{aligned}\phi_{\mu\nu}+\tau_{\phi}\dot{\phi}_{\mu\nu}= & 2\gamma\xi_{\mu\nu}+\gamma_{\phi S}\mathcal{R}_{\langle\mu\rangle\langle\nu\rangle}+\tau_{\phi}\gamma_{\phi S}\mydelta_{\mu\nu\rho\sigma}D\mathcal{R}^{\langle\rho\rangle\langle\sigma\rangle}+\overline{\gamma}\theta\mathcal{R}_{\langle\mu\rangle\langle\nu\rangle}+\widetilde{\gamma}_{\phi}\theta\left(\phi_{\mu\nu}-\gamma_{\phi S}\mathcal{R}_{\langle\mu\rangle\langle\nu\rangle}\right)+2\widetilde{\gamma}\xi_{\mu\nu}\left(\theta\Gamma-2\mathcal{Z}^{\alpha\beta}\mathcal{D}_{\alpha\beta}\right)\\
 & +\widetilde{\gamma}_{\phi S}\mathcal{R}_{\langle\mu\rangle\langle\nu\rangle}\left(\theta\Gamma-2\mathcal{Z}^{\alpha\beta}\mathcal{D}_{\alpha\beta}\right)+\widetilde{\gamma}_{\phi S}\mydelta_{\mu\nu\rho\sigma}D\mathcal{R}^{\langle\rho\rangle\langle\sigma\rangle}+\gamma_{\phi S}\mathcal{T}_{\langle\mu\rangle\langle\nu\rangle}+2\gamma_{\phi pS}\theta\mathcal{R}_{\langle\mu\rangle\langle\nu\rangle}+2\gamma_{\phi p\phi}\theta\xi_{\mu\nu}\\
 & +\gamma_{\phi SS}\mathcal{R}_{\langle\alpha\rangle[\langle\mu\rangle}\mathcal{R}_{\langle\nu\rangle]}^{\,\,\,\,\,\,\alpha}+2\mathcal{R}_{\langle\mu\rangle\langle\nu\rangle}\sum_{i}\gamma_{\phi S\mathfrak{D}_{i}}\mathcal{Z}^{\alpha\beta}\partial_{\epsilon n}^{i}\Omega_{\alpha\beta}+2\gamma_{\phi S\pi}\mathcal{R}_{\langle\alpha\rangle[\langle\mu\rangle}\sigma_{\nu]}^{\,\,\,\,\,\,\alpha}+2\gamma_{\phi S\phi}\mathcal{R}_{\langle\alpha\rangle[\langle\mu\rangle}\xi_{\nu]}^{\,\,\,\,\,\alpha}\\
 & +2\sum_{i}\gamma_{\phi\mathfrak{D}_{i}\phi}\mathcal{Z}^{\rho\sigma}\partial_{\epsilon n}^{i}\Omega_{\rho\sigma}\xi_{\mu\nu}+\sum_{ab}\gamma_{\phi\mathscr{J}_{a}\mathscr{J}_{b}}\nabla_{[\mu}\alpha_{a}\nabla_{\nu]}\alpha_{b}+2\sum_{a}\gamma_{\phi\mathscr{J}_{a}h}\nabla_{[\mu}\alpha_{a}N_{\nu]}+2\sum_{a}\gamma_{\phi\mathscr{J}_{a}q}\nabla_{[\mu}\alpha_{a}M_{\nu]}\\
 & +2\sum_{a}\gamma_{\phi\mathscr{J}_{a}\varpi}\nabla^{\alpha}\alpha_{a}\varXi_{\mu\nu\alpha}+\gamma_{\phi hh}N_{[\mu}N_{\nu]}+2\gamma_{\phi hq}N_{[\mu}M_{\nu]}+2\gamma_{\phi h\varpi}N^{\alpha}\varXi_{\mu\nu\alpha}+\gamma_{\phi qq}M_{[\mu}M_{\nu]}\\
 & +2\gamma_{\phi q\varpi}M^{\alpha}\varXi_{\mu\nu\alpha}+\gamma_{\phi\pi\pi}\sigma_{\alpha[\mu}\sigma_{\nu]}^{\,\,\,\,\alpha}+2\gamma_{\phi\pi\phi}\sigma_{\alpha[\mu}\xi_{\nu]}^{\,\,\,\,\alpha}+\gamma_{\phi\phi\phi}\xi_{\alpha[\mu}\xi_{\nu]}^{\,\,\,\,\alpha}.
\end{aligned}
\label{416}
\end{eqnarray}
where we define $\dot{\phi}_{\mu\nu}=\mydelta_{\mu\nu\rho\sigma}D\phi^{\rho\sigma}$.

\subsubsection{Second-order corrections to the boost heat vector}

Substituting Eqs.~\eqref{90} and Eq.~\eqref{124} into Eq.~\eqref{95} and applying Curie's theorem once more, we find
\begin{eqnarray}
\begin{aligned}
	\langle\hat{q}_{\mu}\left(x\right)\rangle_{2}^{1}= & -\sum_{a}\int d^{4}x_{1}\left(\hat{q}_{\mu}\left(x\right),\hat{\mathscr{J}}_{a\sigma}\left(x_{1}\right)\right)\nabla_{x_{1}}^{\sigma}\alpha_{a}\left(x_{1}\right)+\int d^{4}x_{1}\left(\hat{q}_{\mu}\left(x\right),\hat{h}_{\sigma}\left(x_{1}\right)\right)\beta\left(x_{1}\right)N^{\sigma}\left(x_{1}\right)\\
	& +\int d^{4}x_{1}\left(\hat{q}_{\mu}\left(x\right),\hat{q}_{\sigma}\left(x_{1}\right)\right)\beta\left(x_{1}\right)M^{\sigma}\left(x_{1}\right)-\sum_{a}\lambda_{q\mathscr{J}_{a}}\left(x\right)\nabla_{\mu}\alpha_{a}\left(x\right)-\lambda_{qh}\left(x\right)N_{\mu}\left(x\right)+\lambda\left(x\right)M_{\mu}\left(x\right)
\end{aligned}
\label{417}
\end{eqnarray}
Substitute the correlation functions defined in Eqs.~\eqref{149}-\eqref{151} and~\eqref{155} into Eq.~\eqref{417}, we obtain
\begin{eqnarray}
\begin{aligned}\langle\hat{q}_{\mu}\left(x\right)\rangle_{2}^{1}= & -\frac{1}{3}\sum_{a}\Delta_{\mu\gamma}\left(x\right)\int d^{4}x_{1}\left(\hat{q}^{\lambda}\left(x\right),\hat{\mathscr{J}}_{a\lambda}\left(x_{1}\right)\right)\nabla_{x_{1}}^{\gamma}\alpha_{a}\left(x_{1}\right)+\frac{1}{3}\Delta_{\mu\gamma}\left(x\right)\int d^{4}x_{1}\left(\hat{q}^{\lambda}\left(x\right),\hat{h}_{\lambda}\left(x_{1}\right)\right)\beta\left(x_{1}\right)N^{\gamma}\left(x_{1}\right)\\
	& +\frac{1}{3}\Delta_{\mu\gamma}\left(x\right)\int d^{4}x_{1}\left(\hat{q}^{\lambda}\left(x\right),\hat{q}_{\lambda}\left(x_{1}\right)\right)\beta\left(x_{1}\right)M^{\gamma}\left(x_{1}\right)-\sum_{a}\lambda_{q\mathscr{J}_{a}}\left(x\right)\nabla_{\mu}\alpha_{a}\left(x\right)-\lambda_{qh}\left(x\right)N_{\mu}\left(x\right)+\lambda\left(x\right)M_{\mu}\left(x\right)
\end{aligned}
\label{418}
\end{eqnarray}
Substituting Eqs.~\eqref{308}-\eqref{310} into Eq.~\eqref{418} and expanding the thermodynamic forces around $x_1=x$, we obtain up to the second order in gradients
\begin{eqnarray}
\begin{aligned}
	\langle\hat{q}_{\mu}\left(x\right)\rangle_{2}^{1}= & -\frac{1}{3}\sum_{a}\Delta_{\mu\gamma}\left(x\right)\nabla^{\gamma}\alpha_{a}\left(x\right)\int d^{4}x_{1}\left(\hat{q}^{\lambda}\left(x\right),\frac{\partial\hat{\mathscr{J}}_{a\lambda}\left(x_{1}\right)}{\partial x_{1}^{\tau}}\bigg|_{x_{1}=x}\right)\left(x_{1}-x\right)^{\tau}\\
	& -\frac{1}{3}\sum_{a}\Delta_{\mu\gamma}\left(x\right)\frac{\partial}{\partial x_{1}^{\tau}}\left[\nabla^{\gamma}\alpha_{a}\left(x_{1}\right)\right]\bigg|_{x_{1}=x}\int d^{4}x_{1}\left(\hat{q}^{\lambda}\left(x\right),\hat{\mathscr{J}}_{a\lambda}\left(x_{1}\right)\bigg|_{x_{1}=x}\right)\left(x_{1}-x\right)^{\tau}\\
	& +\frac{1}{3}\Delta_{\mu\gamma}\left(x\right)\beta(x)N^{\gamma}\left(x\right)\int d^{4}x_{1}\left(\hat{q}^{\lambda}\left(x\right),\frac{\partial\hat{h}_{\lambda}\left(x_{1}\right)}{\partial x_{1}^{\tau}}\bigg|_{x_{1}=x}\right)\left(x_{1}-x\right)^{\tau}\\
	& +\frac{1}{3}\Delta_{\mu\gamma}\left(x\right)\frac{\partial}{\partial x_{1}^{\tau}}\left[\beta(x_{1})N^{\gamma}\left(x_{1}\right)\right]\bigg|_{x_{1}=x}\int d^{4}x_{1}\left(\hat{q}^{\lambda}\left(x\right),\hat{h}_{\lambda}\left(x_{1}\right)\bigg|_{x_{1}=x}\right)\left(x_{1}-x\right)^{\tau}\\
	& +\frac{1}{3}\Delta_{\mu\gamma}\left(x\right)\beta\left(x\right)M^{\gamma}\left(x\right)\int d^{4}x_{1}\left(\hat{q}^{\lambda}\left(x\right),\frac{\partial\hat{q}_{\lambda}\left(x_{1}\right)}{\partial x_{1}^{\tau}}\bigg|_{x_{1}=x}\right)\left(x_{1}-x\right)^{\tau}\\
	& +\frac{1}{3}\Delta_{\mu\gamma}\left(x\right)\frac{\partial}{\partial x_{1}^{\tau}}\left[\beta\left(x_{1}\right)M^{\gamma}\left(x_{1}\right)\right]\bigg|_{x_{1}=x}\int d^{4}x_{1}\left(\hat{q}^{\lambda}\left(x\right),\hat{q}_{\lambda}\left(x_{1}\right)\bigg|_{x_{1}=x}\right)\left(x_{1}-x\right)^{\tau}
\end{aligned}
\label{419}
\end{eqnarray}
where we used the following relation to eliminate the first-order terms
\begin{align}
	\lambda_{q\mathscr{J}_{a}}= & -\frac{1}{3}\int d^{4}x_{1}\left(\hat{q}^{\lambda}\left(x\right),\hat{\mathscr{J}}_{a\lambda}\left(x_{1}\right)\bigg|_{x_{1}=x}\right),\label{420}\\
	\lambda_{qh}= & \frac{1}{3}\beta\int d^{4}x_{1}\left(\hat{q}^{\lambda}\left(x\right),\hat{h}_{\lambda}\left(x_{1}\right)\bigg|_{x_{1}=x}\right),\label{421}\\
	\lambda= & -\frac{1}{3}\beta\int d^{4}x_{1}\left(\hat{q}^{\lambda}\left(x\right),\hat{q}_{\lambda}\left(x_{1}\right)\bigg|_{x_{1}=x}\right).\label{422}
\end{align}
By substituting Eqs.~\eqref{314}-\eqref{316} into Eq.~\eqref{419} and invoking Curie's theorem, alongside the orthogonality condition $u^\lambda \hat{q}_\lambda=0$, the first-order approximation $D\beta\simeq\beta\theta\Gamma-2\beta\mathcal{Z}^{\alpha\beta}\mathcal{D}_{\alpha\beta}$, and the relation presented in Eq.~\eqref{328}, we derive
\begin{eqnarray}
\begin{aligned}
	\langle\hat{q}_{\mu}\left(x\right)\rangle_{2}^{1}= & \sum_{a}\widetilde{\lambda}_{qh}n_{a}{w}^{-2}\left[\Gamma {w}\theta+\sum_{a}\delta_{a}n_{a}\theta+\mathcal{K}_{\alpha\beta}\left(\theta S^{\alpha\beta}+u^{\alpha}\partial_{\lambda}S^{\beta\lambda}+S^{\beta\lambda}\partial_{\lambda}u^{\alpha}+u^{\beta}\partial_{\lambda}S^{\lambda\alpha}+S^{\lambda\alpha}\partial_{\lambda}u^{\beta}\right)\right]\nabla_{\mu}\alpha_{a}\\
	&+\sum_{a}\widetilde{\lambda}_{q\mathscr{J}_{a}}\Delta_{\mu\gamma}D\left(\nabla^{\gamma}\alpha_{a}\right)+\widetilde{\lambda}_{qh}N_{\mu}\left(\theta\Gamma-2\mathcal{Z}^{\alpha\beta}\mathcal{D}_{\alpha\beta}\right)+\widetilde{\lambda}_{qh}\Delta_{\mu\gamma}DN^{\gamma}-\widetilde{\lambda}M_{\mu}\left(\theta\Gamma-2\mathcal{Z}^{\alpha\beta}\mathcal{D}_{\alpha\beta}\right)-\widetilde{\lambda}\Delta_{\mu\gamma}DM^{\gamma},
\end{aligned}
\label{423}
\end{eqnarray}
where we define
\begin{align}
	\widetilde{\lambda}_{qh}= & i\frac{d}{d\omega}\lambda_{qh}\left(\omega\right)\bigg|_{\omega=0}=-\frac{1}{6}\frac{d^{2}}{d\omega^{2}}\text{Re}G_{\hat{q}^{\lambda}\hat{h}_{\lambda}}^{R}\left(\omega\right)\bigg|_{\omega=0},\label{424}\\
	\widetilde{\lambda}_{q\mathscr{J}_{a}}= & i\frac{d}{d\omega}\lambda_{qa}\left(\omega\right)\bigg|_{\omega=0}=\frac{T}{6}\frac{d^{2}}{d\omega^{2}}\text{Re}G_{\hat{q}^{\lambda}\hat{\mathscr{J}}_{a\lambda}}^{R}\left(\omega\right)\bigg|_{\omega=0},\label{425}\\
	\widetilde{\lambda}= & i\frac{d}{d\omega}\lambda\left(\omega\right)\bigg|_{\omega=0}=\frac{1}{6}\frac{d^{2}}{d\omega^{2}}\text{Re}G_{\hat{q}^{\lambda}\hat{q}_{\lambda}}^{R}\left(\omega\right)\bigg|_{\omega=0},\label{426}
\end{align}
with the frequency-dependent transport coefficients $\lambda_{qh}\left(\omega\right)$,$\lambda_{q\mathscr{J}_{a}}\left(\omega\right)$, and $\lambda\left(\omega\right)$ are expressed as
\begin{align}
	\lambda_{qh}\left(\omega\right) & =\frac{1}{3}\beta\int d^{4}x_{1}\int_{-\infty}^{t}e^{i\omega\left(t-t_{1}\right)}\left(\hat{q}^{\lambda}\left(\boldsymbol{x},t\right),\hat{h}_{\lambda}\left(\boldsymbol{x}_{1},t_{1}\right)\right),\label{427}\\
	\lambda_{q\mathscr{J}_{a}}\left(\omega\right) & =-\frac{1}{3}\int d^{4}x_{1}\int_{-\infty}^{t}e^{i\omega\left(t-t_{1}\right)}\left(\hat{q}^{\lambda}\left(\boldsymbol{x},t\right),\hat{\mathscr{J}}_{a\lambda}\left(\boldsymbol{x}_{1},t_{1}\right)\right),\label{428}\\
	\lambda\left(\omega\right) & =-\frac{1}{3}\beta\int d^{4}x_{1}\int_{-\infty}^{t}e^{i\omega\left(t-t_{1}\right)}\left(\hat{q}^{\lambda}\left(\boldsymbol{x},t\right),\hat{q}_{\lambda}\left(\boldsymbol{x}_{1},t_{1}\right)\right).\label{429}
\end{align}

Applying Eqs.~\eqref{91} and~\eqref{96} in conjunction with Curie's theorem, the corrections of $q_\mu$ from extended thermodynamic forces are denoted by
\begin{eqnarray}
\langle\hat{q}_{\mu}\rangle_{2}^{2}=\lambda_{qh}\mathcal{H}_{\mu}-\lambda\mathcal{Q}_{\mu}.
\label{430}
\end{eqnarray}

Substituting Eq.~\eqref{90} into Eq.~\eqref{97}, we obtain the corrections from the three-point correlation function to boost heat vector
\begin{equation}
\begin{aligned}
	\langle\hat{q}_{\mu}\left(x\right)\rangle_{2}^{3}= & \int d^{4}x_{1}d^{4}x_{2}\biggl(\hat{q}_{\mu}\left(x\right),\biggl[-\beta\theta\hat{p}^{*}+\beta\mathcal{R}_{\alpha\beta}\hat{S}^{\alpha\beta}+\beta\sum_{i}\left(\hat{\mathfrak{D}}_{i}\partial_{\epsilon n}^{i}\Omega_{\mu\nu}\right)\mathcal{Z}^{\mu\nu}-\sum_{a}\hat{\mathscr{J}}_{a}^{\sigma}\nabla_{\sigma}\alpha_{a}\\
	&+\beta\hat{h}^{\sigma}N_{\sigma}+\beta\hat{q}^{\mu}M_{\mu}+\beta\hat{\pi}^{\mu\nu}\sigma_{\mu\nu}+\beta\hat{\phi}^{\mu\nu}\xi_{\mu\nu}+\hat{\varpi}^{\lambda\alpha\beta}\varXi_{\lambda\alpha\beta}\biggr]_{x_{1}},\biggl[-\beta\theta\hat{p}^{*}+\beta\mathcal{R}_{\alpha\beta}\hat{S}^{\alpha\beta}+\beta\sum_{i}\left(\hat{\mathfrak{D}}_{i}\partial_{\epsilon n}^{i}\Omega_{\mu\nu}\right)\mathcal{Z}^{\mu\nu}\\
	&-\sum_{a}\hat{\mathscr{J}}_{a}^{\sigma}\nabla_{\sigma}\alpha_{a}+\beta\hat{h}^{\sigma}N_{\sigma}+\beta\hat{q}^{\mu}M_{\mu}+\beta\hat{\pi}^{\mu\nu}\sigma_{\mu\nu}+\beta\hat{\phi}^{\mu\nu}\xi_{\mu\nu}+\hat{\varpi}^{\lambda\alpha\beta}\varXi_{\lambda\alpha\beta}\biggr]_{x_{2}}\biggr).
\end{aligned}
\label{431}
\end{equation}
The nonvanishing correlation functions in this case are
\begin{align}
	\left(\hat{q}_{\mu}\left(x\right),\hat{p}^{*}\left(x_{1}\right),\hat{\mathscr{J}}_{a\sigma}\left(x_{2}\right)\right) & =\frac{1}{3}\Delta_{\mu\sigma}\left(x\right)\left(\hat{q}_{\beta}\left(x\right),\hat{p}^{*}\left(x_{1}\right),\hat{\mathscr{J}}_{a}^{\beta}\left(x_{2}\right)\right),\label{432}\\
	\left(\hat{q}_{\mu}\left(x\right),\hat{p}^{*}\left(x_{1}\right),\hat{h}_{\sigma}\left(x_{2}\right)\right) & =\frac{1}{3}\Delta_{\mu\sigma}\left(x\right)\left(\hat{q}_{\beta}\left(x\right),\hat{p}^{*}\left(x_{1}\right),\hat{h}^{\beta}\left(x_{2}\right)\right),\label{433}\\
	\left(\hat{q}_{\mu}\left(x\right),\hat{p}^{*}\left(x_{1}\right),\hat{q}_{\sigma}\left(x_{2}\right)\right) & =\frac{1}{3}\Delta_{\mu\sigma}\left(x\right)\left(\hat{q}_{\beta}\left(x\right),\hat{p}^{*}\left(x_{1}\right),\hat{q}^{\beta}\left(x_{2}\right)\right),\label{434}\\
	\left(\hat{q}_{\mu}\left(x\right),\hat{S}_{\rho\sigma}\left(x_{1}\right),\hat{\mathscr{J}}_{a\alpha}\left(x_{2}\right)\right) & =\frac{1}{3}\mydelta_{\mu\alpha\rho\sigma}\left(x\right)\left(\hat{q}_{\gamma}\left(x\right),\hat{S}^{\gamma\delta}\left(x_{1}\right),\hat{\mathscr{J}}_{a\delta}\left(x_{2}\right)\right),\label{435}\\
	\left(\hat{q}_{\mu}\left(x\right),\hat{S}_{\rho\sigma}\left(x_{1}\right),\hat{h}_{\alpha}\left(x_{2}\right)\right) & =\frac{1}{3}\mydelta_{\mu\alpha\rho\sigma}\left(x\right)\left(\hat{q}_{\gamma}\left(x\right),\hat{S}^{\gamma\delta}\left(x_{1}\right),\hat{h}_{\delta}\left(x_{2}\right)\right),\label{436}\\
	\left(\hat{q}_{\mu}\left(x\right),\hat{S}_{\rho\sigma}\left(x_{1}\right),\hat{q}_{\alpha}\left(x_{2}\right)\right) & =\frac{1}{3}\mydelta_{\mu\alpha\rho\sigma}\left(x\right)\left(\hat{q}_{\gamma}\left(x\right),\hat{S}^{\gamma\delta}\left(x_{1}\right),\hat{q}_{\delta}\left(x_{2}\right)\right),\label{437}\\
	\left(\hat{q}_{\mu}\left(x\right),\hat{S}_{\rho\sigma}\left(x_{1}\right),\hat{\varpi}_{\lambda\alpha\beta}\left(x_{2}\right)\right) & =\myDelta_{\mu\rho\sigma\lambda\alpha\beta}\left(x\right)\left(\hat{q}_{\nu}\left(x\right),\hat{S}_{\gamma\delta}\left(x_{1}\right),\hat{\varpi}^{\nu\gamma\delta}\left(x_{2}\right)\right),\label{438}\\
	\left(\hat{q}_{\mu}\left(x\right),\hat{\mathfrak{D}}_{i}\left(x_{1}\right),\hat{\mathscr{J}}_{a\alpha}\left(x_{2}\right)\right)& =\frac{1}{3}\Delta_{\mu\alpha}\left(x\right)\left(\hat{q}_{\lambda}\left(x\right),\hat{\mathfrak{D}}_{i}\left(x_{1}\right),\hat{\mathscr{J}}_{a}^{\lambda}\left(x_{2}\right)\right),\label{439}\\
	\left(\hat{q}_{\mu}\left(x\right),\hat{\mathfrak{D}}_{i}\left(x_{1}\right),\hat{h}_{\alpha}\left(x_{2}\right)\right) & =\frac{1}{3}\Delta_{\mu\alpha}\left(x\right)\left(\hat{q}_{\lambda}\left(x\right),\hat{\mathfrak{D}}_{i}\left(x_{1}\right),\hat{h}^{\lambda}\left(x_{2}\right)\right),\label{440}\\
	\left(\hat{q}_{\mu}\left(x\right),\hat{\mathfrak{D}}_{i}\left(x_{1}\right),\hat{q}_{\alpha}\left(x_{2}\right)\right) & =\frac{1}{3}\Delta_{\mu\alpha}\left(x\right)\left(\hat{q}_{\lambda}\left(x\right),\hat{\mathfrak{D}}_{i}\left(x_{1}\right),\hat{q}^{\lambda}\left(x_{2}\right)\right),\label{441}\\
	\left(\hat{q}_{\mu}\left(x\right),\hat{\mathscr{J}}_{a\nu}\left(x_{1}\right),\hat{\pi}_{\alpha\beta}\left(x_{2}\right)\right) & =\frac{1}{5}\Delta_{\mu\nu\alpha\beta}\left(x\right)\left(\hat{q}_{\lambda}\left(x\right),\hat{\mathscr{J}}_{a\delta}\left(x_{1}\right),\hat{\pi}^{\lambda\delta}\left(x_{2}\right)\right),\label{442}\\
	\left(\hat{q}_{\mu}\left(x\right),\hat{\mathscr{J}}_{a\nu}\left(x_{1}\right),\hat{\phi}_{\alpha\beta}\left(x_{2}\right)\right) & =\frac{1}{3}\mydelta_{\mu\nu\alpha\beta}\left(x\right)\left(\hat{q}_{\lambda}\left(x\right),\hat{\mathscr{J}}_{a\delta}\left(x_{1}\right),\hat{\phi}^{\lambda\delta}\left(x_{2}\right)\right),\label{443}\\
	\left(\hat{q}_{\mu}\left(x\right),\hat{h}_{\nu}\left(x_{1}\right),\hat{\pi}_{\alpha\beta}\left(x_{2}\right)\right) & =\frac{1}{5}\Delta_{\mu\nu\alpha\beta}\left(x\right)\left(\hat{q}_{\lambda}\left(x\right),\hat{h}_{\delta}\left(x_{1}\right),\hat{\pi}^{\lambda\delta}\left(x_{2}\right)\right),\label{444}\\
	\left(\hat{q}_{\mu}\left(x\right),\hat{h}_{\nu}\left(x_{1}\right),\hat{\phi}_{\alpha\beta}\left(x_{2}\right)\right) & =\frac{1}{3}\mydelta_{\mu\nu\alpha\beta}\left(x\right)\left(\hat{q}_{\lambda}\left(x\right),\hat{h}_{\delta}\left(x_{1}\right),\hat{\phi}^{\lambda\delta}\left(x_{2}\right)\right),\label{445}\\
	\left(\hat{q}_{\mu}\left(x\right),\hat{q}_{\nu}\left(x_{1}\right),\hat{\pi}_{\alpha\beta}\left(x_{2}\right)\right) & =\frac{1}{5}\Delta_{\mu\nu\alpha\beta}\left(x\right)\left(\hat{q}_{\lambda}\left(x\right),\hat{q}_{\delta}\left(x_{1}\right),\hat{\pi}^{\lambda\delta}\left(x_{2}\right)\right),\label{446}\\
	\left(\hat{q}_{\mu}\left(x\right),\hat{q}_{\nu}\left(x_{1}\right),\hat{\phi}_{\alpha\beta}\left(x_{2}\right)\right) & =\frac{1}{3}\mydelta_{\mu\nu\alpha\beta}\left(x\right)\left(\hat{q}_{\lambda}\left(x\right),\hat{q}_{\delta}\left(x_{1}\right),\hat{\phi}^{\lambda\delta}\left(x_{2}\right)\right),\label{447}\\
	\left(\hat{q}_{\mu}\left(x\right),\hat{\phi}_{\nu\sigma}\left(x_{1}\right),\hat{\varpi}_{\lambda\alpha\beta}\left(x_{2}\right)\right) & =\myDelta_{\mu\nu\sigma\lambda\alpha\beta}\left(x\right)\left(\hat{q}_{\rho}\left(x\right),\hat{\phi}_{\gamma\delta}\left(x_{1}\right),\hat{\varpi}^{\rho\gamma\delta}\left(x_{2}\right)\right).\label{448}
\end{align}

The following coefficients are defined as
\begin{align}
	\lambda_{qp\mathscr{J}_{a}} & =\frac{1}{3}\beta\int d^{4}x_{1}d^{4}x_{2}\left(\hat{q}_{\beta}\left(x\right),\hat{p}^{*}\left(x_{1}\right),\hat{\mathscr{J}}_{a}^{\beta}\left(x_{2}\right)\right),\lambda_{qph}=-\frac{1}{3}\beta^{2}\int d^{4}x_{1}d^{4}x_{2}\left(\hat{q}_{\beta}\left(x\right),\hat{p}^{*}\left(x_{1}\right),\hat{h}^{\beta}\left(x_{2}\right)\right),\label{449}\\
	\lambda_{qpq} & =-\frac{1}{3}\beta^{2}\int d^{4}x_{1}d^{4}x_{2}\left(\hat{q}_{\beta}\left(x\right),\hat{p}^{*}\left(x_{1}\right),\hat{q}^{\beta}\left(x_{2}\right)\right),\lambda_{qS\mathscr{J}_{a}}=-\frac{1}{3}\beta\int d^{4}x_{1}d^{4}x_{2}\left(\hat{q}_{\gamma}\left(x\right),\hat{S}^{\gamma\delta}\left(x_{1}\right),\hat{\mathscr{J}}_{a\delta}\left(x_{2}\right)\right),\label{450}\\
	\lambda_{qSh} & =\frac{1}{3}\beta^{2}\int d^{4}x_{1}d^{4}x_{2}\left(\hat{q}_{\gamma}\left(x\right),\hat{S}^{\gamma\delta}\left(x_{1}\right),\hat{h}_{\delta}\left(x_{2}\right)\right),\lambda_{qSq}=\frac{1}{3}\beta^{2}\int d^{4}x_{1}d^{4}x_{2}\left(\hat{q}_{\gamma}\left(x\right),\hat{S}^{\gamma\delta}\left(x_{1}\right),\hat{q}_{\delta}\left(x_{2}\right)\right),\label{451}\\
	\lambda_{qS\varpi} & =\beta\int d^{4}x_{1}d^{4}x_{2}\left(\hat{q}_{\nu}\left(x\right),\hat{S}_{\gamma\delta}\left(x_{1}\right),\hat{\varpi}^{\nu\gamma\delta}\left(x_{2}\right)\right),\lambda_{q\mathfrak{D}_{i}\mathscr{J}_{a}}=-\frac{1}{3}\beta\int d^{4}x_{1}d^{4}x_{2}\left(\hat{q}_{\lambda}\left(x\right),\hat{\mathfrak{D}}_{i}\left(x_{1}\right),\hat{\mathscr{J}}_{a}^{\lambda}\left(x_{2}\right)\right),\label{452}\\
	\lambda_{q\mathfrak{D}_{i}h} & =\frac{1}{3}\beta^{2}\int d^{4}x_{1}d^{4}x_{2}\left(\hat{q}_{\lambda}\left(x\right),\hat{\mathfrak{D}}_{i}\left(x_{1}\right),\hat{h}^{\lambda}\left(x_{2}\right)\right),\lambda_{q\mathfrak{D}_{i}q}=\frac{1}{3}\beta^{2}\int d^{4}x_{1}d^{4}x_{2}\left(\hat{q}_{\lambda}\left(x\right),\hat{\mathfrak{D}}_{i}\left(x_{1}\right),\hat{q}^{\lambda}\left(x_{2}\right)\right),\label{453}\\
	\lambda_{q\mathscr{J}_{a}\pi} & =-\frac{1}{5}\beta\int d^{4}x_{1}d^{4}x_{2}\left(\hat{q}_{\lambda}\left(x\right),\hat{\mathscr{J}}_{a\delta}\left(x_{1}\right),\hat{\pi}^{\lambda\delta}\left(x_{2}\right)\right),\lambda_{q\mathscr{J}_{a}\phi}=-\frac{1}{3}\beta\int d^{4}x_{1}d^{4}x_{2}\left(\hat{q}_{\lambda}\left(x\right),\hat{\mathscr{J}}_{a\delta}\left(x_{1}\right),\hat{\phi}^{\lambda\delta}\left(x_{2}\right)\right),\label{454}\\
	\lambda_{qh\pi} & =\frac{1}{5}\beta^{2}\int d^{4}x_{1}d^{4}x_{2}\left(\hat{q}_{\lambda}\left(x\right),\hat{h}_{\delta}\left(x_{1}\right),\hat{\pi}^{\lambda\delta}\left(x_{2}\right)\right),\lambda_{qh\phi}=\frac{1}{3}\beta^{2}\int d^{4}x_{1}d^{4}x_{2}\left(\hat{q}_{\lambda}\left(x\right),\hat{h}_{\delta}\left(x_{1}\right),\hat{\phi}^{\lambda\delta}\left(x_{2}\right)\right),\label{455}\\
	\lambda_{qq\pi} & =\frac{1}{5}\beta^{2}\int d^{4}x_{1}d^{4}x_{2}\left(\hat{q}_{\lambda}\left(x\right),\hat{q}_{\delta}\left(x_{1}\right),\hat{\pi}^{\lambda\delta}\left(x_{2}\right)\right),\lambda_{qq\phi}=\frac{1}{3}\beta^{2}\int d^{4}x_{1}d^{4}x_{2}\left(\hat{q}_{\lambda}\left(x\right),\hat{q}_{\delta}\left(x_{1}\right),\hat{\phi}^{\lambda\delta}\left(x_{2}\right)\right),\label{456}\\
	\lambda_{q\phi\varpi} & =\beta\int d^{4}x_{1}d^{4}x_{2}\left(\hat{q}_{\rho}\left(x\right),\hat{\phi}_{\gamma\delta}\left(x_{1}\right),\hat{\varpi}^{\rho\gamma\delta}\left(x_{2}\right)\right).\label{457}
\end{align}
From Eqs.~\eqref{431}-\eqref{457} and the symmetry property~\eqref{49}, we obtain
\begin{eqnarray}
\begin{aligned}
\langle\hat{q}_{\mu}\rangle_{2}^{3}= & 2\sum_{a}\lambda_{qp\mathscr{J}_{a}}\theta\nabla_{\mu}\alpha_{a}+2\lambda_{qph}\theta N_{\mu}+2\lambda_{qpq}\theta M_{\mu}+2\sum_{a}\lambda_{qS\mathscr{J}_{a}}\mathcal{R}_{\langle\mu\rangle\langle\alpha\rangle}\nabla^{\alpha}\alpha_{a}+2\lambda_{qSh}\mathcal{R}_{\langle\mu\rangle\langle\alpha\rangle}N^{\alpha}\\
&+2\lambda_{qSq}\mathcal{R}_{\langle\mu\rangle\langle\alpha\rangle}M^{\alpha}+2\lambda_{qS\varpi}\mathcal{R}^{\rho\sigma}\varXi_{\mu\rho\sigma}+2\sum_{i}\mathcal{Z}^{\rho\sigma}\partial_{\epsilon n}^{i}\Omega_{\rho\sigma}\sum_{a}\lambda_{q\mathfrak{D}_{i}\mathscr{J}_{a}}\nabla_{\mu}\alpha_{a}+2\sum_{i}\mathcal{Z}^{\rho\sigma}\partial_{\epsilon n}^{i}\Omega_{\rho\sigma}\lambda_{q\mathfrak{D}_{i}h}N_{\mu}\\
&+2\sum_{i}\mathcal{Z}^{\rho\sigma}\partial_{\epsilon n}^{i}\Omega_{\rho\sigma}\lambda_{q\mathfrak{D}_{i}q}M_{\mu}+2\sum_{a}\lambda_{q\mathscr{J}_{a}\pi}\nabla^{\nu}\alpha_{a}\sigma_{\mu\nu}+2\sum_{a}\lambda_{q\mathscr{J}_{a}\phi}\nabla^{\nu}\alpha_{a}\xi_{\mu\nu}+2\lambda_{qh\pi}N^{\nu}\sigma_{\mu\nu}\\
&+2\lambda_{qh\phi}N^{\nu}\xi_{\mu\nu}+2\lambda_{qq\pi}M^{\nu}\sigma_{\mu\nu}+2\lambda_{qq\phi}M^{\nu}\xi_{\mu\nu}+2\lambda_{q\phi\varpi}\xi^{\nu\sigma}\varXi_{\mu\nu\sigma}.
\end{aligned}
\label{458}
\end{eqnarray}
Combining Eqs.~\eqref{70}, \eqref{92}, \eqref{94}, \eqref{124}, \eqref{423}, \eqref{430}, and~\eqref{458}, we obtain the complete second-order expression for the boost heat vector:
\begin{eqnarray}
\begin{aligned}
q_{\mu}= & -\lambda M_{\mu}+\sum_{a}\lambda_{q\mathscr{J}_{a}}\nabla_{\mu}\alpha_{a}+\lambda_{qh}N_{\mu}+\sum_{a}\widetilde{\lambda}_{qh}n_{a}{w}^{-2}\Bigl[\Gamma {w}\theta+\sum_{a}\delta_{a}n_{a}\theta+\mathcal{K}_{\alpha\beta}\Bigl(\theta S^{\alpha\beta}+u^{\alpha}\partial_{\lambda}S^{\beta\lambda}+S^{\beta\lambda}\partial_{\lambda}u^{\alpha}\\	&+u^{\beta}\partial_{\lambda}S^{\lambda\alpha}+S^{\lambda\alpha}\partial_{\lambda}u^{\beta}\Bigr)\Bigr]\nabla_{\mu}\alpha_{a}+\sum_{a}\widetilde{\lambda}_{q\mathscr{J}_{a}}\Delta_{\mu\gamma}D\left(\nabla^{\gamma}\alpha_{a}\right)+\widetilde{\lambda}_{qh}N_{\mu}\left(\theta\Gamma-2\mathcal{Z}^{\alpha\beta}\mathcal{D}_{\alpha\beta}\right)+\widetilde{\lambda}_{qh}\Delta_{\mu\gamma}DN^{\gamma}\\ &-\widetilde{\lambda}M_{\mu}\left(\theta\Gamma-2\mathcal{Z}^{\alpha\beta}\mathcal{D}_{\alpha\beta}\right)-\widetilde{\lambda}\Delta_{\mu\gamma}DM^{\gamma}+\lambda_{qh}\mathcal{H}_{\mu}-\lambda\mathcal{Q}_{\mu}+2\sum_{a}\lambda_{qp\mathscr{J}_{a}}\theta\nabla_{\mu}\alpha_{a}+2\lambda_{qph}\theta N_{\mu}+2\lambda_{qpq}\theta M_{\mu}\\ &+2\sum_{a}\lambda_{qS\mathscr{J}_{a}}\mathcal{R}_{\langle\mu\rangle\langle\alpha\rangle}\nabla^{\alpha}\alpha_{a}+2\lambda_{qSh}\mathcal{R}_{\langle\mu\rangle\langle\alpha\rangle}N^{\alpha}+2\lambda_{qSq}\mathcal{R}_{\langle\mu\rangle\langle\alpha\rangle}M^{\alpha}+2\lambda_{qS\varpi}\mathcal{R}^{\rho\sigma}\varXi_{\mu\rho\sigma}\\
&+2\sum_{i}\mathcal{Z}^{\rho\sigma}\partial_{\epsilon n}^{i}\Omega_{\rho\sigma}\sum_{a}\lambda_{q\mathfrak{D}_{i}\mathscr{J}_{a}}\nabla_{\mu}\alpha_{a}+2\sum_{i}\mathcal{Z}^{\rho\sigma}\partial_{\epsilon n}^{i}\Omega_{\rho\sigma}\lambda_{q\mathfrak{D}_{i}h}N_{\mu}+2\sum_{i}\mathcal{Z}^{\rho\sigma}\partial_{\epsilon n}^{i}\Omega_{\rho\sigma}\lambda_{q\mathfrak{D}_{i}q}M_{\mu}\\	&+2\sum_{a}\lambda_{q\mathscr{J}_{a}\pi}\nabla^{\nu}\alpha_{a}\sigma_{\mu\nu}+2\sum_{a}\lambda_{q\mathscr{J}_{a}\phi}\nabla^{\nu}\alpha_{a}\xi_{\mu\nu}+2\lambda_{qh\pi}N^{\nu}\sigma_{\mu\nu}+2\lambda_{qh\phi}N^{\nu}\xi_{\mu\nu}+2\lambda_{qq\pi}M^{\nu}\sigma_{\mu\nu}\\ &+2\lambda_{qq\phi}M^{\nu}\xi_{\mu\nu}+2\lambda_{q\phi\varpi}\xi^{\nu\sigma}\varXi_{\mu\nu\sigma}.
\end{aligned}
\label{459}
\end{eqnarray}
To derive a relaxation-type equation for $q_\mu$ from~\eqref{459}, we use the first-order approximation
\begin{eqnarray}	M^{\gamma}\simeq-\lambda^{-1}\left(q^{\gamma}-\sum_{a}\lambda_{q\mathscr{J}_{a}}\nabla^{\gamma}\alpha_{a}-\lambda_{qh}N^{\gamma}\right),	\label{460}
\end{eqnarray}
in the term $-\widetilde{\lambda}\Delta_{\mu\gamma}DM^{\gamma}$ on the right-hand side of Eq.~\eqref{459}. We then have
\begin{equation}
\begin{aligned}
-\widetilde{\lambda}\Delta_{\mu\gamma}DM^{\gamma}\simeq & \widetilde{\lambda}\lambda^{-1}\Delta_{\mu\gamma}Dq^{\gamma}-\widetilde{\lambda}\lambda^{-1}\Delta_{\mu\gamma}\sum_{a}\lambda_{q\mathscr{J}_{a}}D\left(\nabla^{\gamma}\alpha_{a}\right)-\widetilde{\lambda}\lambda^{-1}\beta\theta\sum_{a}\nabla_{\mu}\alpha_{a}\biggl[\left(\frac{\partial\lambda_{q\mathscr{J}_{a}}}{\partial\beta}\Gamma-\sum_{d}\frac{\partial\lambda_{q\mathscr{J}_{a}}}{\partial\alpha_{d}}\delta_{d}-2\frac{\partial\lambda_{q\mathscr{J}_{a}}}{\partial\Omega_{\alpha\beta}}\mathcal{K}_{\alpha\beta}\right)\\
&-2\theta^{-1}\mathcal{Z}^{\alpha\beta}\frac{\partial\lambda_{q\mathscr{J}_{a}}}{\partial S^{\alpha\beta}}\biggr]-\widetilde{\lambda}\lambda^{-1}\Delta_{\mu\gamma}\lambda_{qh}DN^{\gamma}-\widetilde{\lambda}\lambda^{-1}\beta\theta N_{\mu}\biggl[\left(\frac{\partial\lambda_{qh}}{\partial\beta}\Gamma-\sum_{d}\frac{\partial\lambda_{qh}}{\partial\alpha_{d}}\delta_{d}-2\frac{\partial\lambda_{qh}}{\partial\Omega_{\alpha\beta}}\mathcal{K}_{\alpha\beta}\right)-2\theta^{-1}\mathcal{Z}^{\alpha\beta}\frac{\partial\lambda_{qh}}{\partial S^{\alpha\beta}}\biggr]\\
&-\widetilde{\lambda}\lambda^{-2}\left(q_{\mu}-\sum_{a}\lambda_{q\mathscr{J}_{a}}\nabla_{\mu}\alpha_{a}-\lambda_{qh}N_{\mu}\right)\beta\theta\biggl[\left(\frac{\partial\lambda}{\partial\beta}\Gamma-\sum_{d}\frac{\partial\lambda}{\partial\alpha_{d}}\delta_{d}-2\frac{\partial\lambda}{\partial\Omega_{\alpha\beta}}\mathcal{K}_{\alpha\beta}\right)-2\theta^{-1}\mathcal{Z}^{\alpha\beta}\frac{\partial\lambda}{\partial S^{\alpha\beta}}\biggr],
\end{aligned}
\label{461}
\end{equation}
where Eqs.~\eqref{76}-\eqref{78} have been used. Introducing the coefficients
\begin{align}
\tau_{q}= & -\widetilde{\lambda}\lambda^{-1},\label{462}\\
\widetilde{\lambda}_{q}= & \beta\tau_{q}\lambda^{-1}\left[\left(\frac{\partial\lambda}{\partial\beta}\Gamma-\sum_{d}\frac{\partial\lambda}{\partial\alpha_{d}}\delta_{d}-2\frac{\partial\lambda}{\partial\Omega_{\alpha\beta}}\mathcal{K}_{\alpha\beta}\right)-2\theta^{-1}\mathcal{Z}^{\alpha\beta}\frac{\partial\lambda}{\partial S^{\alpha\beta}}\right],\label{463}\\
\overline{\lambda}= & \beta\tau_{q}\left[\left(\frac{\partial\lambda_{qh}}{\partial\beta}\Gamma-\sum_{d}\frac{\partial\lambda_{qh}}{\partial\alpha_{d}}\delta_{d}-2\frac{\partial\lambda_{qh}}{\partial\Omega_{\alpha\beta}}\mathcal{K}_{\alpha\beta}\right)-2\theta^{-1}\mathcal{Z}^{\alpha\beta}\frac{\partial\lambda_{qh}}{\partial S^{\alpha\beta}}\right],\label{464}\\
\widehat{\lambda}^{a}= & \beta\tau_{q}\left[\left(\frac{\partial\lambda_{q\mathscr{J}_{a}}}{\partial\beta}\Gamma-\sum_{d}\frac{\partial\lambda_{q\mathscr{J}_{a}}}{\partial\alpha_{d}}\delta_{d}-2\frac{\partial\lambda_{q\mathscr{J}_{a}}}{\partial\Omega_{\alpha\beta}}\mathcal{K}_{\alpha\beta}\right)-2\theta^{-1}\mathcal{Z}^{\alpha\beta}\frac{\partial\lambda_{q\mathscr{J}_{a}}}{\partial S^{\alpha\beta}}\right],\label{465}
\end{align}
and combining Eqs.~\eqref{459} and \eqref{461}, we arrive at the following relaxation equation for the boost heat vector:
\begin{eqnarray}
\begin{aligned}
q_{\mu}+\tau_{q}\dot{q}_{\mu}= & -\lambda M_{\mu}+\sum_{a}\lambda_{q\mathscr{J}_{a}}\nabla_{\mu}\alpha_{a}+\lambda_{qh}N_{\mu}+\sum_{a}\widetilde{\lambda}_{qh}n_{a}{w}^{-2}\Bigl[\Gamma {w}\theta+\sum_{a}\delta_{a}n_{a}\theta+\mathcal{K}_{\alpha\beta}\Bigl(\theta S^{\alpha\beta}+u^{\alpha}\partial_{\lambda}S^{\beta\lambda}\\
	&+S^{\beta\lambda}\partial_{\lambda}u^{\alpha}+u^{\beta}\partial_{\lambda}S^{\lambda\alpha}+S^{\lambda\alpha}\partial_{\lambda}u^{\beta}\Bigr)\Bigr]\nabla_{\mu}\alpha_{a}+\sum_{a}\widetilde{\lambda}_{q\mathscr{J}_{a}}\Delta_{\mu\gamma}D\left(\nabla^{\gamma}\alpha_{a}\right)+\widetilde{\lambda}_{qh}N_{\mu}\left(\theta\Gamma-2\mathcal{Z}^{\alpha\beta}\mathcal{D}_{\alpha\beta}\right)\\
	&+\widetilde{\lambda}_{qh}\Delta_{\mu\gamma}DN^{\gamma}-\widetilde{\lambda}M_{\mu}\left(\theta\Gamma-2\mathcal{Z}^{\alpha\beta}\mathcal{D}_{\alpha\beta}\right)+\tau_{q}\Delta_{\mu\gamma}\sum_{a}\lambda_{q\mathscr{J}_{a}}D\left(\nabla^{\gamma}\alpha_{a}\right)+\sum_{a}\widehat{\lambda}^{a}\theta\nabla_{\mu}\alpha_{a}+\tau_{q}\lambda_{qh}\Delta_{\mu\gamma}DN^{\gamma}\\
	&+\overline{\lambda}\theta N_{\mu}+\widetilde{\lambda}_{q}\theta\Bigl(q_{\mu}-\sum_{a}\lambda_{qa}\nabla_{\mu}\alpha_{a}-\lambda_{qh}N_{\mu}\Bigr)+\lambda_{qh}\mathcal{H}_{\mu}-\lambda\mathcal{Q}_{\mu}+2\sum_{a}\lambda_{qp\mathscr{J}_{a}}\theta\nabla_{\mu}\alpha_{a}+2\lambda_{qph}\theta N_{\mu}\\
	&+2\lambda_{qpq}\theta M_{\mu}+2\sum_{a}\lambda_{qS\mathscr{J}_{a}}\mathcal{R}_{\langle\mu\rangle\langle\alpha\rangle}\nabla^{\alpha}\alpha_{a}+2\lambda_{qSh}\mathcal{R}_{\langle\mu\rangle\langle\alpha\rangle}N^{\alpha}+2\lambda_{qSq}\mathcal{R}_{\langle\mu\rangle\langle\alpha\rangle}M^{\alpha}+2\lambda_{qS\varpi}\mathcal{R}^{\rho\sigma}\varXi_{\mu\rho\sigma}\\
	&+2\sum_{i}\mathcal{Z}^{\rho\sigma}\partial_{\epsilon n}^{i}\Omega_{\rho\sigma}\sum_{a}\lambda_{q\mathfrak{D}_{i}\mathscr{J}_{a}}\nabla_{\mu}\alpha_{a}+2\sum_{i}\mathcal{Z}^{\rho\sigma}\partial_{\epsilon n}^{i}\Omega_{\rho\sigma}\lambda_{q\mathfrak{D}_{i}h}N_{\mu}+2\sum_{i}\mathcal{Z}^{\rho\sigma}\partial_{\epsilon n}^{i}\Omega_{\rho\sigma}\lambda_{q\mathfrak{D}_{i}q}M_{\mu}\\
	&+2\sum_{a}\lambda_{q\mathscr{J}_{a}\pi}\nabla^{\nu}\alpha_{a}\sigma_{\mu\nu}+2\sum_{a}\lambda_{q\mathscr{J}_{a}\phi}\nabla^{\nu}\alpha_{a}\xi_{\mu\nu}+2\lambda_{qh\pi}N^{\nu}\sigma_{\mu\nu}+2\lambda_{qh\phi}N^{\nu}\xi_{\mu\nu}+2\lambda_{qq\pi}M^{\nu}\sigma_{\mu\nu}\\
	&+2\lambda_{qq\phi}M^{\nu}\xi_{\mu\nu}+2\lambda_{q\phi\varpi}\xi^{\nu\sigma}\varXi_{\mu\nu\sigma}.
\end{aligned}
\label{466}
\end{eqnarray}
where we define $\dot{q}_{\mu}=\Delta_{\mu\gamma}Dq^{\gamma}$.

\subsubsection{Second-order corrections to $\varpi^{\lambda\mu\nu}$}

Substituting Eqs.~\eqref{90} and \eqref{125} into Eq.~\eqref{95} and recalling Curie's theorem we obtain
\begin{eqnarray}
	\langle\hat{\varpi}^{\lambda\mu\nu}\left(x\right)\rangle_{2}^{1}=\int d^{4}x_{1}\left(\hat{\varpi}^{\lambda\mu\nu}\left(x\right),\hat{\varpi}^{\alpha\beta\gamma}\left(x_{1}\right)\right)\varXi_{\alpha\beta\gamma}\left(x_{1}\right)-\varphi\left(x\right)\varXi^{\lambda\mu\nu}\left(x\right).
	\label{467}
\end{eqnarray}
By substituting the two-point correlation function given by Eqs.~\eqref{152} and \eqref{156} into Eq.~\eqref{467}, we obtain
\begin{eqnarray} \langle\hat{\varpi}^{\lambda\mu\nu}\left(x\right)\rangle_{2}^{1}=\myDelta^{\lambda\mu\nu\rho\sigma\delta}\left(x\right)\int d^{4}x_{1}\left(\hat{\varpi}^{\varepsilon\zeta\eta}\left(x\right),\hat{\varpi}_{\varepsilon\zeta\eta}\left(x_{1}\right)\right)\varXi_{\rho\sigma\delta}\left(x_{1}\right)-\varphi\left(x\right)\varXi^{\lambda\mu\nu}\left(x\right).
\label{468}
\end{eqnarray}
Performing a Taylor expansion of the hydrodynamic quantities around $x_1=x$ while retaining linear terms yields:
\begin{eqnarray}
	\hat{\varpi}_{\varepsilon\zeta\eta}\left(x_{1}\right)=\hat{\varpi}_{\varepsilon\zeta\eta}\left(x_{1}\right)\bigg|_{x_{1}=x}+\left(x_{1}-x\right)^{\tau}\frac{\partial\hat{\varpi}_{\varepsilon\zeta\eta}\left(x_{1}\right)}{\partial x_{1}^{\tau}}\bigg|_{x_{1}=x},
	\label{469}
\end{eqnarray}
where
\begin{eqnarray}
	\begin{aligned}
	\hat{\varpi}_{\varepsilon\zeta\eta}\left(x_{1}\right)\bigg|_{x_{1}=x}=&\myDelta_{\alpha\beta\gamma\varepsilon\zeta\eta}\left(x\right)\hat{S}^{\varepsilon\zeta\eta}\left(x_{1}\right),\\
	\frac{\partial\hat{\varpi}_{\varepsilon\zeta\eta}\left(x_{1}\right)}{\partial x_{1}^{\tau}}\bigg|_{x_{1}=x}=&-\left[\hat{\varpi}_{\rho\zeta\eta}\left(x_{1}\right)u_{\varepsilon}\left(x\right)+\hat{\varpi}_{\varepsilon\rho\eta}\left(x_{1}\right)u_{\zeta}\left(x\right)+\hat{\varpi}_{\varepsilon\zeta\rho}\left(x_{1}\right)u_{\eta}\left(x\right)\right]\frac{\partial u^{\rho}\left(x_{1}\right)}{\partial x_{1}}\bigg|_{x_{1}=x}.
	\end{aligned}
	\label{470}
\end{eqnarray}

Substituting Eq.~\eqref{469} into Eq.~\eqref{468} and expanding the thermodynamic force around $x_1=x$, we obtain up to the second order in gradients
\begin{eqnarray}
\begin{aligned}
\langle\hat{\varpi}^{\lambda\mu\nu}\left(x\right)\rangle_{2}^{1}= & \myDelta^{\lambda\mu\nu\rho\sigma\delta}\left(x\right)\varXi_{\rho\sigma\delta}\left(x\right)\int d^{4}x_{1}\left(\hat{\varpi}^{\varepsilon\zeta\eta}\left(x\right),\frac{\partial\hat{\varpi}_{\varepsilon\zeta\eta}\left(x_{1}\right)}{\partial x_{1}^{\tau}}\bigg|_{x_{1}=x}\right)\left(x_{1}-x\right)^{\tau}\\
&+\myDelta^{\lambda\mu\nu\rho\sigma\delta}\left(x\right)\frac{\partial\varXi_{\rho\sigma\delta}\left(x_{1}\right)}{\partial x_{1}^{\tau}}\bigg|_{x_{1}=x}\int d^{4}x_{1}\left(\hat{\varpi}^{\varepsilon\zeta\eta}\left(x\right),\hat{\varpi}_{\varepsilon\zeta\eta}\left(x_{1}\right)\bigg|_{x_{1}=x}\right)\left(x_{1}-x\right)^{\tau},
\end{aligned}
\label{471}
\end{eqnarray}
where we used the following relation to eliminate the first-order term
\begin{eqnarray}
	\varphi=\int d^{4}x_{1}\left(\hat{\varpi}^{\varepsilon\zeta\eta}\left(x\right),\hat{\varpi}_{\varepsilon\zeta\eta}\left(x_{1}\right)\bigg|_{x_{1}=x}\right).
	\label{472}
\end{eqnarray}
Upon substituting Eq.~\eqref{470} into Eq.~\eqref{471} and invoking Curie's theorem, subject to the orthogonality constraint $\hat{\varpi}^{\lambda\mu\nu}u_{\lambda}=0$, we obtain the nonlocal corrections from the two-point correlation function to $\varpi^{\lambda\mu\nu}$
\begin{eqnarray} \langle\hat{\varpi}^{\lambda\mu\nu}\rangle_{2}^{1}=\widetilde{\varphi}\myDelta^{\lambda\mu\nu\rho\sigma\delta}D\varXi_{\rho\sigma\delta},
\label{473}
\end{eqnarray}
where we define
\begin{eqnarray}
	\widetilde{\varphi}=i\frac{d}{d\omega}\varphi\left(\omega\right)\bigg|_{\omega=0}=-\frac{T}{2}\frac{d^{2}}{d\omega^{2}}\text{Re}G_{\hat{\varpi}^{\varepsilon\zeta\eta}\hat{\varpi}_{\varepsilon\zeta\eta}}^{R}\left(\omega\right)\bigg|_{\omega=0},
	\label{474}
\end{eqnarray}
where the frequency-dependent transport coefficients $\varphi\left(\omega\right)$ is expressed as
\begin{eqnarray}
	\varphi\left(\omega\right)=\int d^{4}x_{1}\int_{-\infty}^{t}e^{i\omega\left(t-t_{1}\right)}\left(\hat{\varpi}^{\varepsilon\zeta\eta}\left(\boldsymbol{x},t\right),\hat{\varpi}_{\varepsilon\zeta\eta}\left(\boldsymbol{x}_{1},t_{1}\right)\right).
	\label{475}
\end{eqnarray}
Upon substituting Eq.~\eqref{91} into Eq.~\eqref{96} and applying Curie's theorem, we obtain corrections from extended thermodynamic forces to $\varpi^{\lambda\mu\nu}$
\begin{eqnarray}
\langle\hat{\varpi}^{\lambda\mu\nu}\left(x\right)\rangle_{2}^{2}=\int d^{4}x_{1}\left(\hat{\varpi}^{\lambda\mu\nu}\left(x\right),\hat{C}_{2}\left(x_{1}\right)\right)=0.
\label{476}
\end{eqnarray}
From Eqs.~\eqref{90} and \eqref{97}, we obtain the corrections from the three-point correlation function to $\varpi^{\lambda\mu\nu}$
\begin{equation}
\begin{aligned}
	\left\langle \hat{\varpi}^{\lambda\mu\nu}\left(x\right)\right\rangle _{2}^{3}= & \int d^{4}x_{1}d^{4}x_{2}\biggl(\hat{\varpi}^{\lambda\mu\nu}\left(x\right),\biggl[-\beta\theta\hat{p}^{*}+\beta\mathcal{R}_{\alpha\beta}\hat{S}^{\alpha\beta}+\beta\sum_{i}\left(\hat{\mathfrak{D}}_{i}\partial_{\epsilon n}^{i}\Omega_{\mu\nu}\right)\mathcal{Z}^{\mu\nu}-\sum_{a}\hat{\mathscr{J}}_{a}^{\sigma}\nabla_{\sigma}\alpha_{a}\\
	&+\beta\hat{h}^{\sigma}N_{\sigma}+\beta\hat{q}^{\mu}M_{\mu}+\beta\hat{\pi}^{\mu\nu}\sigma_{\mu\nu}+\beta\hat{\phi}^{\mu\nu}\xi_{\mu\nu}+\hat{\varpi}^{\lambda\alpha\beta}\varXi_{\lambda\alpha\beta}\biggr]_{x_{1}},\biggl[-\beta\theta\hat{p}^{*}+\beta\mathcal{R}_{\alpha\beta}\hat{S}^{\alpha\beta}+\beta\sum_{i}\left(\hat{\mathfrak{D}}_{i}\partial_{\epsilon n}^{i}\Omega_{\mu\nu}\right)\mathcal{Z}^{\mu\nu}\\
	&-\sum_{a}\hat{\mathscr{J}}_{a}^{\sigma}\nabla_{\sigma}\alpha_{a}+\beta\hat{h}^{\sigma}N_{\sigma}+\beta\hat{q}^{\mu}M_{\mu}+\beta\hat{\pi}^{\mu\nu}\sigma_{\mu\nu}+\beta\hat{\phi}^{\mu\nu}\xi_{\mu\nu}+\hat{\varpi}^{\lambda\alpha\beta}\varXi_{\lambda\alpha\beta}\biggr]_{x_{2}}\biggr).
\end{aligned}
\label{477}
\end{equation}
The nonvanishing correlators in Eq.~\eqref{477} are
\begin{align} \left(\hat{\varpi}^{\lambda\mu\nu}\left(x\right),\hat{p}^{*}\left(x_{1}\right),\hat{\varpi}^{\rho\sigma\delta}\left(x_{2}\right)\right) &=\myDelta^{\lambda\mu\nu\rho\sigma\delta}\left(\hat{\varpi}^{\gamma\varepsilon\zeta}\left(x\right),\hat{p}^{*}\left(x_{1}\right),\hat{\varpi}_{\gamma\varepsilon\zeta}\left(x_{2}\right)\right),\label{478}\\	\left(\hat{\varpi}^{\lambda\mu\nu}\left(x\right),\hat{S}^{\rho\sigma}\left(x_{1}\right),\hat{\mathscr{J}}_{a}^{\delta}\left(x_{2}\right)\right) &=\myDelta^{\lambda\mu\nu\rho\sigma\delta}\left(\hat{\varpi}^{\gamma\varepsilon\zeta}\left(x\right),\hat{S}_{\gamma\varepsilon}\left(x_{1}\right),\hat{\mathscr{J}}_{a\zeta}\left(x_{2}\right)\right),\label{479}\\
\left(\hat{\varpi}^{\lambda\mu\nu}\left(x\right),\hat{S}^{\rho\sigma}\left(x_{1}\right),\hat{h}^{\delta}\left(x_{2}\right)\right) &=\myDelta^{\lambda\mu\nu\rho\sigma\delta}\left(\hat{\varpi}^{\gamma\varepsilon\zeta}\left(x\right),\hat{S}_{\gamma\varepsilon}\left(x_{1}\right),\hat{h}_{\zeta}\left(x_{2}\right)\right),\label{480}\\
\left(\hat{\varpi}^{\lambda\mu\nu}\left(x\right),\hat{S}^{\rho\sigma}\left(x_{1}\right),\hat{q}^{\delta}\left(x_{2}\right)\right) &=\myDelta^{\lambda\mu\nu\rho\sigma\delta}\left(\hat{\varpi}^{\gamma\varepsilon\zeta}\left(x\right),\hat{S}_{\gamma\varepsilon}\left(x_{1}\right),\hat{q}_{\zeta}\left(x_{2}\right)\right),\label{481}\\
\left(\hat{\varpi}^{\lambda\mu\nu}\left(x\right),\hat{\mathfrak{D}}_{i}\left(x_{1}\right),\hat{\varpi}^{\rho\sigma\delta}\left(x_{2}\right)\right) &=\myDelta^{\lambda\mu\nu\rho\sigma\delta}\left(\hat{\varpi}^{\gamma\varepsilon\zeta}\left(x\right),\hat{\mathfrak{D}}_{i}\left(x_{1}\right),\hat{\varpi}_{\gamma\varepsilon\zeta}\left(x_{2}\right)\right),\label{482}\\
\left(\hat{\varpi}^{\lambda\mu\nu}\left(x\right),\hat{\mathscr{J}}_{a}^{\rho}\left(x_{1}\right),\hat{\phi}^{\sigma\delta}\left(x_{2}\right)\right) &=\myDelta^{\lambda\mu\nu\rho\sigma\delta}\left(\hat{\varpi}^{\gamma\varepsilon\zeta}\left(x\right),\hat{\mathscr{J}}_{a\gamma}\left(x_{1}\right),\hat{\phi}_{\varepsilon\zeta}\left(x_{2}\right)\right),\label{483}\\
\left(\hat{\varpi}^{\lambda\mu\nu}\left(x\right),\hat{h}^{\rho}\left(x_{1}\right),\hat{\phi}^{\sigma\delta}\left(x_{2}\right)\right) &=\myDelta^{\lambda\mu\nu\rho\sigma\delta}\left(\hat{\varpi}^{\gamma\varepsilon\zeta}\left(x\right),\hat{h}_{\gamma}\left(x_{1}\right),\hat{\phi}_{\varepsilon\zeta}\left(x_{2}\right)\right),\label{484}\\
\left(\hat{\varpi}^{\lambda\mu\nu}\left(x\right),\hat{q}^{\rho}\left(x_{1}\right),\hat{\phi}^{\sigma\delta}\left(x_{2}\right)\right) &=\myDelta^{\lambda\mu\nu\rho\sigma\delta}\left(\hat{\varpi}^{\gamma\varepsilon\zeta}\left(x\right),\hat{q}_{\gamma}\left(x_{1}\right),\hat{\phi}_{\varepsilon\zeta}\left(x_{2}\right)\right),\label{485}
\end{align}
By substituting the correlation functions from Eqs.~\eqref{478}-\eqref{485} into Eq.~\eqref{477}, factoring out the thermodynamic forces evaluated at $x$, and taking into account the symmetry property~\eqref{49}, we define a set of transport coefficients
\begin{align}
	\varphi_{\varpi p\varpi}= & -\beta\int d^{4}x_{1}d^{4}x_{2}\left(\hat{\varpi}^{\gamma\varepsilon\zeta}\left(x\right),\hat{p}^{*}\left(x_{1}\right),\hat{\varpi}_{\gamma\varepsilon\zeta}\left(x_{2}\right)\right),\varphi_{\varpi S\mathscr{J}_{a}}=-\beta\int d^{4}x_{1}d^{4}x_{2}\left(\hat{\varpi}^{\gamma\varepsilon\zeta}\left(x\right),\hat{S}_{\gamma\varepsilon}\left(x_{1}\right),\hat{\mathscr{J}}_{a\zeta}\left(x_{2}\right)\right),\label{486}\\
	\varphi_{\varpi Sh}= & \beta^{2}\int d^{4}x_{1}d^{4}x_{2}\left(\hat{\varpi}^{\gamma\varepsilon\zeta}\left(x\right),\hat{S}_{\gamma\varepsilon}\left(x_{1}\right),\hat{h}_{\zeta}\left(x_{2}\right)\right),\varphi_{\varpi Sq}=\beta^{2}\int d^{4}x_{1}d^{4}x_{2}\left(\hat{\varpi}^{\gamma\varepsilon\zeta}\left(x\right),\hat{S}_{\gamma\varepsilon}\left(x_{1}\right),\hat{q}_{\zeta}\left(x_{2}\right)\right),\label{487}\\
	\varphi_{\varpi\mathfrak{D}_{i}\varpi}= & \beta\int d^{4}x_{1}d^{4}x_{2}\left(\hat{\varpi}^{\gamma\varepsilon\zeta}\left(x\right),\hat{\mathfrak{D}}_{i}\left(x_{1}\right),\hat{\varpi}_{\gamma\varepsilon\zeta}\left(x_{2}\right)\right),\varphi_{\varpi\mathscr{J}_{a}\phi}=-\beta\int d^{4}x_{1}d^{4}x_{2}\left(\hat{\varpi}^{\gamma\varepsilon\zeta}\left(x\right),\hat{\mathscr{J}}_{a\gamma}\left(x_{1}\right),\hat{\phi}_{\varepsilon\zeta}\left(x_{2}\right)\right),\label{488}\\
	\varphi_{\varpi h\phi}= & \beta^{2}\int d^{4}x_{1}d^{4}x_{2}\left(\hat{\varpi}^{\gamma\varepsilon\zeta}\left(x\right),\hat{h}_{\gamma}\left(x_{1}\right),\hat{\phi}_{\varepsilon\zeta}\left(x_{2}\right)\right),\varphi_{\varpi q\phi}=\beta^{2}\int d^{4}x_{1}d^{4}x_{2}\left(\hat{\varpi}^{\gamma\varepsilon\zeta}\left(x\right),\hat{q}_{\gamma}\left(x_{1}\right),\hat{\phi}_{\varepsilon\zeta}\left(x_{2}\right)\right).\label{489}
\end{align}
We finally obtain
\begin{equation}
\begin{aligned}
\left\langle \hat{\varpi}^{\lambda\mu\nu}\left(x\right)\right\rangle _{2}^{3}= & 2\varphi_{\varpi p\varpi}\theta\varXi^{\lambda\mu\nu}+2\sum_{a}\varphi_{\varpi S\mathscr{J}_{a}}\mathcal{R}_{\rho\sigma}\nabla_{\delta}\alpha_{a}\myDelta^{\lambda\mu\nu\rho\sigma\delta}+2\varphi_{\varpi Sh}\mathcal{R}_{\rho\sigma}N_{\delta}\myDelta^{\lambda\mu\nu\rho\sigma\delta}+2\varphi_{\varpi Sq}\mathcal{R}_{\rho\sigma}M_{\delta}\myDelta^{\lambda\mu\nu\rho\sigma\delta}\\
 & +2\sum_{i}\varphi_{\varpi\mathfrak{D}_{i}\varpi}\mathcal{Z}^{\alpha\beta}\partial_{\epsilon n}^{i}\Omega_{\alpha\beta}\varXi^{\lambda\mu\nu}+2\sum_{a}\varphi_{\varpi\mathscr{J}_{a}\phi}\nabla_{\rho}\alpha_{a}\xi_{\sigma\delta}\myDelta^{\lambda\mu\nu\rho\sigma\delta}+2\varphi_{\varpi h\phi}N_{\rho}\xi_{\sigma\delta}\myDelta^{\lambda\mu\nu\rho\sigma\delta}+2\varphi_{\varpi q\phi}M_{\rho}\xi_{\sigma\delta}\myDelta^{\lambda\mu\nu\rho\sigma\delta}.
\end{aligned}
\label{490}
\end{equation}
Combining the corrections from Eqs.~\eqref{70}, \eqref{125}, \eqref{473}, \eqref{476}, \eqref{490} and utilizing Eqs.~\eqref{92} and~\eqref{94}, we derive the complete second-order expression for $\varpi^{\lambda\mu\nu}$
\begin{eqnarray}
\begin{aligned}
\varpi^{\lambda\mu\nu}= & \varphi\varXi^{\lambda\mu\nu}+\widetilde{\varphi}\myDelta^{\lambda\mu\nu\rho\sigma\delta}D\varXi_{\rho\sigma\delta}+2\varphi_{\varpi p\varpi}\theta\varXi^{\lambda\mu\nu}+2\sum_{a}\varphi_{\varpi S\mathscr{J}_{a}}\mathcal{R}_{\rho\sigma}\nabla_{\delta}\alpha_{a}\myDelta^{\lambda\mu\nu\rho\sigma\delta}\\
&+2\varphi_{\varpi Sh}\mathcal{R}_{\rho\sigma}N_{\delta}\myDelta^{\lambda\mu\nu\rho\sigma\delta}+2\varphi_{\varpi Sq}\mathcal{R}_{\rho\sigma}M_{\delta}\myDelta^{\lambda\mu\nu\rho\sigma\delta}+2\sum_{i}\varphi_{\varpi\mathfrak{D}_{i}\varpi}\mathcal{Z}^{\alpha\beta}\partial_{\epsilon n}^{i}\Omega_{\alpha\beta}\varXi^{\lambda\mu\nu}\\
&+2\sum_{a}\varphi_{\varpi\mathscr{J}_{a}\phi}\nabla_{\rho}\alpha_{a}\xi_{\sigma\delta}\myDelta^{\lambda\mu\nu\rho\sigma\delta}+2\varphi_{\varpi h\phi}N_{\rho}\xi_{\sigma\delta}\myDelta^{\lambda\mu\nu\rho\sigma\delta}+2\varphi_{\varpi q\phi}M_{\rho}\xi_{\sigma\delta}\myDelta^{\lambda\mu\nu\rho\sigma\delta}
\end{aligned}
\label{491}
\end{eqnarray}
To derive a relaxation-type equation for $\varpi^{\lambda\mu\nu}$ from Eq.~\eqref{491}, the relation~\eqref{125} is utilized to replace $\varXi_{\lambda\mu\nu}$ with $\varphi^{-1}\varpi^{\lambda\mu\nu}$ in the second term on the right-hand side of Eq.~\eqref{491}. This substitution is valid as the term is already of second order in spacetime gradients. We then have
\begin{equation}
\widetilde{\varphi}\myDelta^{\lambda\mu\nu\rho\sigma\delta}D\varXi_{\rho\sigma\delta}\simeq\widetilde{\varphi}\varphi^{-1}\myDelta^{\lambda\mu\nu\rho\sigma\delta}D\varpi_{\rho\sigma\delta}-\widetilde{\varphi}\varpi^{\lambda\mu\nu}\varphi^{-2}\beta\theta\biggl[\left(\frac{\partial\varphi}{\partial\beta}\Gamma-\sum_{a}\frac{\partial\gamma}{\partial\alpha_{a}}\delta_{a}-2\frac{\partial\gamma}{\partial\Omega_{\alpha\beta}}\mathcal{K}_{\alpha\beta}\right)-2\theta^{-1}\mathcal{Z}^{\alpha\beta}\frac{\partial\varphi}{\partial S^{\alpha\beta}}\biggr],
\label{492}
\end{equation}
where we define
\begin{align}
	\tau_{\varpi}= & -\widetilde{\varphi}\varphi^{-1},\label{493}\\
	\widetilde{\varphi}_{\varpi}= & \tau_{\varpi}\varphi^{-1}\beta\left[\left(\frac{\partial\varphi}{\partial\beta}\Gamma-\sum_{a}\frac{\partial\gamma}{\partial\alpha_{a}}\delta_{a}-2\frac{\partial\gamma}{\partial\Omega_{\alpha\beta}}\mathcal{K}_{\alpha\beta}\right)-2\theta^{-1}\mathcal{Z}^{\alpha\beta}\frac{\partial\varphi}{\partial S^{\alpha\beta}}\right].\label{494}
\end{align}
We finally obtain the following relaxation-type equation for $\varpi^{\lambda\mu\nu}$
\begin{eqnarray}
\begin{aligned}
\varpi^{\lambda\mu\nu}+\tau_{\varpi}\dot{\varpi}^{\lambda\mu\nu}= & \varphi\varXi^{\lambda\mu\nu}+\widetilde{\varphi}_{\varpi}\theta\varpi^{\lambda\mu\nu}+2\varphi_{\varpi p\varpi}\theta\varXi^{\lambda\mu\nu}+2\sum_{a}\varphi_{\varpi S\mathscr{J}_{a}}\mathcal{R}_{\rho\sigma}\nabla_{\delta}\alpha_{a}\myDelta^{\lambda\mu\nu\rho\sigma\delta}\\
 &+2\varphi_{\varpi Sh}\mathcal{R}_{\rho\sigma}N_{\delta}\myDelta^{\lambda\mu\nu\rho\sigma\delta}+2\varphi_{\varpi Sq}\mathcal{R}_{\rho\sigma}M_{\delta}\myDelta^{\lambda\mu\nu\rho\sigma\delta}+2\sum_{i}\varphi_{\varpi\mathfrak{D}_{i}\varpi}\mathcal{Z}^{\alpha\beta}\partial_{\epsilon n}^{i}\Omega_{\alpha\beta}\varXi^{\lambda\mu\nu}\\
 &+2\sum_{a}\varphi_{\varpi\mathscr{J}_{a}\phi}\nabla_{\rho}\alpha_{a}\xi_{\sigma\delta}\myDelta^{\lambda\mu\nu\rho\sigma\delta}+2\varphi_{\varpi h\phi}N_{\rho}\xi_{\sigma\delta}\myDelta^{\lambda\mu\nu\rho\sigma\delta}+2\varphi_{\varpi q\phi}M_{\rho}\xi_{\sigma\delta}\myDelta^{\lambda\mu\nu\rho\sigma\delta},
\end{aligned}
\label{495}
\end{eqnarray}
where we define $\dot{\varpi}^{\lambda\mu\nu}=\myDelta^{\lambda\mu\nu\rho\sigma\delta}D\varpi_{\rho\sigma\delta}$.

\section{Second-order spin hydrodynamics with $\omega\sim\mathcal{O}\left(\partial^{1}\right)$}
\label{section4}

This section discusses the case where the spin chemical potential is a first-order quantity, since the spin chemical potential at global equilibrium can be expressed by the thermal vorticity tensor as $\varpi=-\partial_{[\mu}(u_{\nu]}/T)$. In the previous section, we examined the scenario where both the spin density and the spin chemical potential are zeroth-order quantities. Following exactly the same derivation steps as in the previous section, we adopt the assumption from Ref.~\cite{Tiwari:2024trl} that the gradient ordering of the terms $\mathcal{K}_{\alpha\beta},\mathcal{D}_{\alpha\beta},\mathcal{E}_{\alpha\beta}^{a}$ are $\mathcal{O}(\partial^1)$ and $\mathcal{F}_{\alpha\beta\delta\rho}$ is $\mathcal{O}(\partial^2)$. Thus, the first-order and second-order contributions can be expressed as
\begin{align}
	\hat{C}_{1}\left(x\right)= & -\beta\theta\hat{p}^{*}-\sum_{a}\hat{\mathscr{J}}_{a}^{\sigma}\nabla_{\sigma}\alpha_{a}+\beta\hat{q}^{\mu}M_{\mu}+\beta\hat{\pi}^{\mu\nu}\sigma_{\mu\nu}+\beta\hat{\phi}^{\mu\nu}\xi_{\mu\nu},\label{496}\\
	\hat{C}_{2}\left(x\right)= & \beta\hat{S}^{\alpha\beta}\mathcal{R}_{\alpha\beta}+\beta\sum_{i}\Bigl[\left(\hat{\mathfrak{D}}_{i}\partial_{\epsilon n}^{i}\beta\right)\mathcal{X}+\sum_{a}\left(\hat{\mathfrak{D}_{i}}\partial_{\epsilon n}^{i}\alpha_{a}\right)\mathcal{Y}_{a}+\left(\hat{\mathfrak{D}}_{i}\partial_{\epsilon n}^{i}\Omega_{\mu\nu}\right)\mathcal{Z}^{\mu\nu}\Bigr]+\beta\hat{h}^{\sigma}\mathcal{H}_{\sigma}+\beta\hat{q}^{\mu}\mathcal{Q}_{\mu}+\hat{\varpi}^{\lambda\alpha\beta}\varXi_{\lambda\alpha\beta},\label{497}
\end{align}
where the definitions of all quantities remain consistent with those in the previous section, with the exception of $\mathcal{H}_\sigma$ and $\mathcal{R}_{\alpha\beta}$. The definitions of $\mathcal{H}_\sigma$ and $\mathcal{R}_{\alpha\beta}$ are redefined as follows:
\begin{align}
\mathcal{H}_{\sigma}= & -{w}^{-1}\Bigl(-\frac{1}{2}\beta^{-1}S^{\alpha\beta}\nabla_{\sigma}\Omega_{\alpha\beta}-\nabla_{\sigma}\Pi+\Pi Du_{\sigma}+\Delta_{\sigma\nu}Dh^{\nu}+h^{\mu}\partial_{\mu}u_{\sigma}+h_{\sigma}\theta\nonumber\\
	&+\Delta_{\sigma\nu}\partial_{\mu}\pi^{\mu\nu}+q^{\mu}\partial_{\mu}u_{\sigma}-q_{\sigma}\theta-\Delta_{\sigma\nu}Dq^{\nu}+\Delta_{\sigma\nu}\partial_{\mu}\phi^{\mu\nu}\Bigr),\label{498}\\
	\mathcal{R}_{\alpha\beta}=&\theta\mathcal{K}_{\alpha\beta}+\mathcal{W}_{\alpha\beta}.\label{499}
\end{align}
Next, we will derive the first-order and second-order spin hydrodynamics using the same steps as in the previous section, without elaborating too much on similar processes.

\subsection{First-order spin hydrodynamics}

According to Curie's theorem, we obtain from Eqs.~\eqref{93} and \eqref{496} for the shear-stress tensor to leading order
\begin{eqnarray} \langle\hat{\pi}_{\mu\nu}\left(x\right)\rangle_{1}=\beta\left(x\right)\sigma^{\rho\sigma}\left(x\right)\int d^{4}x_{1}\left(\hat{\pi}_{\mu\nu}\left(x\right),\hat{\pi}_{\rho\sigma}\left(x_{1}\right)\right).
	\label{500}
\end{eqnarray}
The bulk viscous pressure can still be expressed as
\begin{eqnarray}
	\Pi=\langle\hat{p}\rangle-p\left(\epsilon,n_{a},S^{\alpha\beta}\right)=\langle\hat{p}\rangle_{l}+\langle\hat{p}\rangle_{1}-p\left(\epsilon,n_{a},S^{\alpha\beta}\right),
	\label{501}	
\end{eqnarray}
Consequently, to first order in gradients,
\begin{eqnarray}
	\begin{aligned}
		\langle\hat{p}\rangle_{l} & \equiv p\left(\langle\hat{\epsilon}\rangle_{l},\langle\hat{n}_{a}\rangle_{l},\langle\hat{S}^{\alpha\beta}\rangle_{l}\right)\\
		& =p\left(\epsilon-\langle\hat{\epsilon}\rangle_{1},n_{a}-\langle\hat{n}_{a}\rangle_{1},S^{\alpha\beta}-\langle\hat{S}^{\alpha\beta}\rangle_{1}\right)\\
		& =p\left(\epsilon,n_{a},S^{\alpha\beta}\right)-\Gamma\langle\hat{\epsilon}\rangle_{1}-\sum_{a}\delta_{a}\langle\hat{n}_{a}\rangle_{1},
	\end{aligned}
	\label{502}
\end{eqnarray}
where the coefficients $\Gamma$ and $\delta_a$ are defined in Eq.~\eqref{79}. Substituting Eq.~\eqref{502} into Eq.~\eqref{501} for the bulk viscous pressure, we obtain
\begin{eqnarray}
	\Pi =\langle\hat{p}-\Gamma\hat{\epsilon}-\Sigma_{a}\delta_{a}\hat{n}_{a}\rangle_{1}
		=\langle\hat{p}^{*}\rangle_{1},
	\label{503}
\end{eqnarray}
where we have employed the definition of $\hat{p}^*$ given in Eq.~\eqref{81}. Combining Eqs.~\eqref{93} and \eqref{496}, we obtain the first-order correction to the bulk viscous pressure
\begin{eqnarray}
	\Pi\left(x\right)= \langle\hat{p}^{*}\left(x\right)\rangle_{1}=-\beta\left(x\right)\theta\left(x\right)\int d^{4}x_{1}\left(\hat{p}^{*}\left(x\right),\hat{p}^{*}\left(x_{1}\right)\right).
	\label{504}
\end{eqnarray}
Applying Curie’s theorem, we can obtain the remaining dissipative currents,
\begin{eqnarray}
	\begin{aligned}
		\mathscr{J}_{a}^{\mu}\left(x\right)= & \langle\hat{\mathscr{J}}_{a}^{\mu}\left(x\right)\rangle_{1}\\
		= & -\sum_{b}\nabla_{\sigma}\alpha_{b}\left(x\right)\int d^{4}x_{1}\left(\hat{\mathscr{J}}_{a}^{\mu}\left(x\right),\hat{\mathscr{J}}_{b}^{\sigma}\left(x_{1}\right)\right)+\beta\left(x\right)M_{\alpha}\left(x\right)\int d^{4}x_{1}\left(\hat{\mathscr{J}}_{a}^{\mu}\left(x\right),\hat{q}^{\alpha}\left(x_{1}\right)\right),
	\end{aligned}
	\label{505}
\end{eqnarray}
\begin{eqnarray}
	\phi^{\mu\nu}\left(x\right)=\langle\hat{\phi}^{\mu\nu}\left(x\right)\rangle_{1}
		=\beta\left(x\right)\xi_{\rho\sigma}\left(x\right)\int d^{4}x_{1}\left(\hat{\phi}^{\mu\nu}\left(x\right),\hat{\phi}^{\rho\sigma}\left(x_{1}\right)\right),
	\label{506}
\end{eqnarray}
\begin{eqnarray}
	\begin{aligned}
		q^{\mu}\left(x\right)= & \langle\hat{q}^{\mu}\left(x\right)\rangle_{1}\\
		= & -\sum_{a}\nabla_{\sigma}\alpha_{a}\left(x\right)\int d^{4}x_{1}\left(\hat{q}^{\mu}\left(x\right),\hat{\mathscr{J}}_{a}^{\sigma}\left(x_{1}\right)\right)+\beta\left(x\right)M_{\alpha}\left(x\right)\int d^{4}x_{1}\left(\hat{q}^{\mu}\left(x\right),\hat{q}^{\alpha}\left(x_{1}\right)\right).
	\end{aligned}
	\label{507}
\end{eqnarray}
It is important to note that, according to the power counting used in this section, there is no first-order contribution from the dissipative current arising from the spin tensor
\begin{eqnarray}
	\varpi^{\lambda\alpha\beta}\left(x\right)=\langle\hat{\varpi}^{\lambda\alpha\beta}\left(x\right)\rangle_{1}=0.
	\label{508}
\end{eqnarray}
Substituting the two-point correlation functions from Eqs.~\eqref{108}, \eqref{111}, \eqref{113}, \eqref{115}, \eqref{116}, and \eqref{118} into Eqs.~\eqref{500}, \eqref{504}, \eqref{505}-\eqref{507}, we obtain
\begin{align}
\pi_{\mu\nu} & =2\eta\sigma_{\mu\nu},\label{509}\\
\Pi & =-\zeta\theta,\label{510}\\
\phi^{\mu\nu} & =2\gamma\xi^{\mu\nu},\label{511}\\
\mathscr{J}_{a}^{\mu}& =\sum_{b}\chi_{ab}\nabla^{\mu}\alpha_{b}+\chi_{\mathscr{J}_{a}q}M^{\mu},\label{512}\\
q^{\mu} & =-\lambda M^{\mu}+\sum_{a}\lambda_{q\mathscr{J}_{a}}\nabla^{\mu}\alpha_{a},\label{513}
\end{align}
where the first-order transport coefficients defined here are consistent with those in the previous section.

\subsection{Second-order spin hydrodynamics}

This subsection derives all second-order correction contributions to the dissipative currents from three sources using the exact same steps as in the previous section.

\subsubsection{Second-order corrections to the shear stress tensor}

The nonlocal corrections from the two-point correlation function for the shear stress tensor can be obtained through the similar derivation steps as in Eq.~\eqref{166}
\begin{eqnarray}
	\langle\hat{\pi}_{\mu\nu}\rangle_{2}^{1}=2\widetilde{\eta}\Delta_{\mu\nu\rho\sigma}D\sigma^{\rho\sigma}+2\widetilde{\eta}\theta\Gamma\sigma_{\mu\nu},
	\label{514}
\end{eqnarray}
where we have used the definition of $\widetilde{\eta}$ given in Eq.~\eqref{167}.

In the case of the power counting used in this section, since the operator $\hat{C}_{2}$ does not contain symmetric second-rank operators, it can be concluded from Curie's theorem that the second-order correction to the shear stress tensor from the extended thermodynamic forces also vanish
\begin{eqnarray}
	\left\langle \hat{\pi}_{\mu\nu}\left(x\right)\right\rangle _{2}^{2}=\int d^{4}x_{1}\left(\hat{\pi}_{\mu\nu}\left(x\right),\hat{C}_{2}\left(x_{1}\right)\right)=0.
	\label{515}
\end{eqnarray}
Through tedious algebraic calculations similar to those in Eq.~\eqref{192}, the corrections from the three-point correlation function for the shear stress tensor can be obtained as
\begin{eqnarray}
\begin{aligned}
\left\langle \hat{\pi}_{\mu\nu}\right\rangle _{2}^{3}=&2\eta_{\pi p\pi}\theta\sigma_{\mu\nu}+\sum_{ab}\eta_{\pi\mathscr{J}_{a}\mathscr{J}_{b}}\nabla_{\langle\mu}\alpha_{a}\nabla_{\nu\rangle}\alpha_{b}+2\sum_{a}\eta_{\pi\mathscr{J}_{a}q}\nabla_{\langle\mu}\alpha_{a}M_{\nu\rangle}\\
&+\eta_{\pi qq}M_{\langle\mu}M_{\nu\rangle}+\eta_{\pi\pi\pi}\sigma_{\alpha\langle\mu}\sigma_{\nu\rangle}^{\,\,\,\,\alpha}+2\eta_{\pi\pi\phi}\sigma_{\alpha\langle\mu}\xi_{\nu\rangle}^{\,\,\,\,\alpha}+\eta_{\pi\phi\phi}\xi_{\alpha\langle\mu}\xi_{\nu\rangle}^{\,\,\,\,\alpha},
\end{aligned}
\label{516}
\end{eqnarray}
where
\begin{align}
\eta_{\pi p\pi} & =-\frac{1}{5}\beta^{2}\int d^{4}x_{1}d^{4}x_{2}\left(\hat{\pi}_{\gamma\delta}(x),\hat{p}^{*}\left(x_{1}\right),\hat{\pi}^{\gamma\delta}\left(x_{2}\right)\right),\eta_{\pi\mathscr{J}_{a}\mathscr{J}_{b}}=\frac{1}{5}\int d^{4}x_{1}d^{4}x_{2}\left(\hat{\pi}_{\gamma\delta}\left(x\right),\hat{\mathscr{J}}_{a}^{\gamma}\left(x_{1}\right),\hat{\mathscr{J}}_{b}^{\delta}\left(x_{2}\right)\right),\label{517}\\
\eta_{\pi\mathscr{J}_{a}q} & =-\frac{1}{5}\beta\int d^{4}x_{1}d^{4}x_{2}\left(\hat{\pi}_{\gamma\delta}\left(x\right),\hat{\mathscr{J}}_{a}^{\gamma}\left(x_{1}\right),\hat{q}^{\delta}\left(x_{2}\right)\right),\eta_{\pi qq}=\frac{1}{5}\beta^{2}\int d^{4}x_{1}d^{4}x_{2}\left(\hat{\pi}_{\gamma\delta}\left(x\right),\hat{q}^{\gamma}\left(x_{1}\right),\hat{q}^{\delta}\left(x_{2}\right)\right),\label{518}\\
\eta_{\pi\pi\pi} & =\frac{12}{35}\beta^{2}\int d^{4}x_{1}d^{4}x_{2}\left(\hat{\pi}_{\gamma}^{\,\,\,\delta}\left(x\right),\hat{\pi}_{\delta}^{\,\,\,\lambda}\left(x_{1}\right),\hat{\pi}_{\lambda}^{\,\,\,\gamma}\left(x_{2}\right)\right),\eta_{\pi\pi\phi}=-\frac{4}{15}\beta^{2}\int d^{4}x_{1}d^{4}x_{2}\left(\hat{\pi}_{\gamma}^{\,\,\,\delta}\left(x\right),\hat{\pi}_{\delta}^{\,\,\,\lambda}\left(x_{1}\right),\hat{\phi}_{\lambda}^{\,\,\,\gamma}\left(x_{2}\right)\right),\label{519}\\
\eta_{\pi\phi\phi} & =\frac{4}{5}\beta^{2}\int d^{4}x_{1}d^{4}x_{2}\left(\hat{\pi}_{\gamma}^{\,\,\,\delta}\left(x\right),\hat{\phi}_{\delta}^{\,\,\,\lambda}\left(x_{1}\right),\hat{\phi}_{\lambda}^{\,\,\,\gamma}\left(x_{2}\right)\right).\label{520}
\end{align}
The complete second-order expression for the shear stress tensor is
\begin{eqnarray}
\begin{aligned}
\pi_{\mu\nu}=&2\eta\sigma_{\mu\nu}+2\widetilde{\eta}\left(\Delta_{\mu\nu\rho\sigma}D\sigma^{\rho\sigma}+\theta\Gamma\sigma_{\mu\nu}\right)\\
&+2\eta_{\pi p\pi}\theta\sigma_{\mu\nu}+\sum_{ab}\eta_{\pi\mathscr{J}_{a}\mathscr{J}_{b}}\nabla_{\langle\mu}\alpha_{a}\nabla_{\nu\rangle}\alpha_{b}+2\sum_{a}\eta_{\pi\mathscr{J}_{a}q}\nabla_{\langle\mu}\alpha_{a}M_{\nu\rangle}\\
&+\eta_{\pi qq}M_{\langle\mu}M_{\nu\rangle}+\eta_{\pi\pi\pi}\sigma_{\alpha\langle\mu}\sigma_{\nu\rangle}^{\,\,\,\,\alpha}+2\eta_{\pi\pi\phi}\sigma_{\alpha\langle\mu}\xi_{\nu\rangle}^{\,\,\,\,\alpha}+\eta_{\pi\phi\phi}\xi_{\alpha\langle\mu}\xi_{\nu\rangle}^{\,\,\,\,\alpha}.
\end{aligned}
\label{521}
\end{eqnarray}
We use the first-order relation to replace $2\sigma^{\rho\sigma}\to\eta^{-1}\pi^{\rho\sigma}$ in the second term of the right-hand side of Eq.~\eqref{521} to derive the relaxation equation for $\pi_{\mu\nu}$. We then have
\begin{eqnarray}
2\widetilde{\eta}\Delta_{\mu\nu\rho\sigma}D\sigma^{\rho\sigma}\simeq\widetilde{\eta}\eta^{-1}\Delta_{\mu\nu\rho\sigma}D\pi^{\rho\sigma}-\widetilde{\eta}\eta^{-2}\beta\left(\frac{\partial\eta}{\partial\beta}\Gamma-\sum_{a}\frac{\partial\eta}{\partial\alpha_{a}}\delta_{a}\right)\theta\pi_{\mu\nu}.
\label{522}
\end{eqnarray}
Combining Eqs.~\eqref{521} and \eqref{522} and introducing the coefficients
\begin{align}
&\tau_{\pi}=-\widetilde{\eta}\eta^{-1},\label{523}\\
&\widetilde{\eta}_{\pi}=\tau_{\pi}\eta^{-1}\beta\left(\frac{\partial\eta}{\partial\beta}\Gamma-\sum_{a}\frac{\partial\eta}{\partial\alpha_{a}}\delta_{a}\right),\label{524}
\end{align}
we finally obtain the following relaxation equation for the shear-stress tensor,
\begin{eqnarray}
\begin{aligned}
\tau_{\pi}\dot{\pi}_{\mu\nu}+\pi_{\mu\nu}=&2\eta\sigma_{\mu\nu}+\widetilde{\eta}_{\pi}\theta\pi_{\mu\nu}+2\widetilde{\eta}\theta\Gamma\sigma_{\mu\nu}\\
&+2\eta_{\pi p\pi}\theta\sigma_{\mu\nu}+\sum_{ab}\eta_{\pi\mathscr{J}_{a}\mathscr{J}_{b}}\nabla_{\langle\mu}\alpha_{a}\nabla_{\nu\rangle}\alpha_{b}+2\sum_{a}\eta_{\pi\mathscr{J}_{a}q}\nabla_{\langle\mu}\alpha_{a}M_{\nu\rangle}\\
&+\eta_{\pi qq}M_{\langle\mu}M_{\nu\rangle}+\eta_{\pi\pi\pi}\sigma_{\alpha\langle\mu}\sigma_{\nu\rangle}^{\,\,\,\,\alpha}+2\eta_{\pi\pi\phi}\sigma_{\alpha\langle\mu}\xi_{\nu\rangle}^{\,\,\,\,\alpha}+\eta_{\pi\phi\phi}\xi_{\alpha\langle\mu}\xi_{\nu\rangle}^{\,\,\,\,\alpha}.
\end{aligned}
\label{525}
\end{eqnarray}

\subsubsection{Second-order corrections to the bulk viscous pressure}

By following the same derivation steps as in Eq.~\eqref{201}, we can obtain the bulk viscous pressure up to second order
\begin{eqnarray}
\Pi=\langle\hat{p}^{*}\rangle_{1}+\langle\hat{p}^{*}\rangle_{2}+\frac{1}{2}\frac{\partial^{2}p}{\partial\epsilon^{2}}\langle\hat{\epsilon}\rangle_{1}^{2}+\frac{1}{2}\sum_{ab}\frac{\partial^{2}p}{\partial n_{a}\partial n_{b}}\langle\hat{n}_{a}\rangle_{1}\langle\hat{n}_{b}\rangle_{1}+\sum_{a}\frac{\partial^{2}p}{\partial\epsilon\partial n_{a}}\langle\hat{\epsilon}\rangle_{1}\langle\hat{n}_{a}\rangle_{1}-\mathcal{K}_{\alpha\beta}\langle\hat{S}^{\alpha\beta}\rangle_{1}.
\label{526}
\end{eqnarray}
where the first-order correction to energy density, particle densities, and spin density are given by
\begin{align}
\langle\hat{\epsilon}\rangle_{1} & =-\zeta_{\epsilon p}\theta,\label{527}\\
\langle\hat{n}_{a}\rangle_{1} & =-\zeta_{n_{a}p}\theta,\label{528}\\
\langle\hat{S}^{\mu\nu}\rangle_{1} & =-\zeta_{S\phi}\xi^{\mu\nu}.\label{529}
\end{align}
Here, we have used the definitions of the first-order transport coefficients given in Eqs.~\eqref{130},~\eqref{202}, and~\eqref{204}. Using Eqs.~\eqref{527}-\eqref{529}, we can rewrite Eq.~\eqref{526} as
\begin{eqnarray}
\Pi=-\zeta\theta+\langle\hat{p}^{*}\rangle_{2}+\frac{1}{2}\frac{\partial^{2}p}{\partial\epsilon^{2}}\zeta_{\epsilon p}^{2}\theta^{2}+\frac{1}{2}\sum_{ab}\frac{\partial^{2}p}{\partial n_{a}\partial n_{b}}\zeta_{n_{a}p}\zeta_{n_{b}p}\theta^{2}+\sum_{a}\frac{\partial^{2}p}{\partial\epsilon\partial n_{a}}\zeta_{\epsilon p}\zeta_{n_{a}p}\theta^{2}+\mathcal{K}_{\alpha\beta}\zeta_{S\phi}\xi^{\alpha\beta}.
\label{530}
\end{eqnarray}
The nonlocal corrections from the two-point correlation function for the bulk viscous pressure can be obtained through the similar derivation steps as in Eq.~\eqref{246}
\begin{eqnarray}
\langle\hat{p}^{*}\rangle_{2}^{1}=-\widetilde{\zeta}_{p\epsilon}\theta^{2}\widetilde{\Gamma}-\sum_{a}\widetilde{\zeta}_{pn_{a}}\theta^{2}\widetilde{\delta}_{a}-\widetilde{\zeta}\theta^{2}\Gamma-\widetilde{\zeta}D\theta,
\label{531}
\end{eqnarray}
where we have used the definition of $\widetilde{\zeta}$,$\widetilde{\zeta}_{p\epsilon}$,$\widetilde{\zeta}_{pn_{a}}$,$\widetilde{\Gamma}$ and $\widetilde{\delta}_{a}$ given in Eqs.~\eqref{230}, \eqref{231}, \eqref{232}, \eqref{242} and \eqref{243}, respectively.

Substituting Eq.~\eqref{497} into Eq.~\eqref{96} and applying Curie's theorem, we obtain corrections from extended thermodynamic forces to bulk viscous pressure
\begin{eqnarray}
\langle\hat{p}^{*}\rangle_{2}^{2}=&\sum_{i}\zeta_{p\mathfrak{D}_{i}}\left[\left(\partial_{\epsilon n}^{i}\beta\right)\mathcal{X}+\sum_{a}\left(\partial_{\epsilon n}^{i}\alpha_{a}\right)\mathcal{Y}_{a}+\left(\partial_{\epsilon n}^{i}\Omega_{\alpha\beta}\right)\mathcal{Z}^{\alpha\beta}\right],
\label{532}
\end{eqnarray}
where we have used the definition of $\zeta_{p\mathfrak{D}_{i}}$ given in Eq.~\eqref{128}.

Through tedious algebraic calculations similar to those in Eq.~\eqref{279}, the corrections from the three-point correlation function for the bulk viscous pressure can be obtained as
\begin{eqnarray}
\langle\hat{p}^{*}\rangle_{2}^{3}=\zeta_{ppp}\theta^{2}+\sum_{ab}\zeta_{p\mathscr{J}_{a}\mathscr{J}_{b}}\nabla_{\alpha}\alpha_{a}\nabla^{\alpha}\alpha_{b}+2\sum_{a}\zeta_{p\mathscr{J}_{a}q}\nabla^{\sigma}\alpha_{a}M_{\sigma}+\zeta_{pqq}M^{\sigma}M_{\sigma}+\zeta_{p\pi\pi}\sigma^{\rho\sigma}\sigma_{\rho\sigma}+\zeta_{p\phi\phi}\xi^{\rho\sigma}\xi_{\rho\sigma},
\label{533}
\end{eqnarray}
where we define
\begin{align}
\zeta_{ppp} & =\beta^{2}\int d^{4}x_{1}d^{4}x_{2}\left(\hat{p}^{*}\left(x\right),\hat{p}^{*}\left(x_{1}\right),\hat{p}^{*}\left(x_{2}\right)\right),\zeta_{p\mathscr{J}_{a}\mathscr{J}_{b}}=\frac{1}{3}\int d^{4}x_{1}d^{4}x_{2}\left(\hat{p}^{*}\left(x\right),\hat{\mathscr{J}}_{a\gamma}\left(x_{1}\right),\hat{\mathscr{J}}_{b}^{\gamma}\left(x_{2}\right)\right),\label{534}\\
\zeta_{p\mathscr{J}_{a}q} & =-\frac{1}{3}\beta\int d^{4}x_{1}d^{4}x_{2}\left(\hat{p}^{*}\left(x\right),\hat{\mathscr{J}}_{a\gamma}\left(x_{1}\right),\hat{q}^{\gamma}\left(x_{2}\right)\right),\zeta_{pqq}=\frac{1}{3}\beta^{2}\int d^{4}x_{1}d^{4}x_{2}\left(\hat{p}^{*}(x),\hat{q}_{\gamma}(x_{1}),\hat{q}^{\gamma}(x_{2})\right),\label{535}\\
\zeta_{p\pi\pi} & =\frac{1}{5}\beta^{2}\int d^{4}x_{1}d^{4}x_{2}\left(\hat{p}^{*}\left(x\right),\hat{\pi}_{\gamma\delta}\left(x_{1}\right),\hat{\pi}^{\gamma\delta}\left(x_{2}\right)\right),\zeta_{p\phi\phi}=\frac{1}{3}\beta^{2}\int d^{4}x_{1}d^{4}x_{2}\left(\hat{p}^{*}\left(x\right),\hat{\phi}_{\gamma\delta}\left(x_{1}\right),\hat{\phi}^{\gamma\delta}\left(x_{2}\right)\right).\label{536}
\end{align}
The complete second-order expression for the bulk viscous pressure is
\begin{eqnarray}
\begin{aligned}
	\Pi= & -\zeta\theta+\frac{1}{2}\frac{\partial^{2}p}{\partial\epsilon^{2}}\zeta_{\epsilon p}^{2}\theta^{2}+\frac{1}{2}\sum_{ab}\frac{\partial^{2}p}{\partial n_{a}\partial n_{b}}\zeta_{n_{a}p}\zeta_{n_{b}p}\theta^{2}+\sum_{a}\frac{\partial^{2}p}{\partial\epsilon\partial n_{a}}\zeta_{\epsilon p}\zeta_{n_{a}p}\theta^{2}+\mathcal{K}_{\alpha\beta}\zeta_{S\phi}\xi^{\alpha\beta}-\widetilde{\zeta}_{p\epsilon}\theta^{2}\widetilde{\Gamma}\\
	&-\sum_{a}\widetilde{\zeta}_{pn_{a}}\theta^{2}\widetilde{\delta}_{a}-\widetilde{\zeta}\theta^{2}\Gamma-\widetilde{\zeta}D\theta+\sum_{i}\zeta_{p\mathfrak{D}_{i}}\Bigl[\left(\partial_{\epsilon n}^{i}\beta\right)\mathcal{X}+\sum_{a}\left(\partial_{\epsilon n}^{i}\alpha_{a}\right)\mathcal{Y}_{a}+\left(\partial_{\epsilon n}^{i}\Omega_{\alpha\beta}\right)\mathcal{Z}^{\alpha\beta}\Bigr]+\zeta_{ppp}\theta^{2}\\
	&+\sum_{ab}\zeta_{p\mathscr{J}_{a}\mathscr{J}_{b}}\nabla_{\alpha}\alpha_{a}\nabla^{\alpha}\alpha_{b}+2\sum_{a}\zeta_{p\mathscr{J}_{a}q}\nabla^{\sigma}\alpha_{a}M_{\sigma}+\zeta_{pqq}M^{\sigma}M_{\sigma}+\zeta_{p\pi\pi}\sigma^{\rho\sigma}\sigma_{\rho\sigma}+\zeta_{p\phi\phi}\xi^{\rho\sigma}\xi_{\rho\sigma}.
\end{aligned}
\label{537}
\end{eqnarray}
To derive a relaxation-type equation for the bulk viscous pressure, we approximate $\theta\to-\zeta^{-1}\Pi$ in the term $\widetilde{\zeta}D\theta$. Thus, we have
\begin{eqnarray} -\widetilde{\zeta}D\theta\simeq\widetilde{\zeta}\zeta^{-1}D\Pi-\widetilde{\zeta}\zeta^{-2}\beta\left(\frac{\partial\zeta}{\partial\beta}\Gamma-\sum_{a}\frac{\partial\zeta}{\partial\alpha_{a}}\delta_{a}\right)\theta\Pi.
\label{538}
\end{eqnarray}
Combining Eqs.~\eqref{537} and \eqref{538} and introducing the coefficients
\begin{align}
	\dot{\Pi}=&D\Pi,\label{539}\\
	\tau_{\Pi}=&-\widetilde{\zeta}\zeta^{-1},\label{540}\\
	\widetilde{\zeta}_{\Pi}	=&\tau_{\Pi}\zeta^{-1}\beta\left(\frac{\partial\zeta}{\partial\beta}\Gamma-\sum_{a}\frac{\partial\zeta}{\partial\alpha_{a}}\delta_{a}\right),\label{541}
\end{align}
we finally obtain the following relaxation equation for the bulk viscous pressure,
\begin{eqnarray}
\begin{aligned}
	\Pi+\tau_{\Pi}\dot{\Pi}= & -\zeta\theta+\widetilde{\zeta}_{\Pi}\theta\Pi+\Bigl[\frac{1}{2}\frac{\partial^{2}p}{\partial\epsilon^{2}}\zeta_{\epsilon p}^{2}+\frac{1}{2}\sum_{ab}\frac{\partial^{2}p}{\partial n_{a}\partial n_{b}}\zeta_{n_{a}p}\zeta_{n_{b}p}+\sum_{a}\frac{\partial^{2}p}{\partial\epsilon\partial n_{a}}\zeta_{\epsilon p}\zeta_{n_{a}p}\Bigr]\theta^{2}+\zeta_{S\phi}\mathcal{K}_{\alpha\beta}\xi^{\alpha\beta}\\
	&-\Bigl[\Gamma\widetilde{\zeta}+\widetilde{\Gamma}\widetilde{\zeta}_{p\epsilon}+\sum_{a}\widetilde{\zeta}_{pn_{a}}\widetilde{\delta}_{a}\Bigr]\theta^{2}+\sum_{i}\zeta_{p\mathfrak{D}_{i}}\Bigl[\left(\partial_{\epsilon n}^{i}\beta\right)\mathcal{X}+\sum_{a}\left(\partial_{\epsilon n}^{i}\alpha_{a}\right)\mathcal{Y}_{a}+\left(\partial_{\epsilon n}^{i}\Omega_{\alpha\beta}\right)\mathcal{Z}^{\alpha\beta}\Bigr]+\zeta_{ppp}\theta^{2}\\
	&+\sum_{ab}\zeta_{p\mathscr{J}_{a}\mathscr{J}_{b}}\nabla_{\alpha}\alpha_{a}\nabla^{\alpha}\alpha_{b}+2\sum_{a}\zeta_{p\mathscr{J}_{a}q}\nabla^{\sigma}\alpha_{a}M_{\sigma}+\zeta_{pqq}M^{\sigma}M_{\sigma}+\zeta_{p\pi\pi}\sigma^{\rho\sigma}\sigma_{\rho\sigma}+\zeta_{p\phi\phi}\xi^{\rho\sigma}\xi_{\rho\sigma}.
\end{aligned}
\label{542}
\end{eqnarray}

\subsubsection{Second-order corrections to the charge-diffusion currents}

The nonlocal corrections from the two-point correlation function for the charge-diffusion currents can be obtained through
the similar derivation steps as in Eq.~\eqref{329}
\begin{eqnarray}
\langle\hat{\mathscr{J}}_{c\mu}\rangle_{2}^{1}=\widetilde{\chi}_{\mathscr{J}_{c}h}\sum_{a}n_{a}{w}^{-2}\left(\Gamma {w}+\sum_{c}\delta_{c}n_{c}\right)\theta\nabla_{\mu}\alpha_{a}+\sum_{a}\widetilde{\chi}_{ca}\Delta_{\mu\beta}D\left(\nabla^{\beta}\alpha_{a}\right)+\widetilde{\chi}_{\mathscr{J}_{c}q}\theta\Gamma M_{\mu}+\widetilde{\chi}_{\mathscr{J}_{c}q}\Delta_{\mu\beta}DM^{\beta},
\label{543}
\end{eqnarray}
where we have used the definition of $\widetilde{\chi}_{\mathscr{Jc}_{c}h},\widetilde{\chi}_{ca}$, and $\widetilde{\chi}_{\mathscr{J}_{c}q}$ given in Eqs.~\eqref{322}-\eqref{324}.

Substituting Eq.\eqref{497} into Eq.~\eqref{96} and applying Curie's theorem, we obtain corrections from extended thermodynamic forces to charge-diffusion currents
\begin{eqnarray}
\langle\hat{\mathscr{J}}_{c\mu}(x)\rangle_{2}^{2}=\chi_{\mathscr{J}_{c}h}\mathcal{H}_{\mu}+\chi_{\mathscr{J}_{c}q}\mathcal{Q}_{\mu},
\label{544}
\end{eqnarray}
where we have used the definition of $\chi_{\mathscr{J}_{c}h}$ and $\chi_{\mathscr{J}_{c}q}$ given in Eqs.~\eqref{134} and \eqref{135}.

Through tedious algebraic calculations similar to those in Eq.~\eqref{358}, the corrections from the three-point correlation
function for the charge-diffusion currents can be obtained as
\begin{eqnarray}
\begin{aligned}
\langle\hat{\mathscr{J}}_{c\mu}\rangle_{2}^{3}= & 2\sum_{a}\chi_{\mathscr{J}_{c}p\mathscr{J}_{a}}\theta\nabla_{\mu}\alpha_{a}+2\chi_{\mathscr{J}_{c}pq}\theta M_{\mu}+2\sum_{a}\chi_{\mathscr{J}_{c}\mathscr{J}_{a}\pi}\nabla^{\nu}\alpha_{a}\sigma_{\mu\nu}\\
&+2\sum_{a}\chi_{\mathscr{J}_{c}\mathscr{J}_{a}\phi}\nabla^{\nu}\alpha_{a}\xi_{\mu\nu}+2\chi_{\mathscr{J}_{c}q\pi}M^{\nu}\sigma_{\mu\nu}+2\chi_{\mathscr{J}_{c}q\phi}M^{\nu}\xi_{\mu\nu},
\end{aligned}
\label{545}
\end{eqnarray}
where we define
\begin{align}
\chi_{\mathscr{J}_{c}p\mathscr{J}_{a}} & =\frac{1}{3}\beta\int d^{4}x_{1}d^{4}x_{2}\left(\hat{\mathscr{J}}_{c\beta}\left(x\right),\hat{p}^{*}\left(x_{1}\right),\hat{\mathscr{J}}_{a}^{\beta}\left(x_{2}\right)\right),\chi_{\mathscr{J}_{c}pq}=-\frac{1}{3}\beta^{2}\int d^{4}x_{1}d^{4}x_{2}\left(\hat{\mathscr{J}}_{c\beta}\left(x\right),\hat{p}^{*}\left(x_{1}\right),\hat{q}^{\beta}\left(x_{2}\right)\right),\label{546}\\
\chi_{\mathscr{J}_{c}\mathscr{J}_{a}\pi} & =-\frac{1}{5}\beta\int d^{4}x_{1}d^{4}x_{2}\left(\hat{\mathscr{J}}_{c\lambda}\left(x\right),\hat{\mathscr{J}}_{a\delta}\left(x_{1}\right),\hat{\pi}^{\lambda\delta}\left(x_{2}\right)\right),\chi_{\mathscr{J}_{c}\mathscr{J}_{a}\phi}=-\frac{1}{3}\beta\int d^{4}x_{1}d^{4}x_{2}\left(\hat{\mathscr{J}}_{c\lambda}\left(x\right),\hat{\mathscr{J}}_{a\delta}\left(x_{1}\right),\hat{\phi}^{\lambda\delta}\left(x_{2}\right)\right),\label{547}\\
\chi_{\mathscr{J}_{c}q\pi} & =\frac{1}{5}\beta^{2}\int d^{4}x_{1}d^{4}x_{2}\left(\hat{\mathscr{J}}_{c\lambda}\left(x\right),\hat{q}_{\delta}\left(x_{1}\right),\hat{\pi}^{\lambda\delta}\left(x_{2}\right)\right),\chi_{\mathscr{J}_{c}q\phi}=\frac{1}{3}\beta^{2}\int d^{4}x_{1}d^{4}x_{2}\left(\hat{\mathscr{J}}_{c\lambda}\left(x\right),\hat{q}_{\delta}\left(x_{1}\right),\hat{\phi}^{\lambda\delta}\left(x_{2}\right)\right).\label{548}
\end{align}
The complete second-order expression for the charge-diffusion currents are
\begin{eqnarray}
\begin{aligned}\mathscr{J}_{c\mu}= & \sum_{b}\chi_{cb}\nabla_{\mu}\alpha_{b}+\chi_{\mathscr{J}_{c}q}M_{\mu}+\widetilde{\chi}_{\mathscr{J}_{c}h}\sum_{a}n_{a}{w}^{-2}\left(\Gamma {w}+\sum_{a}\delta_{a}n_{a}\right)\theta\nabla_{\mu}\alpha_{a}+\sum_{a}\widetilde{\chi}_{ca}\Delta_{\mu\beta}D\left(\nabla^{\beta}\alpha_{a}\right)\\
	& +\widetilde{\chi}_{\mathscr{J}_{c}q}\theta\Gamma M_{\mu}+\widetilde{\chi}_{\mathscr{J}_{c}q}\Delta_{\mu\beta}DM^{\beta}+\chi_{\mathscr{J}_{c}h}\mathcal{H}_{\mu}+\chi_{\mathscr{J}_{c}q}\mathcal{Q}_{\mu}+2\sum_{a}\chi_{\mathscr{J}_{c}p\mathscr{J}_{a}}\theta\nabla_{\mu}\alpha_{a}+2\chi_{\mathscr{J}_{c}pq}\theta M_{\mu}\\
	& +2\sum_{a}\chi_{\mathscr{J}_{c}\mathscr{J}_{a}\pi}\nabla^{\nu}\alpha_{a}\sigma_{\mu\nu}+2\sum_{a}\chi_{\mathscr{J}_{c}\mathscr{J}_{a}\phi}\nabla^{\nu}\alpha_{a}\xi_{\mu\nu}+2\chi_{\mathscr{J}_{c}q\pi}M^{\nu}\sigma_{\mu\nu}+2\chi_{\mathscr{J}_{c}q\phi}M^{\nu}\xi_{\mu\nu},
\end{aligned}
\label{549}
\end{eqnarray}
To derive a relaxation-type equation for the charge-diffusion currents, we use the first-order approximation 
\begin{eqnarray}
	\nabla^{\beta}\alpha_{a}=\sum_{b}\left(\chi^{-1}\right)_{ab}\left(\mathscr{J}_{b}^{\beta}-\chi_{\mathscr{J}_{b}q}M^{\beta}\right),
	\label{550}
\end{eqnarray}
in the term $\sum_{a}\widetilde{\chi}_{ca}\Delta_{\mu\beta}D\left(\nabla^{\beta}\alpha_{a}\right)$ on the right-hand side of Eq.~\eqref{549}. We then have
\begin{eqnarray}
	\begin{aligned}
		\sum_{a}\widetilde{\chi}_{ca}\Delta_{\mu\beta}D\left(\nabla^{\beta}\alpha_{a}\right)\simeq & \sum_{b}\left(\widetilde{\chi}\chi^{-1}\right)_{cb}\Delta_{\mu\beta}D\mathscr{J}_{b}^{\beta}-\sum_{b}\left(\widetilde{\chi}\chi^{-1}\right)_{cb}M_{\mu}\beta\theta\left(\frac{\partial\chi_{\mathscr{J}_{b}q}}{\partial\beta}\Gamma-\sum_{d}\frac{\partial\chi_{\mathscr{J}_{b}q}}{\partial\alpha_{d}}\delta_{d}\right)\\
		&-\sum_{b}\left(\widetilde{\chi}\chi^{-1}\right)_{cb}\Delta_{\mu\beta}\chi_{\mathscr{J}_{b}q}DM^{\beta}+\sum_{ab}\widetilde{\chi}_{ca}\mathscr{J}_{b\mu}\beta\theta\left(\frac{\partial\left(\chi^{-1}\right)_{ab}}{\partial\beta}\Gamma-\sum_{d}\frac{\partial\left(\chi^{-1}\right)_{ab}}{\partial\alpha_{d}}\delta_{d}\right)\\
		&-\sum_{ab}\widetilde{\chi}_{ca}\chi_{\mathscr{J}_{b}q}M_{\mu}\beta\theta\left(\frac{\partial\left(\chi^{-1}\right)_{ab}}{\partial\beta}\Gamma-\sum_{d}\frac{\partial\left(\chi^{-1}\right)_{ab}}{\partial\alpha_{d}}\delta_{d}\right).
	\end{aligned}
	\label{551}
\end{eqnarray}
Combining Eqs.~\eqref{549} and~\eqref{551} and introducing the coefficients
\begin{align}
	\tau_{\mathscr{J}}^{cb}= & -\left(\widetilde{\chi}\chi^{-1}\right)_{cb}=-\sum_{a}\widetilde{\chi}_{ca}\left(\chi^{-1}\right)_{ab},\label{552}\\
	\widetilde{\chi}_{\mathscr{J}}^{cb}= & \beta\sum_{a}\widetilde{\chi}_{ca}\left(\Gamma\frac{\partial\left(\chi^{-1}\right)_{ab}}{\partial\beta}-\sum_{d}\delta_{d}\frac{\partial\left(\chi^{-1}\right)_{ab}}{\partial\alpha_{d}}\right),\label{553}\\
	\overline{\chi}^{c}= & \beta\sum_{b}\tau_{\mathscr{J}}^{cb}\left(\Gamma\frac{\partial\chi_{\mathscr{J}_{b}q}}{\partial\beta}-\sum_{d}\delta_{d}\frac{\partial\chi_{\mathscr{J}_{b}q}}{\partial\alpha_{d}}\right),\label{554}
\end{align}
we finally obtain the following relaxation equation for the charge-diffusion currents,
\begin{eqnarray}
\begin{aligned}
\mathscr{J}_{c\mu}+\sum_{b}\tau_{\mathscr{J}}^{cb}\dot{\mathscr{J}}_{b\mu}= & \sum_{b}\chi_{cb}\nabla_{\mu}\alpha_{b}+\chi_{\mathscr{J}_{c}q}M_{\mu}+\widetilde{\chi}_{\mathscr{J}_{c}h}\sum_{a}n_{a}{w}^{-2}\left(\Gamma {w}+\sum_{a}\delta_{a}n_{a}\right)\theta\nabla_{\mu}\alpha_{a}+\overline{\chi}^{c}\theta M_{\mu}\\
&+\sum_{b}\tau_{\mathscr{J}}^{cb}\chi_{\mathscr{J}_{b}q}\Delta_{\mu\beta}DM^{\beta}+\sum_{b}\widetilde{\chi}_{\mathscr{J}}^{cb}\theta\mathscr{J}_{b\mu}-\sum_{b}\widetilde{\chi}_{\mathscr{J}}^{cb}\chi_{\mathscr{J}_{b}q}\theta M_{\mu}+\widetilde{\chi}_{\mathscr{J}_{c}q}\theta\Gamma M_{\mu}\\
&+\widetilde{\chi}_{\mathscr{J}_{c}q}\Delta_{\mu\beta}DM^{\beta}+\chi_{\mathscr{J}_{c}h}\mathcal{H}_{\mu}+\chi_{\mathscr{J}_{c}q}\mathcal{Q}_{\mu}+2\sum_{a}\chi_{\mathscr{J}_{c}p\mathscr{J}_{a}}\theta\nabla_{\mu}\alpha_{a}+2\chi_{\mathscr{J}_{c}pq}\theta M_{\mu}\\
&+2\sum_{a}\chi_{\mathscr{J}_{c}\mathscr{J}_{a}\pi}\nabla^{\nu}\alpha_{a}\sigma_{\mu\nu}+2\sum_{a}\chi_{\mathscr{J}_{c}\mathscr{J}_{a}\phi}\nabla^{\nu}\alpha_{a}\xi_{\mu\nu}+2\chi_{\mathscr{J}_{c}q\pi}M^{\nu}\sigma_{\mu\nu}+2\chi_{\mathscr{J}_{c}q\phi}M^{\nu}\xi_{\mu\nu}.
\end{aligned}
\label{555}
\end{eqnarray}

\subsubsection{Second-order corrections to the rotational stress tensor}

The nonlocal corrections from the two-point correlation function for the rotational stress tensor can be obtained through the similar derivation steps as in Eq.~\eqref{377}
\begin{eqnarray}
\langle\hat{\phi}_{\mu\nu}\rangle_{2}^{1}=2\widetilde{\gamma}\mydelta_{\mu\nu\rho\sigma}D\xi^{\rho\sigma}+2\widetilde{\gamma}\theta\Gamma\xi_{\mu\nu},
\label{556}
\end{eqnarray}
where the definition of $\widetilde{\gamma}$ is the same as that in Eq.~\eqref{374}.

Substituting Eq.~\eqref{497} into Eq.~\eqref{96} and applying Curie's theorem, we obtain corrections from extended thermodynamic forces to the rotational stress tensor
\begin{eqnarray}
\langle\hat{\phi}_{\mu\nu}\rangle_{2}^{2}=\gamma_{\phi S}\mathcal{R}_{\langle\mu\rangle\langle\nu\rangle}.
\label{557}
\end{eqnarray}
where the definition of $\gamma_{\phi S}$ is the same as that in Eq.~\eqref{131}.

Similarly, the corrections from the three-point correlation function for the rotational stress tensor can be obtained as
\begin{eqnarray}
\begin{aligned}
	\langle\hat{\phi}_{\mu\nu}\rangle_{2}^{3}= & 2\gamma_{\phi p\phi}\theta\xi_{\mu\nu}+\sum_{ab}\gamma_{\phi\mathscr{J}_{a}\mathscr{J}_{b}}\nabla_{[\mu}\alpha_{a}\nabla_{\nu]}\alpha_{b}+2\sum_{a}\gamma_{\phi\mathscr{J}_{a}q}\nabla_{[\mu}\alpha_{a}M_{\nu]}\\
	& +\gamma_{\phi qq}M_{[\mu}M_{\nu]}+\gamma_{\phi\pi\pi}\sigma_{\alpha[\mu}\sigma_{\nu]}^{\,\,\,\,\alpha}+2\gamma_{\phi\pi\phi}\sigma_{\alpha[\mu}\xi_{\nu]}^{\,\,\,\,\alpha}+\gamma_{\phi\phi\phi}\xi_{\alpha[\mu}\xi_{\nu]}^{\,\,\,\,\alpha},
\end{aligned}
\label{558}
\end{eqnarray}
where
\begin{align}
	\gamma_{\phi p\phi} & =-\frac{1}{3}\beta^{2}\int d^{4}x_{1}d^{4}x_{2}\left(\hat{\phi}_{\gamma\delta}\left(x\right),\hat{p}^{*}\left(x_{1}\right),\hat{\phi}^{\gamma\delta}\left(x_{2}\right)\right),\gamma_{\phi\mathscr{J}_{a}\mathscr{J}_{b}}=\frac{1}{3}\int d^{4}x_{1}d^{4}x_{2}\left(\hat{\phi}_{\gamma\delta}\left(x\right),\hat{\mathscr{J}}_{a}^{\gamma}\left(x_{1}\right),\hat{\mathscr{J}}_{b}^{\delta}\left(x_{2}\right)\right),\label{559}\\
	\gamma_{\phi\mathscr{J}_{a}q} & =-\frac{1}{3}\beta\int d^{4}x_{1}d^{4}x_{2}\left(\hat{\phi}_{\gamma\delta}\left(x\right),\hat{\mathscr{J}}_{a}^{\gamma}\left(x_{1}\right),\hat{q}^{\delta}\left(x_{2}\right)\right),\gamma_{\phi qq}=\frac{1}{3}\beta^{2}\int d^{4}x_{1}d^{4}x_{2}\left(\hat{\phi}_{\gamma\delta}\left(x\right),\hat{q}^{\gamma}\left(x_{1}\right),\hat{q}^{\delta}\left(x_{2}\right)\right),\label{560}\\
	\gamma_{\phi\pi\pi} & =-\frac{4}{15}\beta^{2}\int d^{4}x_{1}d^{4}x_{2}\left(\hat{\phi}_{\lambda}^{\,\,\,\,\delta}\left(x\right),\hat{\pi}_{\delta}^{\,\,\,\,\eta}\left(x_{1}\right),\hat{\pi}_{\eta}^{\,\,\,\,\lambda}\left(x_{2}\right)\right),\gamma_{\phi\pi\phi}=\frac{4}{5}\beta^{2}\int d^{4}x_{1}d^{4}x_{2}\left(\hat{\phi}_{\lambda}^{\,\,\,\,\delta}\left(x\right),\hat{\pi}_{\delta}^{\,\,\,\,\eta}\left(x_{1}\right),\hat{\phi}_{\eta}^{\,\,\,\,\lambda}\left(x_{2}\right)\right),\label{561}\\
	\gamma_{\phi\phi\phi} & =-\frac{4}{3}\beta^{2}\int d^{4}x_{1}d^{4}x_{2}\left(\hat{\phi}_{\lambda}^{\,\,\,\,\delta}\left(x\right),\hat{\phi}_{\delta}^{\,\,\,\,\eta}\left(x_{1}\right),\hat{\phi}_{\eta}^{\,\,\,\,\lambda}\left(x_{2}\right)\right).\label{562}
\end{align}
The complete second-order expression for the rotational stress tensor is
\begin{eqnarray}
\begin{aligned}
	\phi_{\mu\nu}=&2\gamma\xi_{\mu\nu}+2\widetilde{\gamma}\left[\mydelta_{\mu\nu\rho\sigma}D\xi^{\rho\sigma}+\theta\Gamma\xi_{\mu\nu}\right]+\gamma_{\phi S}\mathcal{R}_{\langle\mu\rangle\langle\nu\rangle}\\
	&+2\gamma_{\phi p\phi}\theta\xi_{\mu\nu}+\sum_{ab}\gamma_{\phi\mathscr{J}_{a}\mathscr{J}_{b}}\nabla_{[\mu}\alpha_{a}\nabla_{\nu]}\alpha_{b}+2\sum_{a}\gamma_{\phi\mathscr{J}_{a}q}\nabla_{[\mu}\alpha_{a}M_{\nu]}\\
	&+\gamma_{\phi qq}M_{[\mu}M_{\nu]}+\gamma_{\phi\pi\pi}\sigma_{\alpha[\mu}\sigma_{\nu]}^{\,\,\,\,\alpha}+2\gamma_{\phi\pi\phi}\sigma_{\alpha[\mu}\xi_{\nu]}^{\,\,\,\,\alpha}+\gamma_{\phi\phi\phi}\xi_{\alpha[\mu}\xi_{\nu]}^{\,\,\,\,\alpha}.
\end{aligned}
\label{563}
\end{eqnarray}
The evolution equation for the rotational stress tensor can be obtained by replacing $2\xi^{\rho\sigma}\sim\gamma^{-1}\phi^{\rho\sigma}$ in the term $D\xi^{\rho\sigma}$, and we obtain
\begin{eqnarray}
2\widetilde{\gamma}\mydelta_{\mu\nu\rho\sigma}D\xi^{\rho\sigma}\simeq\widetilde{\gamma}\gamma^{-1}\mydelta_{\mu\nu\rho\sigma}D\phi^{\rho\sigma}-\widetilde{\gamma}\gamma^{-2}\beta\left(\frac{\partial\gamma}{\partial\beta}\Gamma-\sum_{a}\frac{\partial\gamma}{\partial\alpha_{a}}\delta_{a}\right)\theta\phi_{\mu\nu}.\label{564}
\end{eqnarray}
Combining Eqs.~\eqref{563} and \eqref{564} and introducing the coefficients
\begin{align}
\dot{\phi}_{\mu\nu}&=\mydelta_{\mu\nu\rho\sigma}D\phi^{\rho\sigma},\label{565}\\
\tau_{\phi}&=-\widetilde{\gamma}\gamma^{-1},\label{566}\\
\widetilde{\gamma}_{\phi}	&=\tau_{\phi}\gamma^{-1}\beta\left(\frac{\partial\gamma}{\partial\beta}\Gamma-\sum_{a}\frac{\partial\gamma}{\partial\alpha_{a}}\delta_{a}\right),\label{567}
\end{align}
we finally obtain the following relaxation equation for the rotational stress tensor,
\begin{eqnarray}
\begin{aligned}
	\phi_{\mu\nu}+\tau_{\phi}\dot{\phi}_{\mu\nu}=&2\gamma\xi_{\mu\nu}+\widetilde{\gamma}_{\phi}\theta\phi_{\mu\nu}+2\widetilde{\gamma}\theta\Gamma\xi_{\mu\nu}+\gamma_{\phi S}\mathcal{R}_{\langle\mu\rangle\langle\nu\rangle}\\
	& +2\gamma_{\phi p\phi}\theta\xi_{\mu\nu}+\sum_{ab}\gamma_{\phi\mathscr{J}_{a}\mathscr{J}_{b}}\nabla_{[\mu}\alpha_{a}\nabla_{\nu]}\alpha_{b}+2\sum_{a}\gamma_{\phi\mathscr{J}_{a}q}\nabla_{[\mu}\alpha_{a}M_{\nu]}\\
	&+\gamma_{\phi qq}M_{[\mu}M_{\nu]}+\gamma_{\phi\pi\pi}\sigma_{\alpha[\mu}\sigma_{\nu]}^{\,\,\,\,\alpha}+2\gamma_{\phi\pi\phi}\sigma_{\alpha[\mu}\xi_{\nu]}^{\,\,\,\,\alpha}+\gamma_{\phi\phi\phi}\xi_{\alpha[\mu}\xi_{\nu]}^{\,\,\,\,\alpha}.
\end{aligned}
\label{568}
\end{eqnarray}
It can be noted from symmetry that term $\gamma_{\phi qq}M_{[\mu}M_{\nu]}$ does not exist.

\subsubsection{Second-order corrections to the boost heat vector}

The nonlocal corrections from the two-point correlation function for the charge-diffusion currents can be obtained
through the similar derivation steps as in Eq.~\eqref{423}
\begin{eqnarray}
\langle\hat{q}_{\mu}\rangle_{2}^{1}=\widetilde{\lambda}_{qh}\sum_{a}n_{a}{w}^{-2}\left(\Gamma {w}+\sum_{d}\delta_{d}n_{d}\right)\theta\nabla_{\mu}\alpha_{a}+\sum_{a}\widetilde{\lambda}_{q\mathscr{J}_{a}}\Delta_{\mu\gamma}D\left(\nabla^{\gamma}\alpha_{a}\right)-\widetilde{\lambda}\theta\Gamma M_{\mu}-\widetilde{\lambda}\Delta_{\mu\gamma}DM^{\gamma}.
\label{569}
\end{eqnarray}
where the definitions of $\widetilde{\lambda}_{qh},\widetilde{\lambda}_{q\mathscr{J}_{a}}$, and $\widetilde{\lambda}$ are the same as those in Eqs.~\eqref{424}-\eqref{426}.

Substituting Eq.~\eqref{497} into Eq.~\eqref{96} and applying Curie’s theorem, we obtain corrections from extended thermodynamic forces to the boost heat vector
\begin{eqnarray}
\langle\hat{q}_{\mu}\rangle_{2}^{2}=\lambda_{qh}\mathcal{H}_{\mu}-\lambda\mathcal{Q}_{\mu},
\label{570}
\end{eqnarray}
where the definitions of $\lambda_{qh}$ and $\lambda$ are the same as those in Eqs.~\eqref{136} and \eqref{138}.

Similarly, the corrections from the three-point correlation function for the boost heat vector can be obtained
as
\begin{equation}
\langle\hat{q}_{\mu}\rangle_{2}^{3}=2\sum_{a}\lambda_{qp\mathscr{J}_{a}}\theta\nabla_{\mu}\alpha_{a}+2\lambda_{qpq}\theta M_{\mu}+2\sum_{a}\lambda_{q\mathscr{J}_{a}\pi}\nabla^{\nu}\alpha_{a}\sigma_{\mu\nu}+2\sum_{a}\lambda_{q\mathscr{J}_{a}\phi}\nabla^{\nu}\alpha_{a}\xi_{\mu\nu}+2\lambda_{qq\pi}M^{\nu}\sigma_{\mu\nu}+2\lambda_{qq\phi}M^{\nu}\xi_{\mu\nu},
\label{571}
\end{equation}
where
\begin{align}
	\lambda_{qp\mathscr{J}_{a}} & =\frac{1}{3}\beta\int d^{4}x_{1}d^{4}x_{2}\left(\hat{q}_{\beta}\left(x\right),\hat{p}^{*}\left(x_{1}\right),\hat{\mathscr{J}}_{a}^{\beta}\left(x_{2}\right)\right),\lambda_{qpq}=-\frac{1}{3}\beta^{2}\int d^{4}x_{1}d^{4}x_{2}\left(\hat{q}_{\beta}\left(x\right),\hat{p}^{*}\left(x_{1}\right),\hat{q}^{\beta}\left(x_{2}\right)\right),\label{572}\\
	\lambda_{q\mathscr{J}_{a}\pi} & =-\frac{1}{5}\beta\int d^{4}x_{1}d^{4}x_{2}\left(\hat{q}_{\lambda}\left(x\right),\hat{\mathscr{J}}_{a\delta}\left(x_{1}\right),\hat{\pi}^{\lambda\delta}\left(x_{2}\right)\right),\lambda_{q\mathscr{J}_{a}\phi}=-\frac{1}{3}\beta\int d^{4}x_{1}d^{4}x_{2}\left(\hat{q}_{\lambda}\left(x\right),\hat{\mathscr{J}}_{a\delta}\left(x_{1}\right),\hat{\phi}^{\lambda\delta}\left(x_{2}\right)\right),\label{573}\\
	\lambda_{qq\pi} & =\frac{1}{5}\beta^{2}\int d^{4}x_{1}d^{4}x_{2}\left(\hat{q}_{\lambda}\left(x\right),\hat{q}_{\delta}\left(x_{1}\right),\hat{\pi}^{\lambda\delta}\left(x_{2}\right)\right),\lambda_{qq\phi}=\frac{1}{3}\beta^{2}\int d^{4}x_{1}d^{4}x_{2}\left(\hat{q}_{\lambda}\left(x\right),\hat{q}_{\delta}\left(x_{1}\right),\hat{\phi}^{\lambda\delta}\left(x_{2}\right)\right).\label{574}
\end{align}
The complete second-order expression for the boost heat vector is
\begin{eqnarray}
\begin{aligned}
	q_{\mu}=&\sum_{a}\lambda_{q\mathscr{J}_{a}}\nabla_{\mu}\alpha_{a}-\lambda M_{\mu}+\widetilde{\lambda}_{qh}\sum_{a}n_{a}{w}^{-2}\left(\Gamma {w}+\sum_{a}\delta_{a}n_{a}\right)\theta\nabla_{\mu}\alpha_{a}+\sum_{a}\widetilde{\lambda}_{q\mathscr{J}_{a}}\Delta_{\mu\gamma}D\left(\nabla^{\gamma}\alpha_{a}\right)\\
	&-\widetilde{\lambda}\theta\Gamma M_{\mu}-\widetilde{\lambda}\Delta_{\mu\gamma}DM^{\gamma}+\lambda_{qh}\mathcal{H}_{\mu}-\lambda\mathcal{Q}_{\mu}+2\sum_{a}\lambda_{qp\mathscr{J}_{a}}\theta\nabla_{\mu}\alpha_{a}+2\lambda_{qpq}\theta M_{\mu}\\
	&+2\sum_{a}\lambda_{q\mathscr{J}_{a}\pi}\nabla^{\nu}\alpha_{a}\sigma_{\mu\nu}+2\sum_{a}\lambda_{q\mathscr{J}_{a}\phi}\nabla^{\nu}\alpha_{a}\xi_{\mu\nu}+2\lambda_{qq\pi}M^{\nu}\sigma_{\mu\nu}+2\lambda_{qq\phi}M^{\nu}\xi_{\mu\nu}.
\end{aligned}
\label{575}
\end{eqnarray}
In order to derive a relaxation-type equation for $q_\mu$, we use the first-order approximation
\begin{eqnarray}
M^{\gamma}=-\lambda^{-1}\left(q^{\gamma}-\sum_{a}\lambda_{q\mathscr{J}_{a}}\nabla^{\gamma}\alpha_{a}\right),
	\label{576}
\end{eqnarray}
in the term $-\widetilde{\lambda}\Delta_{\mu\gamma}DM^{\gamma}$ on the right-hand side of Eq.~\eqref{575}. We then have
\begin{eqnarray}
	\begin{aligned}
		-\widetilde{\lambda}\Delta_{\mu\gamma}DM^{\gamma}\simeq & \widetilde{\lambda}\lambda^{-1}\Delta_{\mu\gamma}Dq^{\gamma}-\widetilde{\lambda}\lambda^{-1}\Delta_{\mu\gamma}\sum_{a}\lambda_{q\mathscr{J}_{a}}D\left(\nabla^{\gamma}\alpha_{a}\right)-\widetilde{\lambda}\lambda^{-1}\sum_{a}\nabla_{\mu}\alpha_{a}\beta\theta\left(\frac{\partial\lambda_{q\mathscr{J}_{a}}}{\partial\beta}\Gamma-\sum_{d}\frac{\partial\lambda_{q\mathscr{J}_{a}}}{\partial\alpha_{d}}\delta_{d}\right)\\
		&-\widetilde{\lambda}\lambda^{-2}q_{\mu}\beta\theta\left(\frac{\partial\lambda}{\partial\beta}\Gamma-\sum_{d}\frac{\partial\lambda}{\partial\alpha_{d}}\delta_{d}\right)+\widetilde{\lambda}\lambda^{-2}\sum_{a}\lambda_{q\mathscr{J}_{a}}\nabla_{\mu}\alpha_{a}\beta\theta\left(\frac{\partial\lambda}{\partial\beta}\Gamma-\sum_{d}\frac{\partial\lambda}{\partial\alpha_{d}}\delta_{d}\right).
	\end{aligned}
	\label{577}
\end{eqnarray}
Combining Eqs.~\eqref{575} and \eqref{577} and introducing the coefficients
\begin{align}
	\tau_{q}= & -\widetilde{\lambda}\lambda^{-1},\label{578}\\
	\widetilde{\lambda}_{q}= & \beta\tau_{q}\lambda^{-1}\left(\frac{\partial\lambda}{\partial\beta}\Gamma-\sum_{d}\frac{\partial\lambda}{\partial\alpha_{d}}\delta_{d}\right),\label{579}\\
	\overline{\lambda}^{a}= & \beta\tau_{q}\left(\frac{\partial\lambda_{q\mathscr{J}_{a}}}{\partial\beta}\Gamma-\sum_{d}\frac{\partial\lambda_{q\mathscr{J}_{a}}}{\partial\alpha_{d}}\delta_{d}\right),\label{580}
\end{align}
we finally obtain the following relaxation equation for the boost heat vector,
\begin{eqnarray}
	\begin{aligned}
		q_{\mu}+\tau_{q}\dot{q}_{\mu}=&\sum_{a}\lambda_{q\mathscr{J}_{a}}\nabla_{\mu}\alpha_{a}-\lambda M_{\mu}+\sum_{a}\widetilde{\lambda}_{qh}n_{a}{w}^{-2}\left(\Gamma {w}+\sum_{a}\delta_{a}n_{a}\right)\theta\nabla_{\mu}\alpha_{a}+\sum_{a}\widetilde{\lambda}_{q\mathscr{J}_{a}}\Delta_{\mu\gamma}D\left(\nabla^{\gamma}\alpha_{a}\right)\\
		&-\widetilde{\lambda}\theta\Gamma M_{\mu}+\tau_{q}\Delta_{\mu\gamma}\sum_{a}\lambda_{q\mathscr{J}_{a}}D\left(\nabla^{\gamma}\alpha_{a}\right)+\sum_{a}\overline{\lambda}^{a}\theta\nabla_{\mu}\alpha_{a}+\widetilde{\lambda}_{q}\theta q_{\mu}-\sum_{a}\widetilde{\lambda}_{q}\lambda_{q\mathscr{J}_{a}}\theta\nabla_{\mu}\alpha_{a}\\
		&+\lambda_{qh}\mathcal{H}_{\mu}-\lambda\mathcal{Q}_{\mu}+2\sum_{a}\lambda_{qp\mathscr{J}_{a}}\theta\nabla_{\mu}\alpha_{a}+2\lambda_{qpq}\theta M_{\mu}+2\sum_{a}\lambda_{q\mathscr{J}_{a}\pi}\nabla^{\nu}\alpha_{a}\sigma_{\mu\nu}\\
		&+2\sum_{a}\lambda_{q\mathscr{J}_{a}\phi}\nabla^{\nu}\alpha_{a}\xi_{\mu\nu}+2\lambda_{qq\pi}M^{\nu}\sigma_{\mu\nu}+2\lambda_{qq\phi}M^{\nu}\xi_{\mu\nu}.
	\end{aligned}
	\label{581}
\end{eqnarray}

\subsubsection{Second-order corrections to $\varpi^{\lambda\mu\nu}$}

Since the operator $\hat{C}_1$ does not contain a third-rank tensor part, the nonlocal correction for $\varpi^{\lambda\mu\nu}$ from the two-point correlation function vanishes due to Curie's theorem
\begin{eqnarray}
\langle\hat{\varpi}^{\lambda\mu\nu}\rangle_{2}^{1}=0.
\label{5582}
\end{eqnarray}
Substituting Eq.~\eqref{497} into Eq.~\eqref{96} and applying Curie’s theorem, we obtain corrections from extended thermodynamic forces to $\varpi^{\lambda\mu\nu}$
\begin{eqnarray}
\langle\hat{\varpi}^{\lambda\mu\nu}\rangle_{2}^{2}=\varphi\varXi^{\lambda\mu\nu},
\label{583}
\end{eqnarray}
where the definition of $\varphi$ is the same as that in Eq.~\eqref{139}.

Similarly, the corrections from the three-point correlation function for $\varpi^{\lambda\mu\nu}$ can be obtained as
\begin{eqnarray}
\left\langle \hat{\varpi}^{\lambda\mu\nu}\right\rangle _{2}^{3}=2\sum_{a}\varphi_{\varpi\mathscr{J}_{a}\phi}\nabla_{\rho}\alpha_{a}\xi_{\sigma\delta}\myDelta^{\lambda\mu\nu\rho\sigma\delta}+2\varphi_{\varpi q\phi}M_{\rho}\xi_{\sigma\delta}\myDelta^{\lambda\mu\nu\rho\sigma\delta},
\label{584}
\end{eqnarray}
where
\begin{align}
	\varphi_{\varpi\mathscr{J}_{a}\phi}=-\beta\int d^{4}x_{1}d^{4}x_{2}\left(\hat{\varpi}^{\gamma\varepsilon\zeta}\left(x\right),\hat{\mathscr{J}}_{a\gamma}\left(x_{1}\right),\hat{\phi}_{\varepsilon\zeta}\left(x_{2}\right)\right),\varphi_{\varpi q\phi}=\beta^{2}\int d^{4}x_{1}d^{4}x_{2}\left(\hat{\varpi}^{\gamma\varepsilon\zeta}\left(x\right),\hat{q}_{\gamma}\left(x_{1}\right),\hat{\phi}_{\varepsilon\zeta}\left(x_{2}\right)\right).\label{585}
\end{align}
The complete second-order expression for $\varpi^{\lambda\mu\nu}$ is
\begin{eqnarray}
\varpi^{\lambda\mu\nu}=\varphi\varXi^{\lambda\mu\nu}+2\sum_{a}\varphi_{\varpi\mathscr{J}_{a}\phi}\nabla_{\rho}\alpha_{a}\xi_{\sigma\delta}\myDelta^{\lambda\mu\nu\rho\sigma\delta}+2\varphi_{\varpi q\phi}M_{\rho}\xi_{\sigma\delta}\myDelta^{\lambda\mu\nu\rho\sigma\delta}.
\label{586}
\end{eqnarray}

\section{Conclusion}
\label{section5}

In this study, we have derived a novel formulation of relativistic canonical-like second-order spin hydrodynamics for two distinct power counting schemes using Zubarev’s nonequilibrium statistical operator formalism. Our analysis focuses on a multicomponent quantum system within the hydrodynamic regime, characterized by the energy-momentum tensor with symmetric and antisymmetric parts, conserved charge currents, and the totally antisymmetric spin tensor. We have successfully derived second-order expressions for the shear-stress tensor, bulk-viscous pressure, charge-diffusion currents, rotational stress tensor, boost heat vector, and spin tensor-related dissipative flux. The formal expressions for all second-order transport coefficients were derived in terms of two- and three-point equilibrium correlation functions, which can be computed using standard thermal field theory methods ~\cite{itzykson1991statistical}.

In the first power counting scheme  $\omega_{\mu\nu}\sim\mathcal{O}\left(\partial^{0}\right)$, the constitutive relations for first-order spin hydrodynamics are described by Eqs.~\eqref{120}-\eqref{125}.  Notably, while the shear stress tensor and the spin tensor-related dissipative flux  $\varpi^{\lambda\alpha\beta}$  exhibit no cross correlations, the remaining four dissipative currents—bulk viscous pressure, rotational stress tensor, charge-diffusion current, and boost heat vector—do involve cross correlations. The presence of these cross-coupling effects indicates that different physical processes are coupled, leading to the emergence of new transport phenomena. Consequently, additional transport coefficients must be introduced to fully describe these novel behaviors. 

In contrast, the second power counting scheme $\omega_{\mu\nu}\sim\mathcal{O}\left(\partial^{1}\right)$,introduces more complex thermodynamic forces, including $\mathcal{R}_{\mu\nu},\mathcal{Z}^{\mu\nu},N_{\sigma}$, and $\varXi_{\lambda\alpha\beta}$ ,  besides the usual $\theta,\sigma_{\mu\nu},\nabla_{\mu}\alpha_{a},M_{\mu}$, and $\xi_{\mu\nu}$.  The introduction of these new thermodynamic forces results in additional terms in the physical equations, indicating a greater richness in the physical phenomena of the current system compared to the second power counting scheme. Second-order terms involve interactions between dissipative currents and thermodynamic forces, arising from both three-point and two-point correlations, as well as nonlocal effects. Some of the first-order transport coefficients in the first power counting scheme become second-order transport coefficients in the second power counting scheme. As demonstrated in our previous work ~\cite{She:2021lhe}, for systems where the spin chemical potential is the leading-order in gradient expansion, entropy current analysis reveals that if $S_{(0)}^{\mu}$  is to correspond to a reversible ideal spin fluid, then the first-order correction of $T^{[\mu\nu]}$ must vanish. Consequently, the antisymmetric part of the energy-momentum tensor can only appear at the second or higher order in gradient expansion. This finding necessitates the exclusion of contributions from both  $q^\mu$ and $\phi^{\mu\nu}$ thereby removing the need for the assumption that $\xi_{\mu\nu}$ is a first-order tensor in the first power counting scheme.

Our findings differ from those of Ref. ~\cite{Tiwari:2024trl}, which retains a term proportional to $M_{[\mu}M_{\nu]}$ when discussing the phenomenological form of second-order spin hydrodynamics. According to symmetry, this term should vanish. Similarly, Ref. ~\cite{Tiwari:2024trl} retains a term proportional to $\nabla_{[\mu}\alpha\nabla_{\nu]}\alpha$ when discussing the rotational stress tensor. For a single current, this term should also vanish. However, for multiple currents, cross terms need to be retained. To facilitate subsequent calculations of transport coefficients in spin fluids using thermal field theory, we plan to reformulate the three-point correlation function as a retarded three-point Green’s function in future work.

The novel formulation of relativistic canonical-like second-order spin hydrodynamics presented in this study significantly extends our understanding of multicomponent quantum systems in the hydrodynamic regime. The introduction of cross-coupling effects and additional thermodynamic forces enriches the physical phenomena described by the model, necessitating the development of new transport coefficients. Future research should focus on the detailed computation of these transport coefficients using thermal field theory methods and exploring the implications of the second-order terms for the behavior of spin fluids. Additionally, further investigation is needed to reconcile the differences between our findings and those of previous studies, particularly in the context of symmetry considerations and the formulation of the rotational stress tensor.

In conclusion, this work provides a robust framework for the study of relativistic canonical-like second-order spin hydrodynamics, offering new insights into the transport phenomena of multicomponent quantum systems. The findings underscore the importance of considering higher-order terms and cross-coupling effects in the modeling of such systems, paving the way for future advancements in the field.

\section*{Acknowledgments}

The authors would like to thank Jin Hu and Matteo Buzzegoli for valuable discussions and contributions. The referee's constructive criticism has been instrumental in shaping this research. This research was partly funded by the Startup Research Fund of Henan Academy of Sciences (No. 231820058) and the 2024 Henan Province International Science and Technology Cooperation Projects (No. 242102521068).  The research of D.H. is support in part by the National Key Research and Development
Program of China under Contract No. 2022YFA1604900. Additionally, he received partial support from the National Natural Science Foundation of China (NSFC) under Grants No.12435009 and No. 12275104.

\bibliography{rotation.bib}

\end{document}